\documentclass[journal,comsoc]{IEEEtran}
\usepackage{amsmath,amsfonts}
\usepackage{algorithmic}
\usepackage{algorithm}
\usepackage[dvipsnames, table]{xcolor}
\usepackage{array}
\usepackage[caption=false,font=footnotesize,labelfont=rm,textfont=rm]{subfig}
\usepackage{textcomp}
\usepackage{stfloats}
\usepackage{url}
\usepackage{bbding}
\usepackage{color}
\usepackage{fontawesome5}
\usepackage{verbatim}
\usepackage{hyperref}
\usepackage{pifont}
\usepackage{subfig}
\usepackage{multirow}
\usepackage{pythonhighlight}
\usepackage{threeparttable}
\usepackage{bm}
\usepackage{colortbl}
\usepackage{xcolor}
\usepackage{hyperref}
\definecolor{lightgray}{rgb}{0.94,0.94,0.94}
\definecolor{colorl}{rgb}{1,0.88,0.70}
\hypersetup{
hidelinks,
colorlinks=true,
linkcolor=red,
citecolor=green,
urlcolor = magenta
}
\usepackage{colortbl}
\definecolor{colorh}{rgb}{1,0.60,0.20}
\definecolor{colorm}{rgb}{1,0.72,0.30}
\definecolor{colorl}{rgb}{1,0.88,0.70}

\usepackage{orcidlink}
\usepackage{graphicx}
\usepackage{cite}
\usepackage{microtype}
\usepackage{graphicx}
\usepackage{subfig}
\usepackage{graphicx}
\usepackage{float}
\usepackage{multirow}
\usepackage{bm}
\usepackage{booktabs}
\usepackage{amsmath}
\usepackage{amssymb}
\usepackage{mathtools}
\usepackage{amsthm}
\usepackage{tikz}
\usepackage{pgfplots}
\pgfplotsset{compat=1.18}
\usepgfplotslibrary{groupplots}
\usetikzlibrary{arrows.meta}
\usetikzlibrary{calc}
\usetikzlibrary{decorations.markings}
\def\BibTeX{{\rm B\kern-.05em{\sc i\kern-.025em b}\kern-.08em
    T\kern-.1667em\lower.7ex\hbox{E}\kern-.125emX}}
\usepackage{balance}
\begin{document}
\raggedbottom
\title{BiCRVC: An Efficient Bidirectional Neural Video Compression Framework via Coupled\\Representation Coding}
\author{
Wei Jiang~\orcidlink{0000-0001-9169-1924},
Junru Li~\orcidlink{0000-0001-7603-8599},~\IEEEmembership{Member,~IEEE},
Kai Zhang~\orcidlink{0000-0002-6627-0009},~\IEEEmembership{Senior Member,~IEEE}, 
Li Zhang\textsuperscript{\faEnvelope}~\orcidlink{0000-0003-2118-4876},~\IEEEmembership{Senior Member,~IEEE}
\thanks{
Wei Jiang, Junru Li, Kai Zhang, and Li Zhang are with ByteDance. Li Zhang is the corresponding author 
(e-mail: \texttt{\{jiangwei.lvc, lijunru, zhangkai.video, lizhang.idm\}@bytedance.com}).}
}\markboth{Preprint, under review}%
{How to Use the IEEEtran \LaTeX \ Templates}

\maketitle

\begin{abstract}
Neural video compression (NVC) has achieved strong compression performance, but practical
random-access coding still faces two technical challenges: existing bidirectional NVCs (BVCs)
usually require costly motion-first decoding, and reliable motion estimation is 
difficult under long-range bidirectional prediction.
To address these issues, we present BiCRVC, an efficient bidirectional neural video compression framework based on coupled representation coding.
Instead of coding motion and frame information with two separate codecs, BiCRVC transforms the motion representation and the current-frame latent into a unified latent representation for entropy coding.
This design enables motion and frame information to be decoded from the same bitstream with one unified codec, while still reconstructing motion-aligned contexts for frame decoding.
To improve motion accuracy, we introduce multi-candidate motion estimation (MCME), which combines multi-scale motion estimation and parallel accumulated motion estimation to better handle diverse and long-range motions.
To reduce motion coding overhead, we further propose bidirectional motion feature propagation (BMFP), which reuses previously decoded motion features at both the encoder and decoder as temporal priors for conditional motion coding.
In addition, coupled distortion training and random GOP structure training are used to encourage joint motion-frame coding and improve adaptation to hierarchical random-access structures.
Experiments show that BiCRVC achieves better compression performance than state-of-the-art BVCs while 
providing about 30$\times$ faster 1080p decoding than recent BVCs.
\end{abstract}


\section{Introduction}
\begin{figure}[t]
  \centering
  \includegraphics[width=0.9\columnwidth]{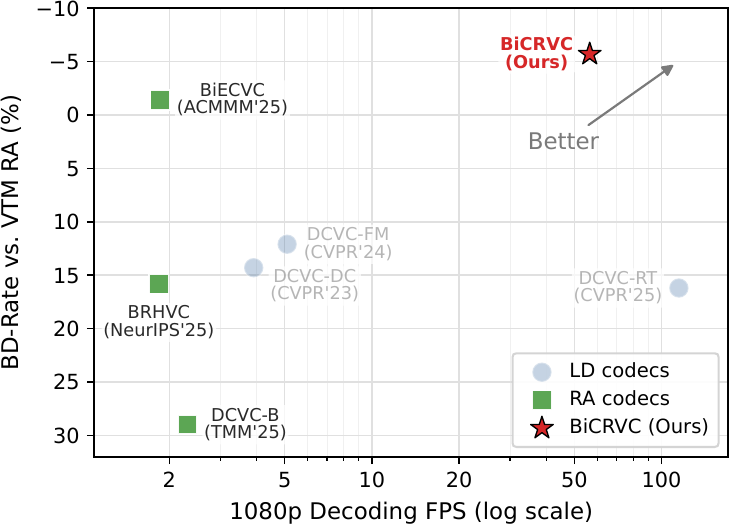}
  \caption{BD-Rate and 1080p decoding FPS comparison on 64-frame sequences with intra-period 64 (IP64/FN64). The vertical axis is inverted, so a higher position indicates better coding efficiency.}
  \label{fig:fps_bd_rate}
\end{figure}
Neural video compression (NVC)~\cite{lu2019dvc, hu2021fvc,lin2020mlvc,lu2024deep,zhai2024hybrid,chen2024bcanf,guo2023learning,vctmentzer2022,rippel2021elf,yang2020learning,yang2021hierarchical,li2025ustc,sheng2024vnvc}
has attracted increasing attention because it can learn spatial and temporal redundancy from data instead of relying on handcrafted coding tools~\cite{wiegand2003overview, sullivan2012overview,bross2021overview}.
Recent NVC~\cite{li2023neural,jiang2025ecvc,li2024neural,tang2025neural,bian2025augmented}
has surpassed H.266/VVC~\cite{bross2021overview} and the emerging ECM reference software~\cite{ye2025ecm}
under the low-delay configuration, where each frame is coded using only past reconstructed frames as references.
However, most existing NVCs~\cite{jiang2025ecvc,tang2025neural,bian2025augmented,li2023neural,li2024neural,lu2020end,tang2026nvc,jia2025practical,hu2020improving,hu2022fvc,ho2022canf,guo2023learning,vctmentzer2022,zhai2024hybrid,jiang2024lvc,chen2025hytip,sheng2024spatial,sheng2025prediction,chen2024maskcrt}
focus on forward prediction for low-delay coding.
Bidirectional NVCs (BVCs)~\cite{alexandre2023hierarchical,chen2024bcanf,jiang2025biecvc,zhai2025llbvc} are important for random-access applications such as video-on-demand and latency-tolerant streaming, but they remain less mature than low-delay NVCs.
\par
Practical random-access NVC faces two key technical challenges.
First, existing BVCs usually rely on explicit motion coding and require costly motion-first decoding.
Since motion must be decoded before frame reconstruction, previous bidirectional codecs often use separate motion and frame codecs, which increases decoding complexity and limits parallelism.
Second, motion estimation is more difficult in hierarchical bidirectional coding.
Forward prediction usually uses adjacent references with relatively small motion, whereas random-access coding may predict a frame from distant past and future references, leading to larger and more diverse motions.
Implicit alignment, as used by efficient low-delay codecs such as DCVC-RT~\cite{jia2025practical}, is therefore often insufficient for accurate bidirectional context construction.
\par
\begin{figure*}
  \centering
  \includegraphics[width=0.95\textwidth]{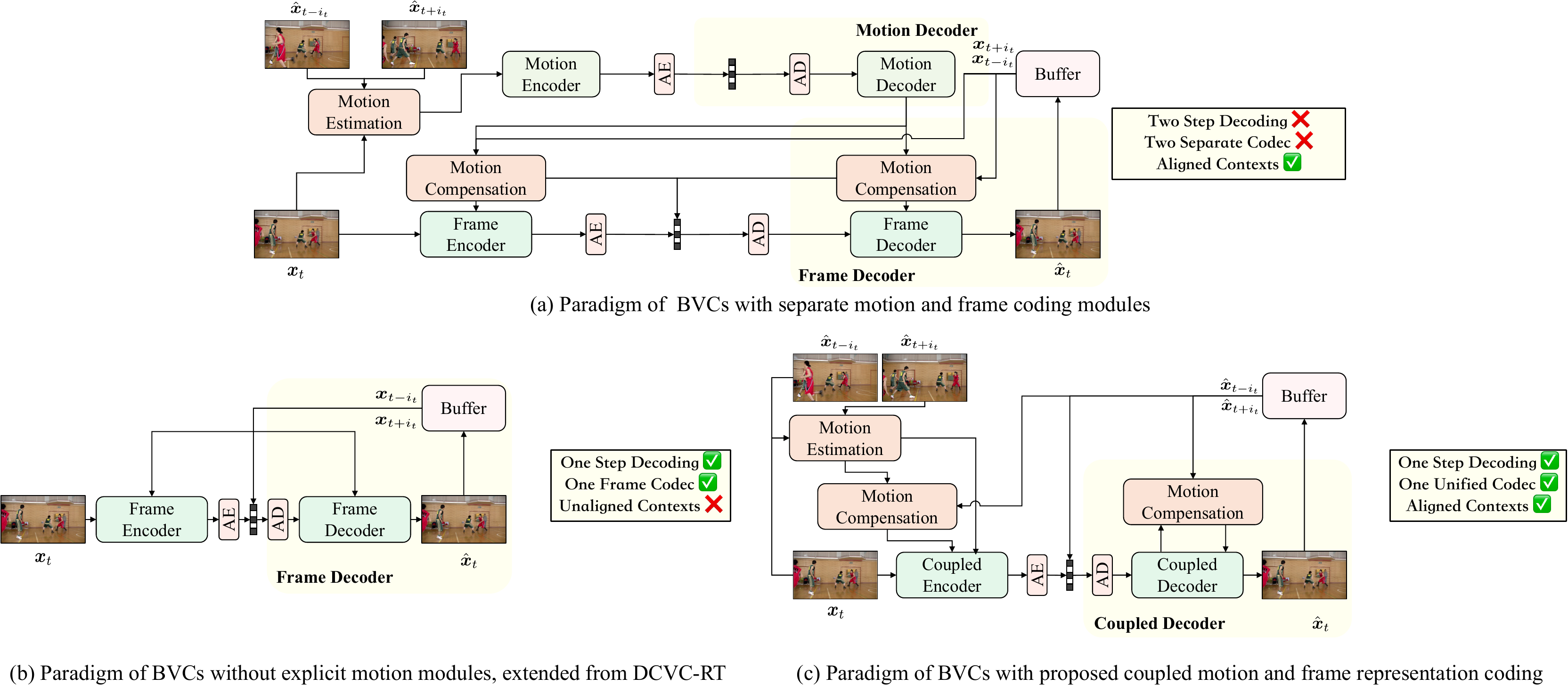}
  \caption{Comparison of different bidirectional video compression architectures: (a) existing BVCs use two separate codecs for motion and frame information, (b) a direct extension of DCVC-RT uses one frame codec without explicit motion coding, and (c) BiCRVC uses one unified codec for coupled motion and frame representation coding.}
  \label{fig:arch_compare}
\end{figure*} 
To achieve practical decoding complexity while maintaining competitive compression performance,
we propose BiCRVC, an efficient bidirectional neural video compression framework based on coupled representation coding.
BiCRVC jointly transforms the motion representation and the current-frame latent into a unified latent representation for entropy coding.
As illustrated in Figure~\ref{fig:arch_compare}, this design decodes motion and frame information from the same bitstream with one unified codec, while still reconstructing motion-aligned contexts for frame decoding.
\par
To improve compression performance, BiCRVC further enhances motion estimation and reduces motion coding overhead.
Since motion accuracy is important for context alignment and frame reconstruction in bidirectional compression, we propose multi-candidate motion estimation (MCME), which combines multi-scale motion estimation (MSME) and parallel accumulated motion estimation (PAME).
MSME estimates motion candidates at different input scales to cover different motion ranges, while PAME decomposes difficult long-range motion estimation into shorter-range estimates and accumulates them with reduced serial dependency.
The resulting motion candidates are fused according to warping errors defined on aligned backward-flow coordinates.
To reduce motion coding overhead, BiCRVC introduces bidirectional motion feature propagation (BMFP), which reuses previously decoded motion features at both the encoder and decoder as temporal priors for conditional motion coding.
\par
Finally, we design an improved training strategy to better exploit bidirectional prediction.
Coupled distortion training jointly supervises frame reconstruction and motion reconstruction, encouraging joint motion-frame coding and exploiting their redundancy.
Random GOP structure training exposes short clips to different hierarchical coding layers, helping the model learn diverse reference patterns.
With these techniques, BiCRVC achieves better compression performance than state-of-the-art BVCs~\cite{sheng2025bi,liuneural,jiang2025biecvc} while providing about 30$\times$ faster 1080p decoding than recent BVCs.
As summarized in Fig.~\ref{fig:fps_bd_rate}, BiCRVC offers a favorable tradeoff between coding efficiency and decoding throughput on 64-frame sequences with intra-period 64 (IP64/FN64), achieving lower BD-Rate than real-time low-delay codecs while being much faster than existing random-access neural codecs.
\par
Our contributions are summarized as follows:
\begin{itemize}
\item We propose BiCRVC, an efficient bidirectional neural video compression framework based on coupled representation coding, which jointly codes motion and frame representations with one unified codec while preserving motion-aligned contexts.
\item We introduce MCME to improve motion accuracy for bidirectional compression by combining MSME for different motion ranges and PAME for long-range motion modeling with reduced serial dependency.
\item We propose BMFP to reduce motion coding overhead by reusing previously decoded motion features as temporal priors at both the encoder and decoder.
\item We design a training strategy with coupled distortion training and random GOP structure training. Experiments show that BiCRVC achieves better compression performance than state-of-the-art BVCs with about 30$\times$ faster 1080p decoding than recent BVCs.
\end{itemize} 
\begin{figure*}[!t]
  \centering
  \footnotesize
  \resizebox{0.98\textwidth}{!}{
  \begin{tikzpicture}[x=0.48cm,y=0.7cm]
    \tikzset{
      gopnode/.style={draw, rounded corners=1pt, minimum width=0.70cm, minimum height=0.50cm, inner sep=1pt, font=\scriptsize},
      intranode/.style={gopnode, fill=gray!30},
      layera/.style={gopnode, fill=blue!12},
      layerb/.style={gopnode, fill=green!12},
      layerc/.style={gopnode, fill=yellow!20},
      layerd/.style={gopnode, fill=orange!18},
      layere/.style={gopnode, fill=red!12}
    }

    \draw[->, thick] (-1.0,-0.8) -- (33.2,-0.8) node[right] {POC};
    \foreach \l in {0,1,2,3,4,5} {
      \node[left] at (-0.8,\l) {Layer \l};
    }

    \foreach \p in {0,32} {
      \node[intranode] (p\p) at (\p,0) {\p};
    }
    \node[layera] (p16) at (16,1) {16};
    \foreach \p in {8,24} {
      \node[layerb] (p\p) at (\p,2) {\p};
    }
    \foreach \p in {4,12,20,28} {
      \node[layerc] (p\p) at (\p,3) {\p};
    }
    \foreach \p in {2,6,10,14,18,22,26,30} {
      \node[layerd] (p\p) at (\p,4) {\p};
    }
    \foreach \p in {1,3,5,7,9,11,13,15,17,19,21,23,25,27,29,31} {
      \node[layere] (p\p) at (\p,5) {\p};
    }

    \foreach \a/\b/\c in {16/0/32,8/0/16,24/16/32,4/0/8,12/8/16,20/16/24,28/24/32,2/0/4,6/4/8,10/8/12,14/12/16,18/16/20,22/20/24,26/24/28,30/28/32,1/0/2,3/2/4,5/4/6,7/6/8,9/8/10,11/10/12,13/12/14,15/14/16,17/16/18,19/18/20,21/20/22,23/22/24,25/24/26,27/26/28,29/28/30,31/30/32} {
      \draw[gray!85, line width=0.45pt, postaction={decorate},
        decoration={markings, mark=at position 0.55 with {\arrow{Stealth[length=1.5mm,width=1.0mm]}}}]
        (p\b.north) -- (p\a.south);
      \draw[gray!85, line width=0.45pt, postaction={decorate},
        decoration={markings, mark=at position 0.55 with {\arrow{Stealth[length=1.5mm,width=1.0mm]}}}]
        (p\c.north) -- (p\a.south);
    }
  \end{tikzpicture}}
  \caption{Illustration of the GOP32 structure used in BiCRVC. POCs 0 and 32 are intra frames with layer index 0, while the remaining frames are B frames organized hierarchically according to their temporal references. The arrows indicate the dependency from a reference frame to the current coded frame.}
  \label{fig:gop32}
\end{figure*}
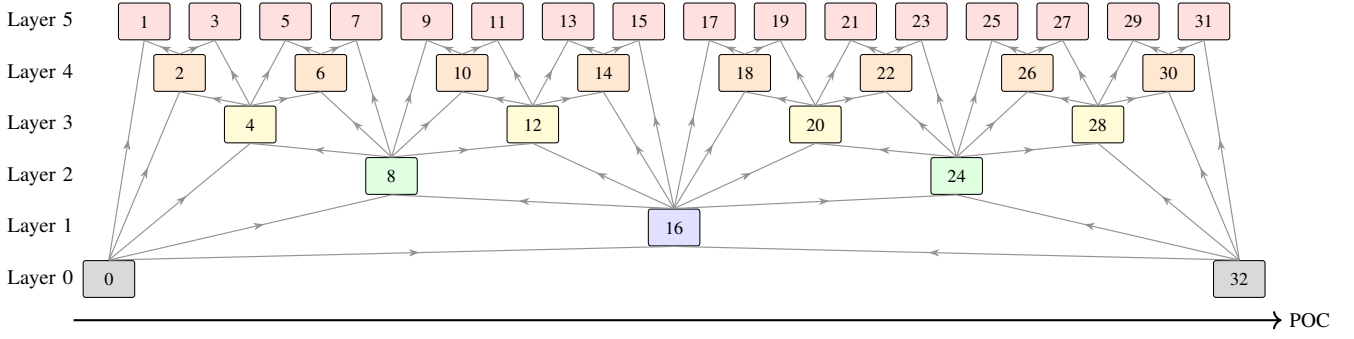
\section{Related Works}
\subsection{Neural Video Compression with Forward Prediction}
Forward-prediction codecs encode each inter frame from previously reconstructed frames and have been extensively studied in learned video compression.
A typical early design, represented by DVC~\cite{lu2020end}, estimates optical flow~\cite{ranjan2017optical}, performs motion compensation, and transmits motion and residual information.
This predictive pipeline has been strengthened by more effective motion compensation~\cite{agustsson2020scale,shi2022alphavc,hu2022coarse}, feature-domain residual coding~\cite{hu2022fvc}, and selective entropy coding~\cite{shi2022alphavc}.
Nevertheless, explicitly coding residuals still limits how much temporal prior information can be exploited.
\par
Conditional coding~\cite{li2021deep} offers a more flexible alternative by reconstructing frames under decoded temporal conditions rather than predefined residuals.
Following this idea, later methods improve temporal prior propagation~\cite{sheng2022temporal}, context alignment and motion transmission~\cite{li2023neural,Qi_2023_CVPR,bian2025augmented}, and non-local or look-ahead temporal modeling~\cite{jiang2025ecvc,liao2025ehvc}.
Entropy models are also enhanced with spatial-temporal priors, quadtree structures, and quality-aware conditions~\cite{li2022hybrid,li2023neural,qi2024long}.
For long coding chains, context refreshment~\cite{li2024neural} reduces accumulated errors.
\par
Although these methods achieve strong compression performance, their computational complexity remains high and limits practical deployment.
To improve practicality, DCVC-RT~\cite{jia2025practical} adopts implicit motion alignment and removes explicit motion estimation and compensation.
It enables real-time encoding and decoding while still outperforming VTM~\cite{bross2021overview} under low-delay scenarios.
\subsection{Neural Video Compression with Bidirectional Prediction}
Bidirectional coding requires the model to exploit two-sided temporal references, which makes learned bidirectional compression more difficult than forward prediction~\cite{lu2020end,li2021deep,jiang2025ecvc}.
Existing methods mainly differ in how they build the prediction signal from the past and future references.
One line of work first synthesizes the target-frame prediction through interpolation and then codes the remaining information.
Early interpolation-based codecs follow this design~\cite{wu2018video}, while later methods improve feature-level residual coding and model expressiveness~\cite{djelouah2019neural,alexandre2023hierarchical}.
\par
Another line of work keeps explicit bidirectional motion modeling, which is useful when the current frame has large displacement from both references.
Existing methods estimate bidirectional motion with optical flow and compress both the warped prediction and the motion information~\cite{yang2024adaptive}, reuse low-delay models with MEMC~\cite{pourreza2021extending}, or support P-frame and B-frame coding in a unified conditional framework~\cite{yang2024ucvc}.
Recent methods further extend temporal propagation and context modeling to bidirectional coding~\cite{sheng2025bi}, harmonize hierarchical bidirectional references~\cite{liuneural}, and introduce richer bidirectional contexts with more reference frames, linear attention, and context gating~\cite{jiang2025biecvc}.
Among them, BiECVC~\cite{jiang2025biecvc} is the first BVC to surpass VTM under the random-access configuration on all evaluated datasets.
\par
Although existing bidirectional methods achieve strong performance, their gains are often obtained at the cost of high model complexity.
At the same time, practical low-delay models such as DCVC-RT~\cite{jia2025practical} do not generalize well to random-access coding, where implicit motion estimation is less effective for large motion.
Therefore, it remains an open problem to design a bidirectional video compression model that achieves both high compression efficiency and practical decoding complexity.
\section{Method}
\subsection{Background}
Bidirectional neural video compression encodes the sequence $\{\boldsymbol{x}_t \in \mathbb{R}^{H \times W \times 3}\}_{t=0}^{N-1}$
using information from both past and future frames. Here, $t$ denotes the picture order count (POC), $H$ and $W$ denote the frame height and width, and $N$ denotes the number of frames.
For coding, the sequence is partitioned into groups of pictures (GOPs) under a predefined structure. Frames are coded as either intra frames or inter frames. Intra frames are coded independently, while B frames use temporal references from both past and future frames.\par
Within each GOP, frames are organized hierarchically through binary partitioning. Specifically, a B frame $\boldsymbol{x}_t$ references the past and future frames $\boldsymbol{x}_{t-i_t}$ and $\boldsymbol{x}_{t+i_t}$, where $i_t$ denotes the reference distance at POC $t$.
This dependency forms multiple layers within the GOP. The layer index of each intra frame is set to 0, and the layer index $\ell_t$ of a B frame $\boldsymbol{x}_t$ is defined as $\max(\ell_{t-i_t}, \ell_{t+i_t})+1$. 
Different quality factors can then be assigned to different layers to enable hierarchical coding. In the current H.266/VVC
VTM standard~\cite{bross2021overview}, the GOP size is set to 32, as illustrated in Fig.~\ref{fig:gop32}.\par
Most current BVC methods~\cite{sheng2025bi,zhai2025llbvc, jiang2025biecvc, liuneural} follow this bidirectional hierarchical
structure and rely on optical flow networks~\cite{ranjan2017optical} to model temporal correspondence between the current and reference frames.
However, under random access settings, large motion and long temporal distance make vanilla optical-flow-based prediction
less reliable, leading to noticeable alignment errors and inferior coding efficiency compared with VTM. Moreover, explicit
motion estimation and motion transmission remain necessary in this regime and cannot be removed as in low-delay settings
such as DCVC-RT~\cite{jia2025practical} without severe performance degradation. These operations further increase model complexity and decoding
latency, since decoding has to wait for motion decoding and reference reconstruction to finish, thereby limiting parallelism.
\subsection{Overview}
\begin{figure*}
  \centering
  \includegraphics[width=\textwidth]{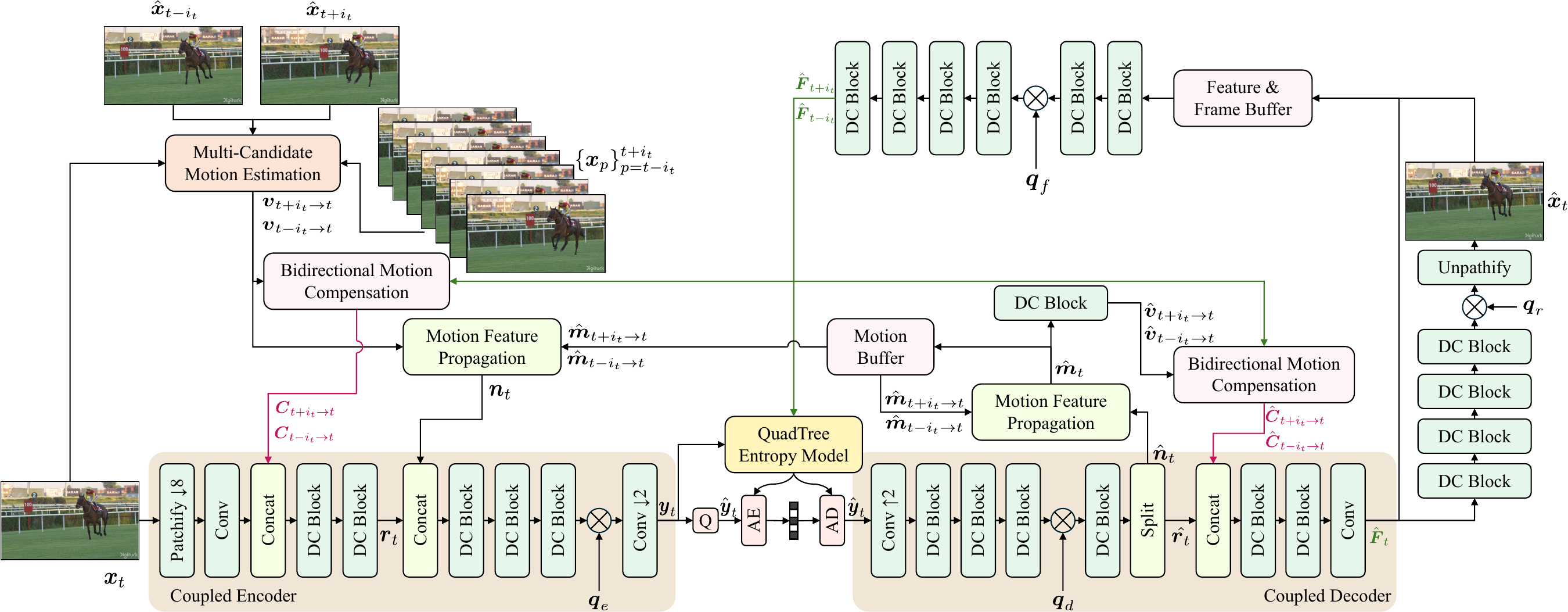}
  \caption{Overall framework of BiCRVC. The quantization parameters ($\boldsymbol{q}_e, \boldsymbol{q}_d, \boldsymbol{q}_r, \boldsymbol{q}_f$) control variable-rate compression. $\boldsymbol{x}_t$ and $\hat{\boldsymbol{x}}_t$ denote the current frame and its reconstructed frame, respectively. $\hat{\boldsymbol{x}}_{t-i_t}$ and $\hat{\boldsymbol{x}}_{t+i_t}$ denote the forward and backward reference frames, respectively, and $\hat{\boldsymbol{F}}_{t-i_t}$ and $\hat{\boldsymbol{F}}_{t+i_t}$ denote their corresponding reference features. $\boldsymbol{v}_{t\rightarrow t-i_t}$ and $\boldsymbol{v}_{t\rightarrow t+i_t}$ denote the backward motion vectors from the current frame to the two reference frames, respectively, while $\hat{\boldsymbol{v}}_{t\rightarrow t-i_t}$ and $\hat{\boldsymbol{v}}_{t\rightarrow t+i_t}$ denote their reconstructed versions. Q denotes quantization. AE and AD denote arithmetic encoding and arithmetic decoding, respectively.}
  \label{fig:overall}
\end{figure*}

To improve large-motion modeling and decoding parallelism, we propose BiCRVC, an efficient bidirectional neural video compression framework based on coupled representation coding. The overall framework is shown in Figure~\ref{fig:overall}.
BiCRVC builds on DCVC-RT~\cite{jia2025practical}, a practical low-delay neural video codec.
It keeps the low-resolution single-scale feature propagation and coding design of DCVC-RT, instead of the multi-scale design used in earlier DCVC-DC-based methods~\cite{li2023neural,jiang2025ecvc}.
Specifically, the input frame is directly patchified to $1/8$ resolution for conditional coding, and the propagated features are also maintained at $1/8$ resolution.
This greatly reduces computational cost compared with full-resolution or progressively downsampled methods~\cite{li2023neural,jiang2025ecvc,jiang2025biecvc,sheng2025bi,liuneural}.
Following DCVC-RT, BiCRVC further uses scale-based selective coding.
For variable-rate coding, BiCRVC adopts module-bank-based rate control~\cite{jia2025practical} with 64 quality levels.
For hierarchical quality control, it uses different quality layers trained with distinct quality factors, following DCVC-B~\cite{sheng2025bi} and BiECVC~\cite{jiang2025biecvc}.
For effective entropy coding, BiCRVC also adopts the quadtree context model (QCM)~\cite{li2023neural}, which is widely used in recent neural video codecs including DCVC-B~\cite{sheng2025bi}, BRHVC~\cite{liuneural}, and BiECVC~\cite{jiang2025biecvc}.
\par
\begin{figure*}
  \centering
  \includegraphics[width=0.9\textwidth]{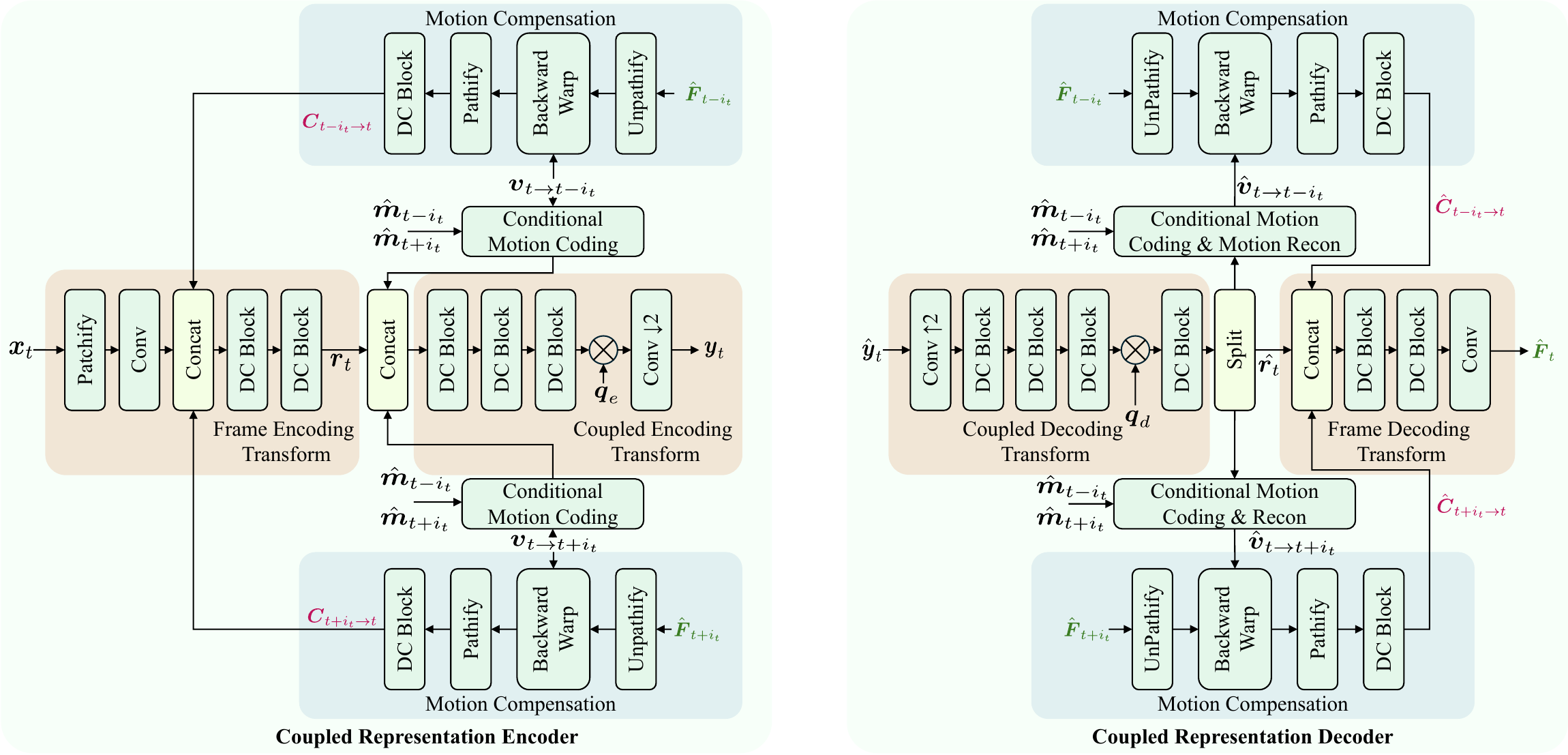}
  \caption{Architecture of the proposed coupled representation coding.}
  \label{fig:crc}
\end{figure*}
Bidirectional random-access coding requires both effective motion compensation and efficient parallel decoding.
Existing methods struggle to achieve both at the same time.
DCVC-RT~\cite{jia2025practical} is efficient, but its implicit motion alignment is not sufficient for large motions under random-access settings.
Previous bidirectional video codecs~\cite{sheng2025bi,liuneural,jiang2025biecvc} rely on explicit motion, but require motion-first serial decoding.
To address this issue, BiCRVC adopts coupled representation coding (CRC) to jointly code motion and frame representations.
This design supports bidirectional motion compensation and enables parallel motion and frame decoding.
\par
Motion compensation accuracy is also critical to bidirectional coding efficiency.
To better handle large and long-term motions, BiCRVC employs multi-candidate motion estimation (MCME), which combines multi-scale motion estimation and parallel accumulated motion estimation.
MSME improves adaptation to different motion ranges, while PAME enhances long-term motion modeling with better parallelism.
Their motion candidates are fused to obtain the final motion vector.
\par
\begin{figure*}
  \centering
  \includegraphics[width=0.97\textwidth]{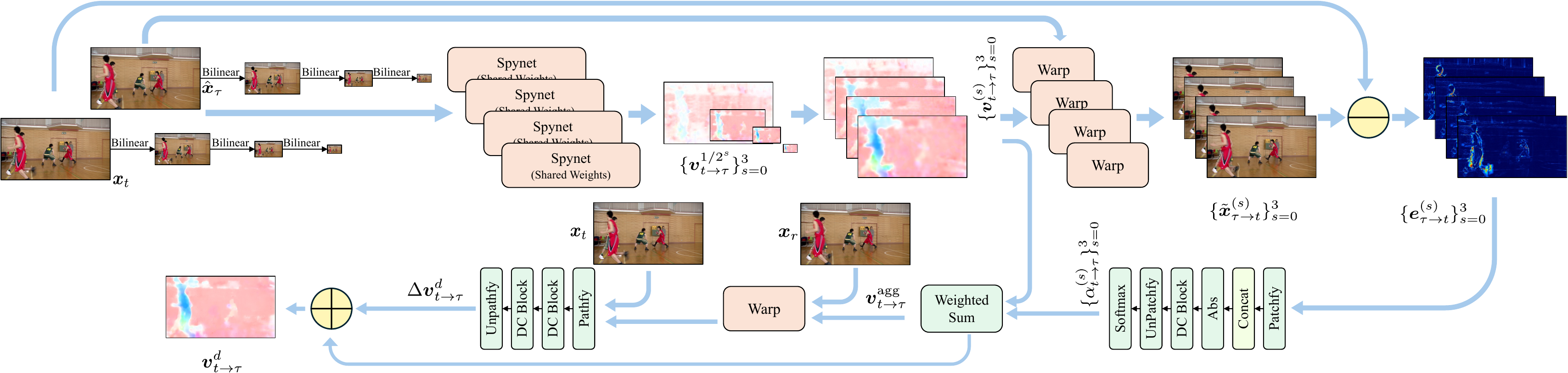}
  \caption{Architecture of the proposed multi-scale motion estimation.}
  \label{fig:msme}
\end{figure*}
To reduce motion transmission overhead while preserving accurate motion compensation,
BiCRVC further introduces bidirectional motion feature propagation (BMFP).
Previously decoded motion features are reused at both the encoder and decoder as temporal priors for conditional motion coding.
\par
To fully exploit the proposed framework, BiCRVC also uses a carefully designed training strategy.
It promotes joint motion-frame coding, exploits motion-frame redundancy, and improves reconstruction quality for decoder-side motion compensation.
Training with random GOP structures further helps the model learn complex 
motion dependencies, reference patterns, and hierarchical quality even with short training sequences.
\subsection{Coupled Representation Coding}
Existing BVCs~\cite{sheng2025bi,zhai2025llbvc, jiang2025biecvc, liuneural} typically use two separate codecs for motion and frame information.
At the encoder, motion is first estimated and encoded. The decoded motion vectors are then used to align the reference frames before frame coding.
The same dependency also appears at the decoder, where motion must be decoded before frame decoding, as illustrated in Figure~\ref{fig:arch_compare}.
This design can improve compression efficiency through better context alignment, but it also introduces a two-stage decoding pipeline, limits parallelism, slows decoding, and requires two separate codecs for motion and frame coding.\par
In contrast, the low-delay-oriented DCVC-RT removes explicit motion modules and lets the encoder and decoder implicitly learn temporal dynamics from the reference and current-frame features.
This design works well for low-delay coding, where temporal correlations are mostly local.
Directly adopting DCVC-RT for random-access coding enables one-step decoding with one frame codec.
However, the temporal distance between the current frame and its references is often much larger in this setting.
Large motions are therefore harder to model with the limited receptive field of convolutional networks, leading to clear performance degradation.
These observations show that accurate motion modeling is critical to bidirectional coding efficiency.
\par
Considering the strengths and limitations of both designs, BiCRVC aims to enable one-step decoding with one unified codec while maintaining effective bidirectional context modeling.
In previous BVCs, motion vectors are also used to align the priors for entropy modeling, which forces motion to be decoded before the current frame.
Using unaligned contexts as entropy priors can improve speed, but separate motion transmission still requires an additional motion codec.
Moreover, the complexity of a standalone motion codec can be comparable to that of the frame codec, leading to higher overall complexity.
\par
These observations raise a natural question: can motion and frame coding be handled by one unified codec while maintaining effective context modeling?
\par
To this end, we propose coupled representation coding, whose architecture is shown in Figure~\ref{fig:crc}.
It consists of a coupled representation encoder and a coupled representation decoder.
The coupled representation encoder consists of a frame encoding transform (FET) and a coupled encoding transform (CET).
In FET, the input frame is first patchified ($\mathcal{P}$) to obtain the $1/8$-resolution representation for conditional coding.
At the encoder, the estimated uncompressed backward optical flows $\boldsymbol{v}_{t\rightarrow t-i_t}$ and $\boldsymbol{v}_{t\rightarrow t+i_t}$ are directly used as motion vectors.
This design accelerates encoding and enables the coupled representation decoder.
For encoder-side motion compensation (MC), the propagated features $\hat{\boldsymbol{F}}_{t-i_t}$ and $\hat{\boldsymbol{F}}_{t+i_t}$ serve as unaligned references.
Since $\hat{\boldsymbol{F}}_{t-i_t}$ and $\hat{\boldsymbol{F}}_{t+i_t}$ are also at $1/8$ resolution, they are unpatchified ($\mathcal{P}^{-1}$) and aligned by backward flow warping ($\mathcal{W}$) with optical flows $\boldsymbol{v}_{t\rightarrow t-i_t}$ and $\boldsymbol{v}_{t\rightarrow t+i_t}$, respectively.
The aligned references are then patchified into $\boldsymbol{C}_{t-i_t\rightarrow t}$ and $\boldsymbol{C}_{t+i_t\rightarrow t}$ to match the resolution of the patchified frame representation.
The contexts $\boldsymbol{C}_{t-i_t\rightarrow t}$ and $\boldsymbol{C}_{t+i_t\rightarrow t}$ are used as priors for conditional coding of the current frame, resulting in the latent $\boldsymbol{r}_t$.
The latent $\boldsymbol{r}_t$ and optical flows $\boldsymbol{v}_{t\rightarrow t-i_t}$ and $\boldsymbol{v}_{t\rightarrow t+i_t}$ are then fed into CET for unified transmission.
For clarity, we first describe the basic coupled representation coding pipeline without BMFP, and leave the BMFP-enhanced motion representation to Section~\ref{sec:bmfp}.
Specifically, the two optical flows are first patchified to $1/8$ resolution and concatenated along the channel dimension as $\boldsymbol{n}_t$, which matches the resolution of $\boldsymbol{r}_t$.
The latent $\boldsymbol{r}_t$ is concatenated with $\boldsymbol{n}_t$ as the input to CET.
CET then produces the coupled latent representation $\boldsymbol{y}_t$, which jointly contains the frame latent and motion representation.
The encoder-side process is formulated as
\begin{equation}
\begin{aligned}
\boldsymbol{C}_{t-i_t\rightarrow t}
&= \mathcal{P}\!\left(\mathcal{W}\!\left(\mathcal{P}^{-1}(\hat{\boldsymbol{F}}_{t-i_t}), \boldsymbol{v}_{t\rightarrow t-i_t}\right)\right), \\
\boldsymbol{C}_{t+i_t\rightarrow t}
&= \mathcal{P}\!\left(\mathcal{W}\!\left(\mathcal{P}^{-1}(\hat{\boldsymbol{F}}_{t+i_t}), \boldsymbol{v}_{t\rightarrow t+i_t}\right)\right), \\
\boldsymbol{r}_t
&= \mathrm{FET}\!\left(\mathcal{P}(\boldsymbol{x}_t), \boldsymbol{C}_{t-i_t\rightarrow t}, \boldsymbol{C}_{t+i_t\rightarrow t}\right), \\
\boldsymbol{n}_t
&= \mathrm{Concat}\!\left(
\mathcal{P}(\boldsymbol{v}_{t\rightarrow t-i_t}),
\mathcal{P}(\boldsymbol{v}_{t\rightarrow t+i_t})
\right), \\
\boldsymbol{y}_t
&= \mathrm{CET}\!\left(\mathrm{Concat}\!\left(\boldsymbol{r}_t, \boldsymbol{n}_t\right)\right).
\end{aligned}
\end{equation}
$\boldsymbol{y}_t$ is quantized as $\hat{\boldsymbol{y}}_t$ and entropy coded into a unified bitstream.
At the coupled representation decoder, the frame latent and motion representation are recovered from this unified bitstream.
Specifically, $\hat{\boldsymbol{y}}_t$ is first transformed by the coupled representation decoder and 
then split into the decoded latent $\hat{\boldsymbol{r}}_t$ and the decoded patchified motion representation $\hat{\boldsymbol{n}}_t$ along the channel dimension.
The decoded motion representation $\hat{\boldsymbol{n}}_t$ is further split into two parts and unpatchified ($\mathcal{P}^{-1}$) to obtain the decoded optical flows $\hat{\boldsymbol{v}}_{t\rightarrow t-i_t}$ and $\hat{\boldsymbol{v}}_{t\rightarrow t+i_t}$.
These decoded optical flows are used to align the propagated features $\hat{\boldsymbol{F}}_{t-i_t}$ and $\hat{\boldsymbol{F}}_{t+i_t}$ by backward flow warping ($\mathcal{W}$).
The aligned references are then patchified into $\hat{\boldsymbol{C}}_{t-i_t\rightarrow t}$ and $\hat{\boldsymbol{C}}_{t+i_t\rightarrow t}$.
Finally, $\hat{\boldsymbol{r}}_t$, $\hat{\boldsymbol{C}}_{t-i_t\rightarrow t}$, and $\hat{\boldsymbol{C}}_{t+i_t\rightarrow t}$ are concatenated and fed into 
the frame decoding transform to generate the frame feature 
$\hat{\boldsymbol{F}}_t$, which is further used by the 
reconstruction generation network (REC)
to reconstruct $\hat{\boldsymbol{x}}_t$.
The decoder side process is formulated as
\begin{equation}
\begin{aligned}
\hat{\boldsymbol{r}}_t, \hat{\boldsymbol{n}}_t
&= \mathrm{Split}\!\left(\mathrm{CRD}(\hat{\boldsymbol{y}}_t)\right), \\
\hat{\boldsymbol{v}}_{t\rightarrow t-i_t}
&= \mathcal{P}^{-1}\!\left(\mathrm{Split}_1(\hat{\boldsymbol{n}}_t)\right), \\
\hat{\boldsymbol{v}}_{t\rightarrow t+i_t}
&= \mathcal{P}^{-1}\!\left(\mathrm{Split}_2(\hat{\boldsymbol{n}}_t)\right), \\
\hat{\boldsymbol{C}}_{t-i_t\rightarrow t}
&= \mathcal{P}\!\left(\mathcal{W}\!\left(\mathcal{P}^{-1}(\hat{\boldsymbol{F}}_{t-i_t}), \hat{\boldsymbol{v}}_{t\rightarrow t-i_t}\right)\right), \\
\hat{\boldsymbol{C}}_{t+i_t\rightarrow t}
&= \mathcal{P}\!\left(\mathcal{W}\!\left(\mathcal{P}^{-1}(\hat{\boldsymbol{F}}_{t+i_t}), \hat{\boldsymbol{v}}_{t\rightarrow t+i_t}\right)\right), \\
\hat{\boldsymbol{F}}_t
&= \mathrm{FDT}\!\left(\mathrm{Concat}\!\left(\hat{\boldsymbol{r}}_t, \hat{\boldsymbol{C}}_{t-i_t\rightarrow t}, \hat{\boldsymbol{C}}_{t+i_t\rightarrow t}\right)\right), \\
\hat{\boldsymbol{x}}_t
&= \mathrm{REC}\!\left(\hat{\boldsymbol{F}}_t\right).
\end{aligned}
\end{equation}
\par
By jointly coding motion and frame representations through concatenation, BiCRVC uses one unified codec, enables one-step decoding, and exploits the correlation between motion and frame coding.
This design has the potential to improve coding efficiency while reducing decoding complexity.
Moreover, coupled representation coding provides a \textit{scalable} framework.
It can be extended to jointly transmit \textit{other compact side information}, 
such as context-adaptive weights, so that these priors can be made available at both the encoder and decoder 
through coupled transmission.
\begin{figure*}
  \centering
  \includegraphics[width=0.92\textwidth]{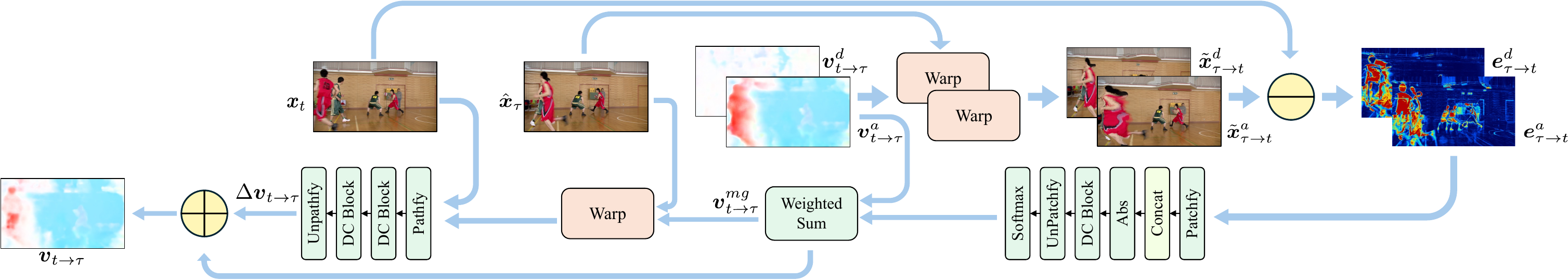}
  \caption{Architecture of the proposed multi-candidate motion estimation.}
  \label{fig:mcme_arch}
\end{figure*}
\subsection{Multi-Candidate Motion Estimation}
In bidirectional video compression, the quality of motion estimation has a significant impact on coding efficiency.
Existing BVCs~\cite{sheng2025bi,jiang2025biecvc} directly use optical flows estimated by SPyNet~\cite{ranjan2017optical}, which can be unreliable for large motions.
To address this issue, BiCRVC employs multi-candidate motion estimation (MCME), as shown in Fig.~\ref{fig:mcme_arch}.
MCME improves motion estimation flexibility and robustness to complex motion patterns by merging multiple motion candidates, each of which is suited to a specific motion range.
\subsubsection{Multi-Scale Motion Estimation}
In random-access scenarios, the motion between the current frame and its references can be very large, and some sequences may contain fast object or camera motion.
These factors make the motion field difficult to estimate reliably.
Convolutional optical flow networks are often limited in this case because their effective receptive fields grow slowly with network depth.
For example, SPyNet~\cite{ranjan2017optical} stacks $5\times 5$ and $7\times 7$ convolutional kernels, but its ability to capture long-range correlations remains limited, especially for high-resolution frames.
One possible remedy is to introduce global attention for motion estimation.
However, full-frame attention brings high computational and memory costs, and its practical benefit for dense motion estimation is not guaranteed.
This raises a practical question: 
\textit{how can an optical flow network estimate more accurate 
motion fields under a limited effective receptive field?}
To address this issue, we propose multi-scale motion estimation (MSME).
The architecture is illustrated in Fig.~\ref{fig:msme}.
The key idea is to change the input scale rather than the network itself.
Although the receptive field of an optical flow network is fixed in its input coordinate system, its coverage in the original image coordinate system depends on the input resolution.
When the input frames are downsampled, the same receptive field corresponds to a larger region after being projected back to the original resolution.
As a result, the network can estimate larger displacements without increasing the depth or complexity of the motion estimator.
Specifically, let $\tau\in\{t-i_t,t+i_t\}$ denote a reference POC.
Given the current frame $\boldsymbol{x}_t\in \mathbb{R}^{H\times W\times 3}$ and a reference frame
$\boldsymbol{x}_\tau \in \mathbb{R}^{H\times W\times 3}$, we first build multi-scale frame pyramids
$\{\boldsymbol{x}_t^{1/2^s}\}_{s=0}^3$ and $\{\boldsymbol{x}_{\tau}^{1/2^s}\}_{s=0}^3$
by downsampling the frames with scale factor $1/2^s$.
At each scale, the current frame and the reference frame are fed into the optical flow network to estimate a scale-specific backward motion field
$\boldsymbol{v}_{t\rightarrow \tau}^{1/2^s}$.
To apply these motion fields at the original resolution, 
we bilinearly upsample them to $H\times W$ and rescale their magnitudes accordingly, obtaining motion candidates
$\{\boldsymbol{v}_{t\rightarrow \tau}^{(s)}\}_{s=0}^3$.
This simple operation keeps the additional cost low while making each scale provide a motion candidate in the original image coordinate system.
\par
\begin{figure*}[t]
\centering
\setlength{\tabcolsep}{2pt}
\renewcommand{\arraystretch}{1.5}
\scriptsize
\begin{tabular}{c !{\vrule width 0.5pt} c c c c !{\vrule width 0.5pt} c}
\multicolumn{1}{c}{\textbf{Input Frames}} &
\multicolumn{4}{c !{\vrule width 0.5pt}}{\textbf{Multi-Scale Motions, Warped Frames and Warped Errors}} &
\multicolumn{1}{c}{\textbf{Output Results}} \\

\textbf{Ref.} $\hat{\boldsymbol{x}}_0$ &
$\boldsymbol{v}_{1\rightarrow0}^{(1)}$ &
$\boldsymbol{v}_{1\rightarrow0}^{(2)}$ &
$\boldsymbol{v}_{1\rightarrow0}^{(4)}$ &
$\boldsymbol{v}_{1\rightarrow0}^{(8)}$ &
$\boldsymbol{v}_{1\rightarrow0}^{d}$ \\

\includegraphics[width=0.115\textwidth]{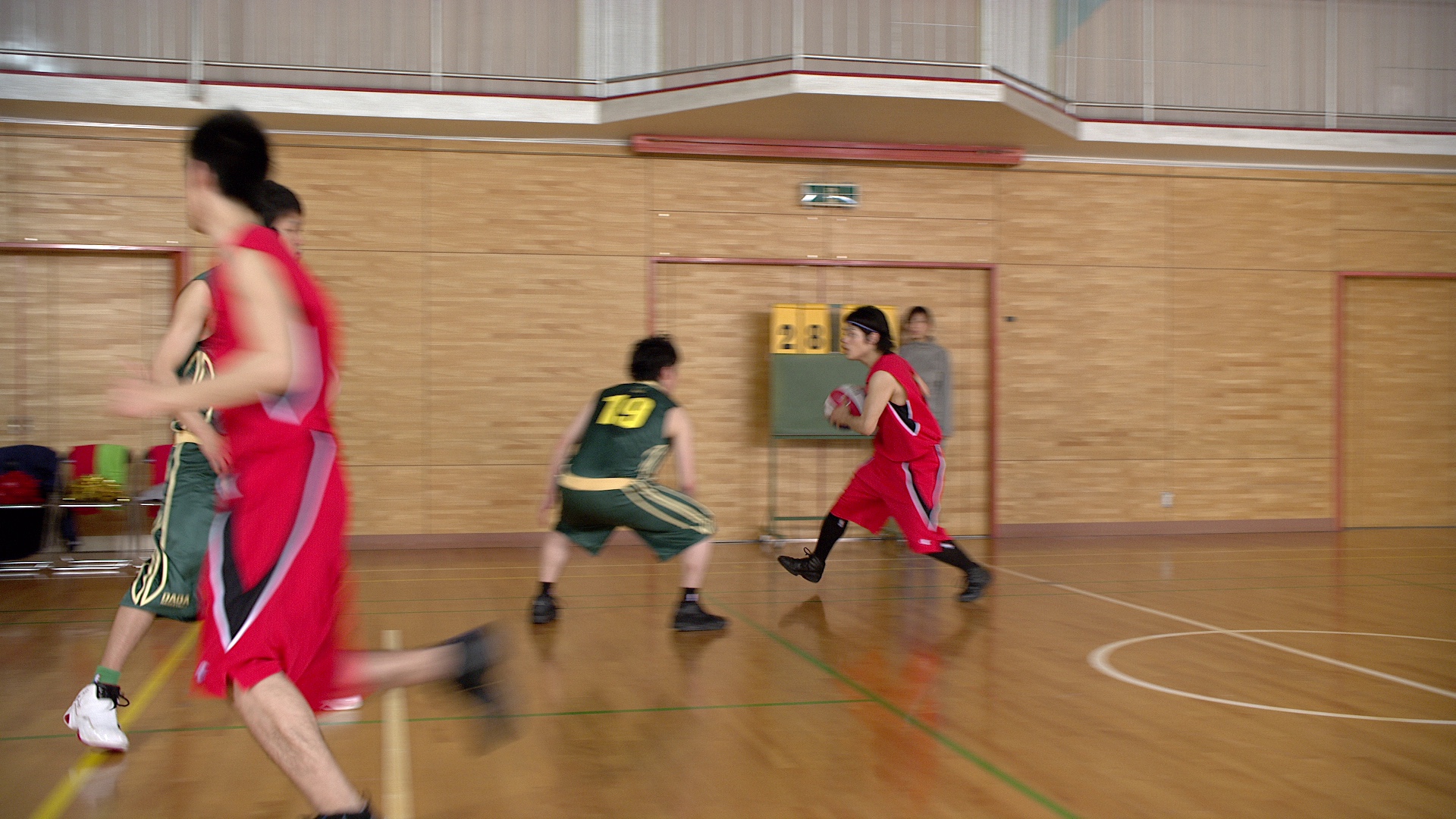} &
\includegraphics[width=0.115\textwidth]{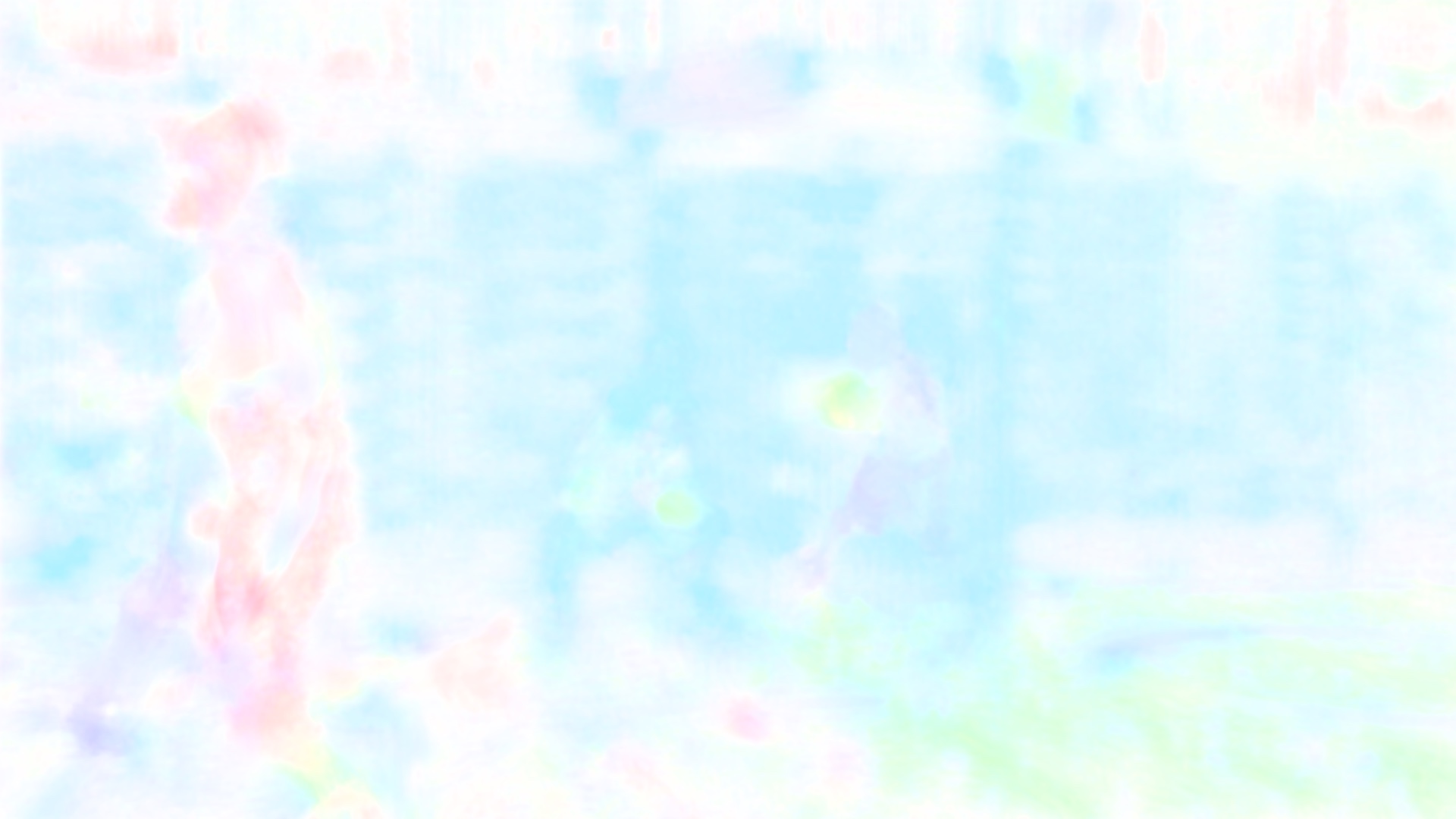} &
\includegraphics[width=0.115\textwidth]{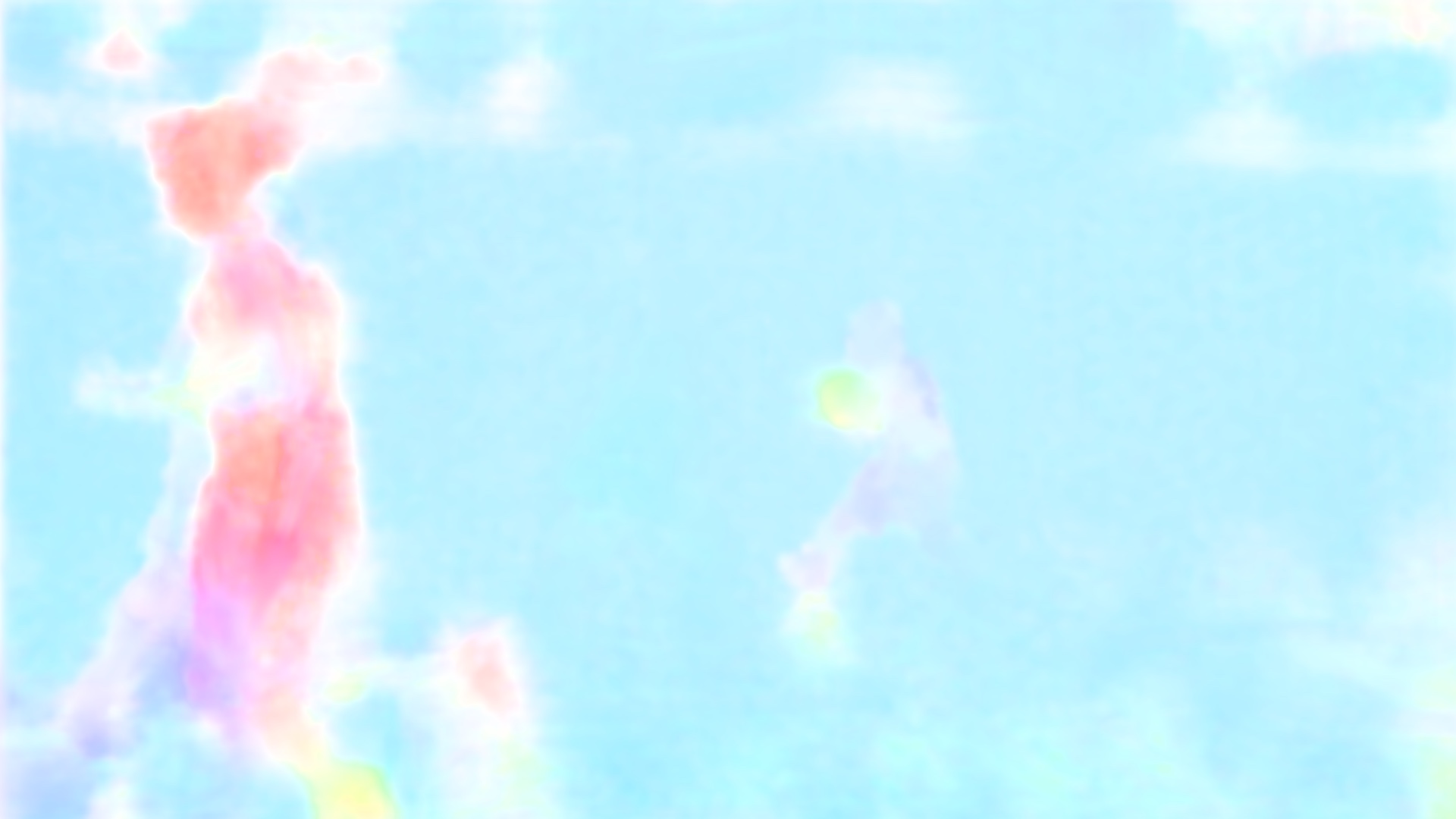} &
\includegraphics[width=0.115\textwidth]{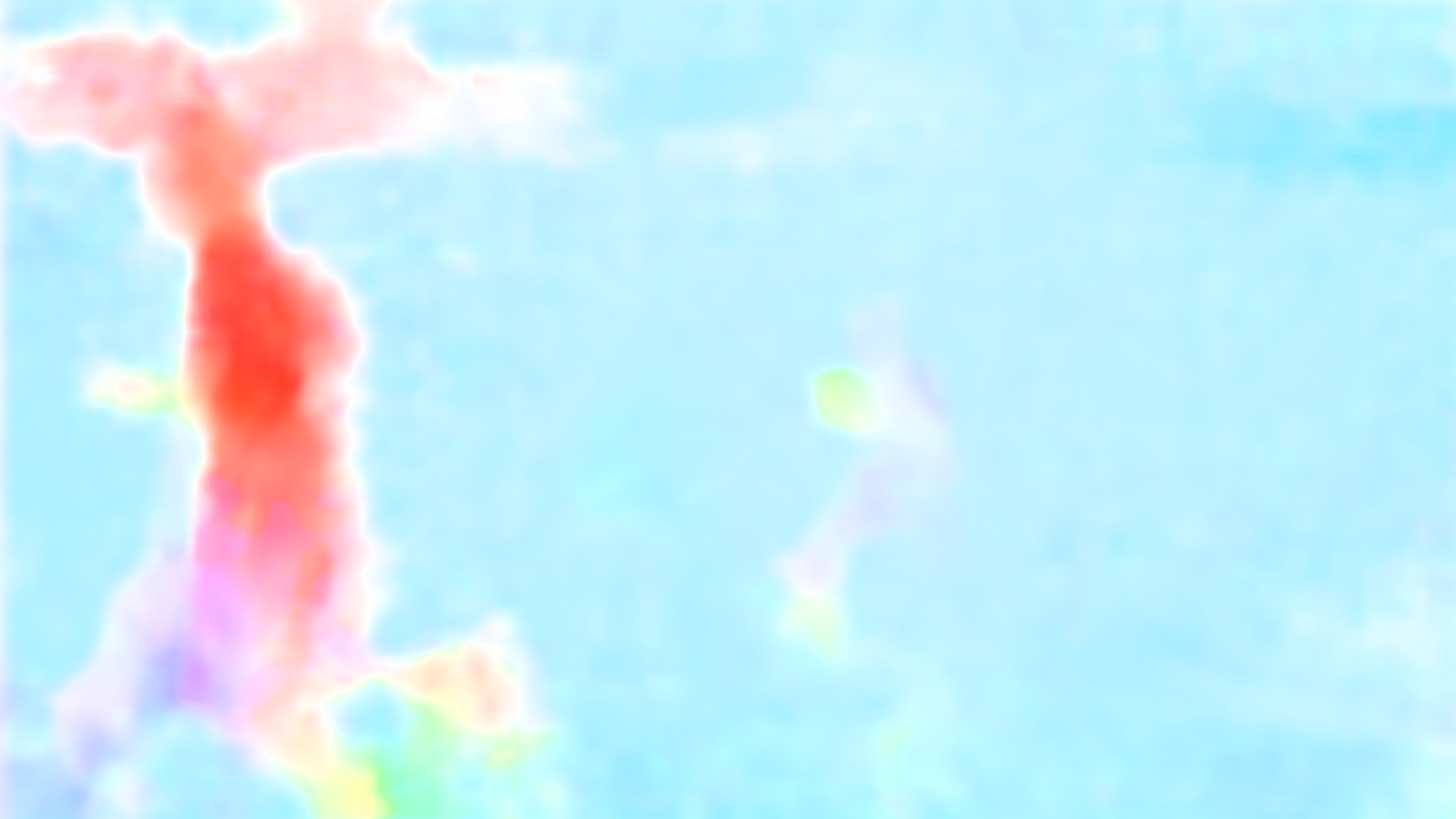} &
\includegraphics[width=0.115\textwidth]{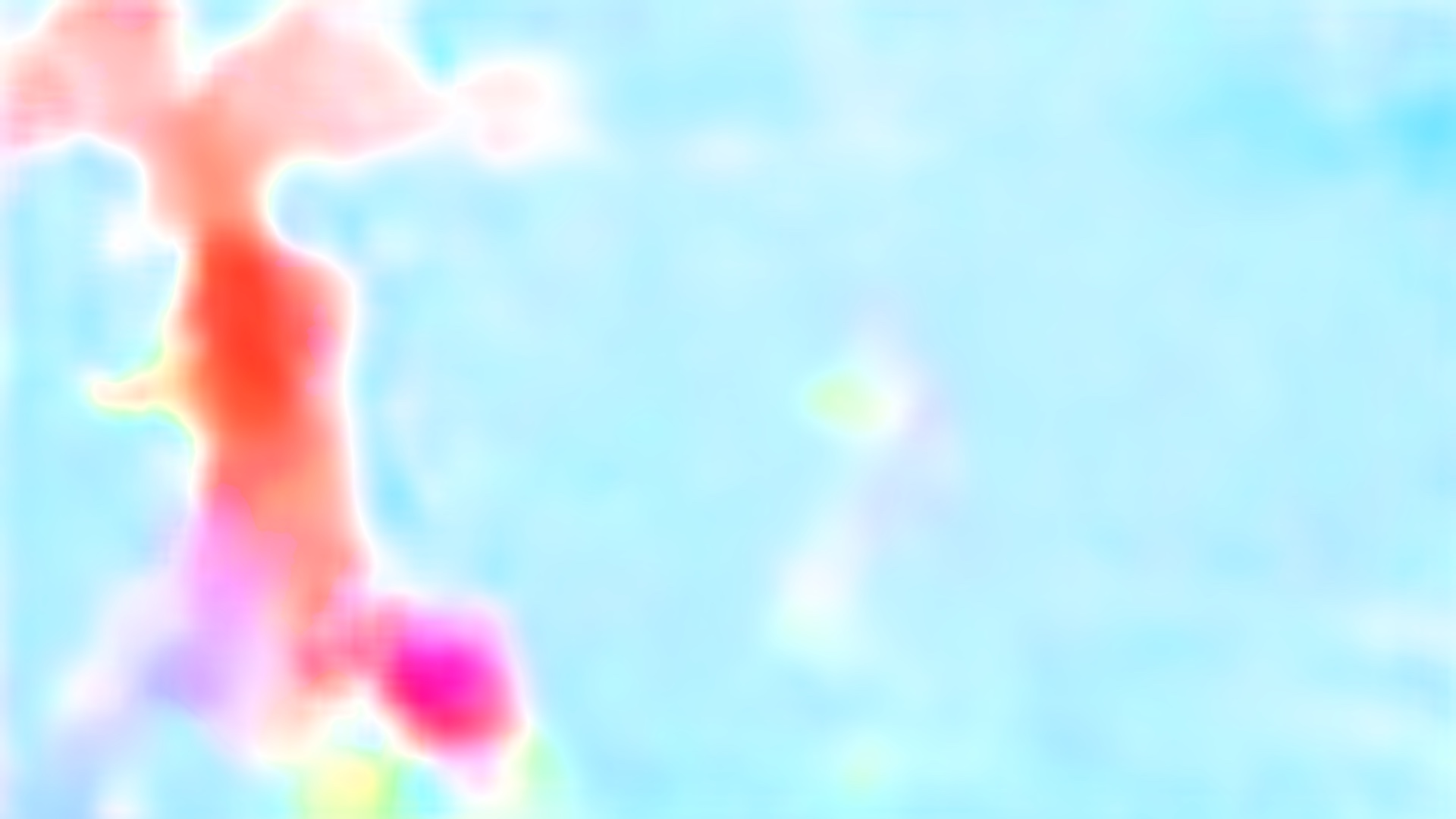} &
\includegraphics[width=0.115\textwidth]{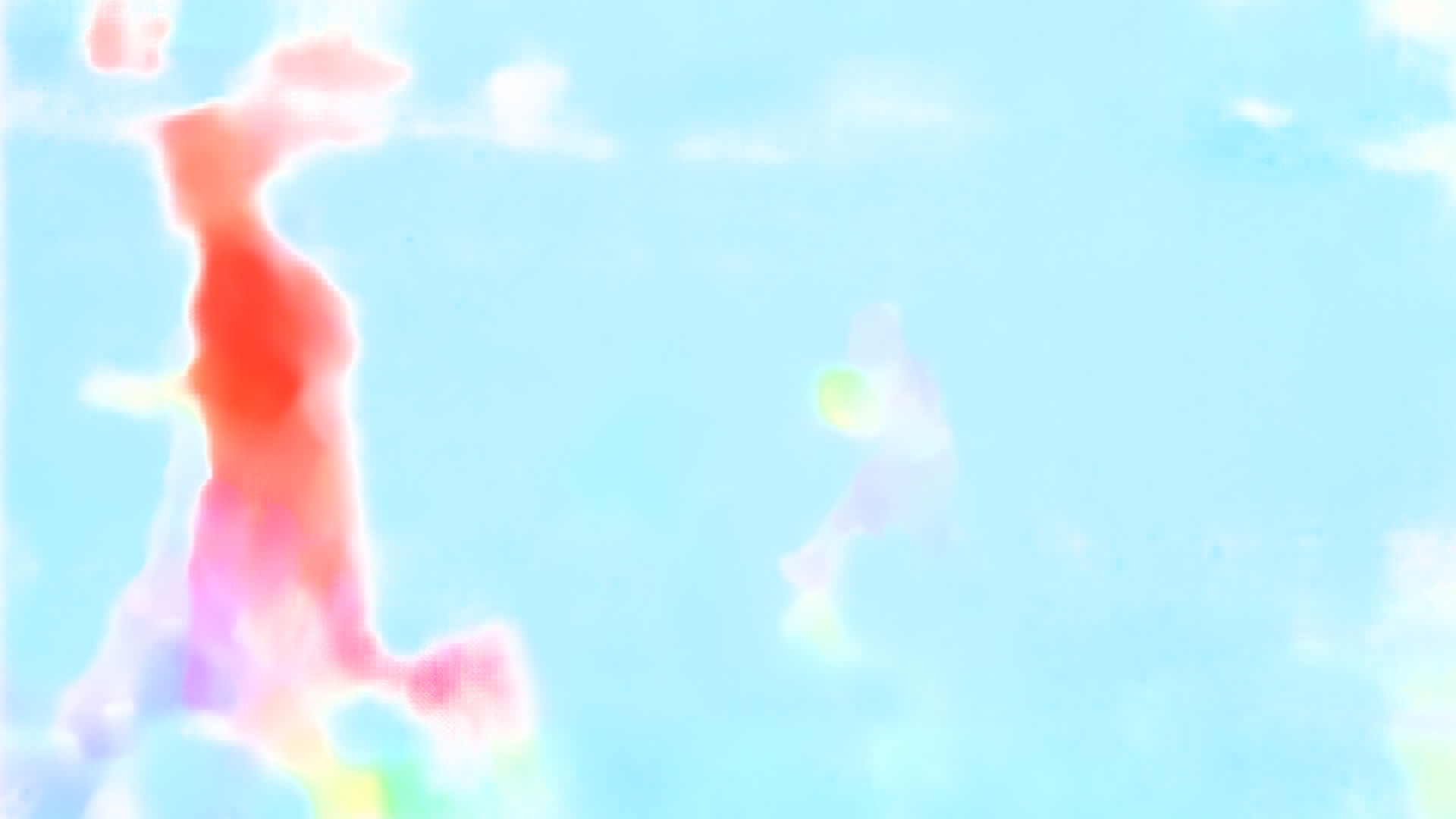} \\

\textbf{Cur.} $\boldsymbol{x}_1$ &
$\hat{\boldsymbol{x}}_{0\rightarrow1}^{(1)}\ 23.53\,\mathrm{dB}$ &
$\hat{\boldsymbol{x}}_{0\rightarrow1}^{(2)}\ 25.51\,\mathrm{dB}$ &
$\hat{\boldsymbol{x}}_{0\rightarrow1}^{(4)}\ 27.18\,\mathrm{dB}$ &
$\hat{\boldsymbol{x}}_{0\rightarrow1}^{(8)}\ 26.79\,\mathrm{dB}$ &
$\hat{\boldsymbol{x}}_{0\rightarrow1}^{d}\ 28.19\,\mathrm{dB}$ \\

\includegraphics[width=0.115\textwidth]{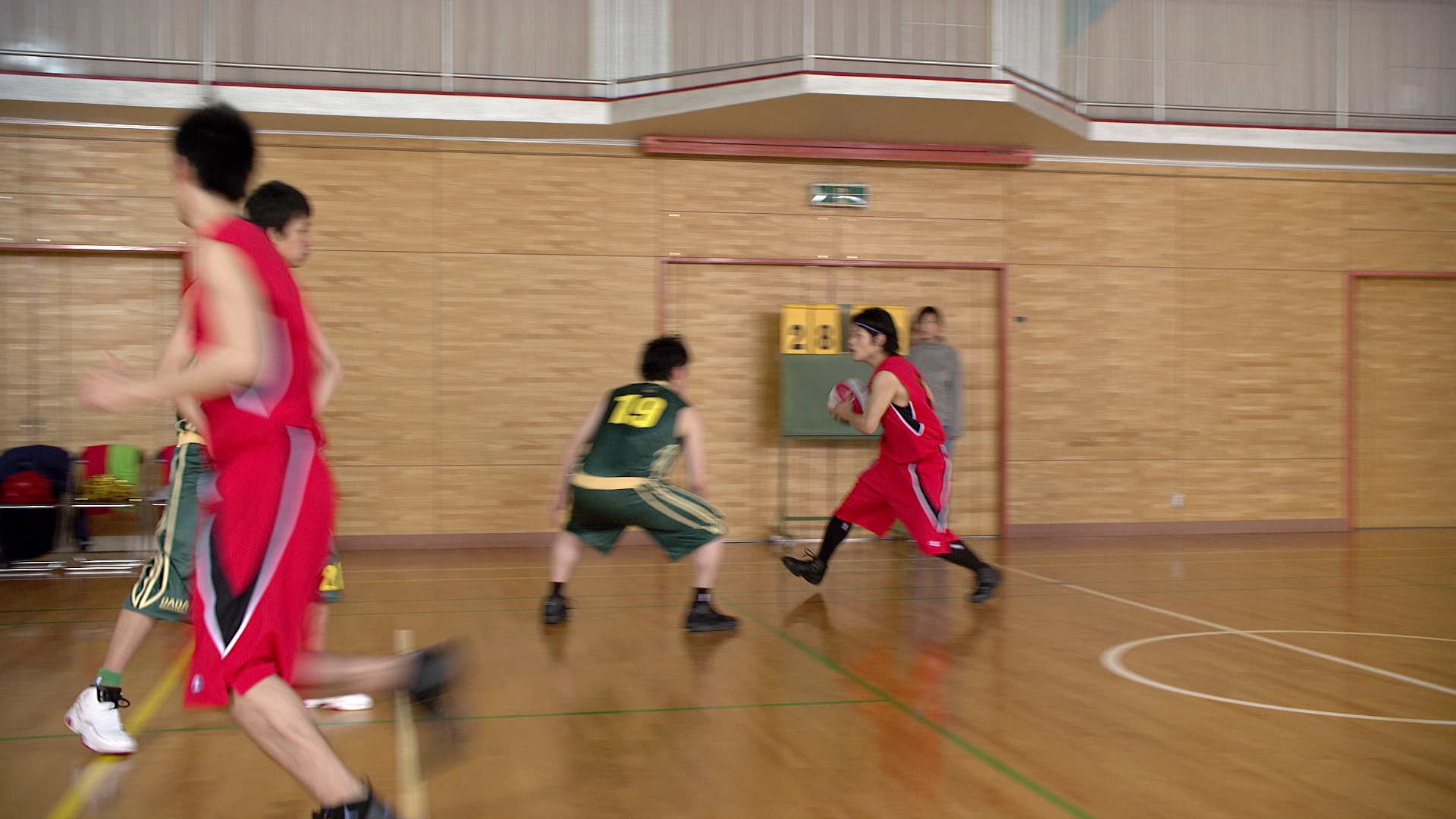} &
\includegraphics[width=0.115\textwidth]{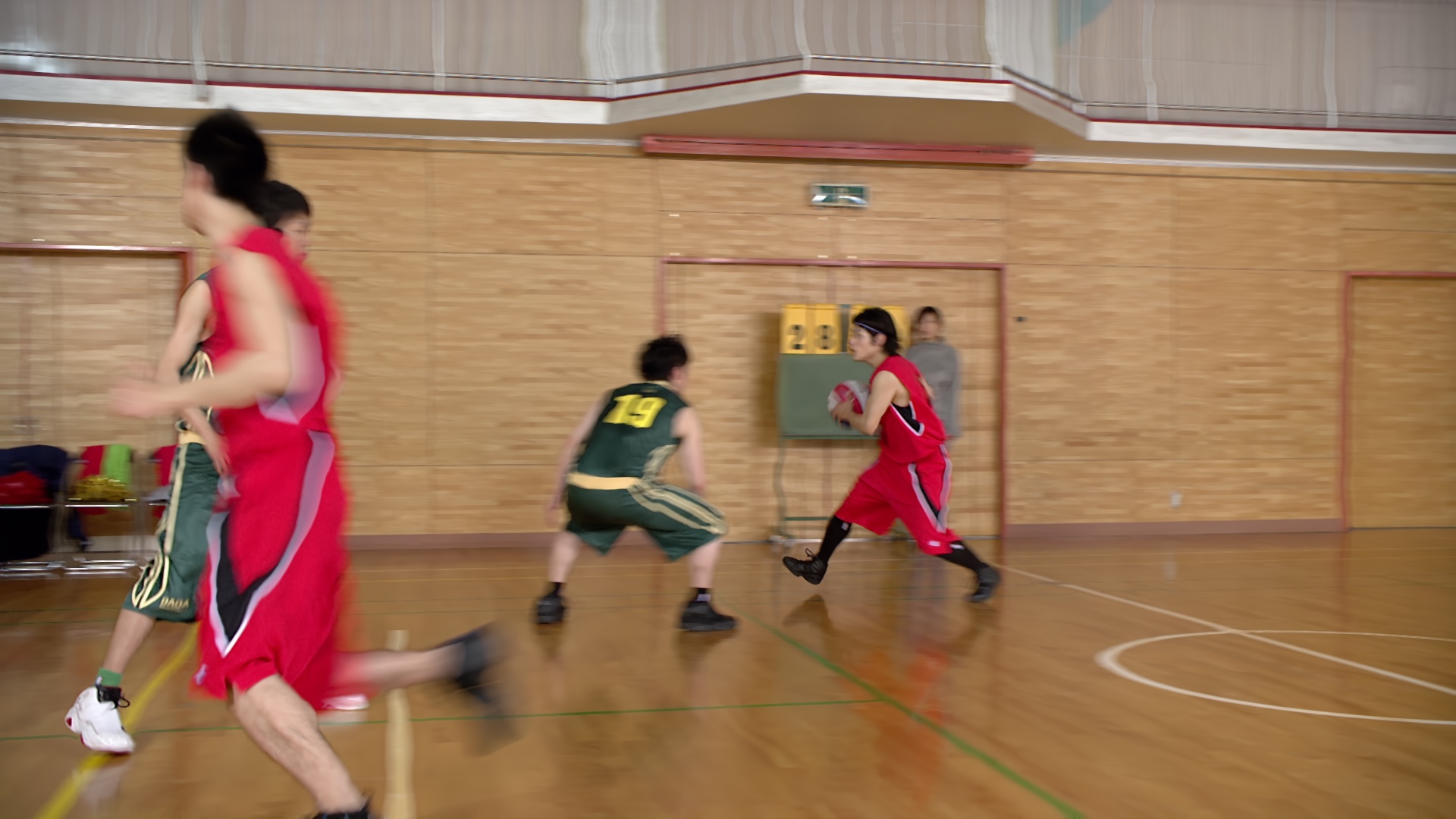} &
\includegraphics[width=0.115\textwidth]{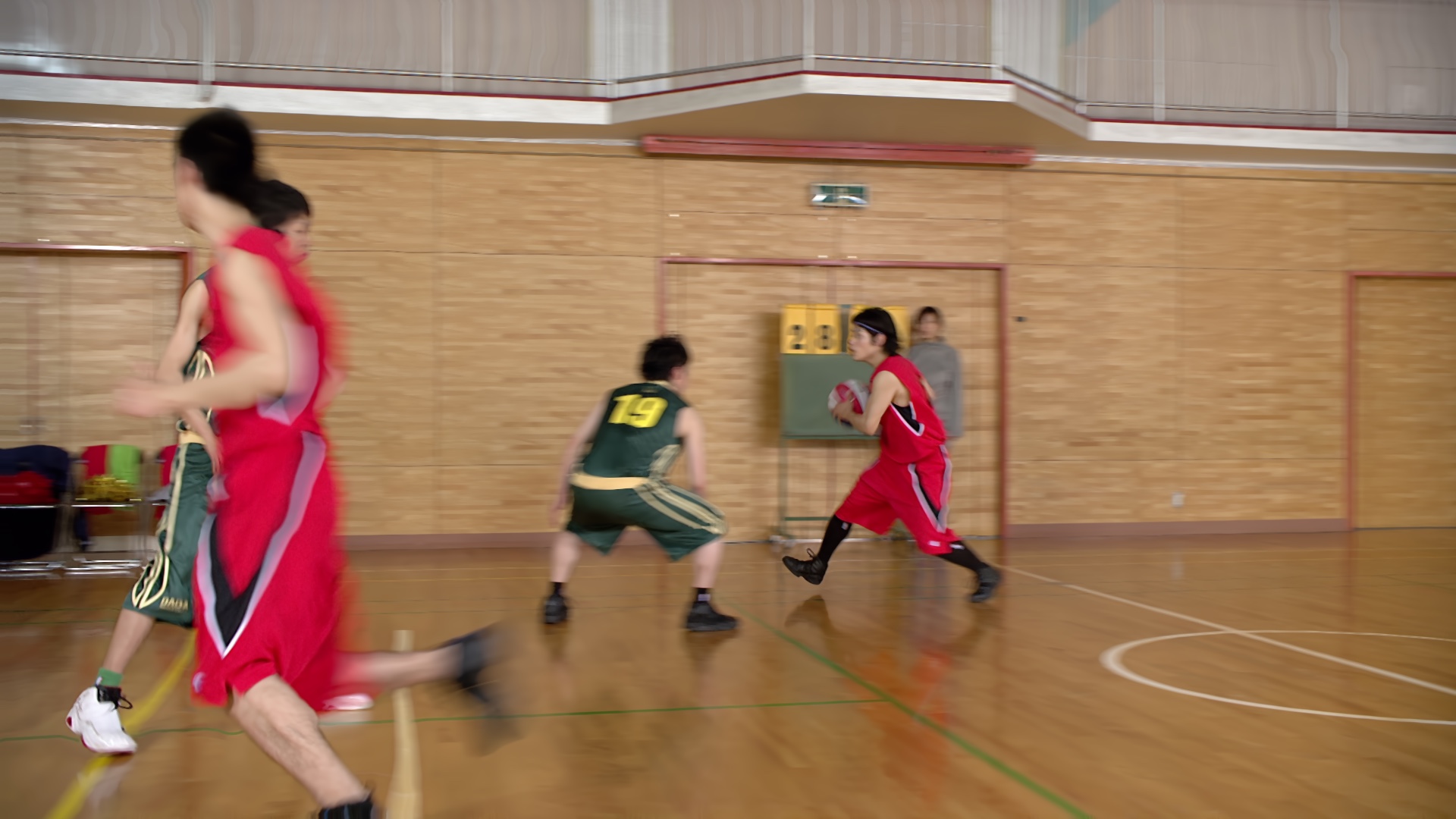} &
\includegraphics[width=0.115\textwidth]{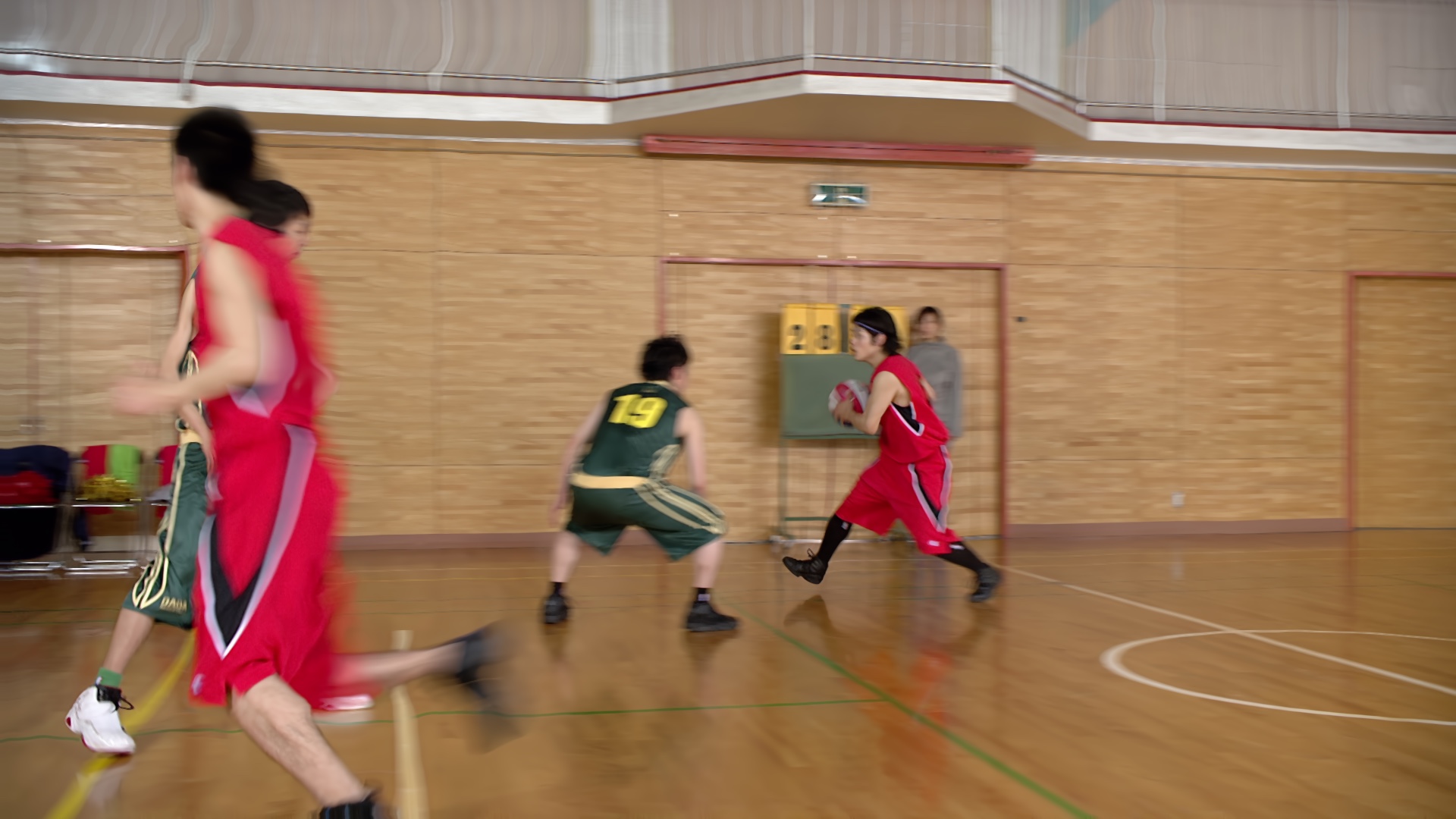} &
\includegraphics[width=0.115\textwidth]{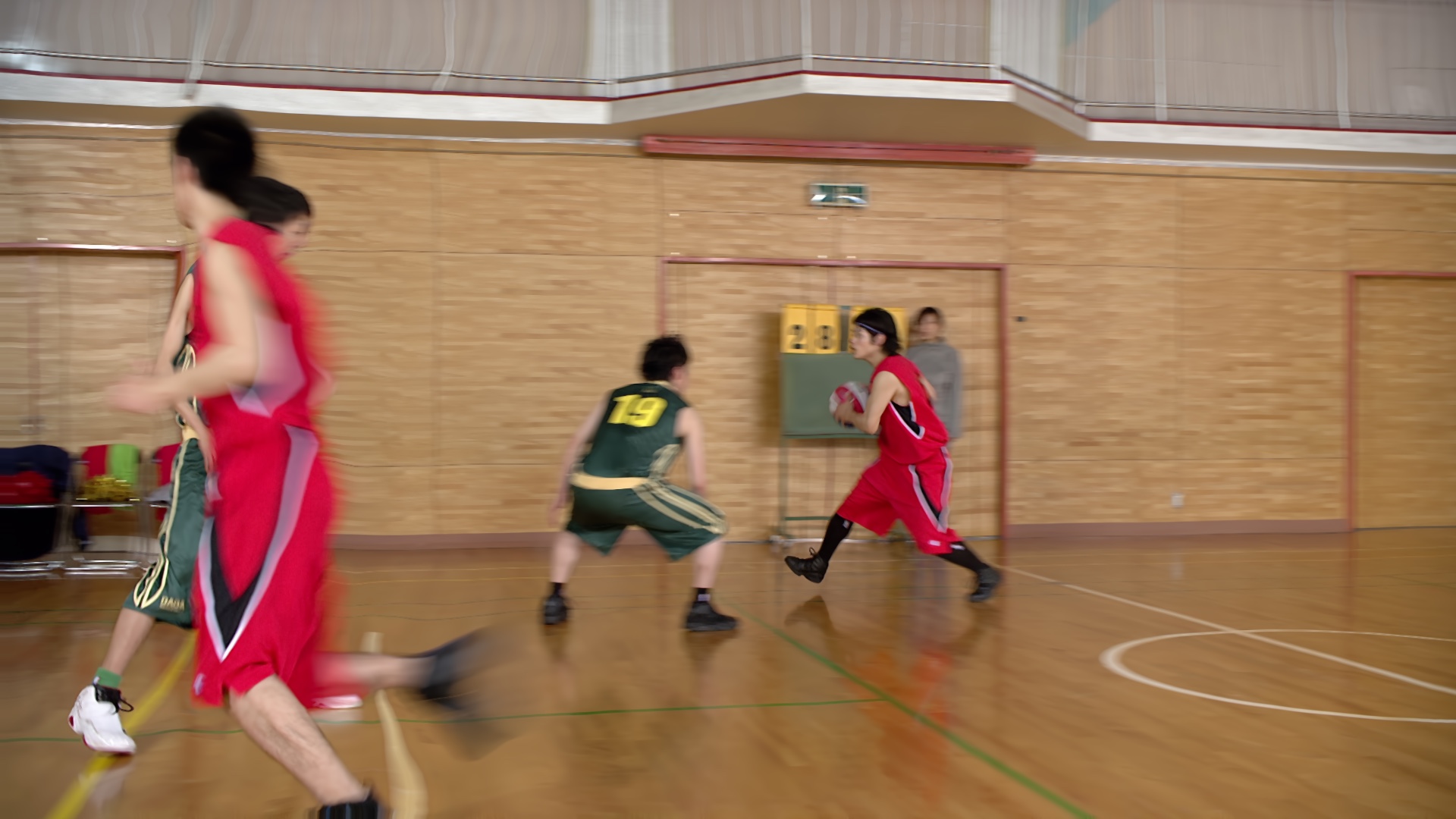} &
\includegraphics[width=0.115\textwidth]{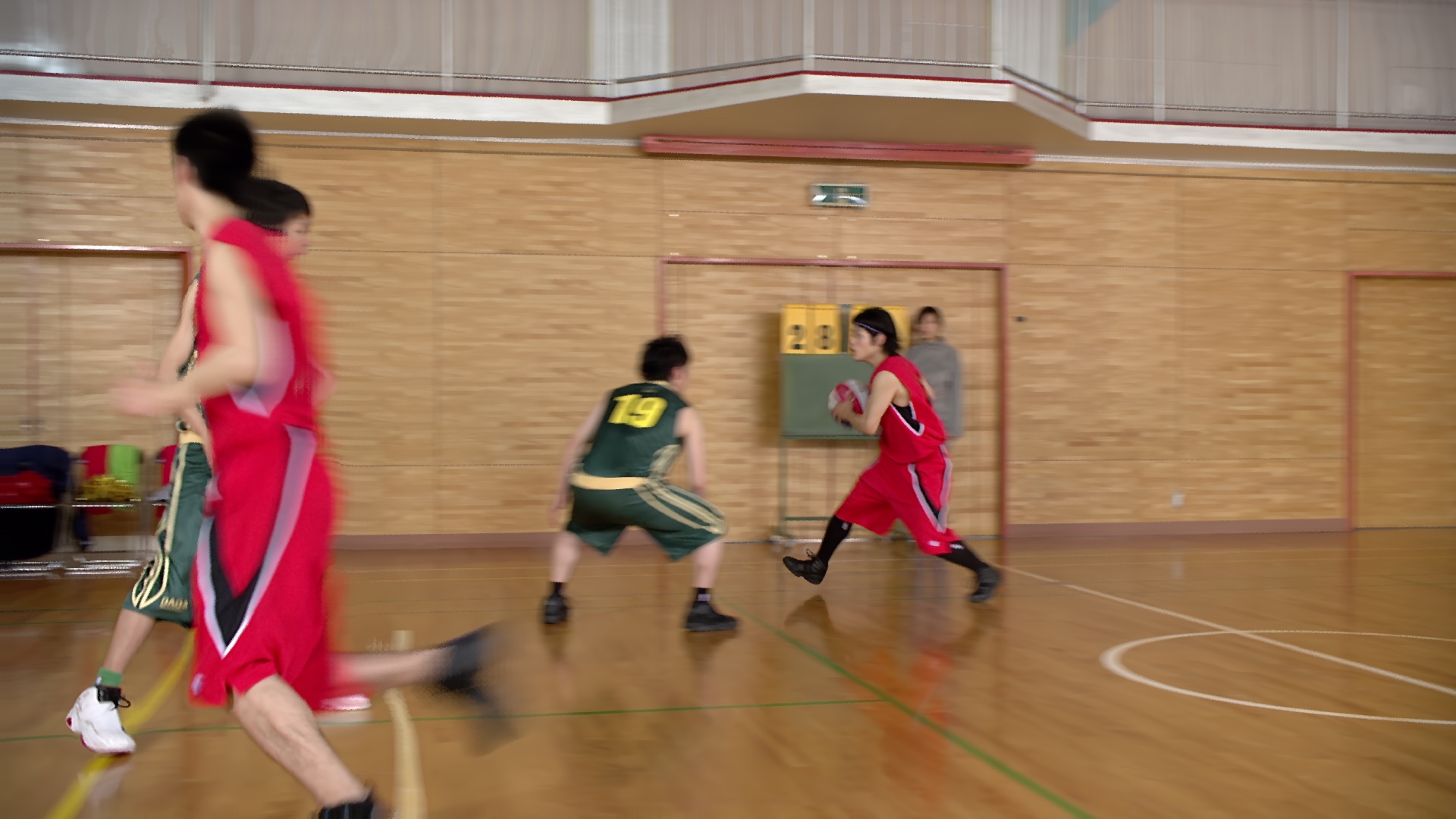} \\

 &
$e_{0\rightarrow1}^{(1)}$ &
$e_{0\rightarrow1}^{(2)}$ &
$e_{0\rightarrow1}^{(4)}$ &
$e_{0\rightarrow1}^{(8)}$ &
$e_{0\rightarrow1}^{d}$ \\

 &
\includegraphics[width=0.115\textwidth]{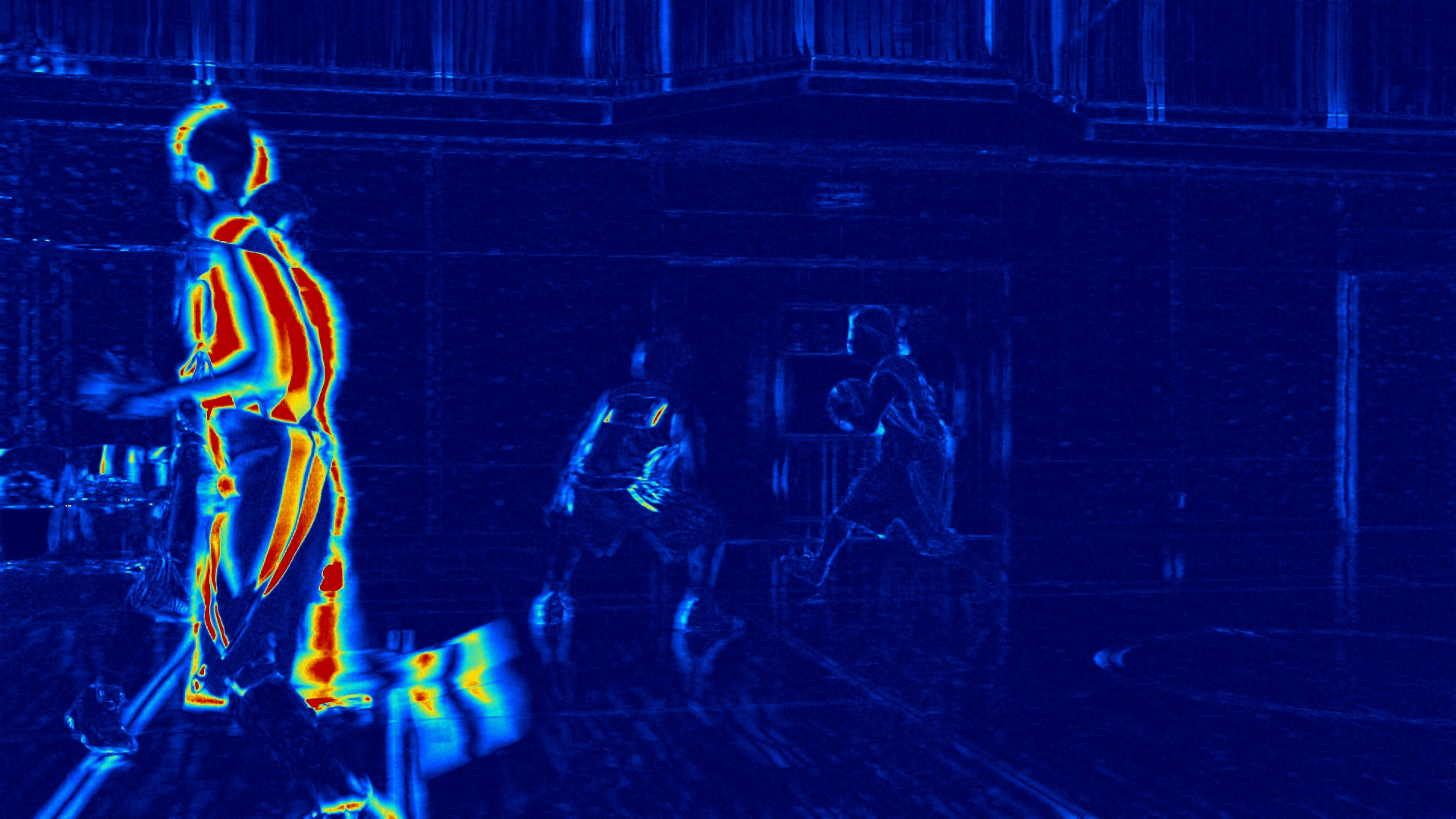} &
\includegraphics[width=0.115\textwidth]{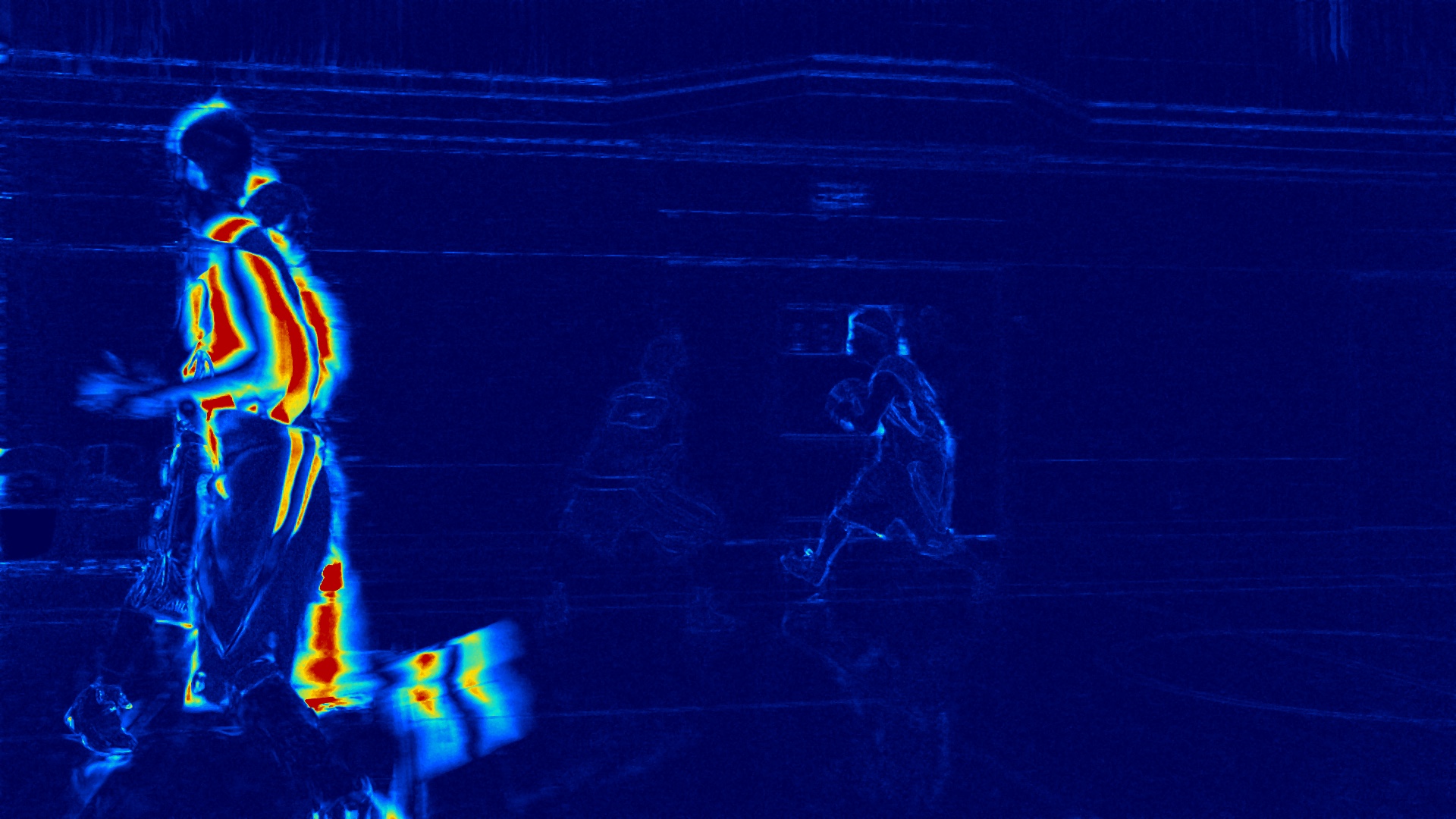} &
\includegraphics[width=0.115\textwidth]{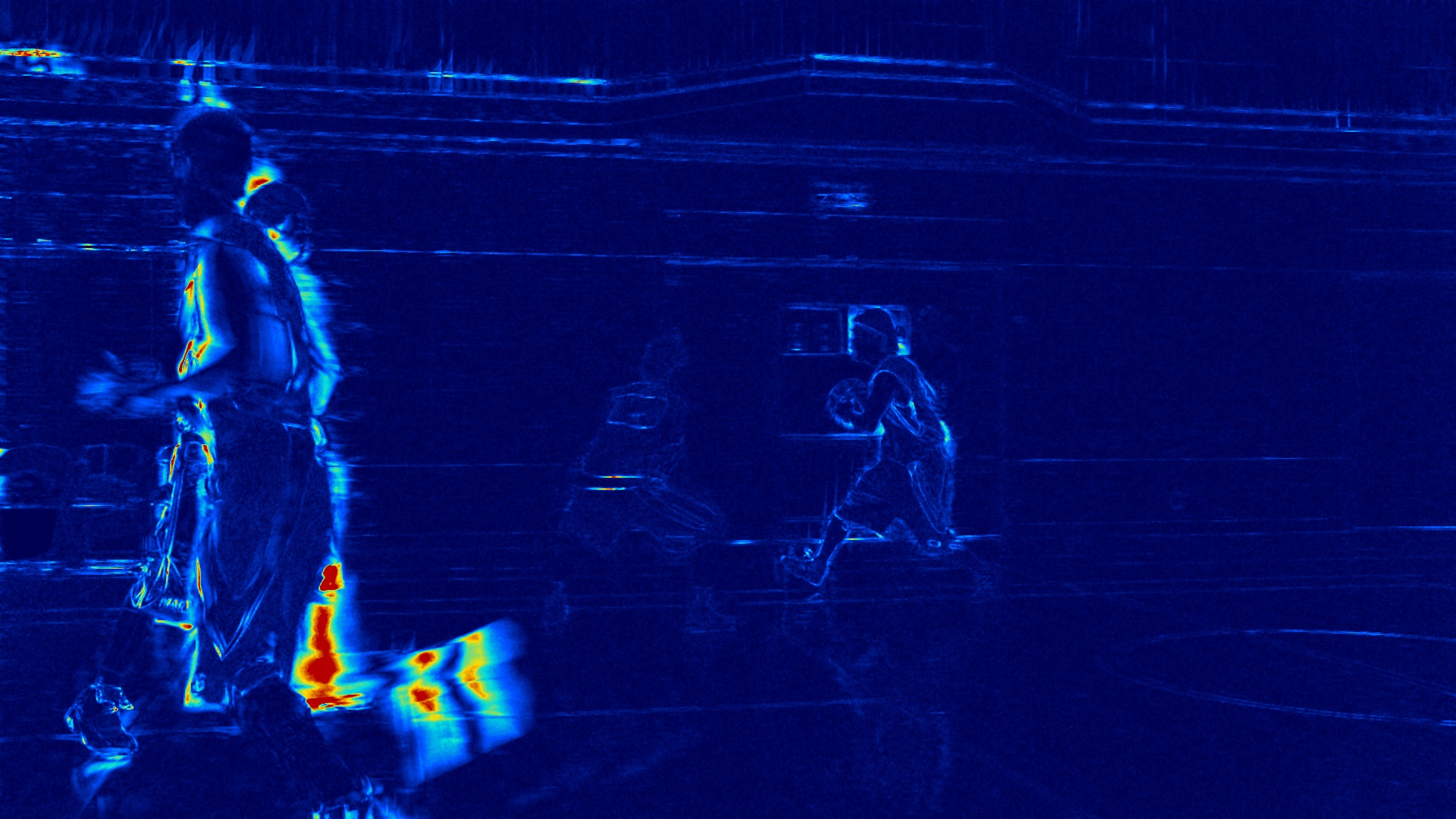} &
\includegraphics[width=0.115\textwidth]{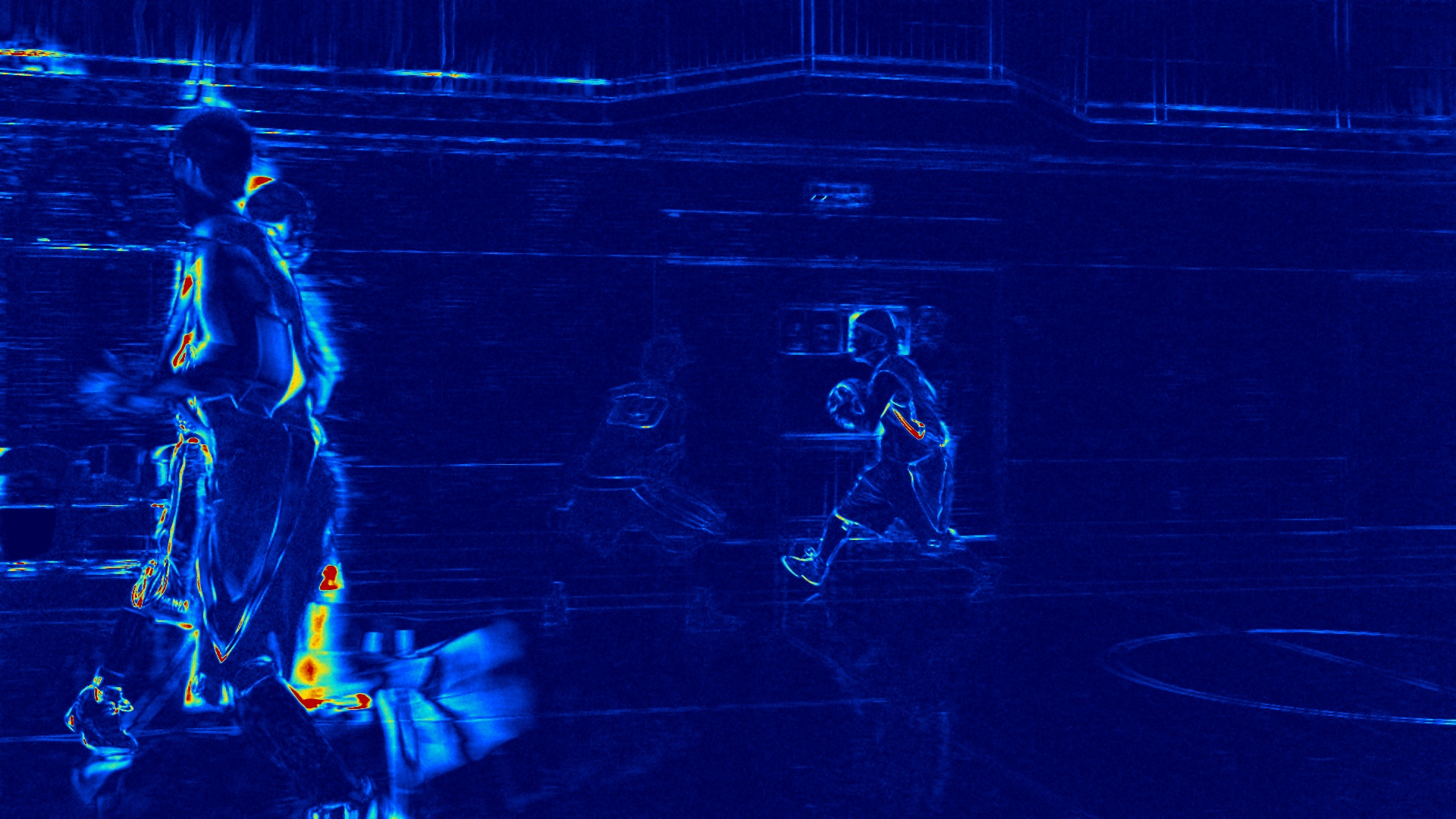} &
\includegraphics[width=0.115\textwidth]{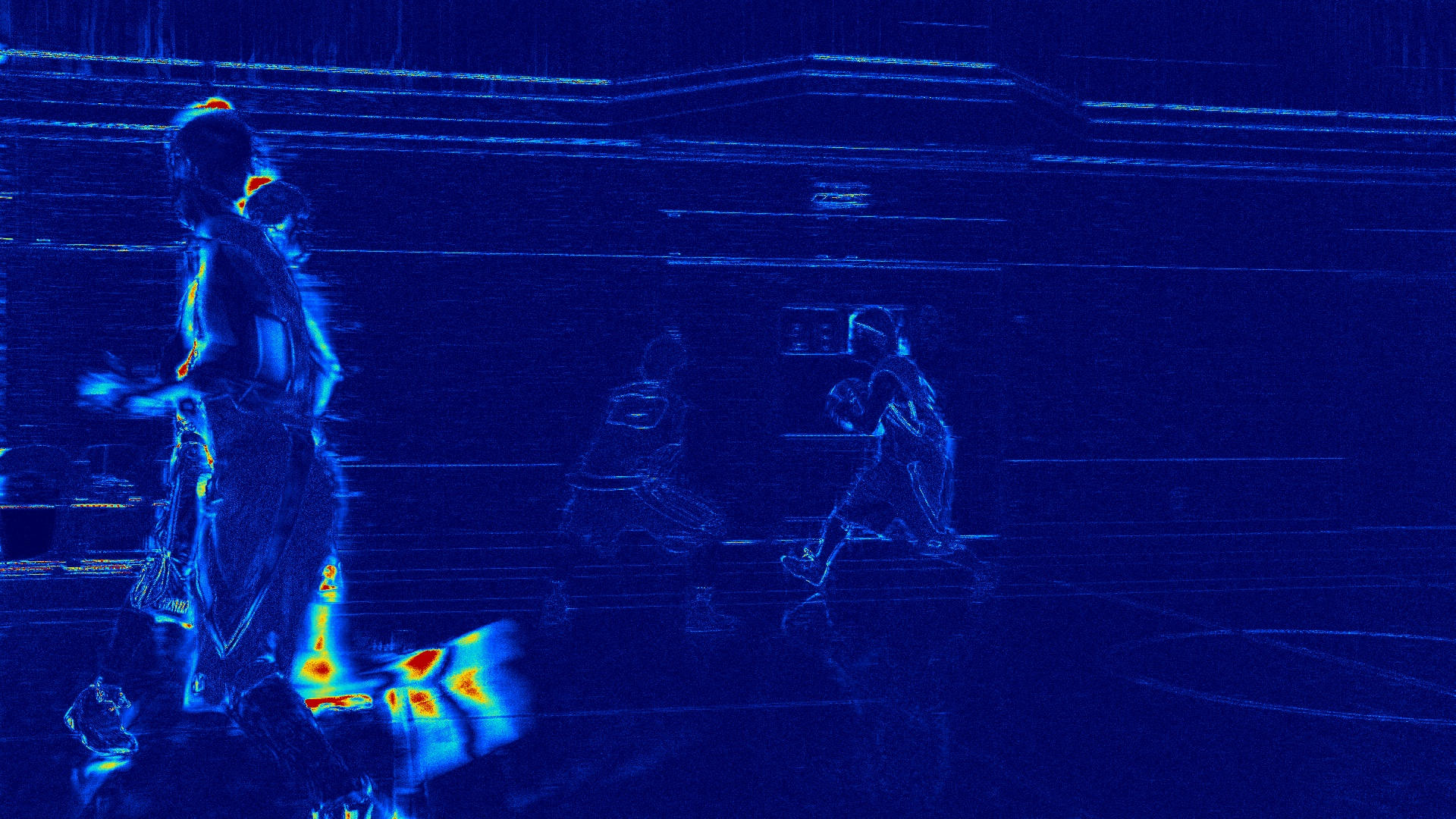} \\
\noalign{\vskip 2pt \color{gray!35}\hrule height 0.35pt \vskip 2pt}

\textbf{Ref.} $\hat{\boldsymbol{x}}_0$ &
$\boldsymbol{v}_{2\rightarrow0}^{(1)}$ &
$\boldsymbol{v}_{2\rightarrow0}^{(2)}$ &
$\boldsymbol{v}_{2\rightarrow0}^{(4)}$ &
$\boldsymbol{v}_{2\rightarrow0}^{(8)}$ &
$\boldsymbol{v}_{2\rightarrow0}^{d}$ \\

\includegraphics[width=0.115\textwidth]{figures_jpg/motion_vis/pptx/BasketballDrive/frame/im00001.jpg} &
\includegraphics[width=0.115\textwidth]{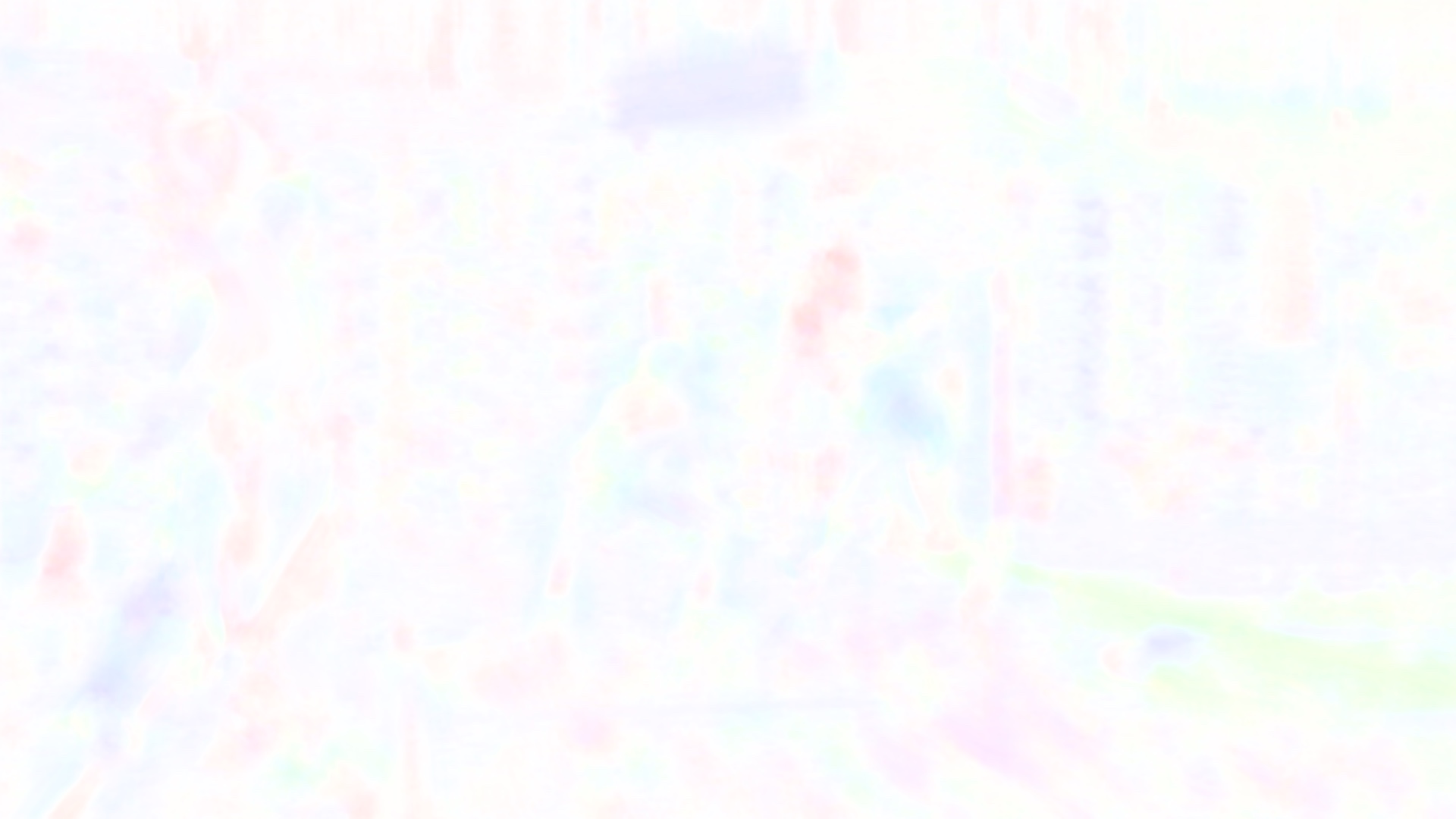} &
\includegraphics[width=0.115\textwidth]{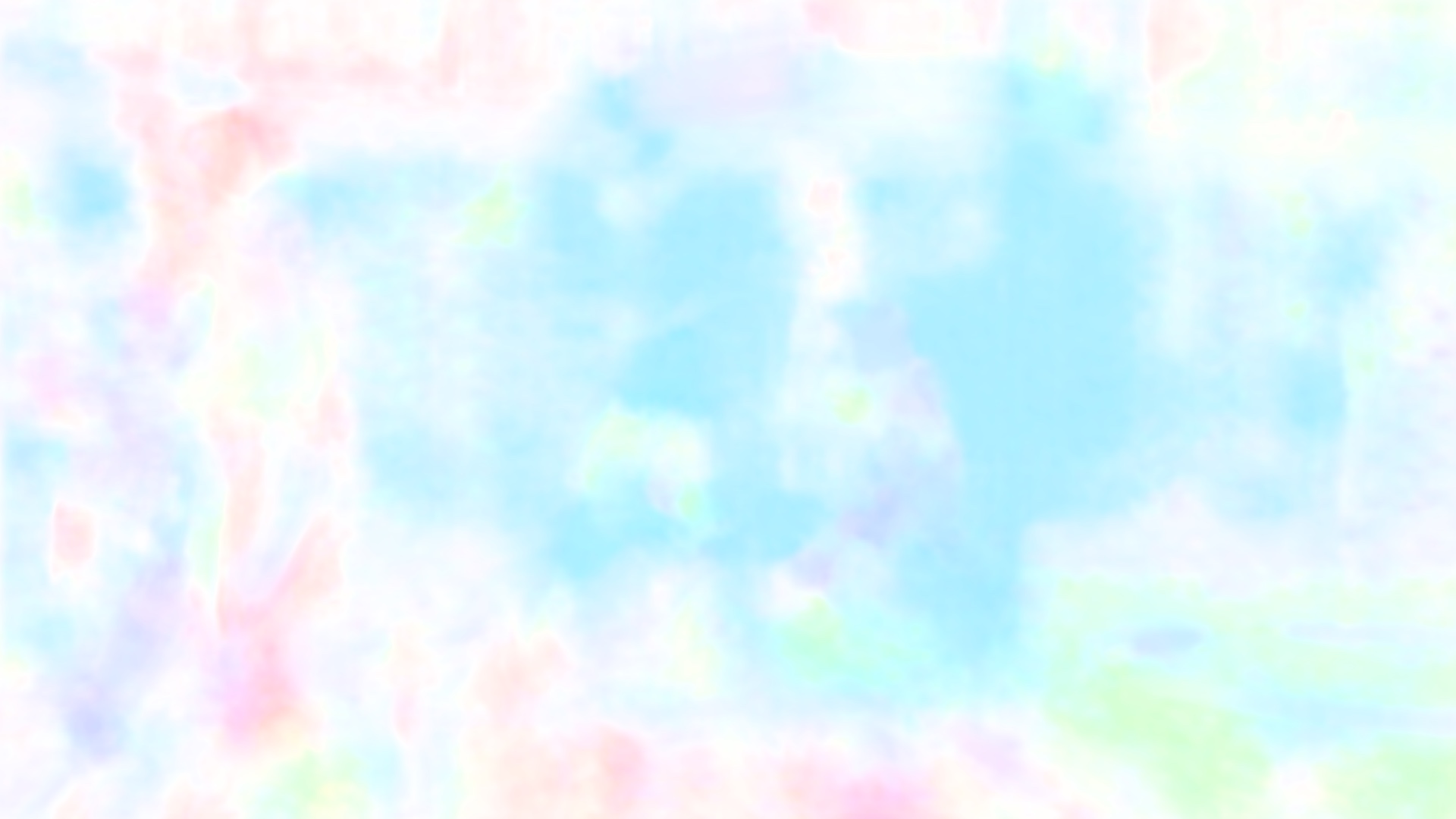} &
\includegraphics[width=0.115\textwidth]{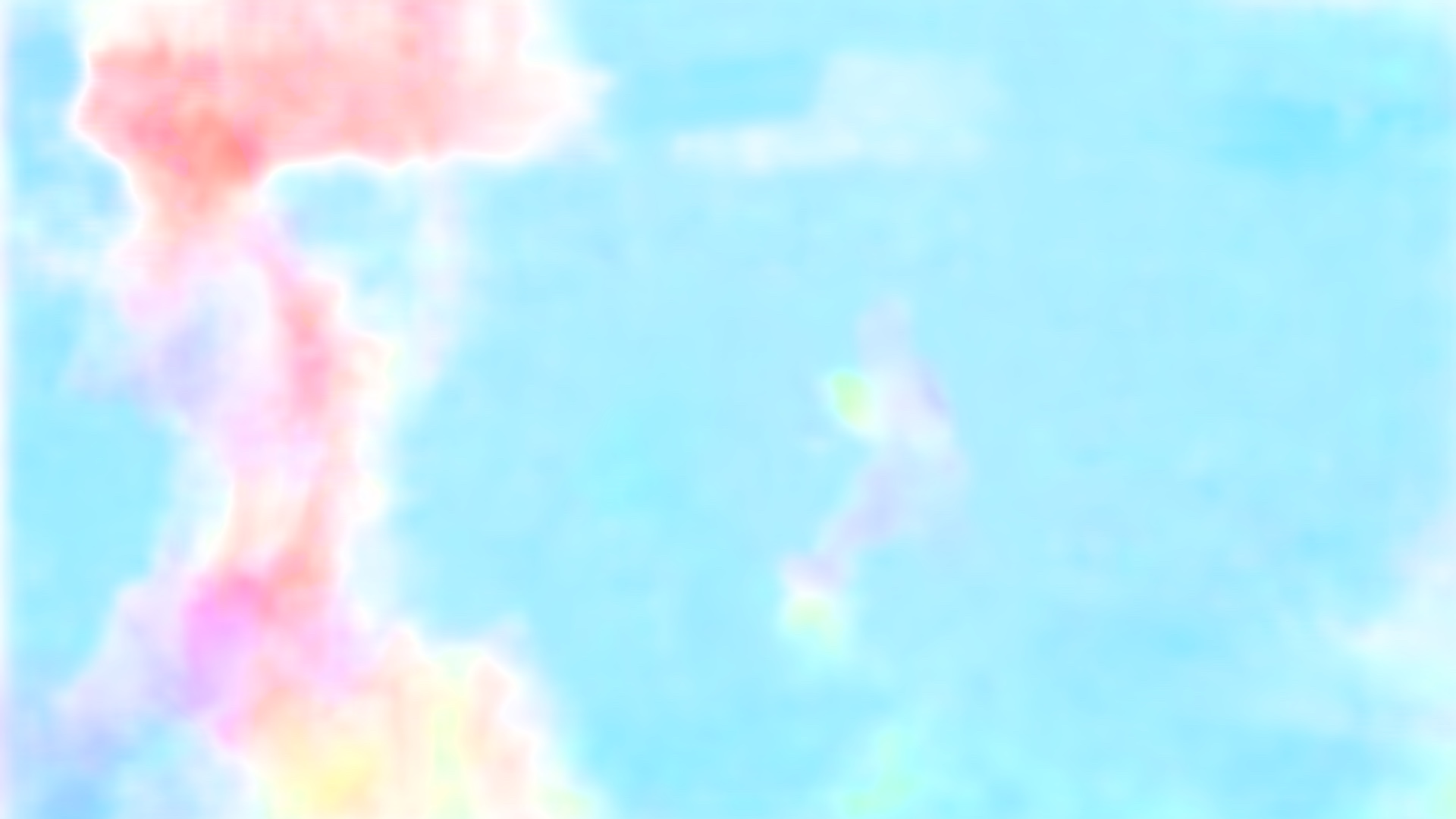} &
\includegraphics[width=0.115\textwidth]{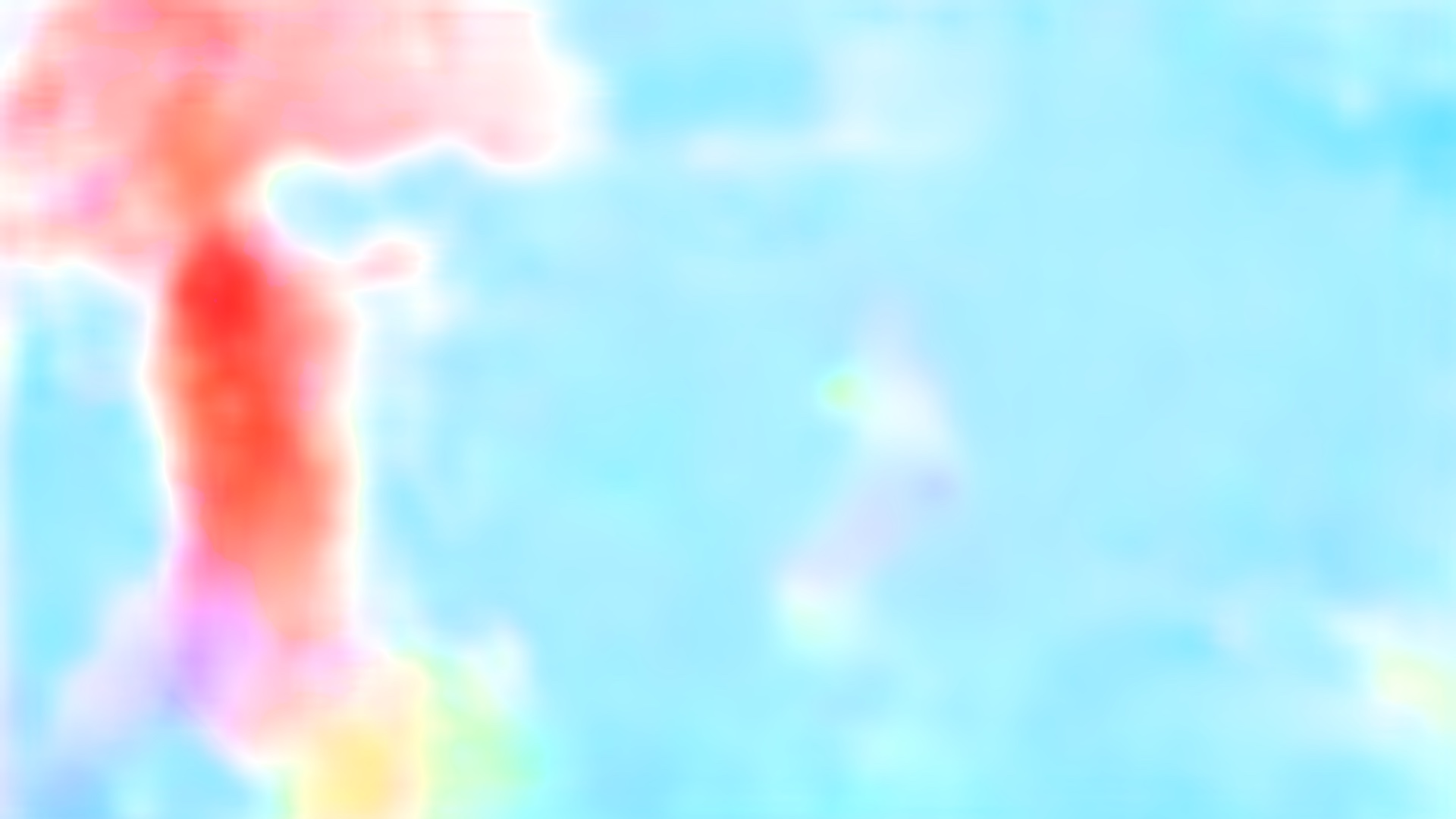} &
\includegraphics[width=0.115\textwidth]{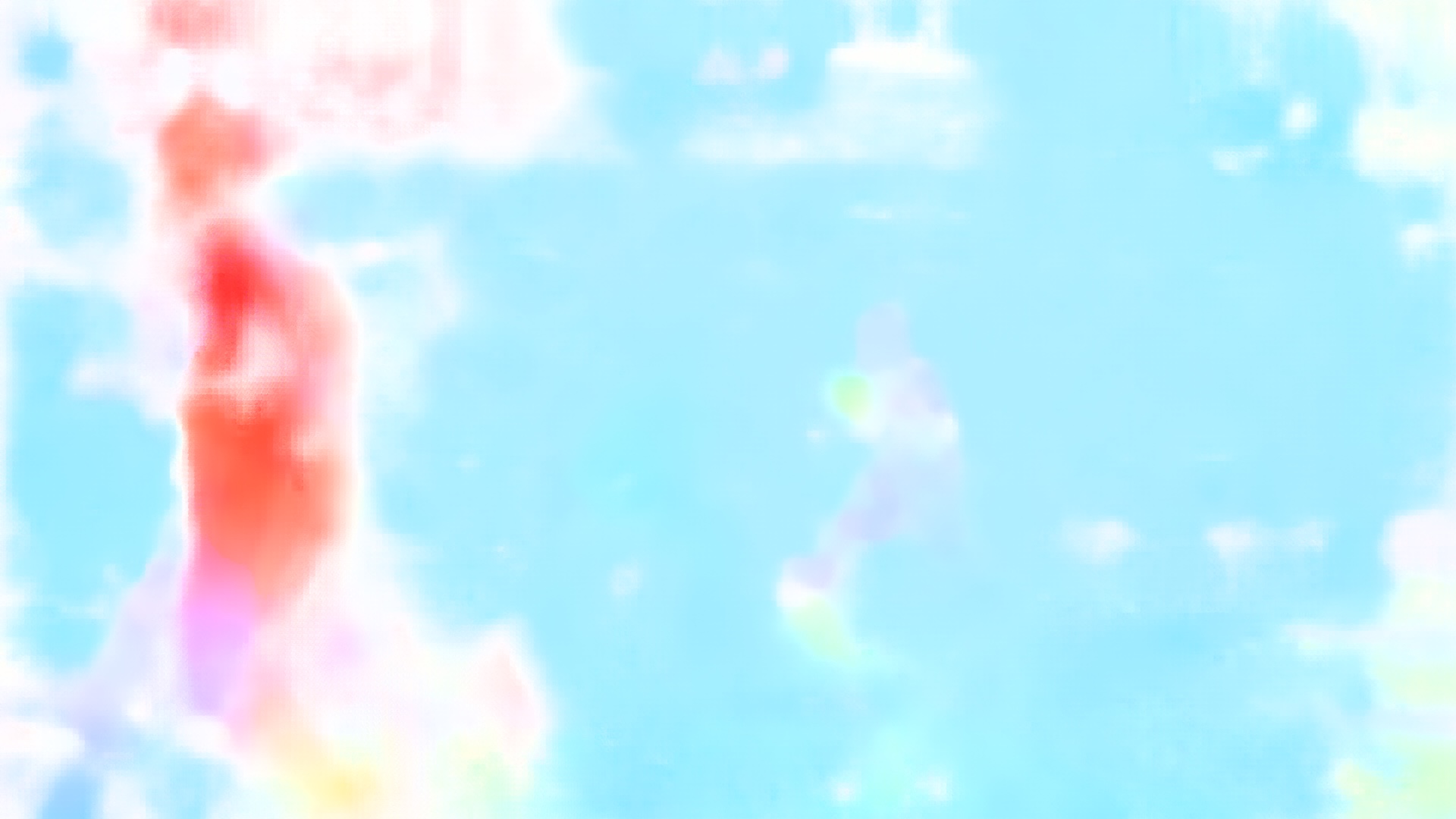} \\

\textbf{Cur.} $\boldsymbol{x}_2$ &
$\hat{\boldsymbol{x}}_{0\rightarrow2}^{(1)}\ 19.04\,\mathrm{dB}$ &
$\hat{\boldsymbol{x}}_{0\rightarrow2}^{(2)}\ 20.42\,\mathrm{dB}$ &
$\hat{\boldsymbol{x}}_{0\rightarrow2}^{(4)}\ 21.87\,\mathrm{dB}$ &
$\hat{\boldsymbol{x}}_{0\rightarrow2}^{(8)}\ 23.06\,\mathrm{dB}$ &
$\hat{\boldsymbol{x}}_{0\rightarrow2}^{d}\ 23.27\,\mathrm{dB}$ \\

\includegraphics[width=0.115\textwidth]{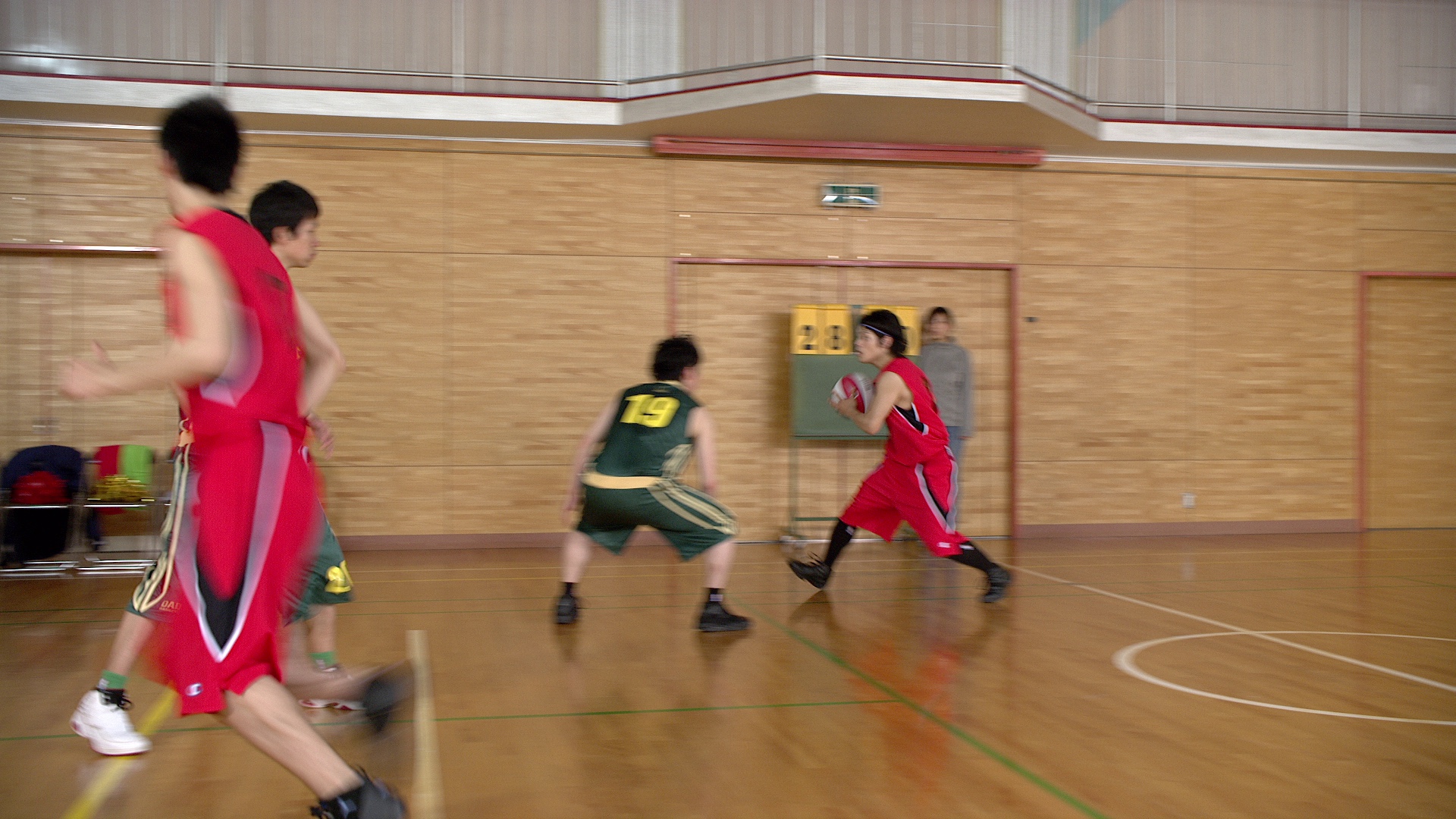} &
\includegraphics[width=0.115\textwidth]{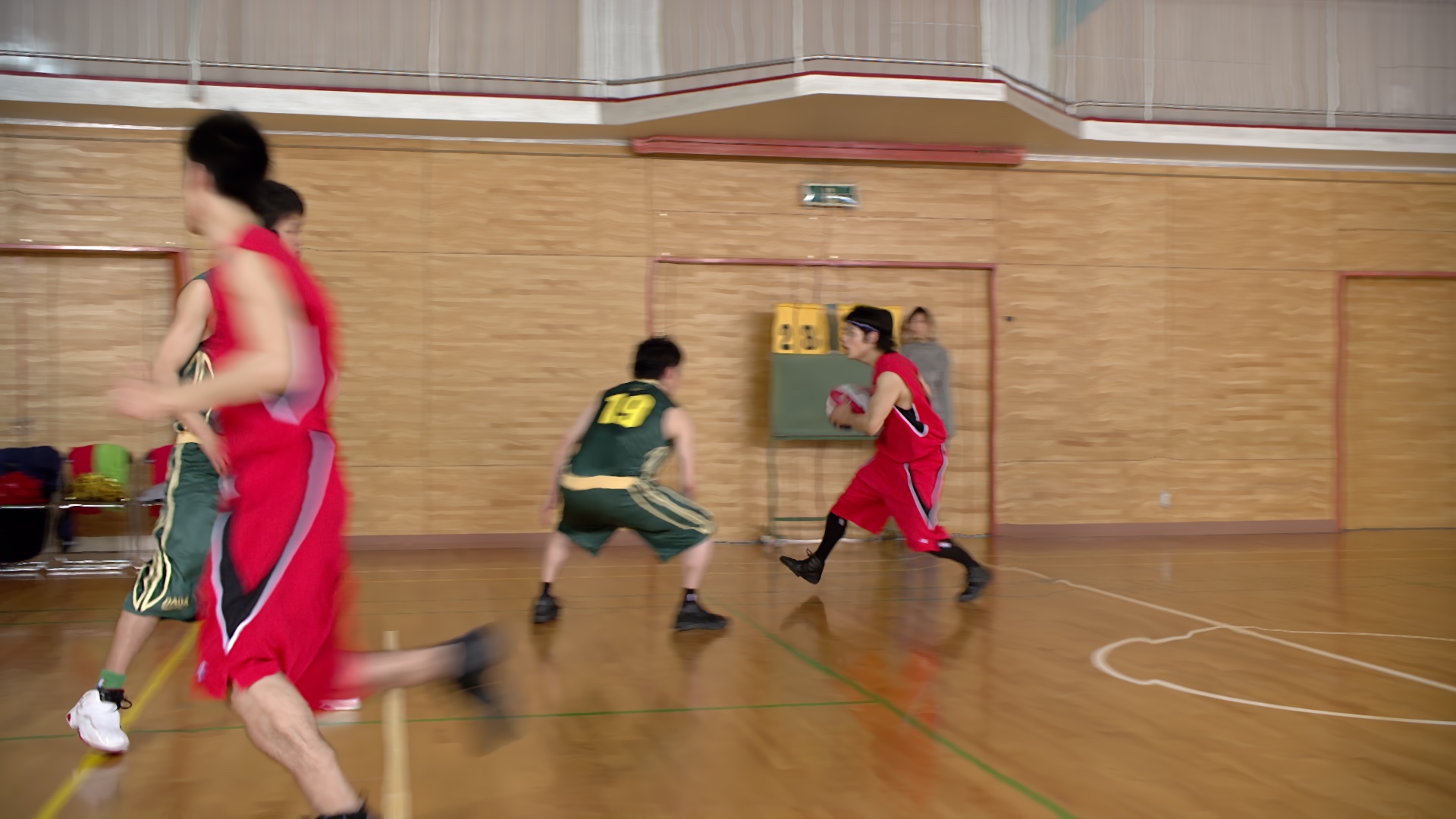} &
\includegraphics[width=0.115\textwidth]{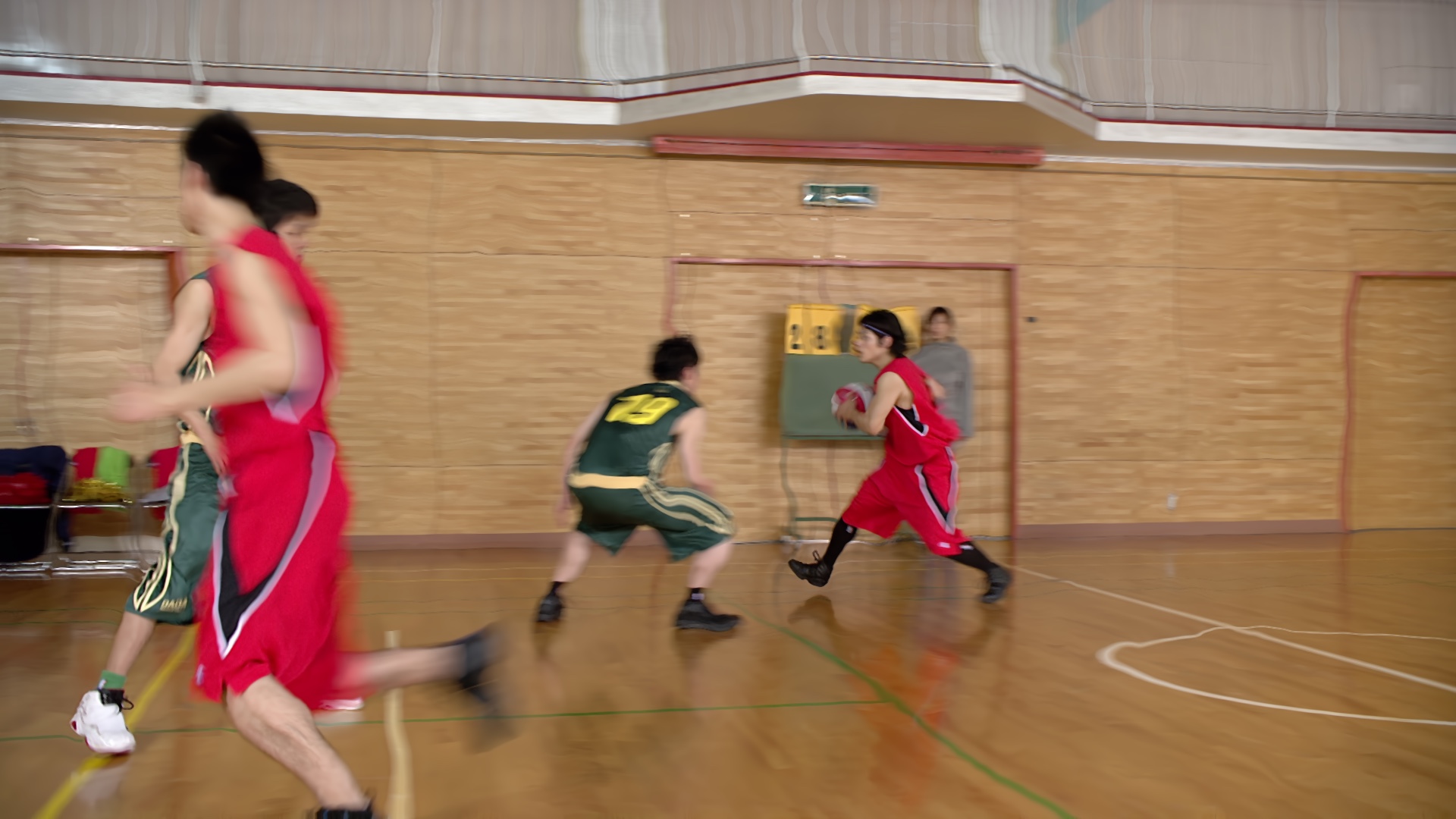} &
\includegraphics[width=0.115\textwidth]{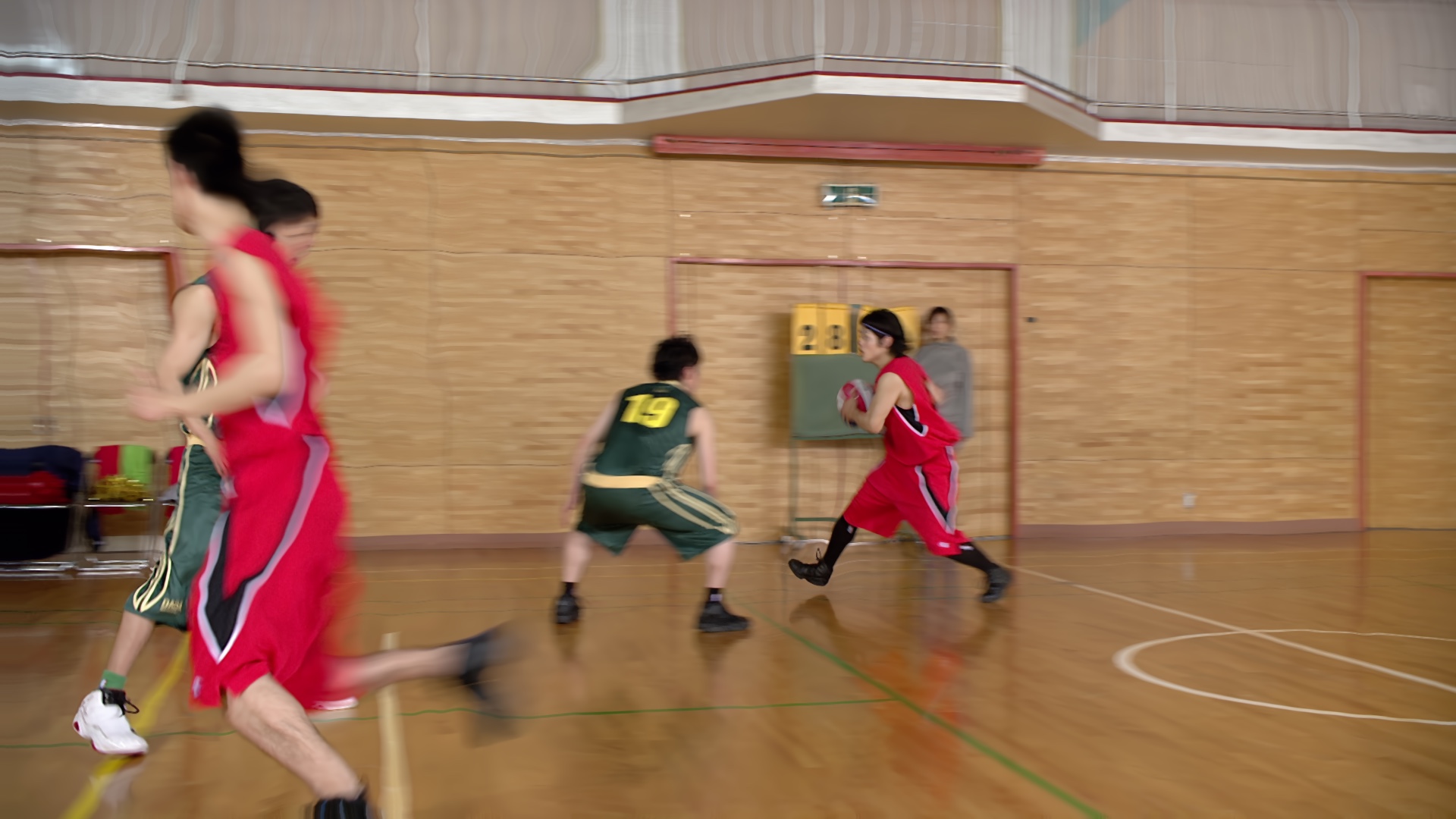} &
\includegraphics[width=0.115\textwidth]{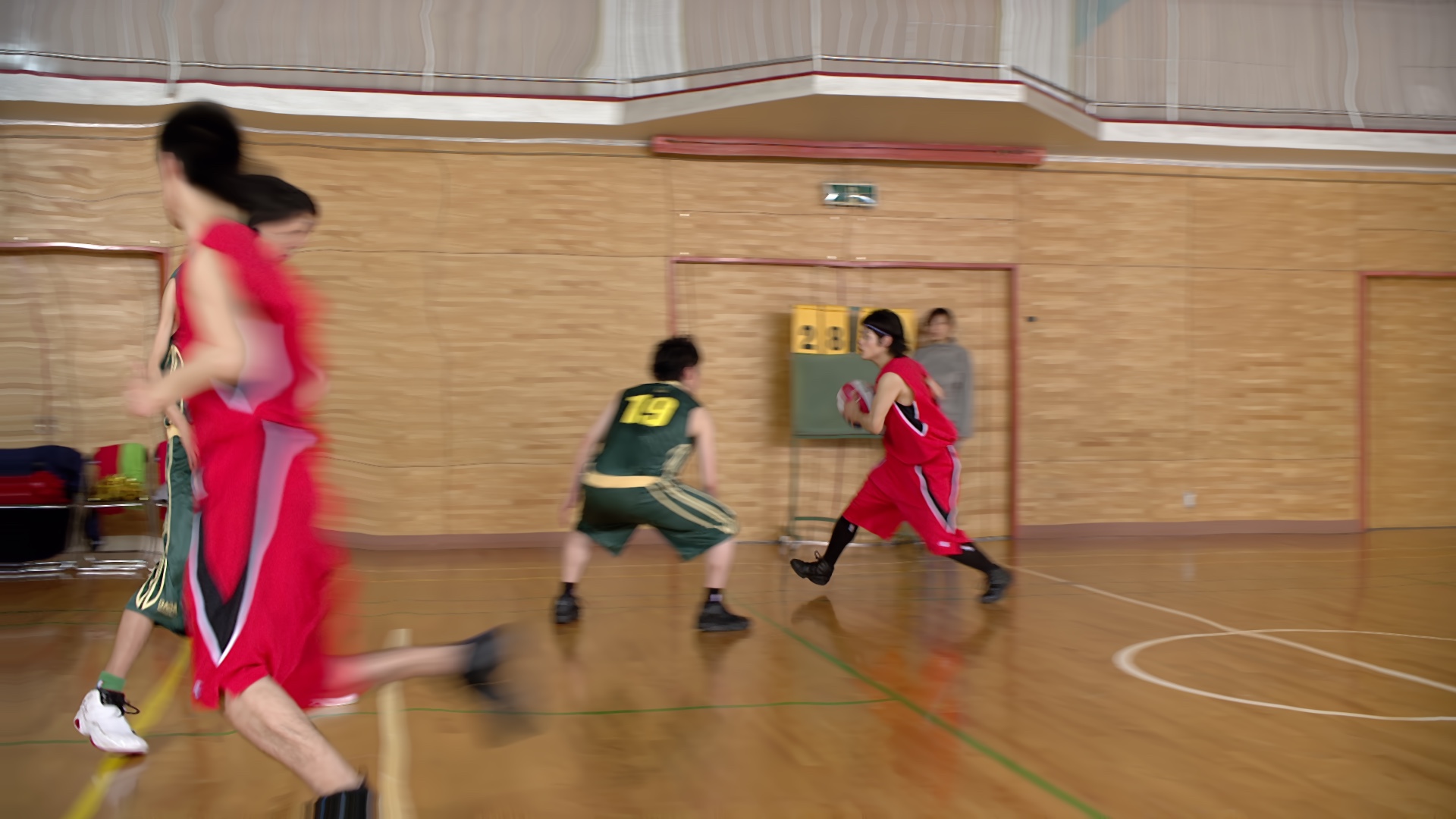} &
\includegraphics[width=0.115\textwidth]{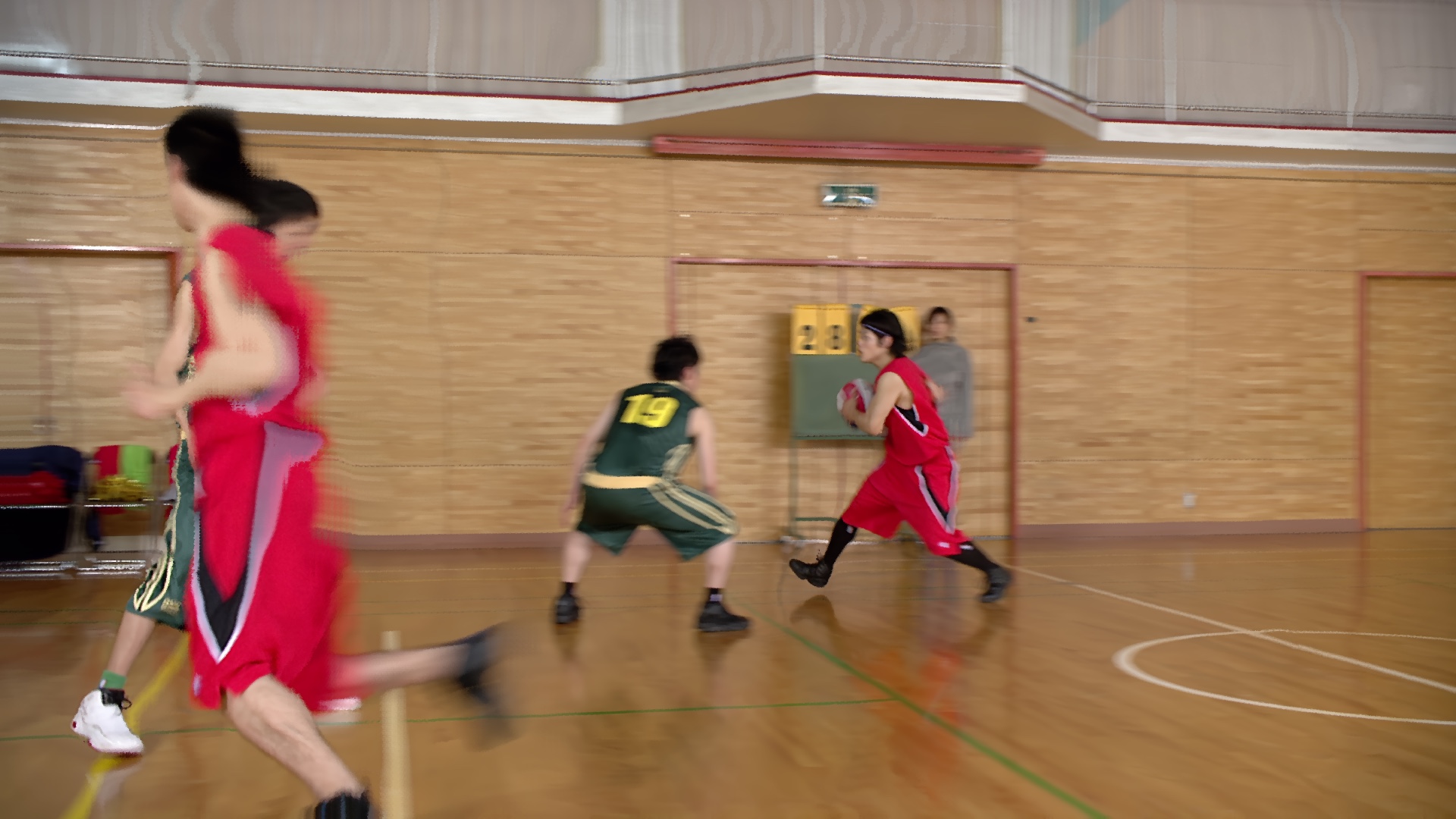} \\

 &
$e_{0\rightarrow2}^{(1)}$ &
$e_{0\rightarrow2}^{(2)}$ &
$e_{0\rightarrow2}^{(4)}$ &
$e_{0\rightarrow2}^{(8)}$ &
$e_{0\rightarrow2}^{d}$ \\

 &
\includegraphics[width=0.115\textwidth]{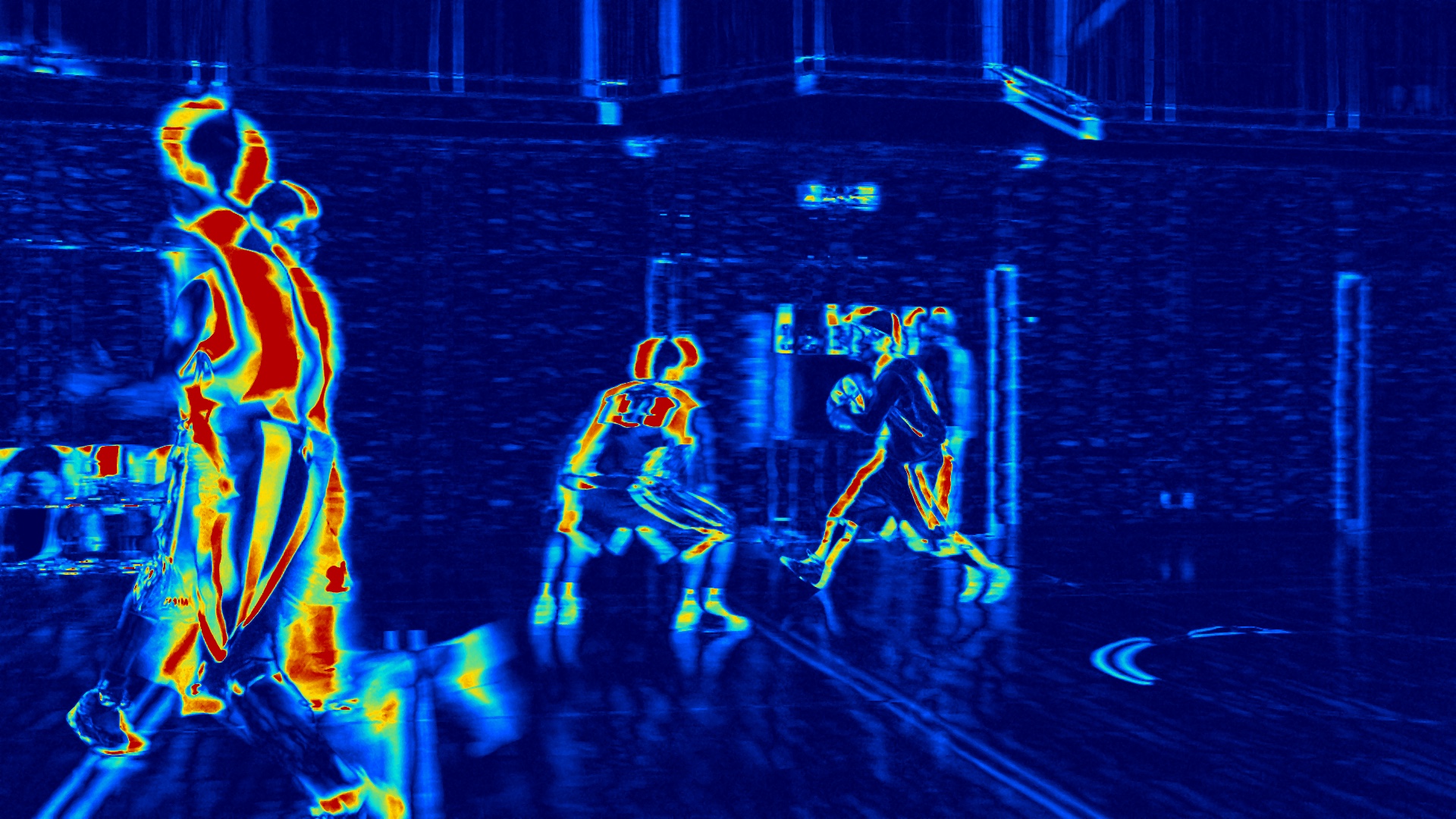} &
\includegraphics[width=0.115\textwidth]{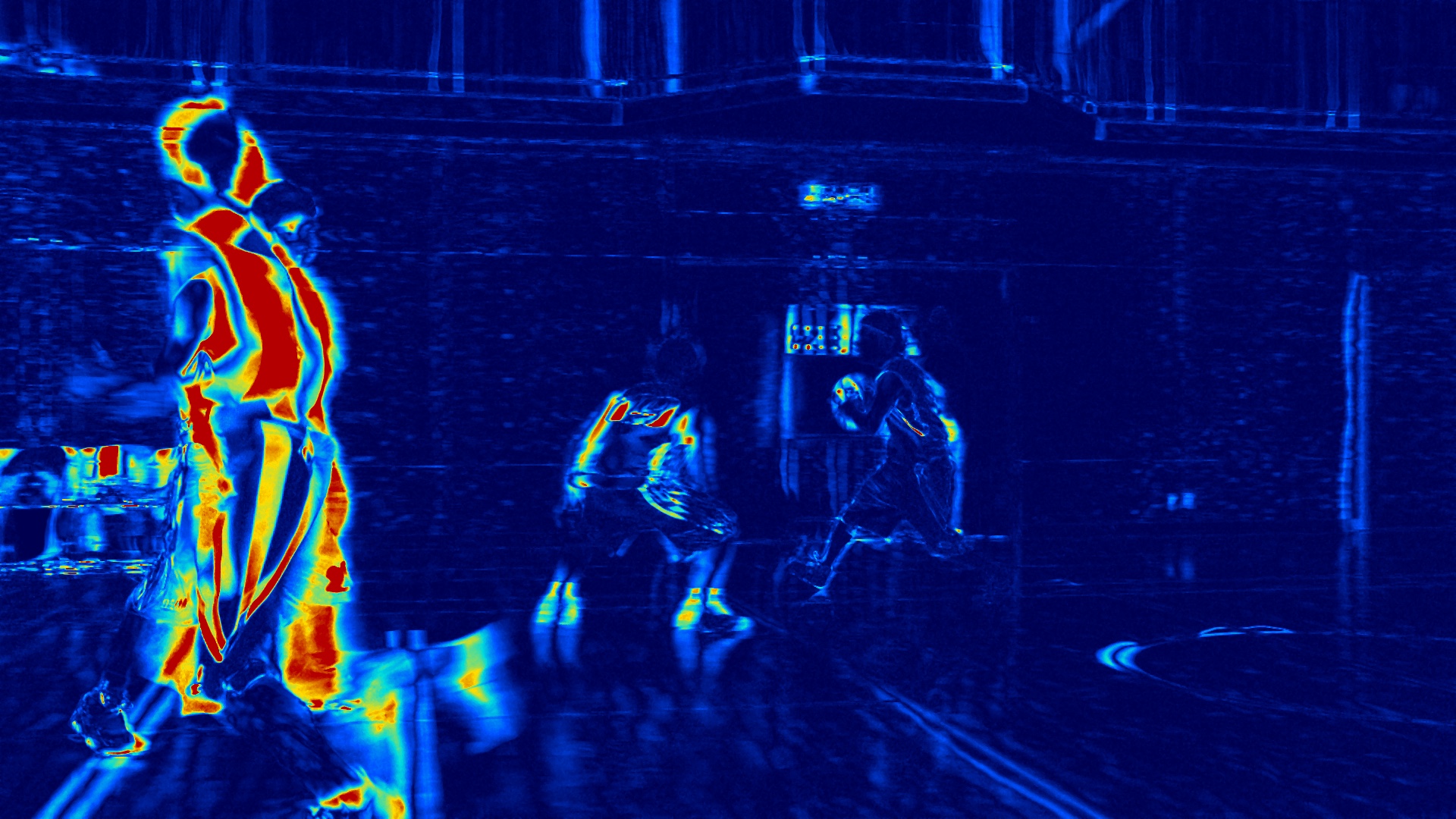} &
\includegraphics[width=0.115\textwidth]{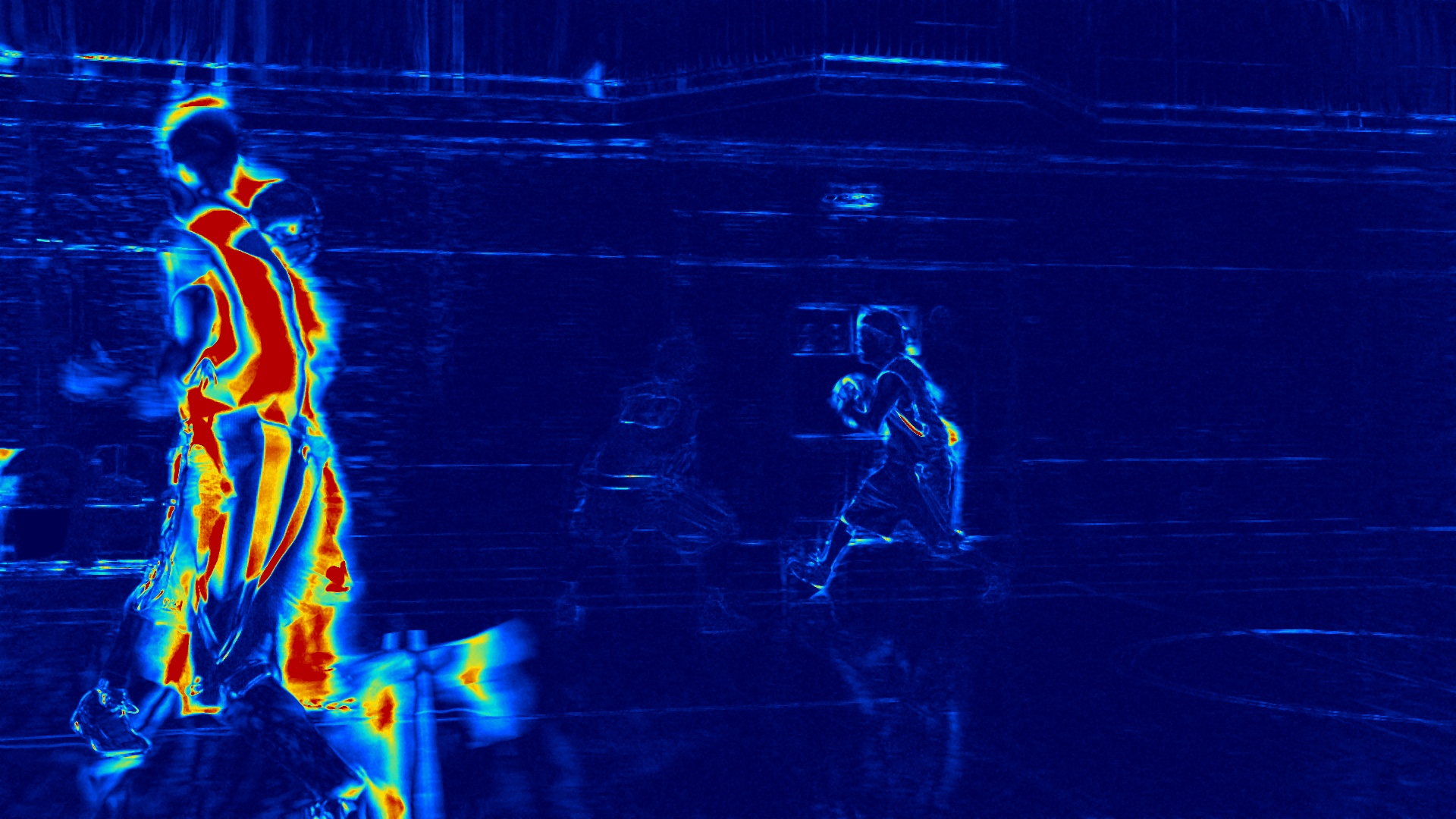} &
\includegraphics[width=0.115\textwidth]{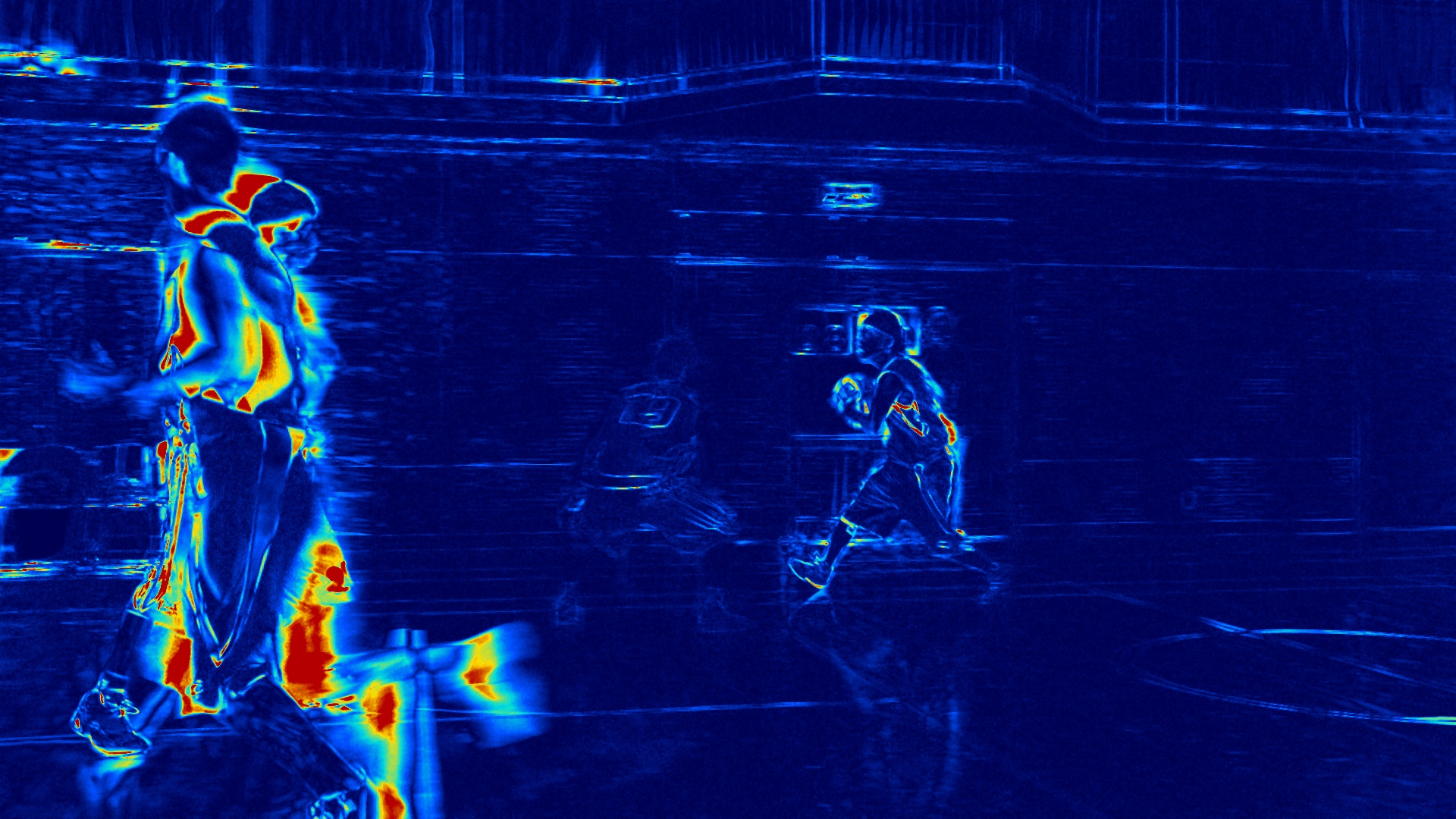} &
\includegraphics[width=0.115\textwidth]{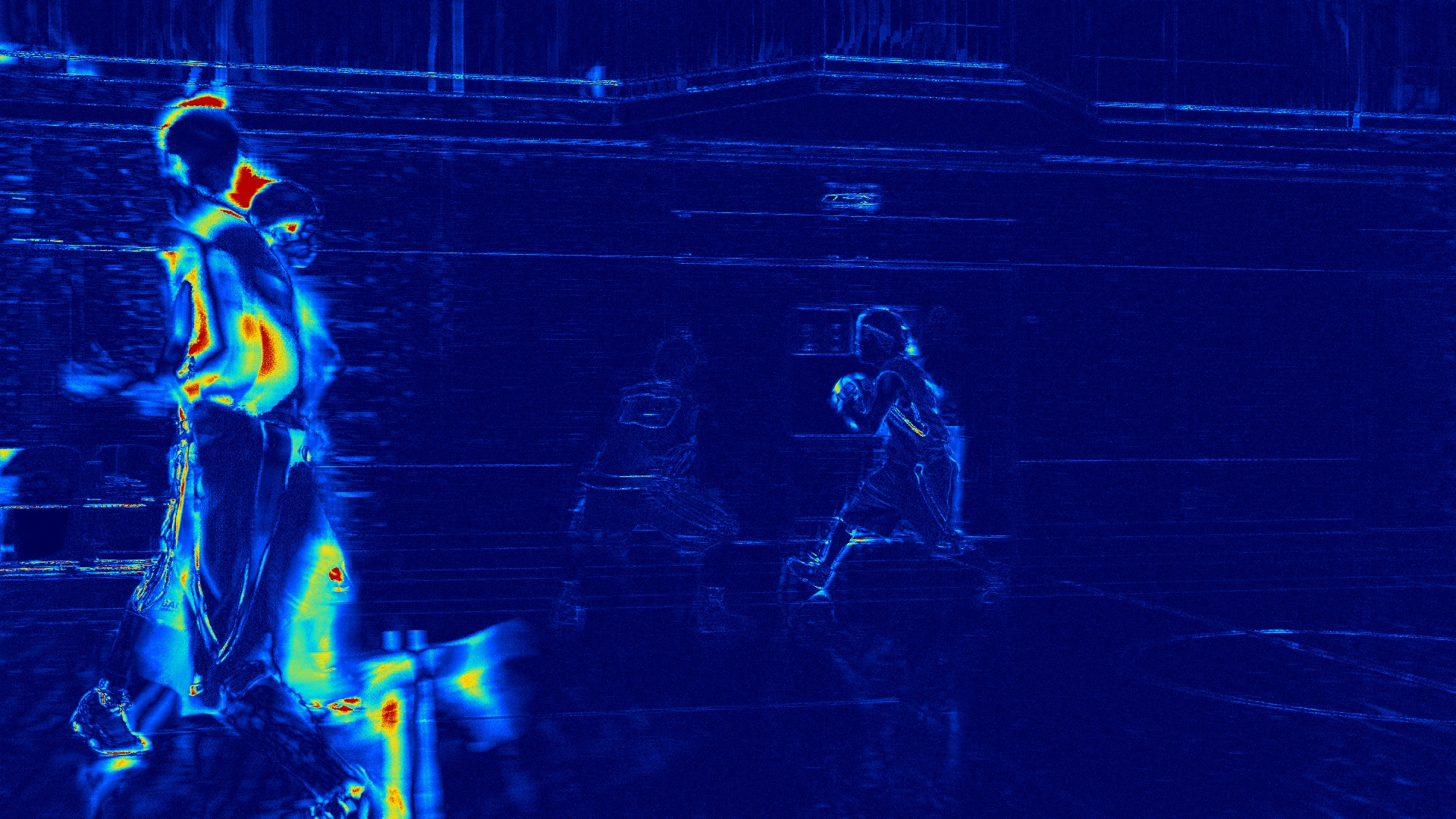} \\
\end{tabular}
\caption{Visualization of multi-scale motion estimation for forward prediction from POC 0 to POC 1 and POC 2 on BasketballDrive. The PSNR values are computed between each warped frame and the target frame, rather than the reference frame.}
\label{fig:basketballdrive_msme_vis}
\end{figure*}

Different scales are reliable for different motion patterns.
Fine-scale candidates preserve local details and are usually more suitable for small motions, while coarse-scale candidates can better handle large displacements.
A single global selection among scales is therefore suboptimal, because the best scale may vary across spatial locations.
We instead perform pixel-wise candidate fusion according to the alignment quality of each candidate.
This motivation is also visualized in Fig.~\ref{fig:basketballdrive_msme_vis}, where different scales show different warping quality and their aggregation produces a more accurate motion field.
\par

\begin{figure*}[t]
\centering
\setlength{\tabcolsep}{2pt}
\renewcommand{\arraystretch}{1.5}
\scriptsize
\begin{tabular}{c !{\vrule width 0.5pt} c c c c c !{\vrule width 0.5pt} c !{\vrule width 0.5pt} c}
\multicolumn{1}{c}{\textbf{Input Frames}} &
\multicolumn{5}{c !{\vrule width 0.5pt}}{\textbf{Multi-Scale Motion Estimation}} &
\multicolumn{1}{c !{\vrule width 0.5pt}}{\parbox[c][2.8em][c]{0.13\textwidth}{\centering\textbf{Parallel Accumulated}\\\textbf{Motion Estimation}}} &
\multicolumn{1}{c}{\parbox[c][2.8em][c]{0.115\textwidth}{\centering\textbf{Output Results}}} \\

\textbf{Ref.} $\hat{\boldsymbol{x}}_0$ &
$\boldsymbol{v}_{8\rightarrow0}^{(1)}$ &
$\boldsymbol{v}_{8\rightarrow0}^{(2)}$ &
$\boldsymbol{v}_{8\rightarrow0}^{(4)}$ &
$\boldsymbol{v}_{8\rightarrow0}^{(8)}$ &
$\boldsymbol{v}_{8\rightarrow0}^{d}$ &
$\boldsymbol{v}_{8\rightarrow0}^{a}$ &
$\boldsymbol{v}_{8\rightarrow0}$ \\

\includegraphics[width=0.115\textwidth]{figures_jpg/motion_vis/pptx/BasketballDrive/frame/im00001.jpg} &
\includegraphics[width=0.115\textwidth]{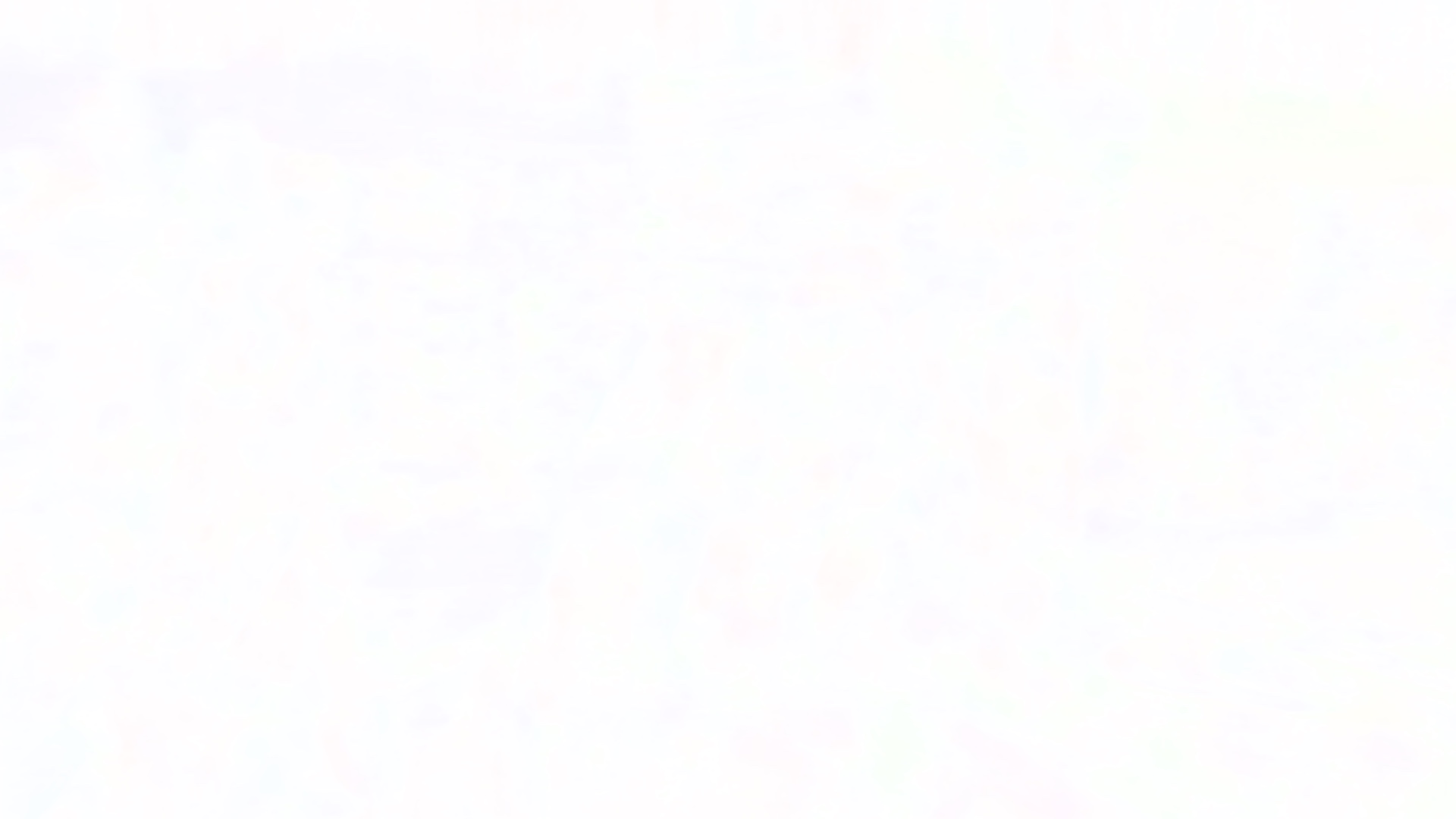} &
\includegraphics[width=0.115\textwidth]{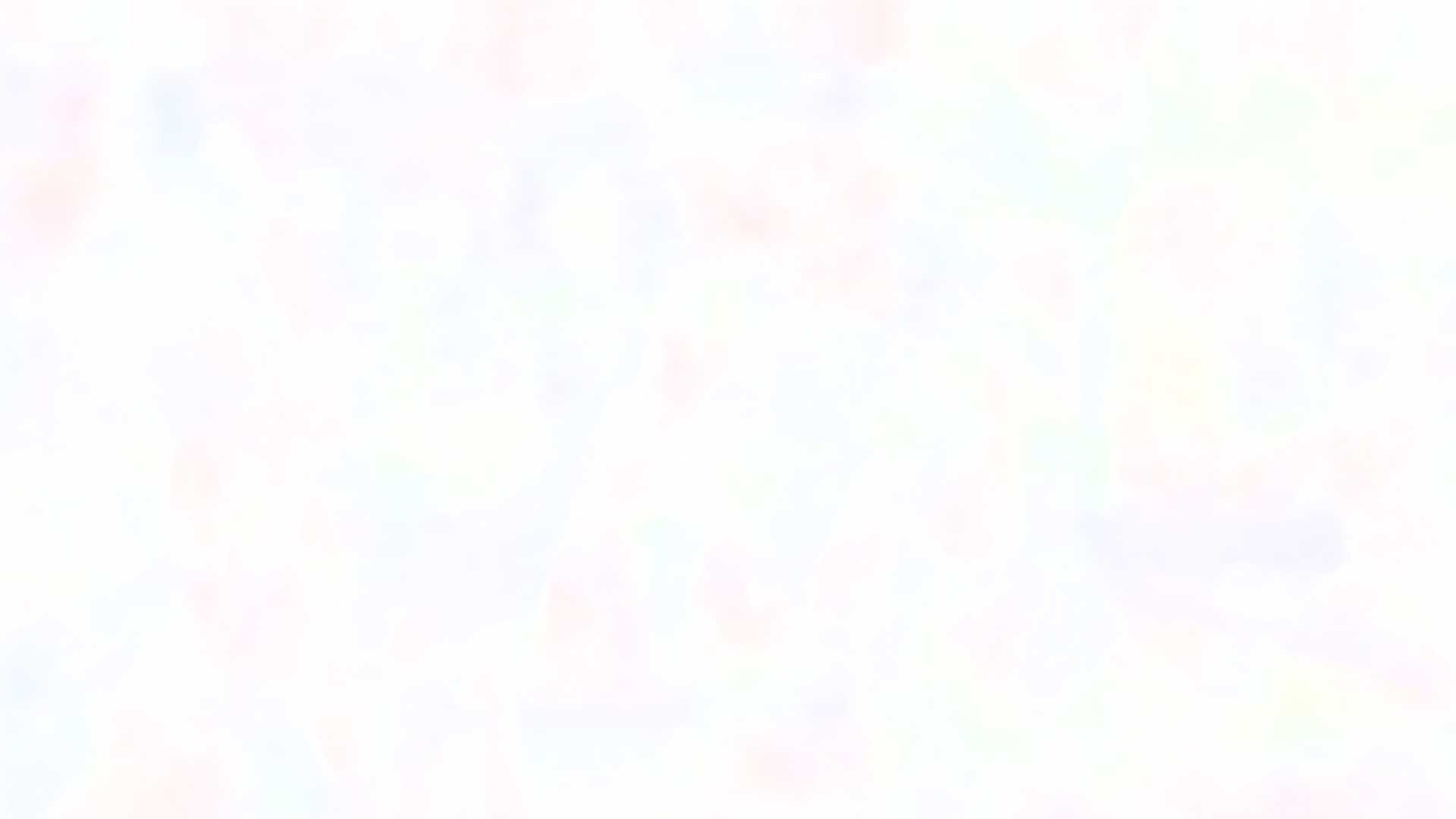} &
\includegraphics[width=0.115\textwidth]{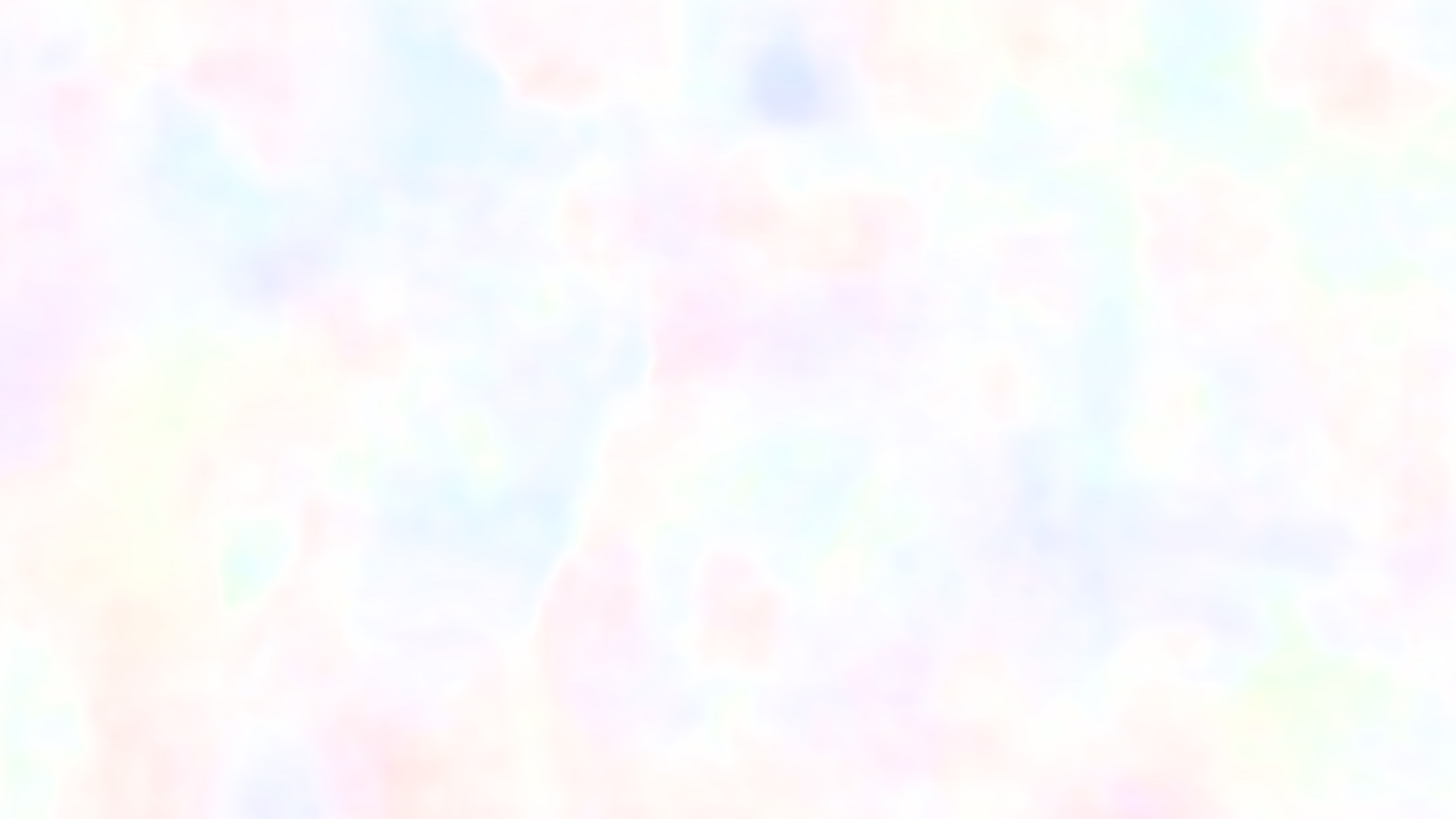} &
\includegraphics[width=0.115\textwidth]{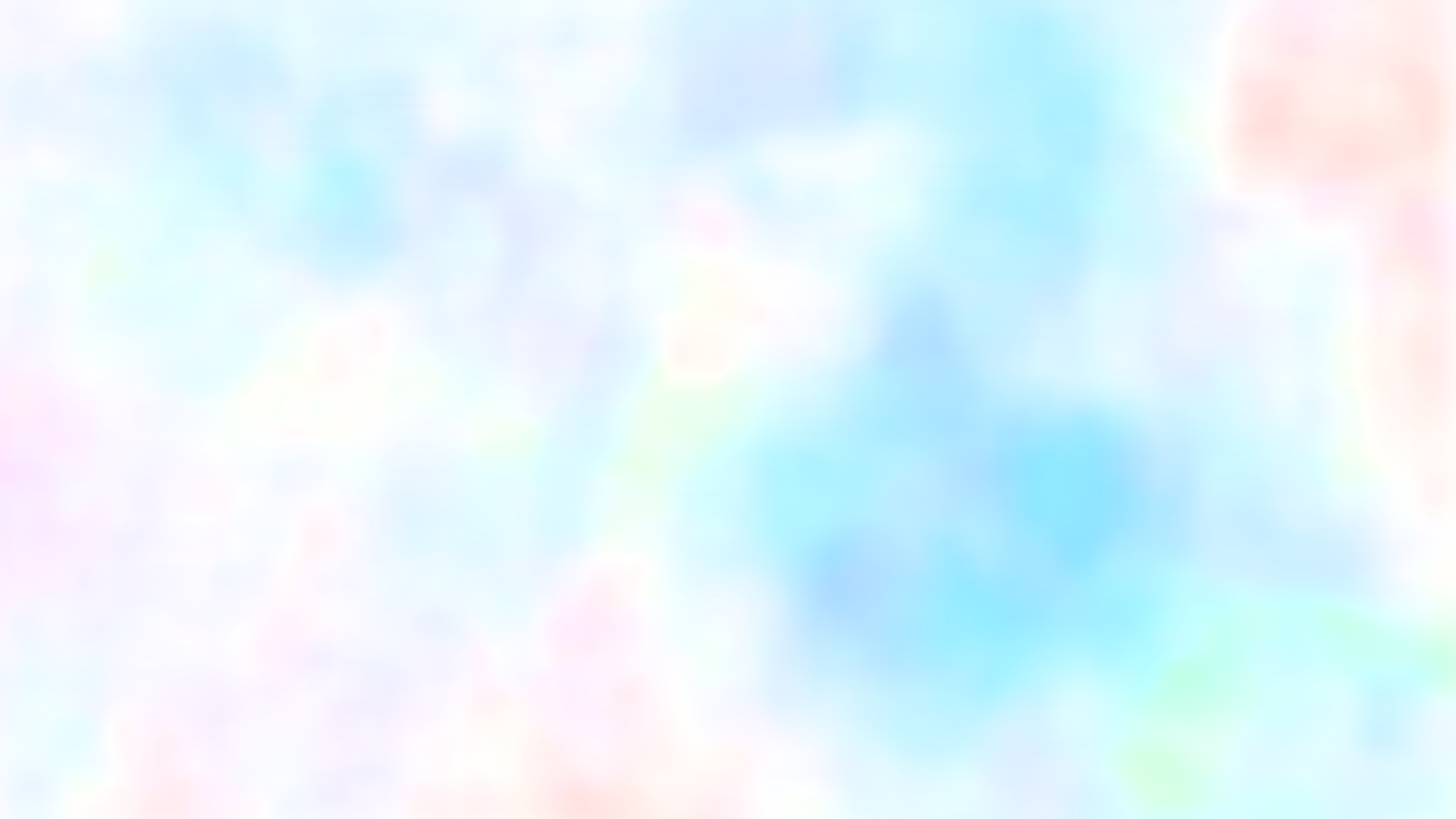} &
\includegraphics[width=0.115\textwidth]{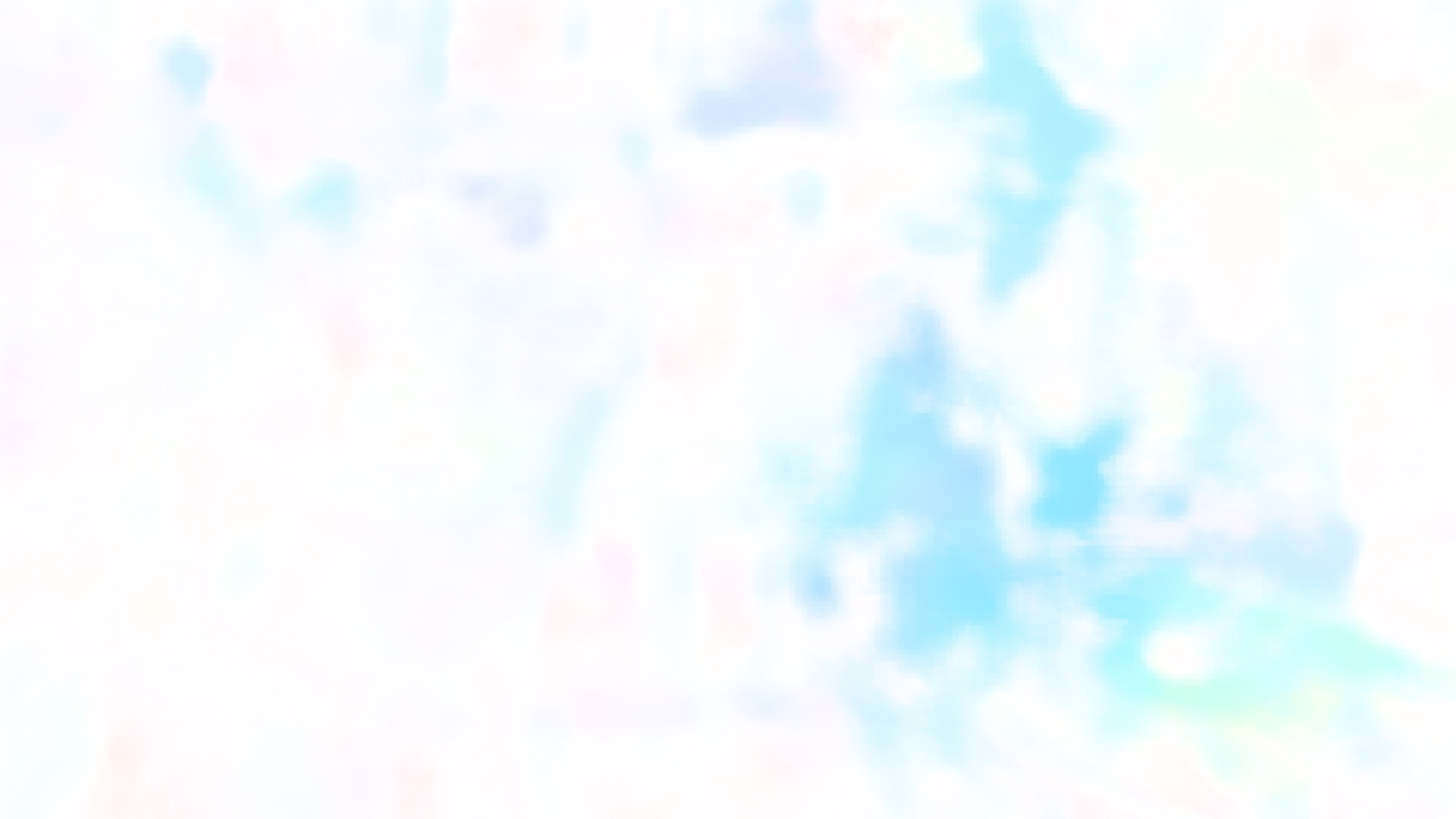} &
\includegraphics[width=0.115\textwidth]{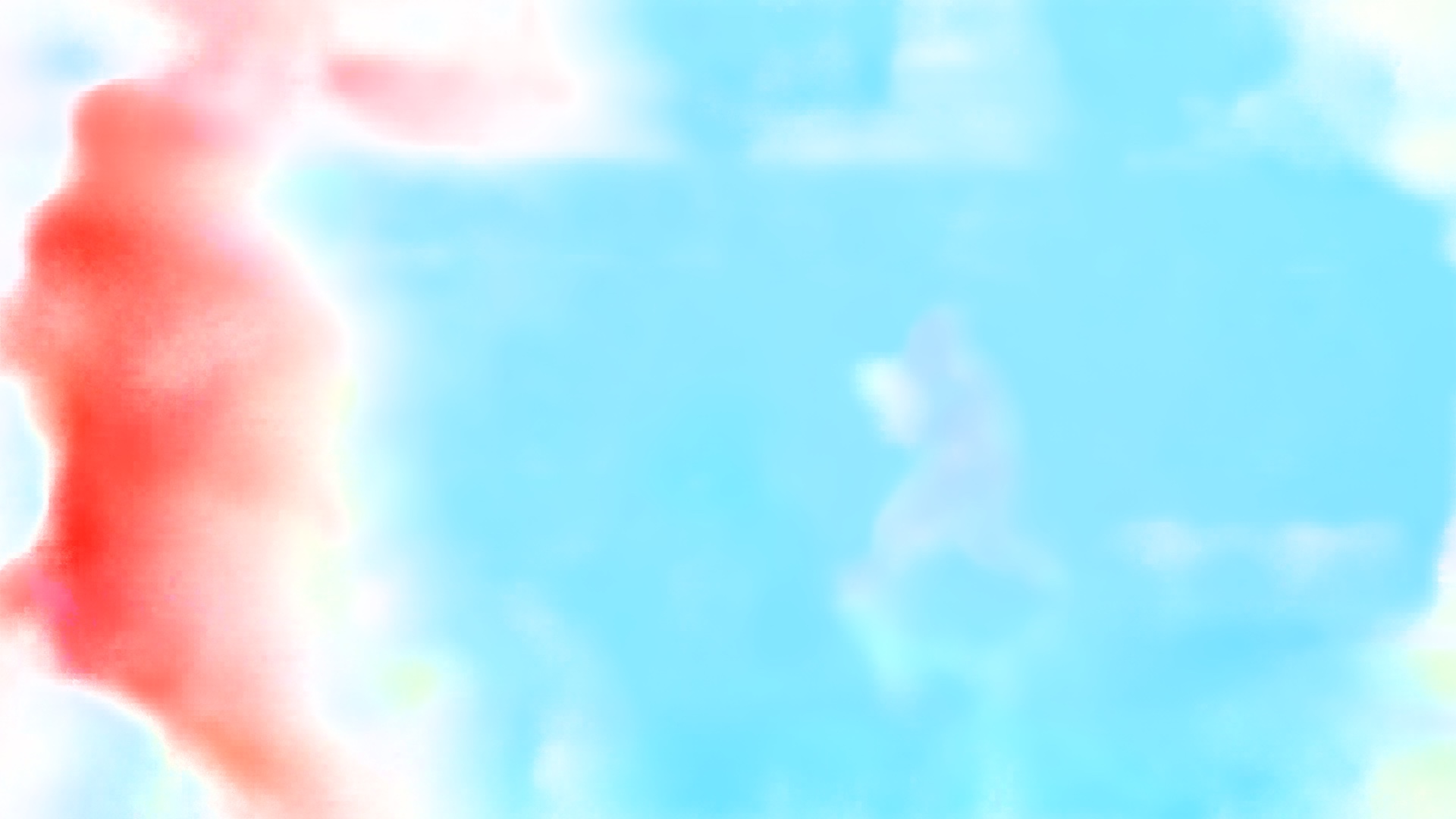} &
\includegraphics[width=0.115\textwidth]{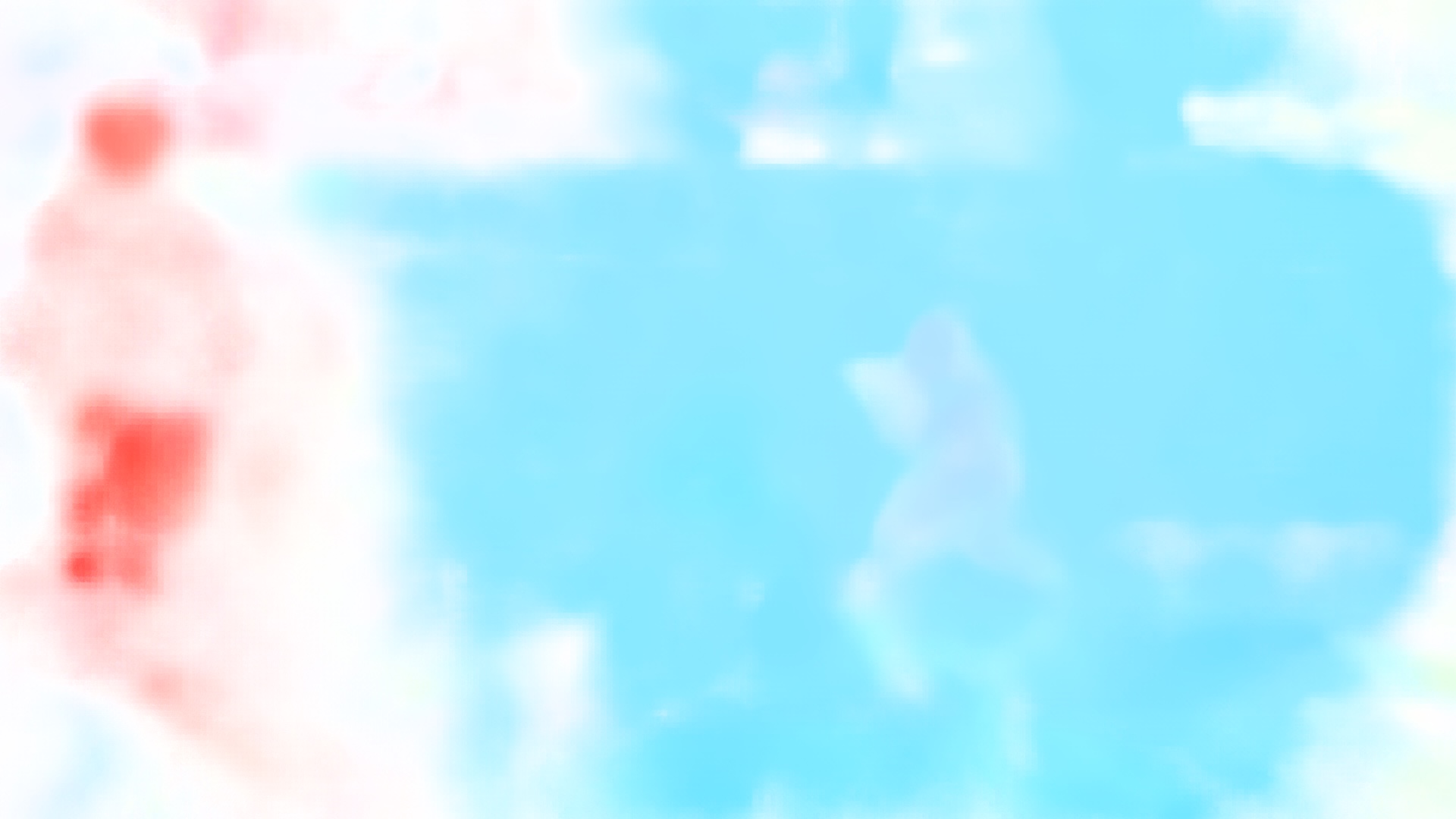} \\

\textbf{Cur.} $\boldsymbol{x}_{8}$ &
$\hat{\boldsymbol{x}}_{0\rightarrow8}^{(1)}\ 15.54\,\mathrm{dB}$ &
$\hat{\boldsymbol{x}}_{0\rightarrow8}^{(2)}\ 15.64\,\mathrm{dB}$ &
$\hat{\boldsymbol{x}}_{0\rightarrow8}^{(4)}\ 15.76\,\mathrm{dB}$ &
$\hat{\boldsymbol{x}}_{0\rightarrow8}^{(8)}\ 16.04\,\mathrm{dB}$ &
$\hat{\boldsymbol{x}}_{0\rightarrow8}^{d}\ 16.07\,\mathrm{dB}$ &
$\hat{\boldsymbol{x}}_{0\rightarrow8}^{a}\ 17.34\,\mathrm{dB}$ &
$\hat{\boldsymbol{x}}_{0\rightarrow8}\ 17.54\,\mathrm{dB}$ \\

\includegraphics[width=0.115\textwidth]{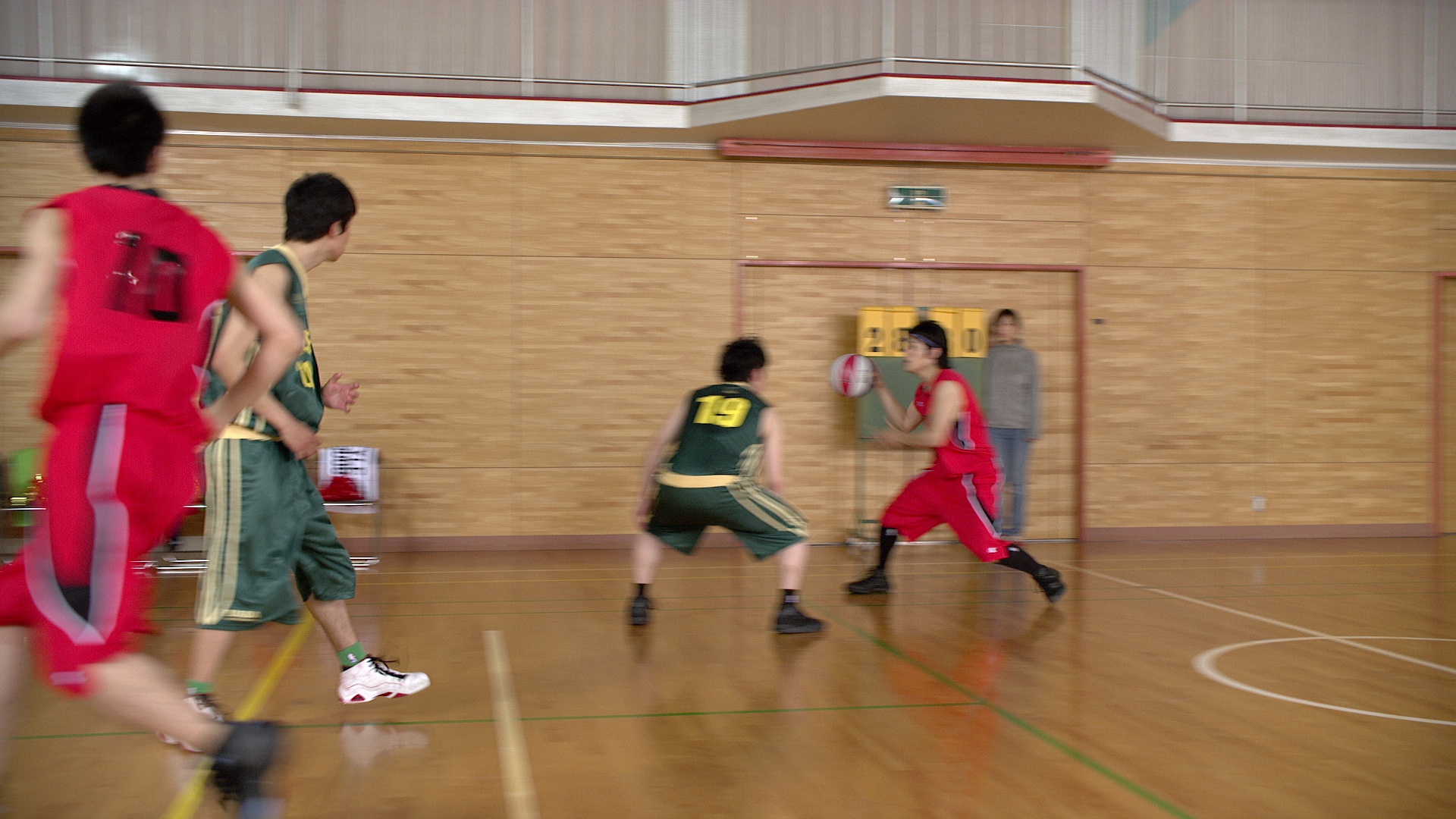} &
\includegraphics[width=0.115\textwidth]{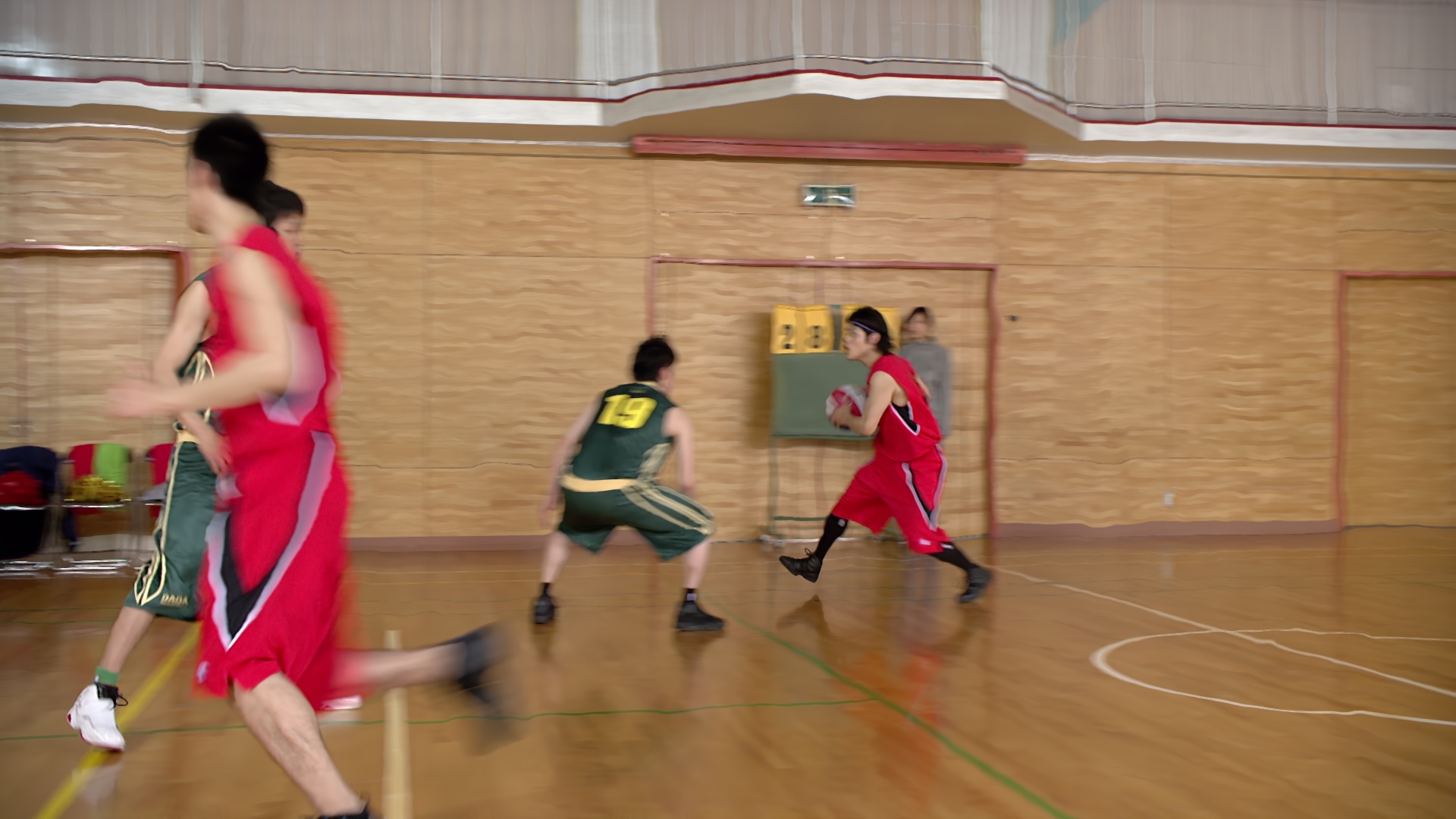} &
\includegraphics[width=0.115\textwidth]{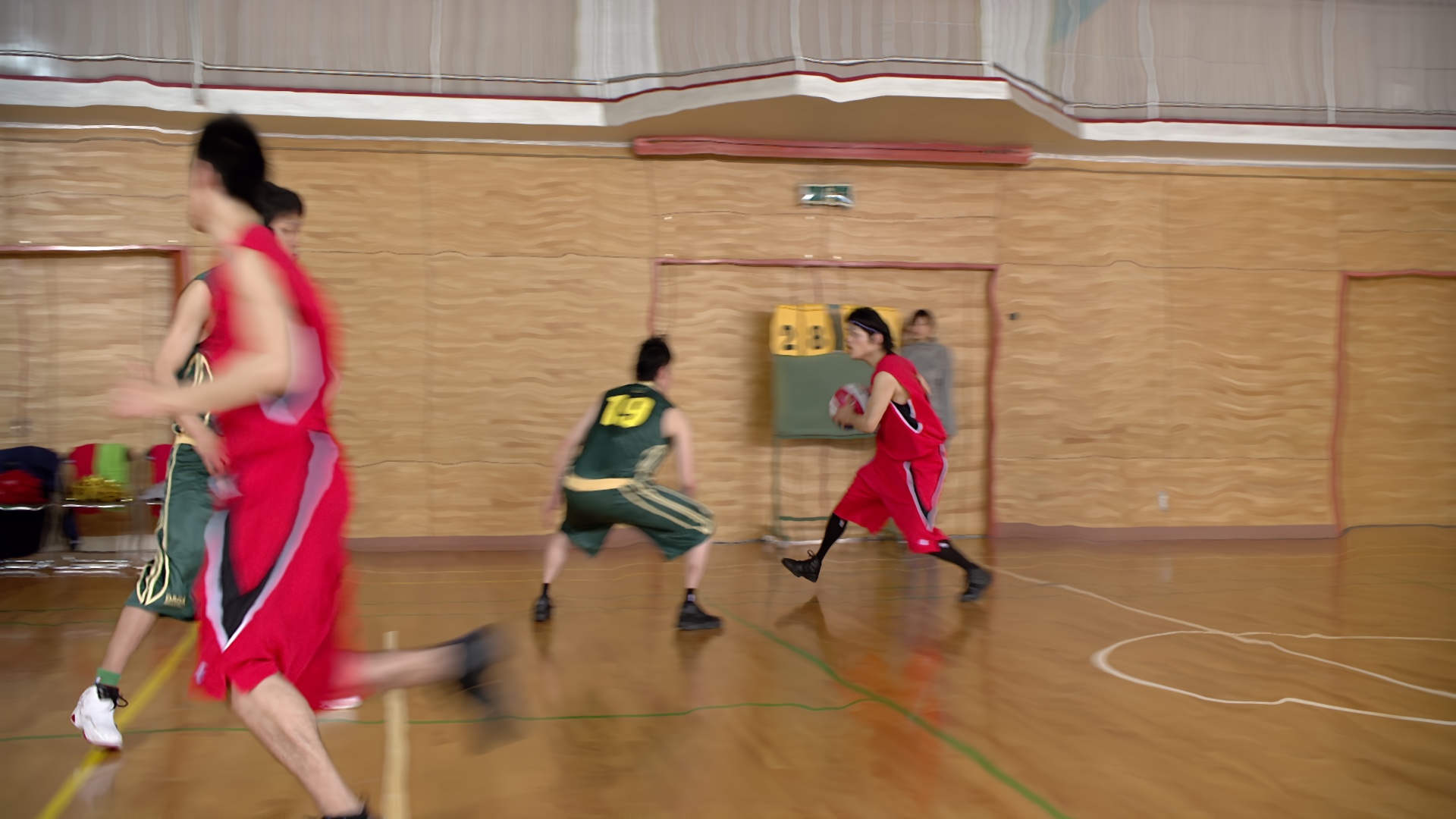} &
\includegraphics[width=0.115\textwidth]{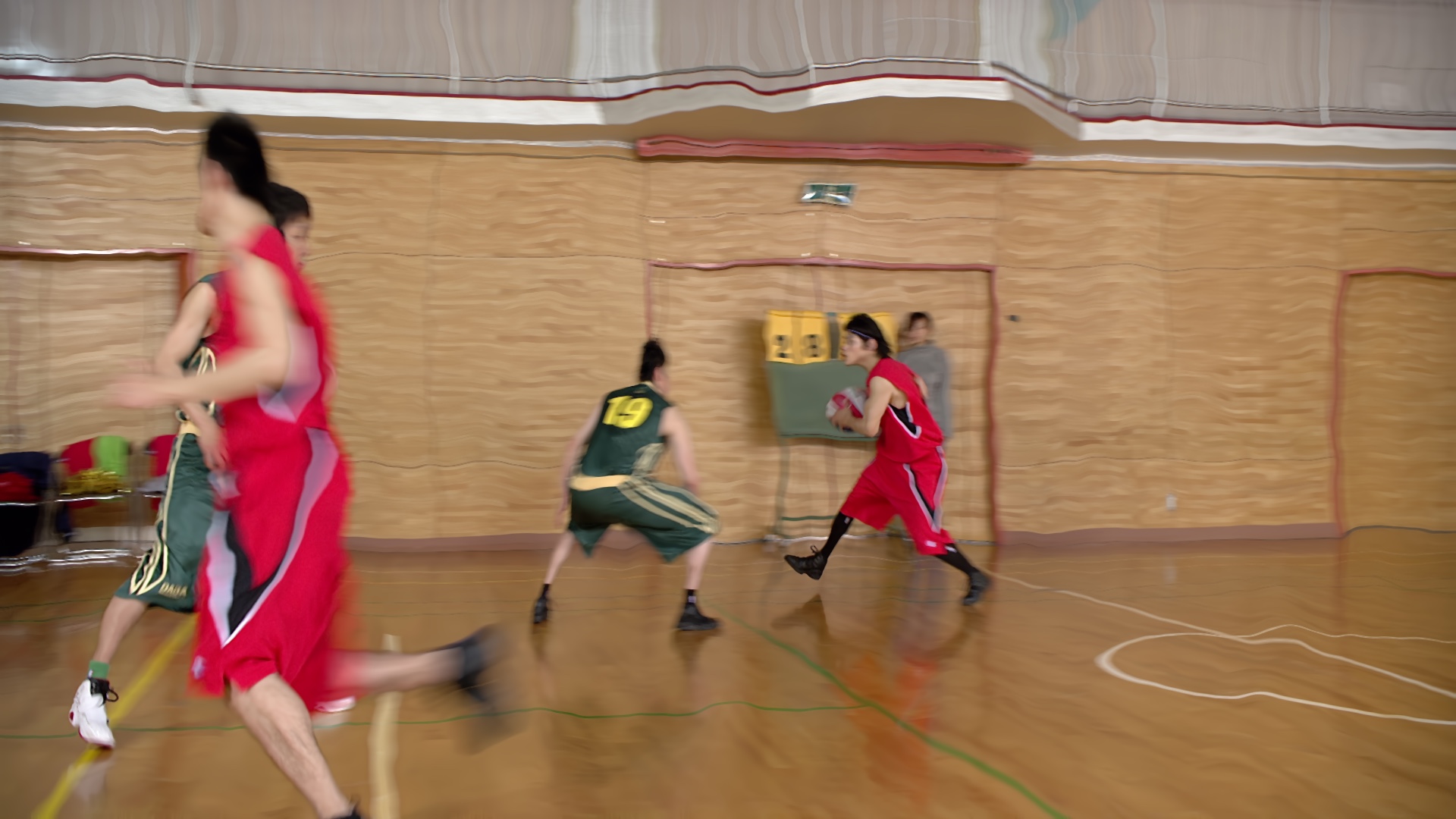} &
\includegraphics[width=0.115\textwidth]{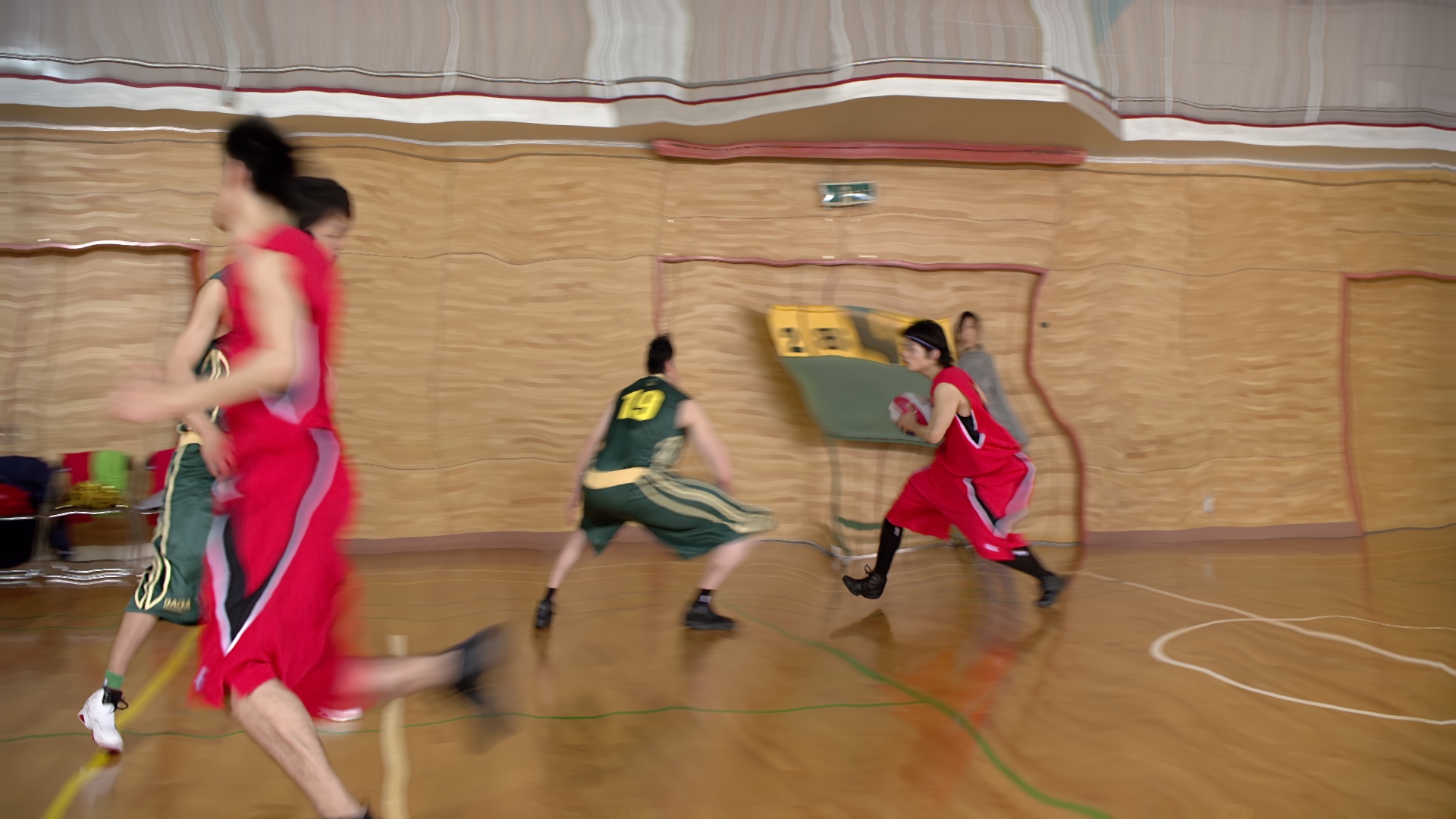} &
\includegraphics[width=0.115\textwidth]{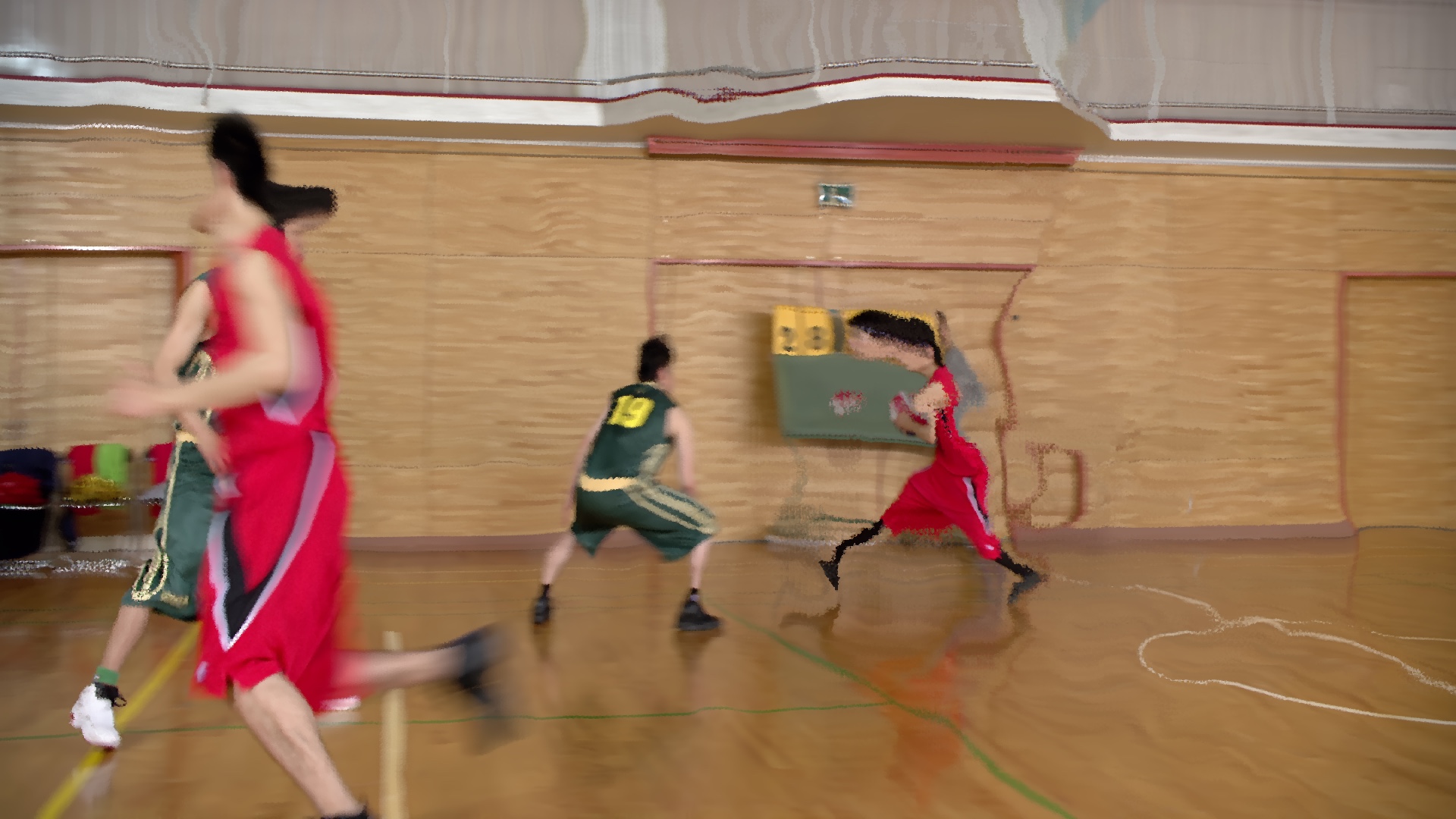} &
\includegraphics[width=0.115\textwidth]{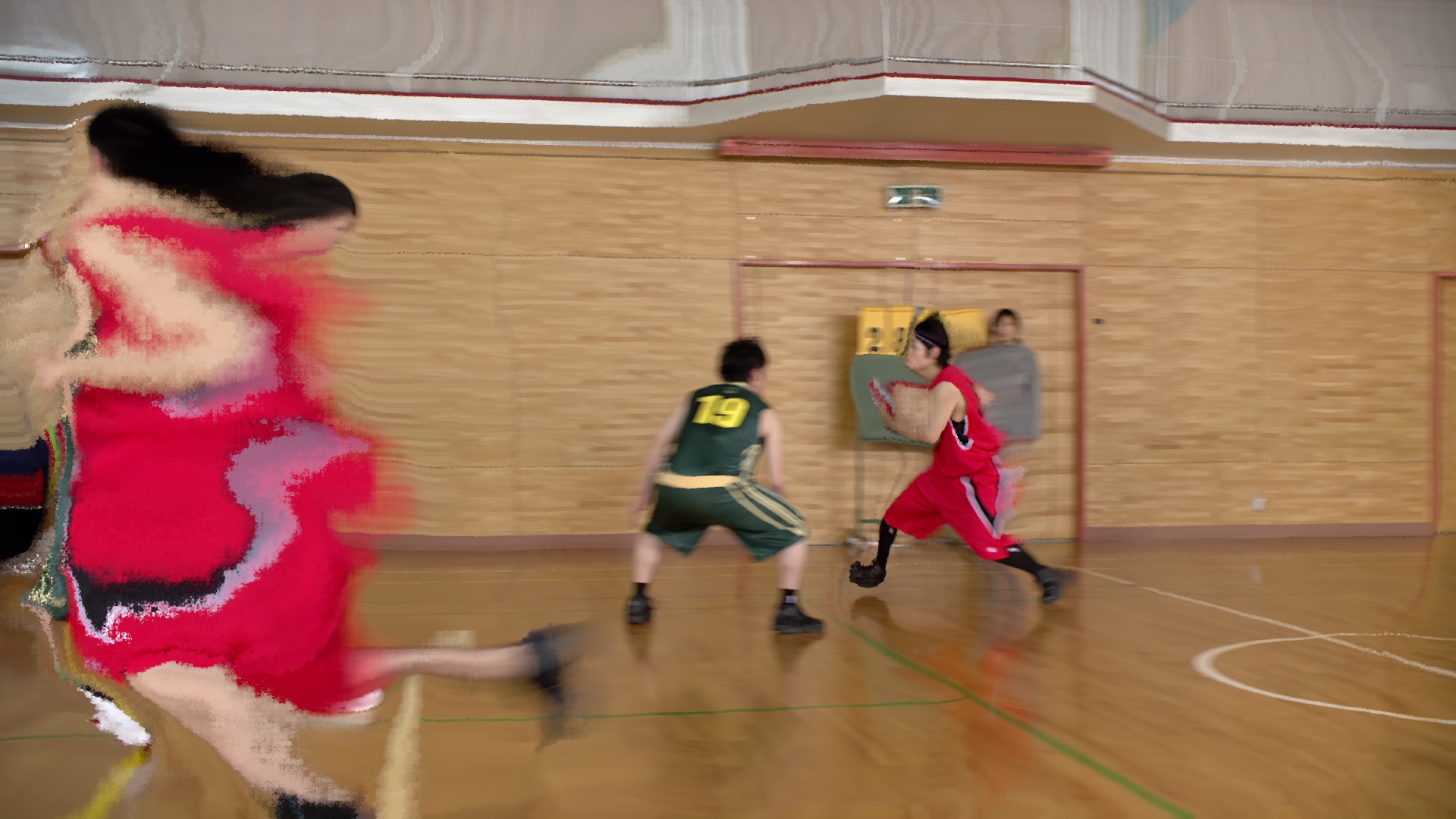} &
\includegraphics[width=0.115\textwidth]{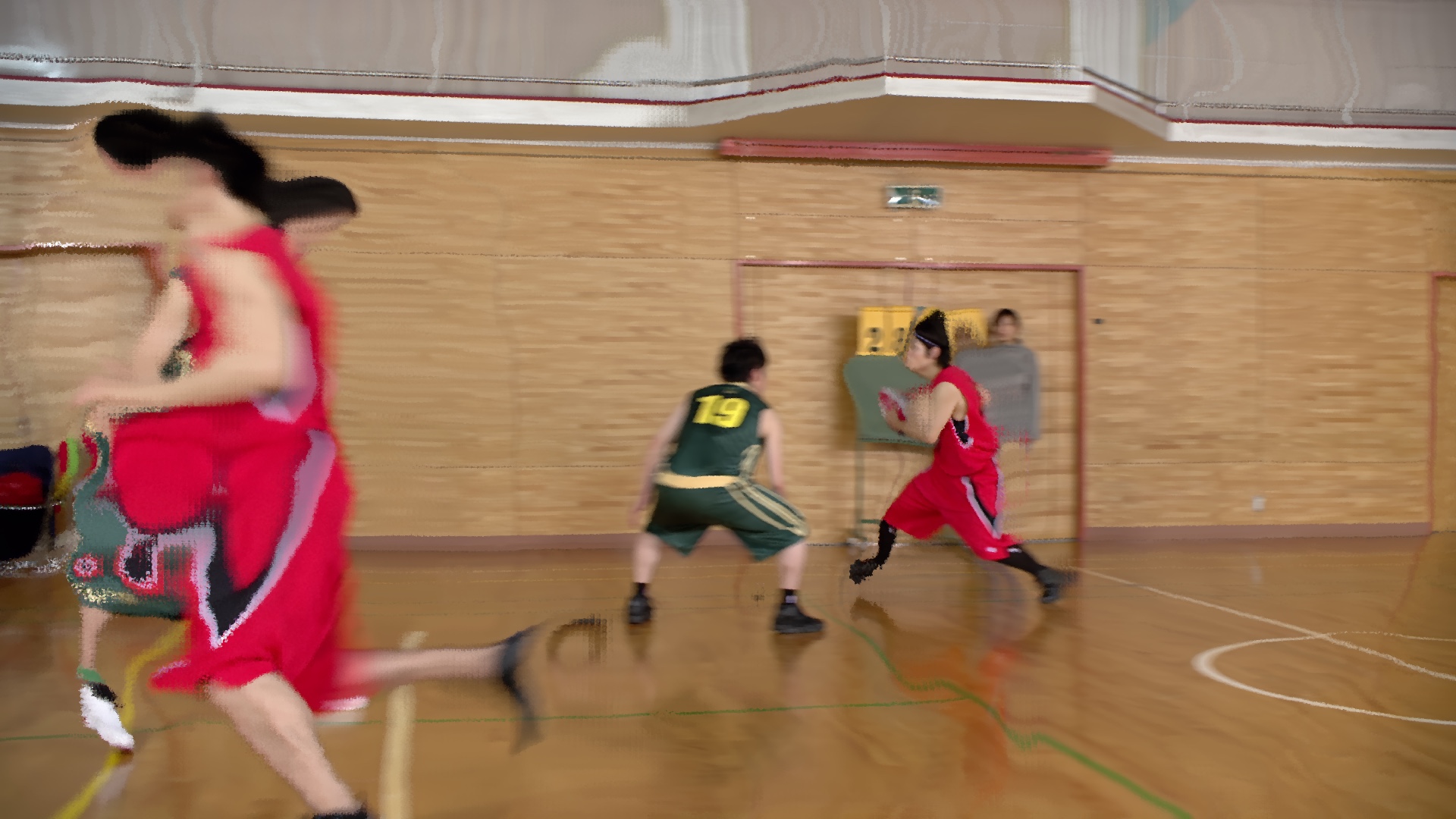} \\

 &
$\boldsymbol{e}_{0\rightarrow8}^{(1)}$ &
$\boldsymbol{e}_{0\rightarrow8}^{(2)}$ &
$\boldsymbol{e}_{0\rightarrow8}^{(4)}$ &
$\boldsymbol{e}_{0\rightarrow8}^{(8)}$ &
$\boldsymbol{e}_{0\rightarrow8}^{d}$ &
$\boldsymbol{e}_{0\rightarrow8}^{a}$ &
$\boldsymbol{e}_{0\rightarrow8}$ \\

 &
\includegraphics[width=0.115\textwidth]{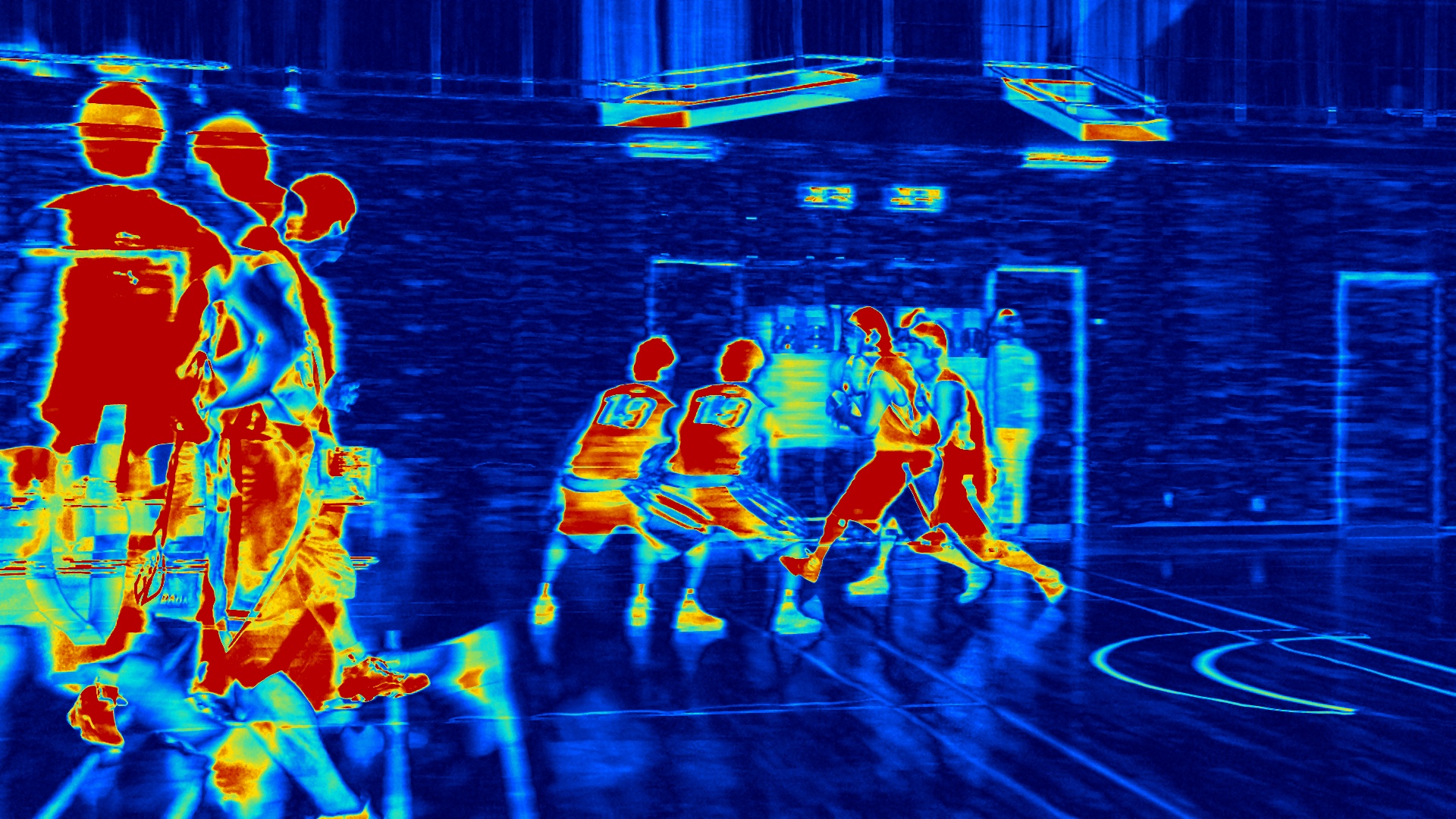} &
\includegraphics[width=0.115\textwidth]{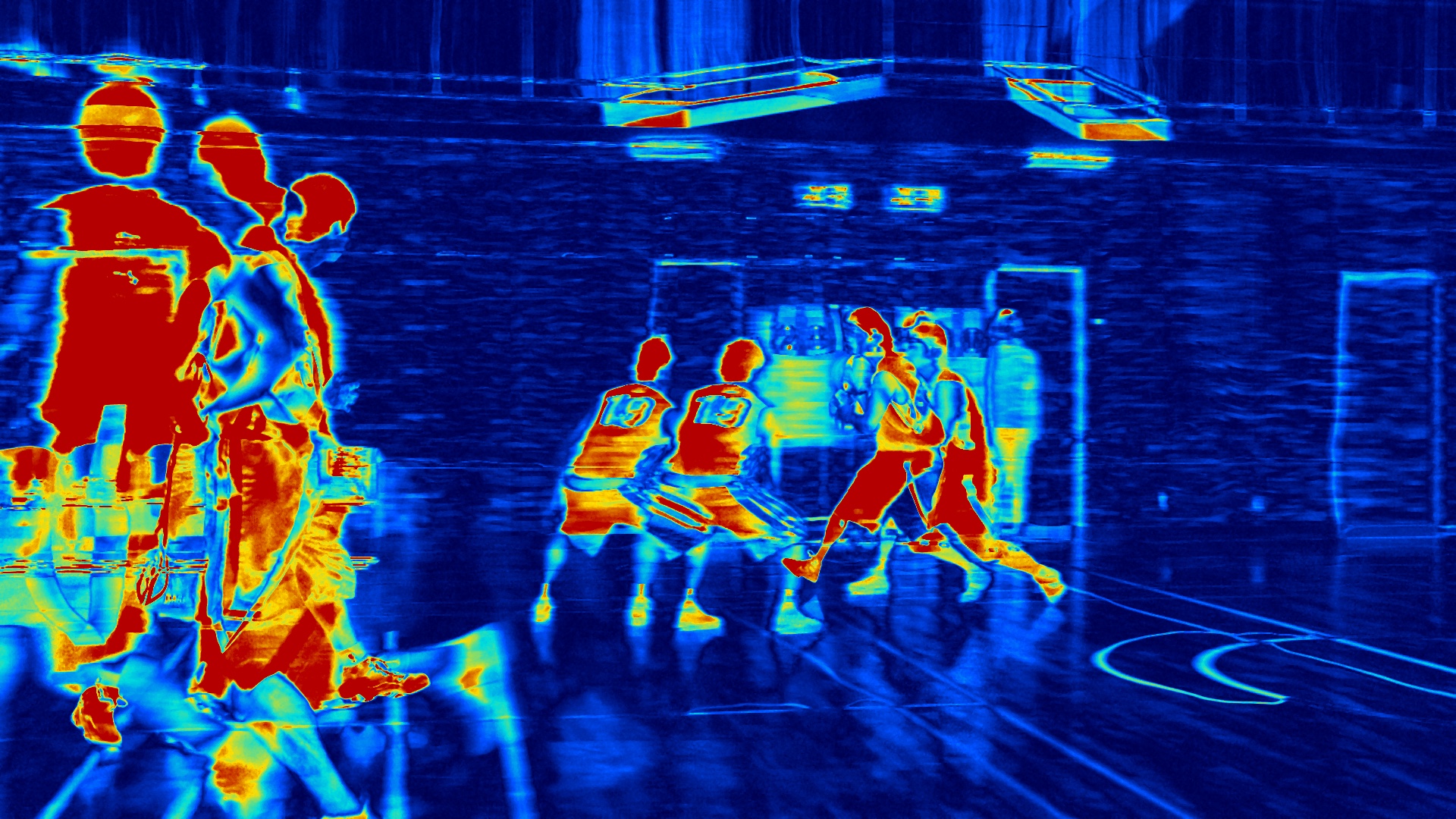} &
\includegraphics[width=0.115\textwidth]{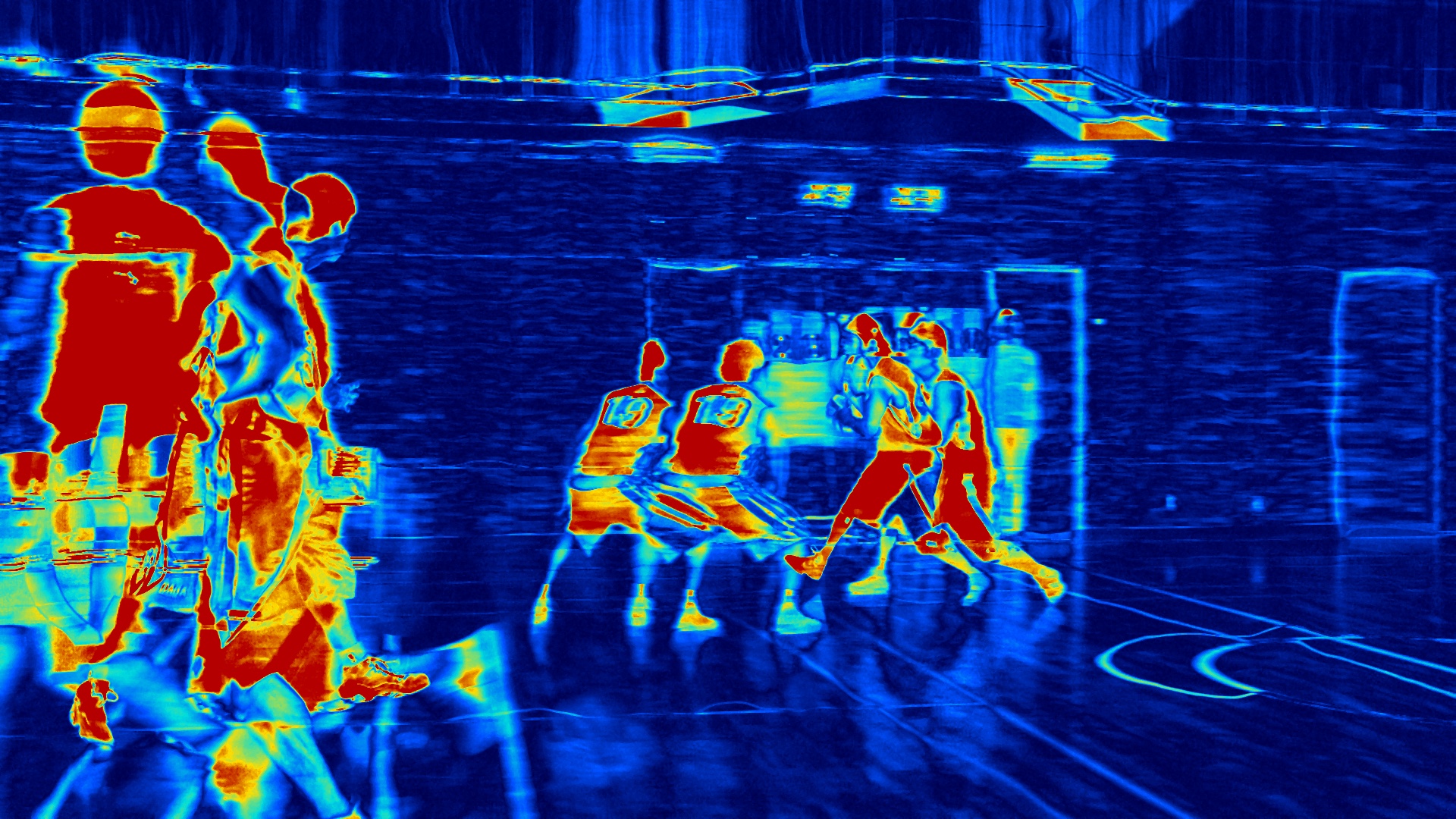} &
\includegraphics[width=0.115\textwidth]{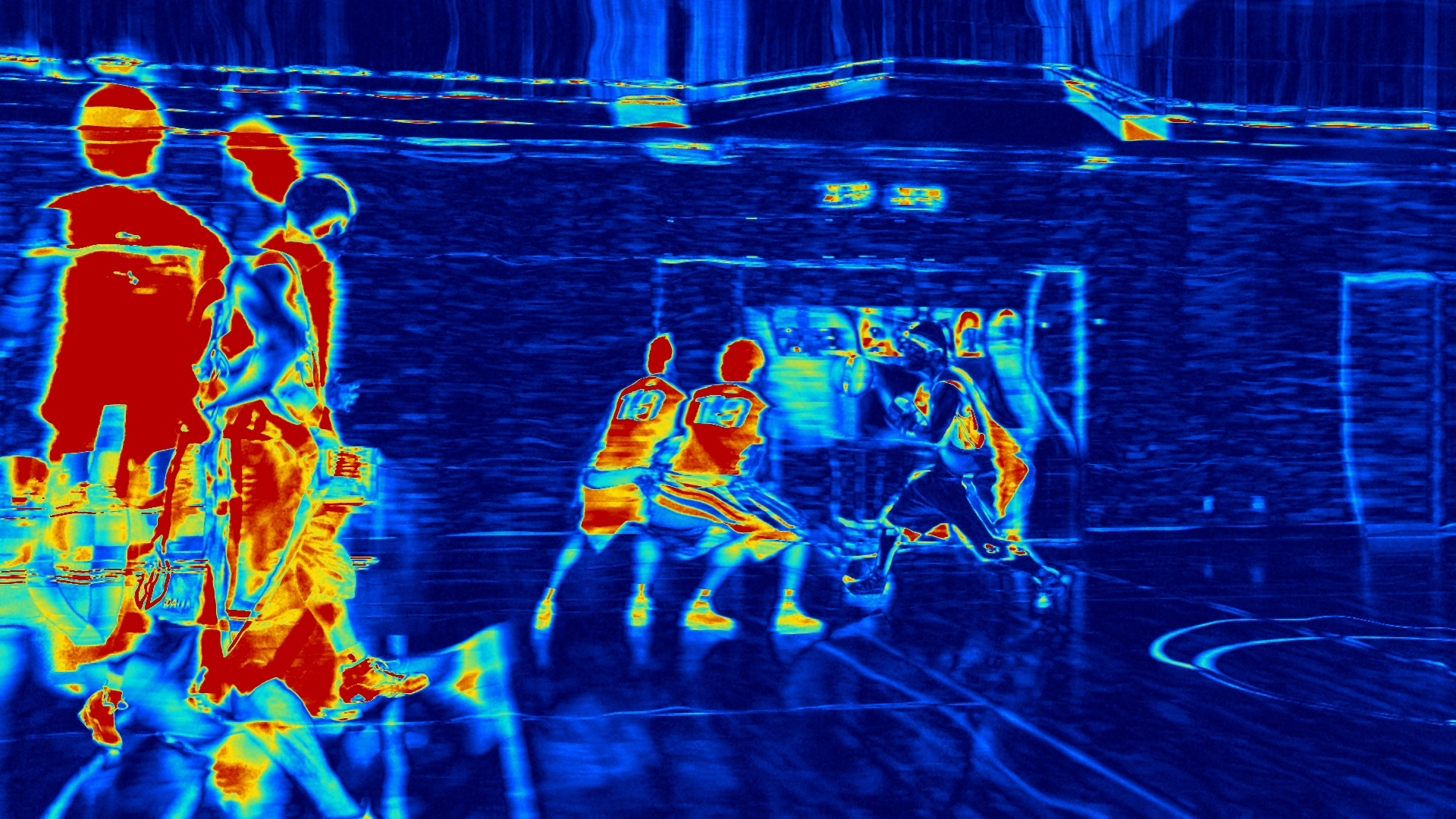} &
\includegraphics[width=0.115\textwidth]{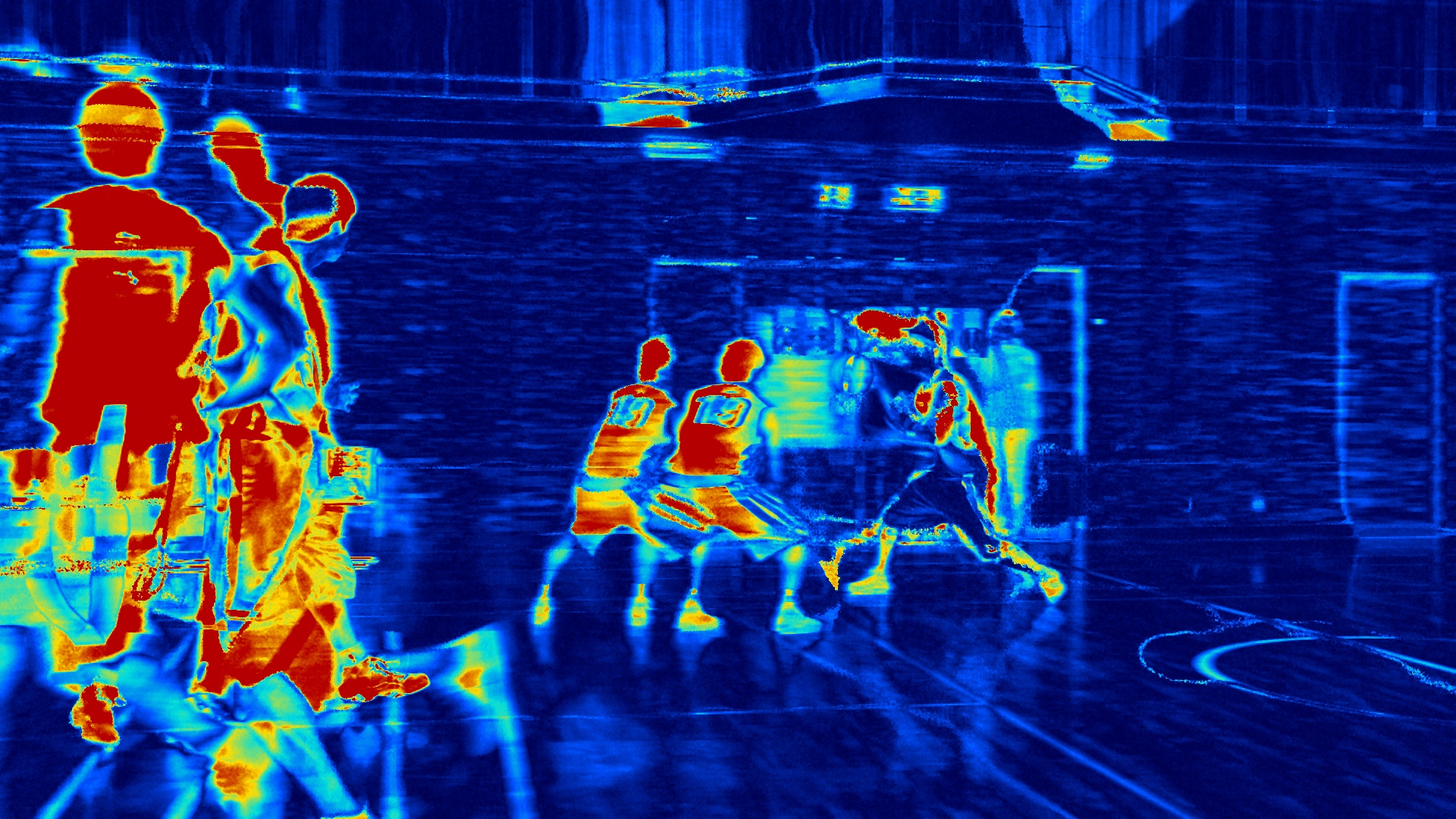} &
\includegraphics[width=0.115\textwidth]{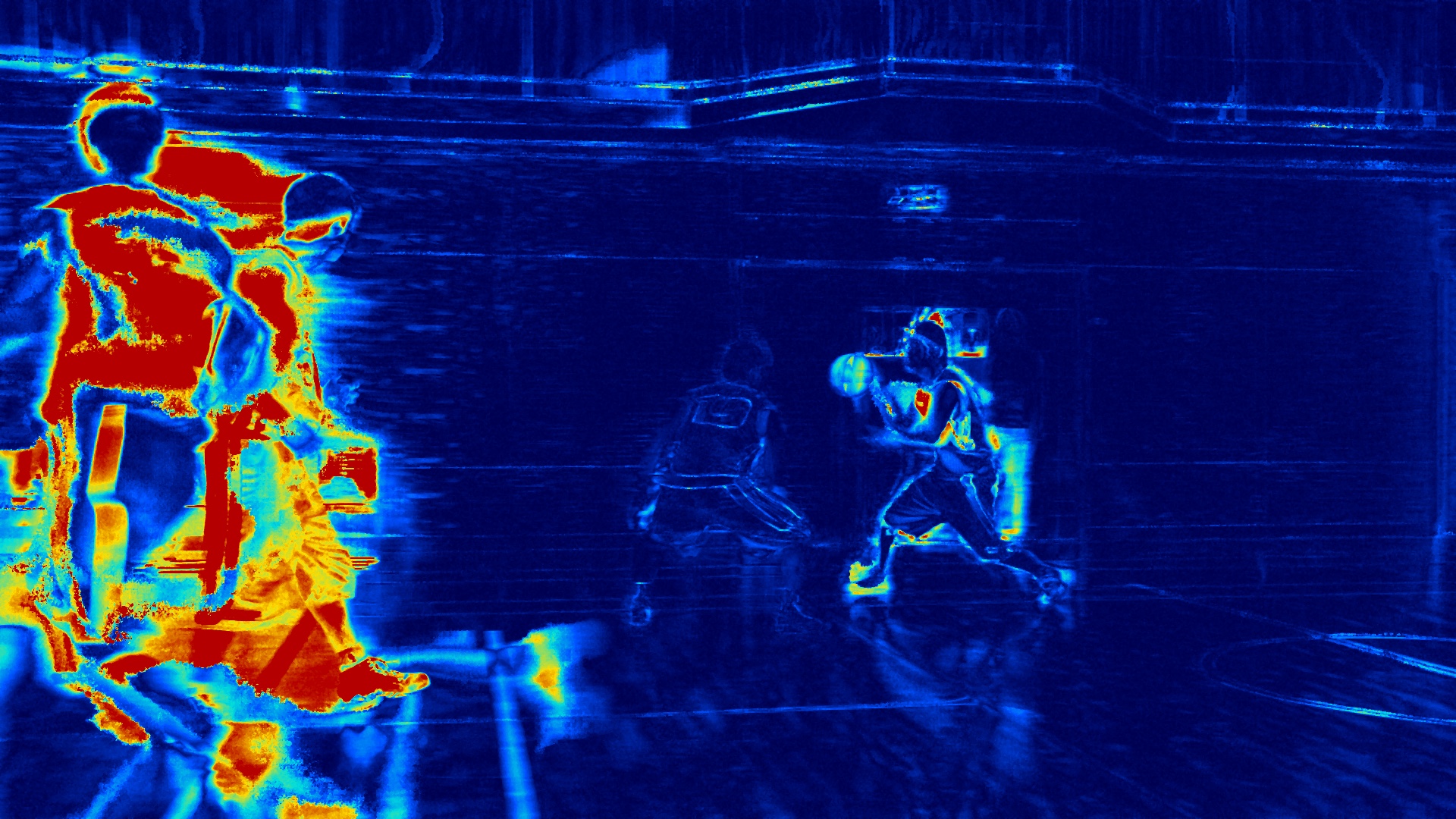} &
\includegraphics[width=0.115\textwidth]{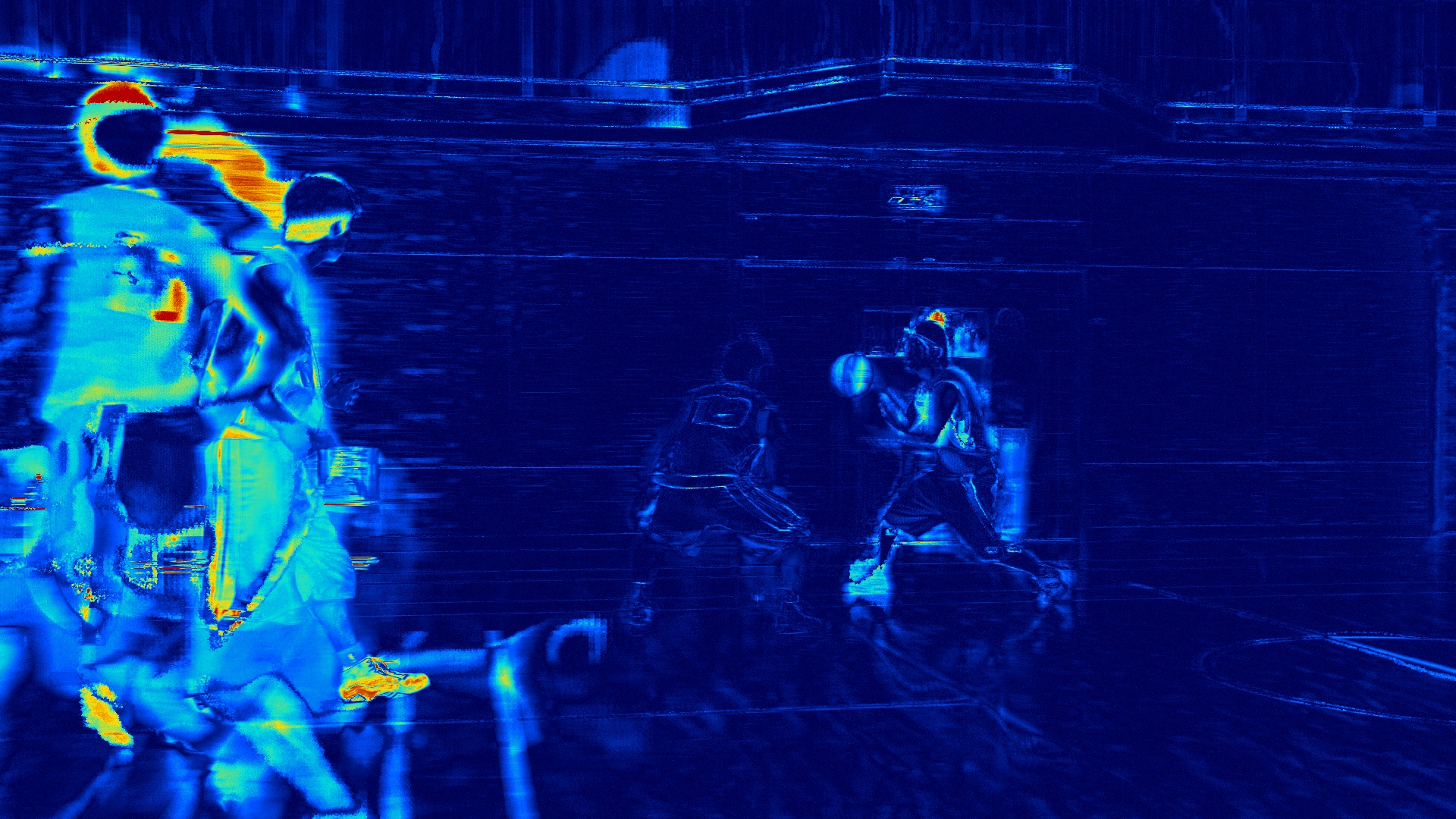} \\
\noalign{\vskip 2pt \color{gray!35}\hrule height 0.35pt \vskip 2pt}

\textbf{Ref.} $\hat{\boldsymbol{x}}_0$ &
$\boldsymbol{v}_{16\rightarrow0}^{(1)}$ &
$\boldsymbol{v}_{16\rightarrow0}^{(2)}$ &
$\boldsymbol{v}_{16\rightarrow0}^{(4)}$ &
$\boldsymbol{v}_{16\rightarrow0}^{(8)}$ &
$\boldsymbol{v}_{16\rightarrow0}^{d}$ &
$\boldsymbol{v}_{16\rightarrow0}^{a}$ &
$\boldsymbol{v}_{16\rightarrow0}$ \\

\includegraphics[width=0.115\textwidth]{figures_jpg/motion_vis/pptx/BasketballDrive/frame/im00001.jpg} &
\includegraphics[width=0.115\textwidth]{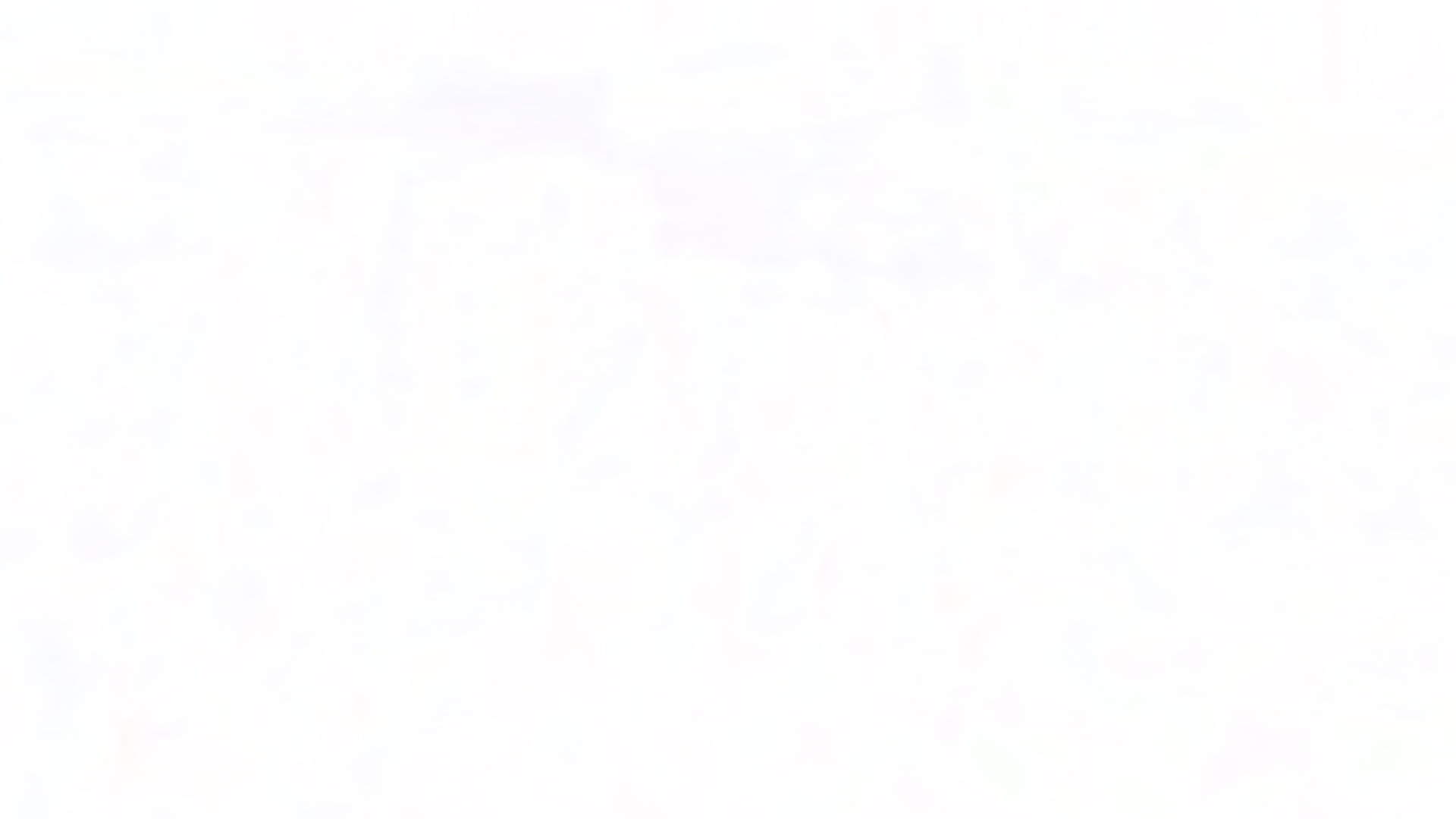} &
\includegraphics[width=0.115\textwidth]{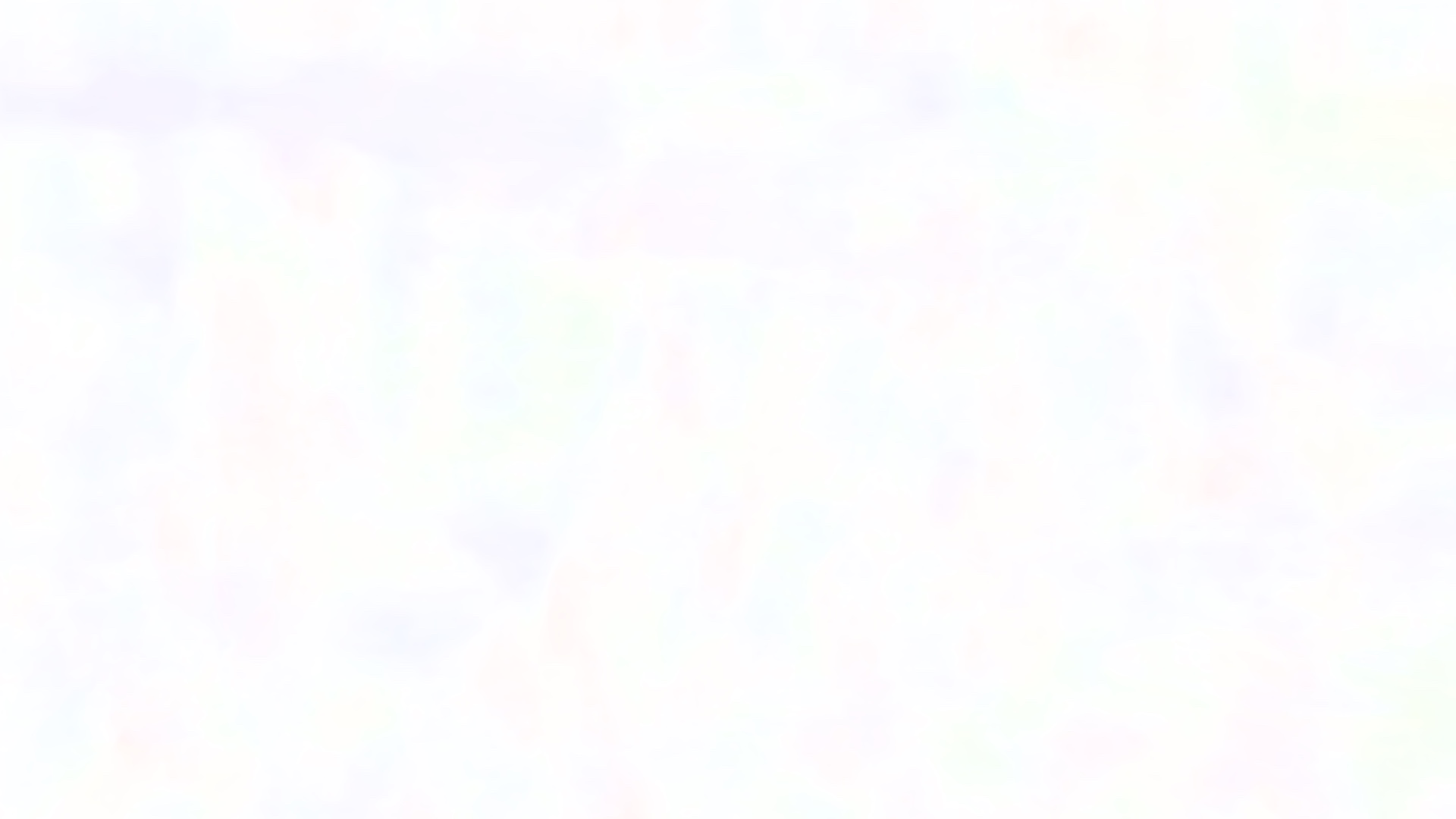} &
\includegraphics[width=0.115\textwidth]{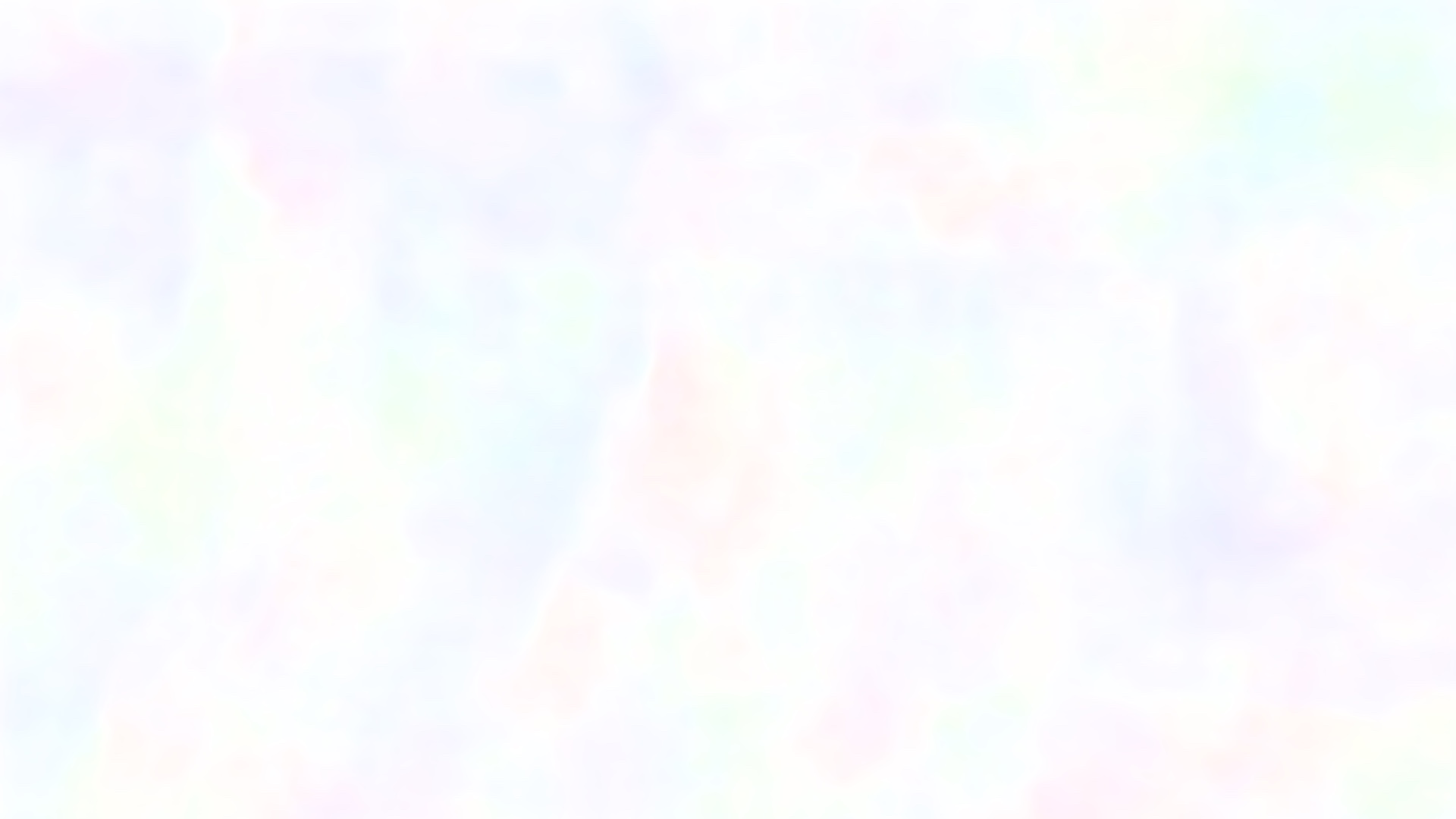} &
\includegraphics[width=0.115\textwidth]{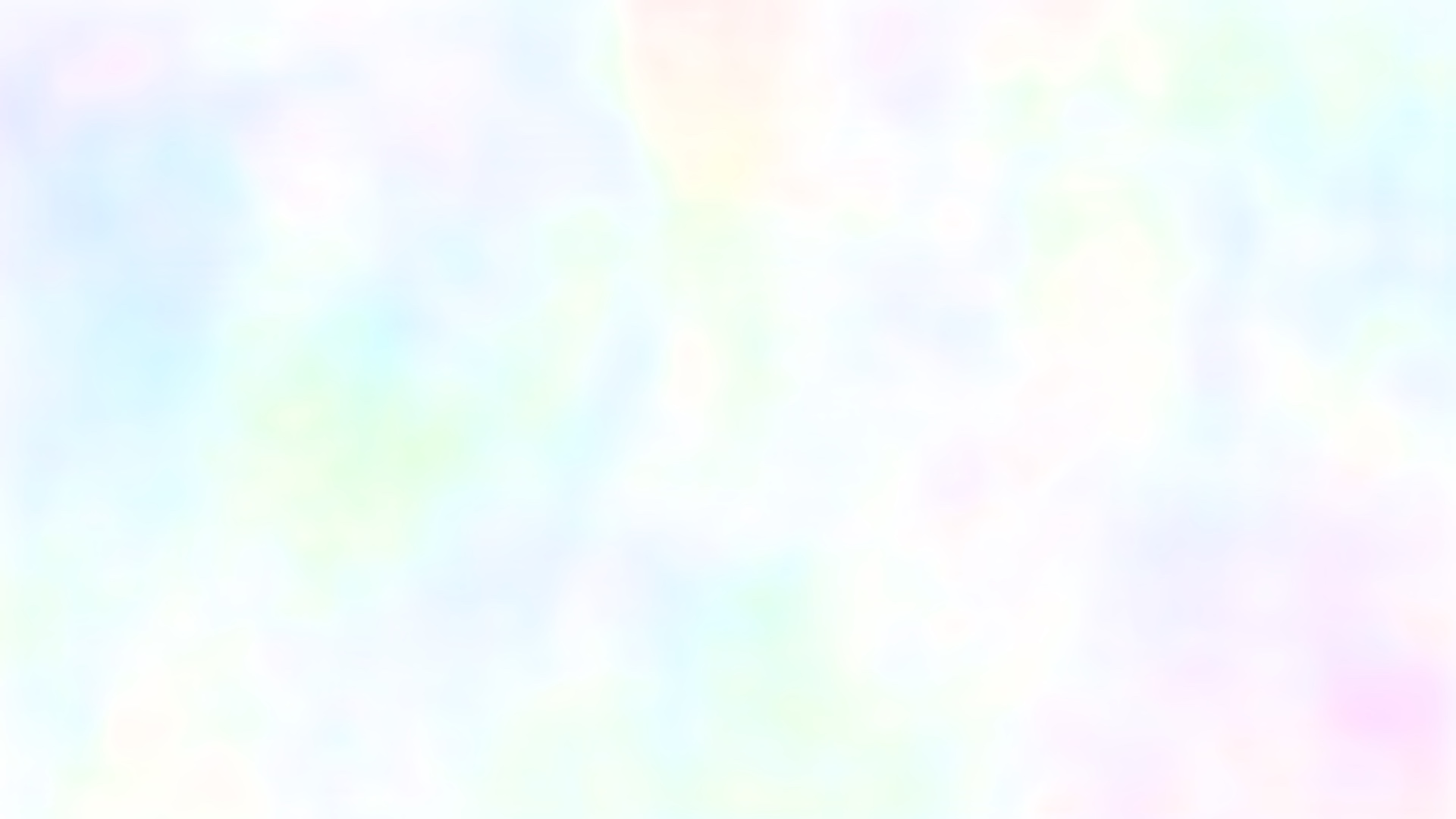} &
\includegraphics[width=0.115\textwidth]{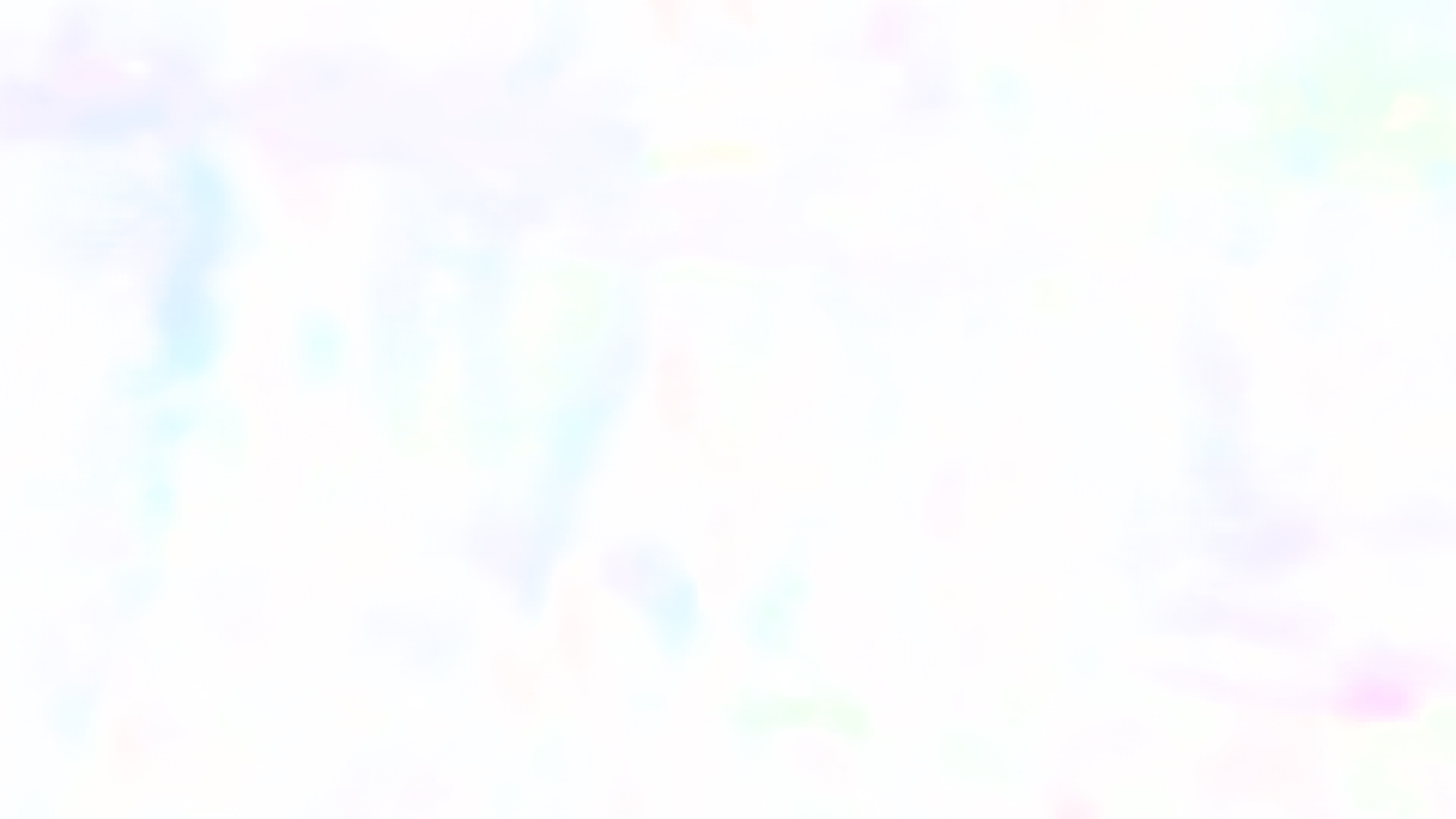} &
\includegraphics[width=0.115\textwidth]{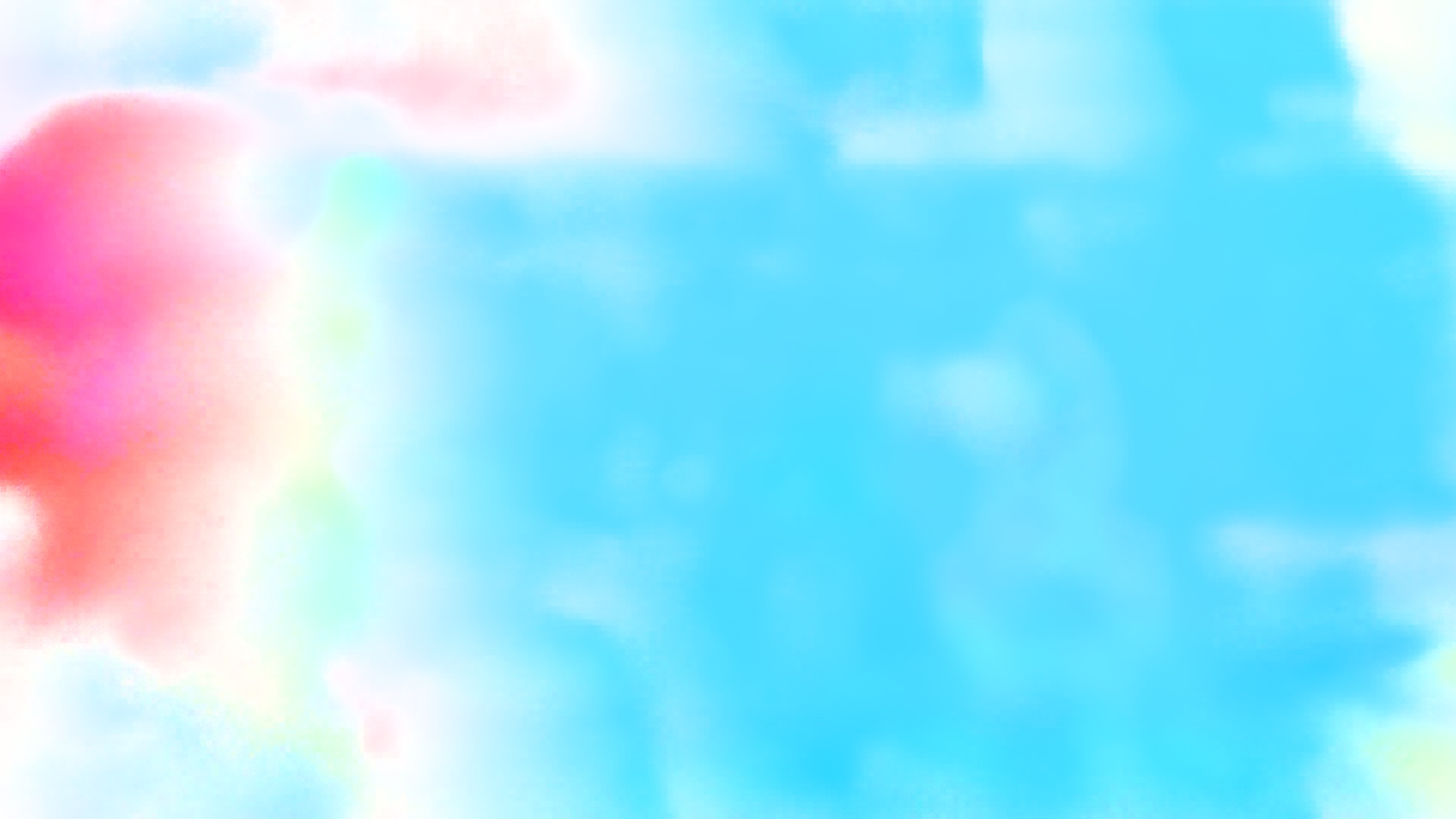} &
\includegraphics[width=0.115\textwidth]{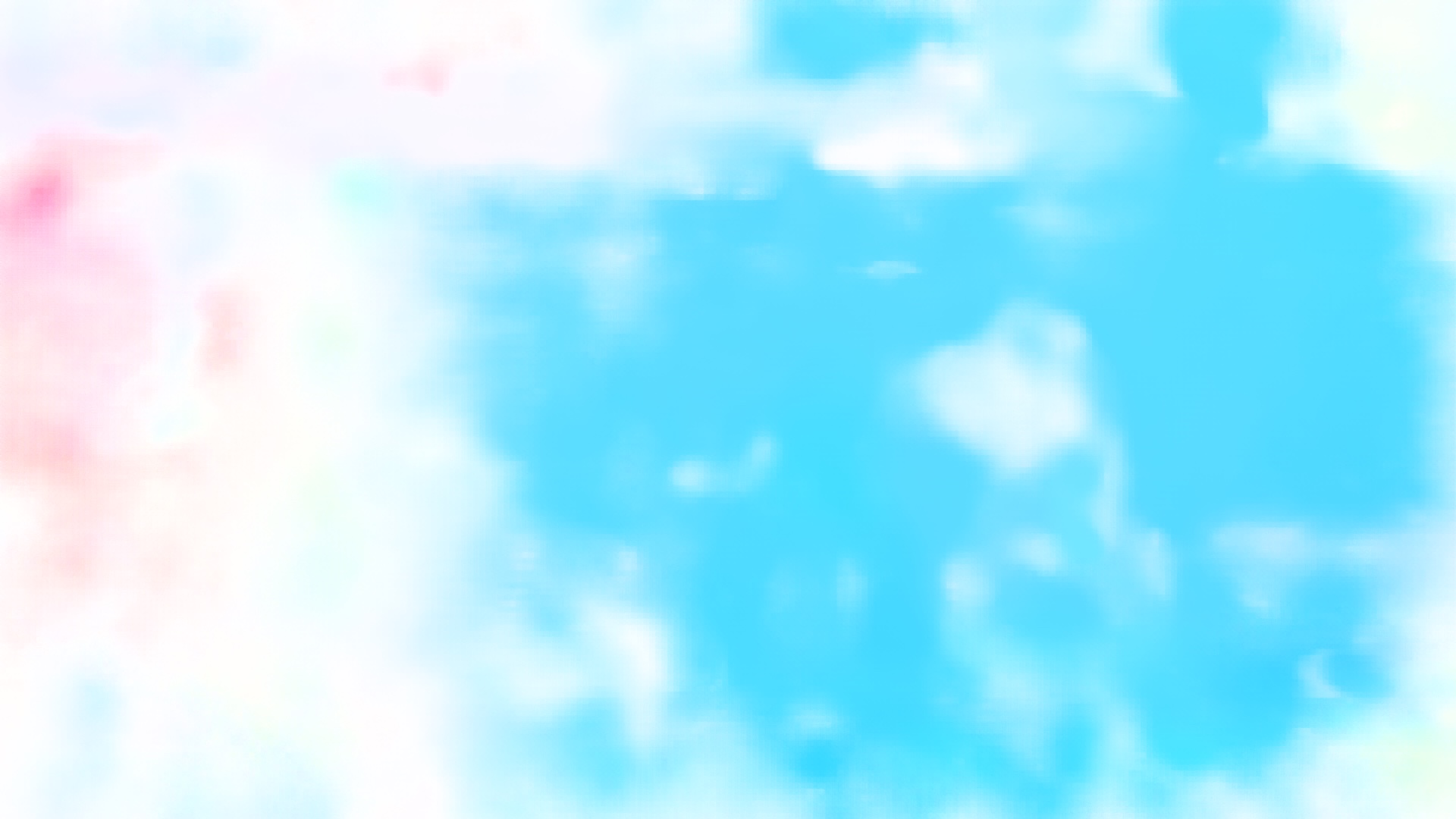} \\

\textbf{Cur.} $\boldsymbol{x}_{16}$ &
$\hat{\boldsymbol{x}}_{0\rightarrow16}^{(1)}\ 15.26\,\mathrm{dB}$ &
$\hat{\boldsymbol{x}}_{0\rightarrow16}^{(2)}\ 15.61\,\mathrm{dB}$ &
$\hat{\boldsymbol{x}}_{0\rightarrow16}^{(4)}\ 15.72\,\mathrm{dB}$ &
$\hat{\boldsymbol{x}}_{0\rightarrow16}^{(8)}\ 15.65\,\mathrm{dB}$ &
$\hat{\boldsymbol{x}}_{0\rightarrow16}^{d}\ 15.76\,\mathrm{dB}$ &
$\hat{\boldsymbol{x}}_{0\rightarrow16}^{a}\ 16.31\,\mathrm{dB}$ &
$\hat{\boldsymbol{x}}_{0\rightarrow16}\ 17.04\,\mathrm{dB}$ \\

\includegraphics[width=0.115\textwidth]{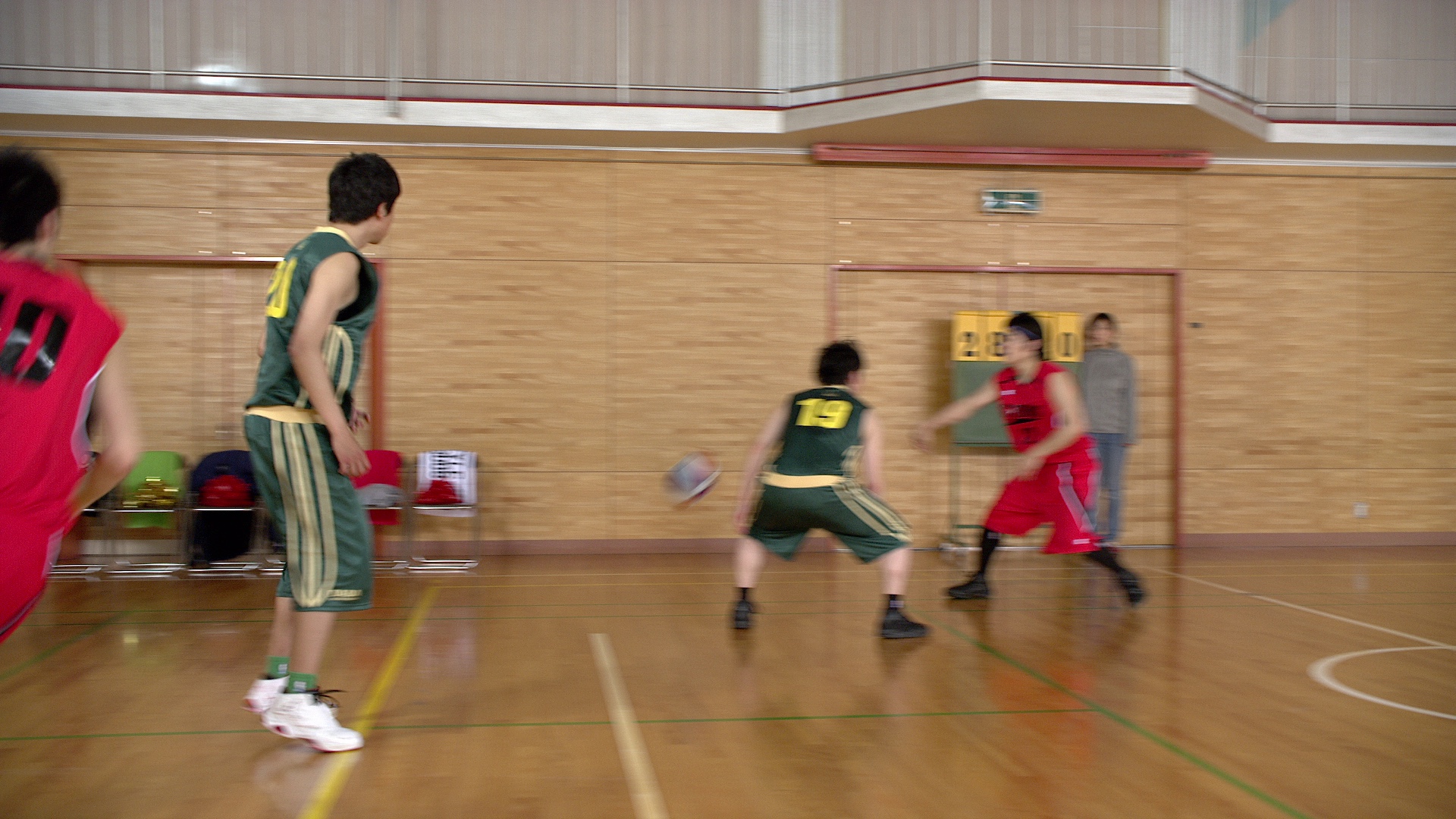} &
\includegraphics[width=0.115\textwidth]{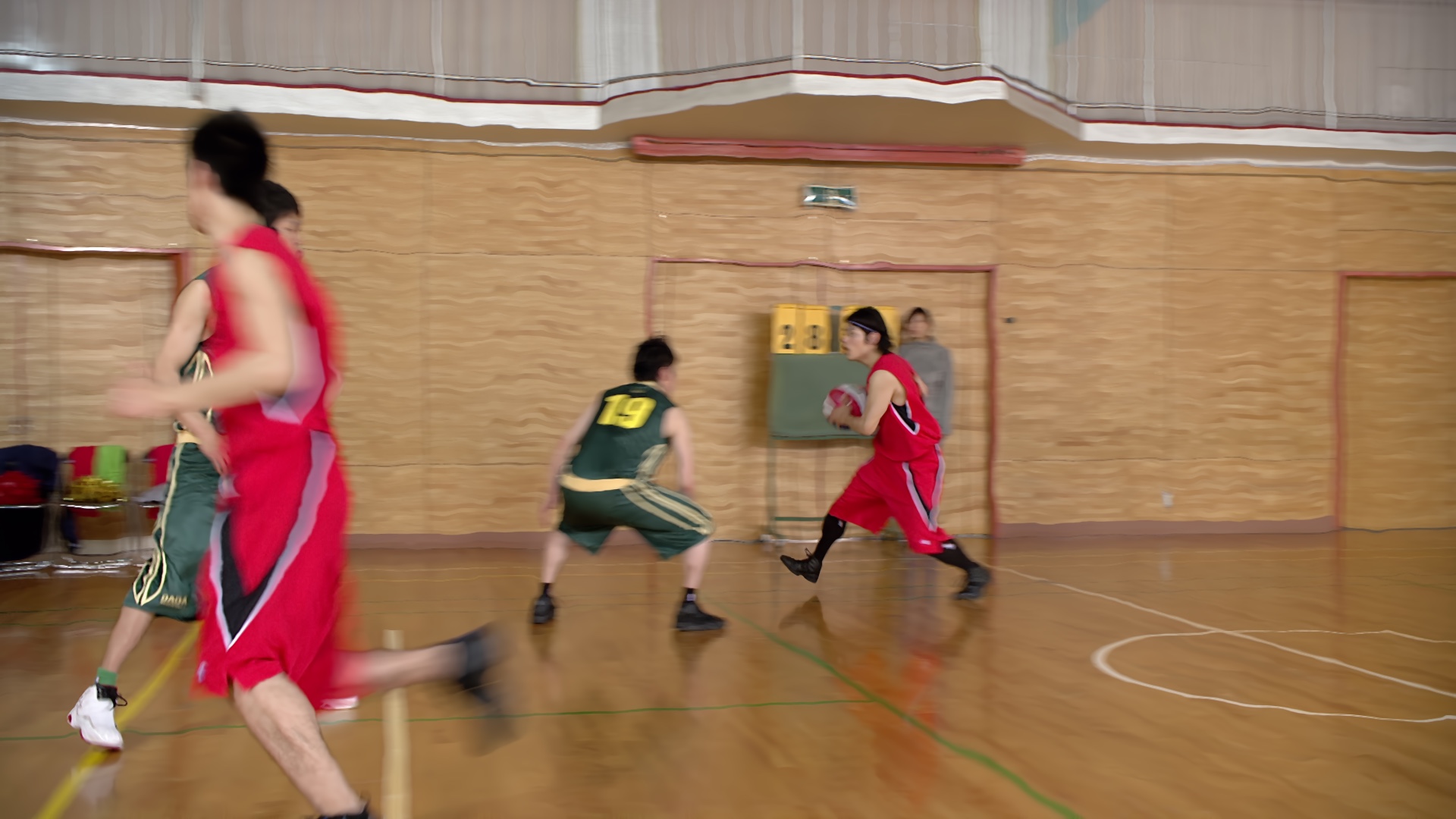} &
\includegraphics[width=0.115\textwidth]{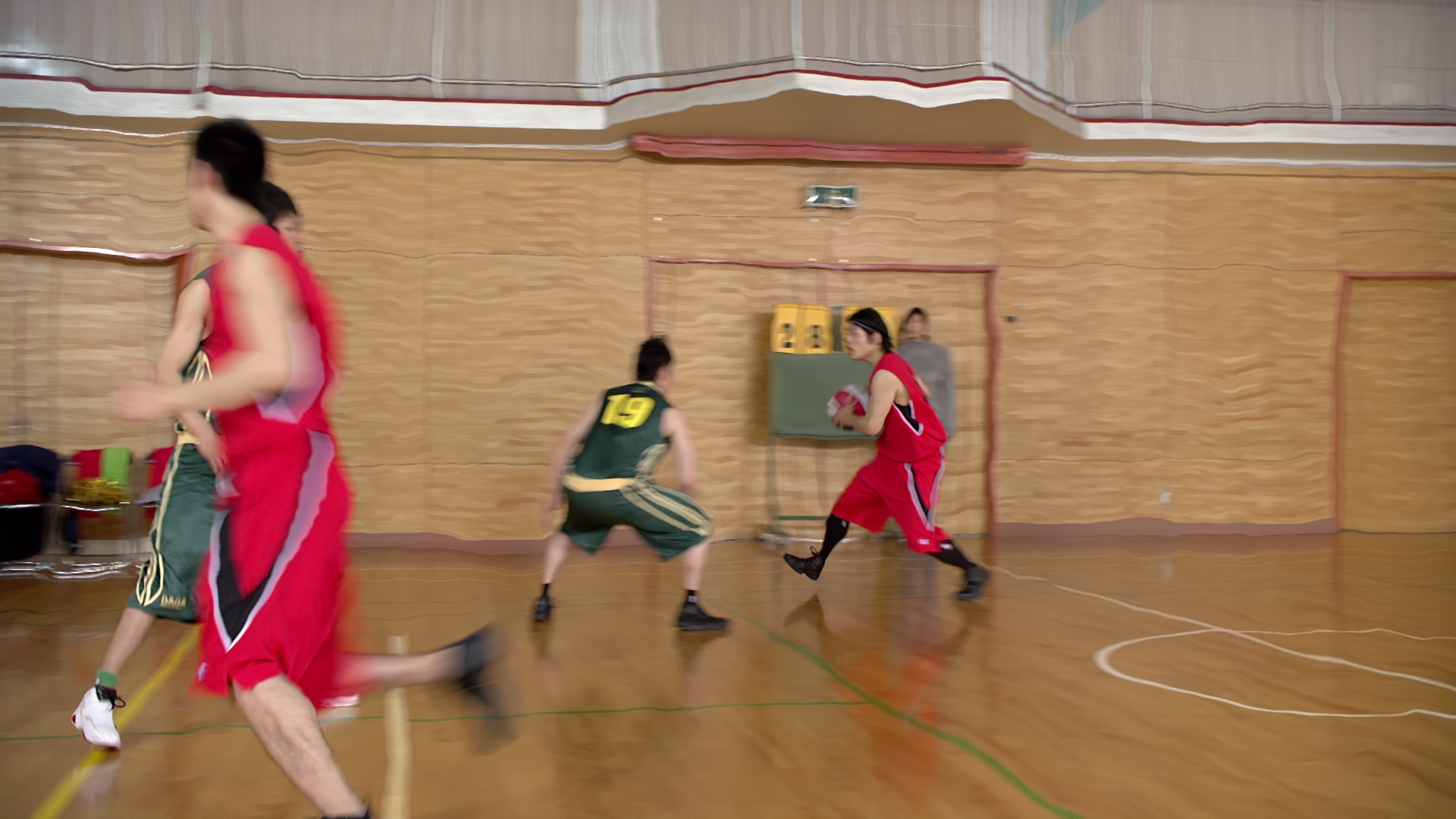} &
\includegraphics[width=0.115\textwidth]{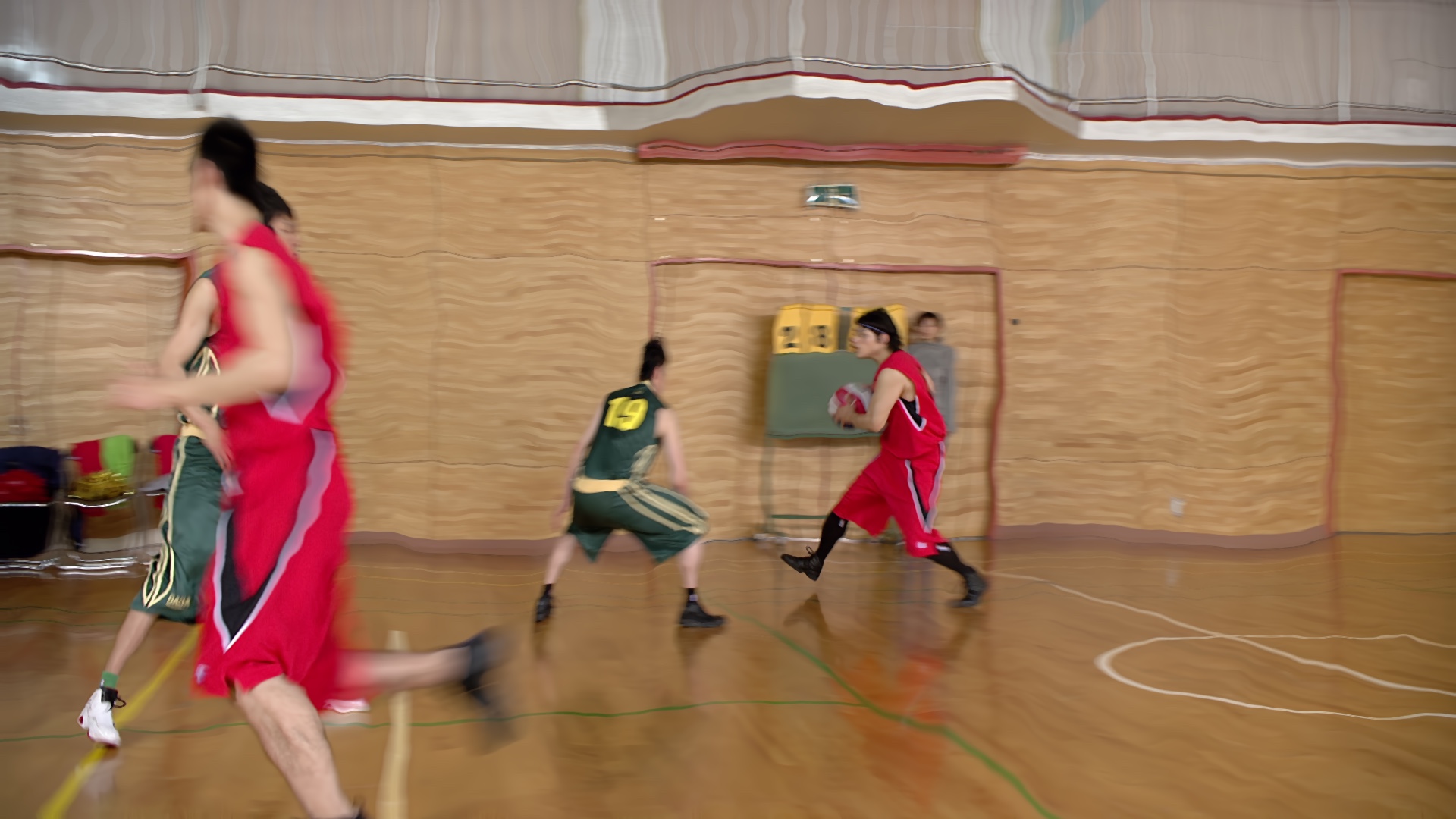} &
\includegraphics[width=0.115\textwidth]{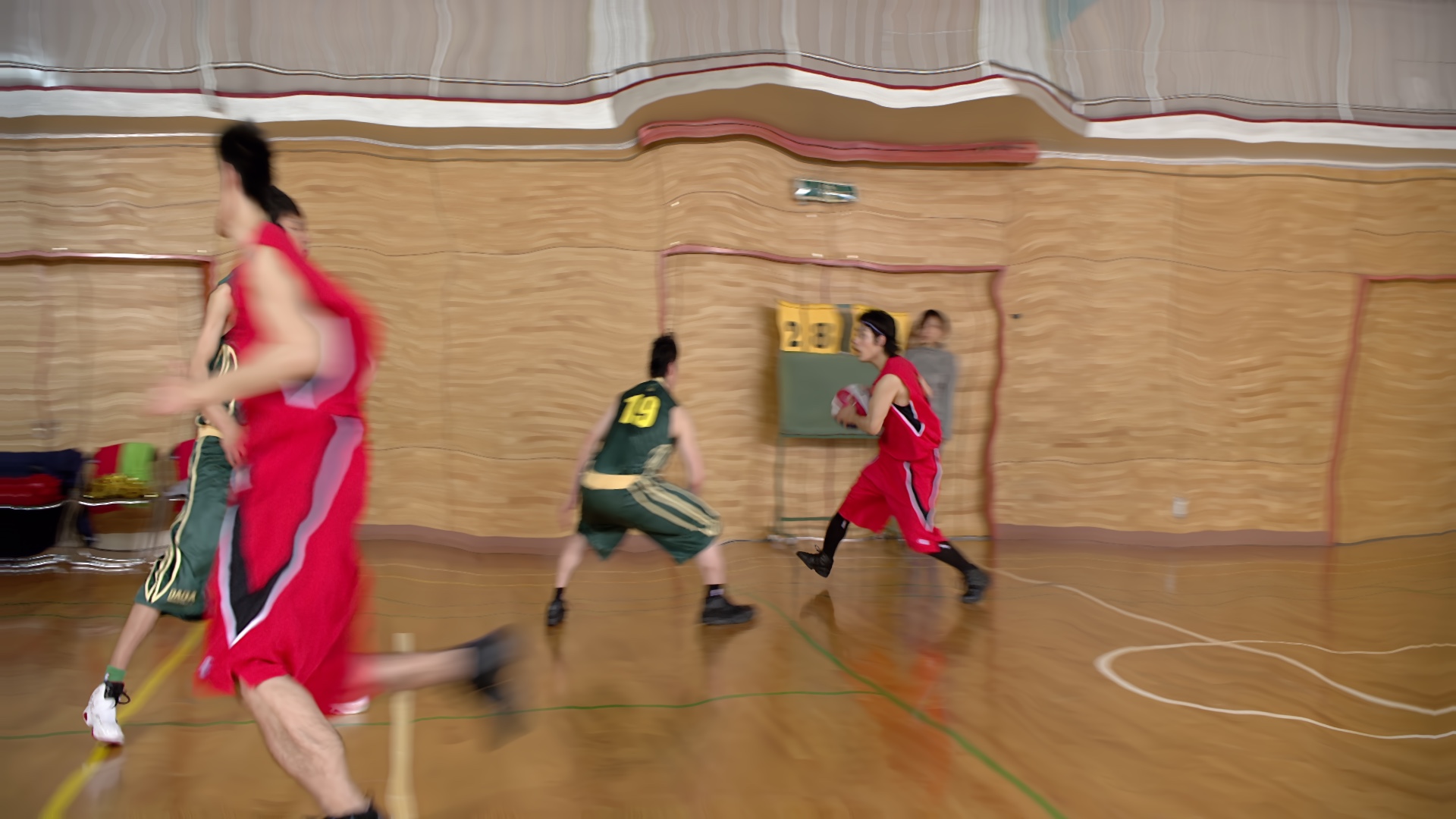} &
\includegraphics[width=0.115\textwidth]{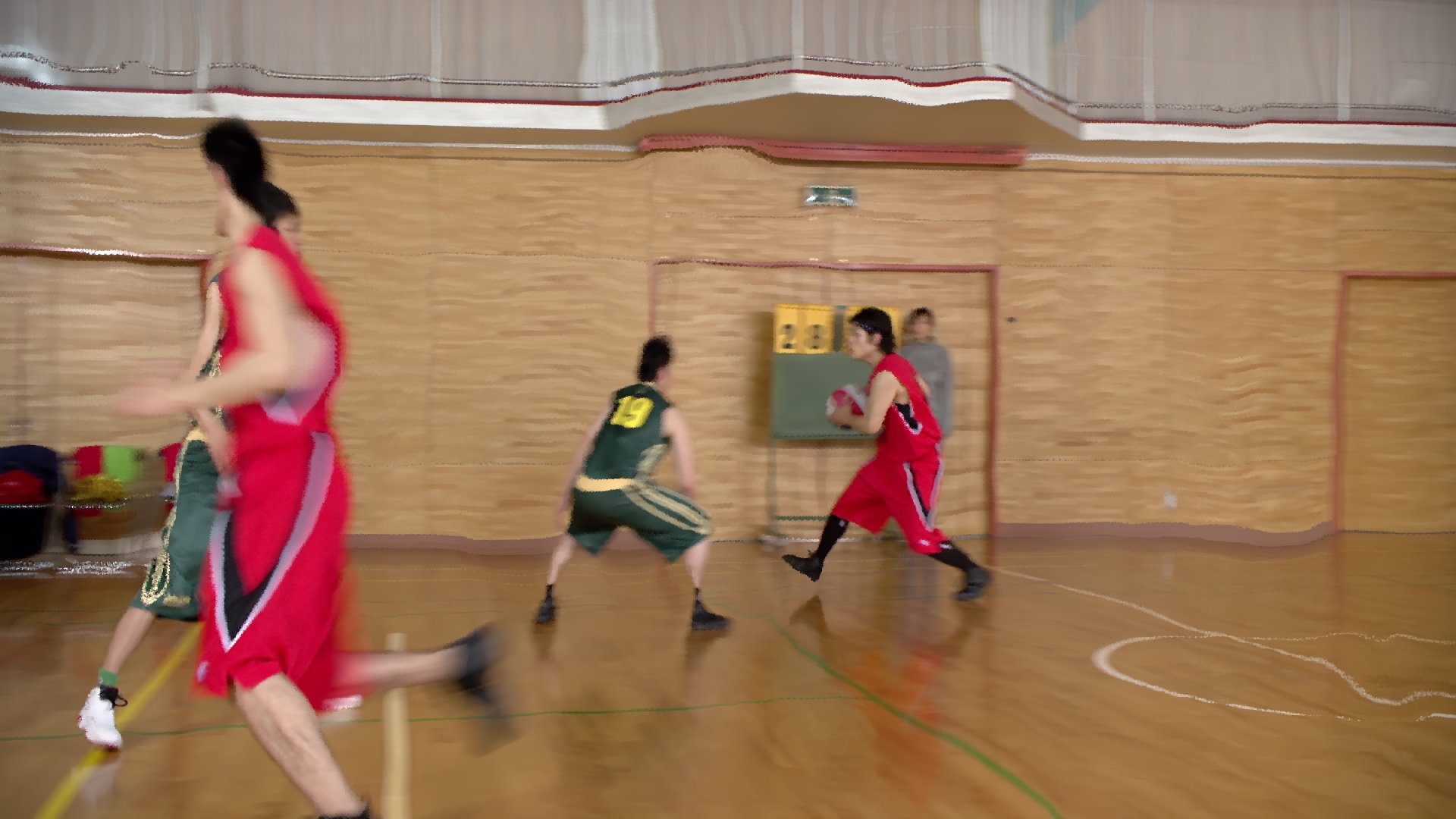} &
\includegraphics[width=0.115\textwidth]{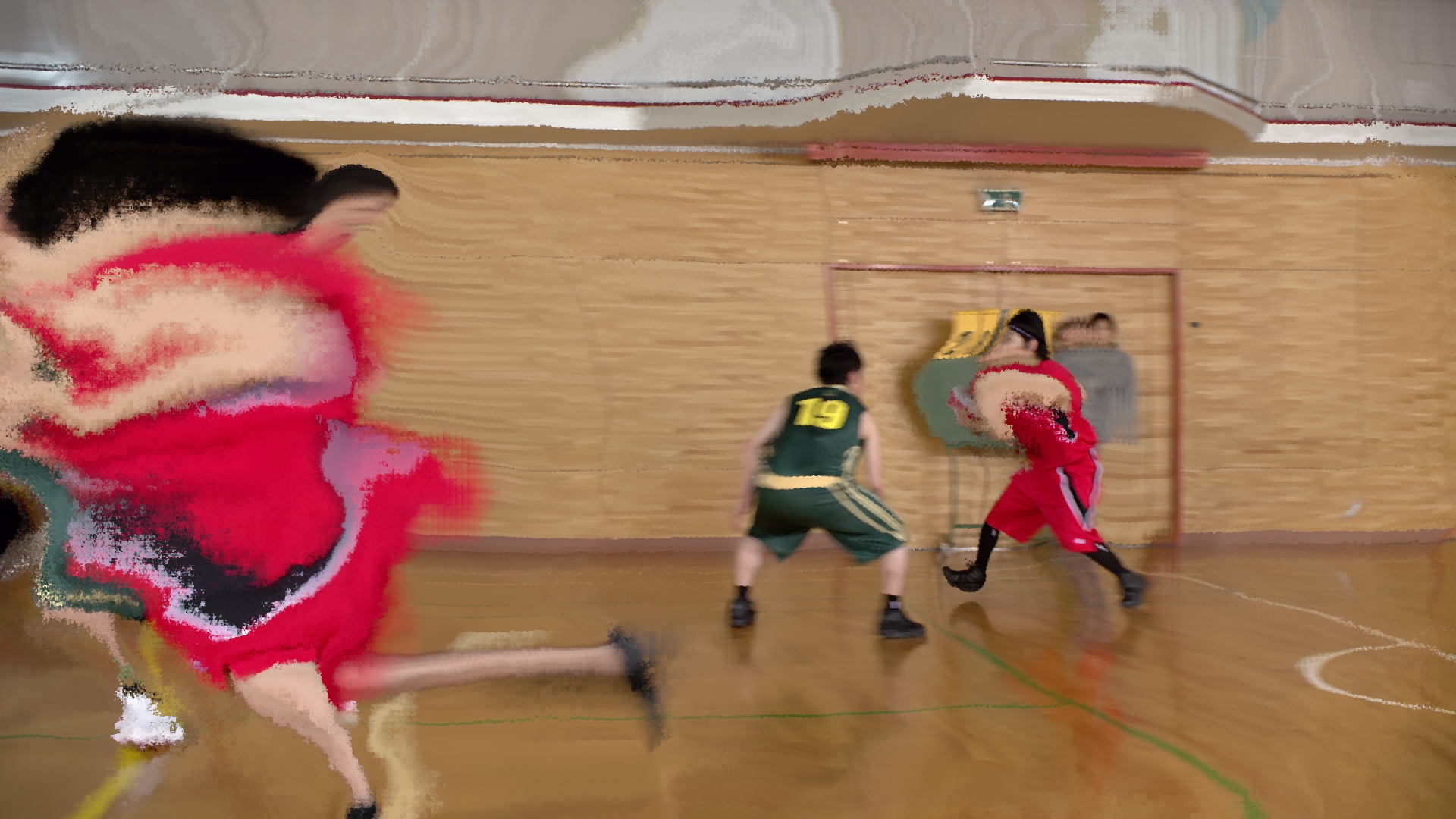} &
\includegraphics[width=0.115\textwidth]{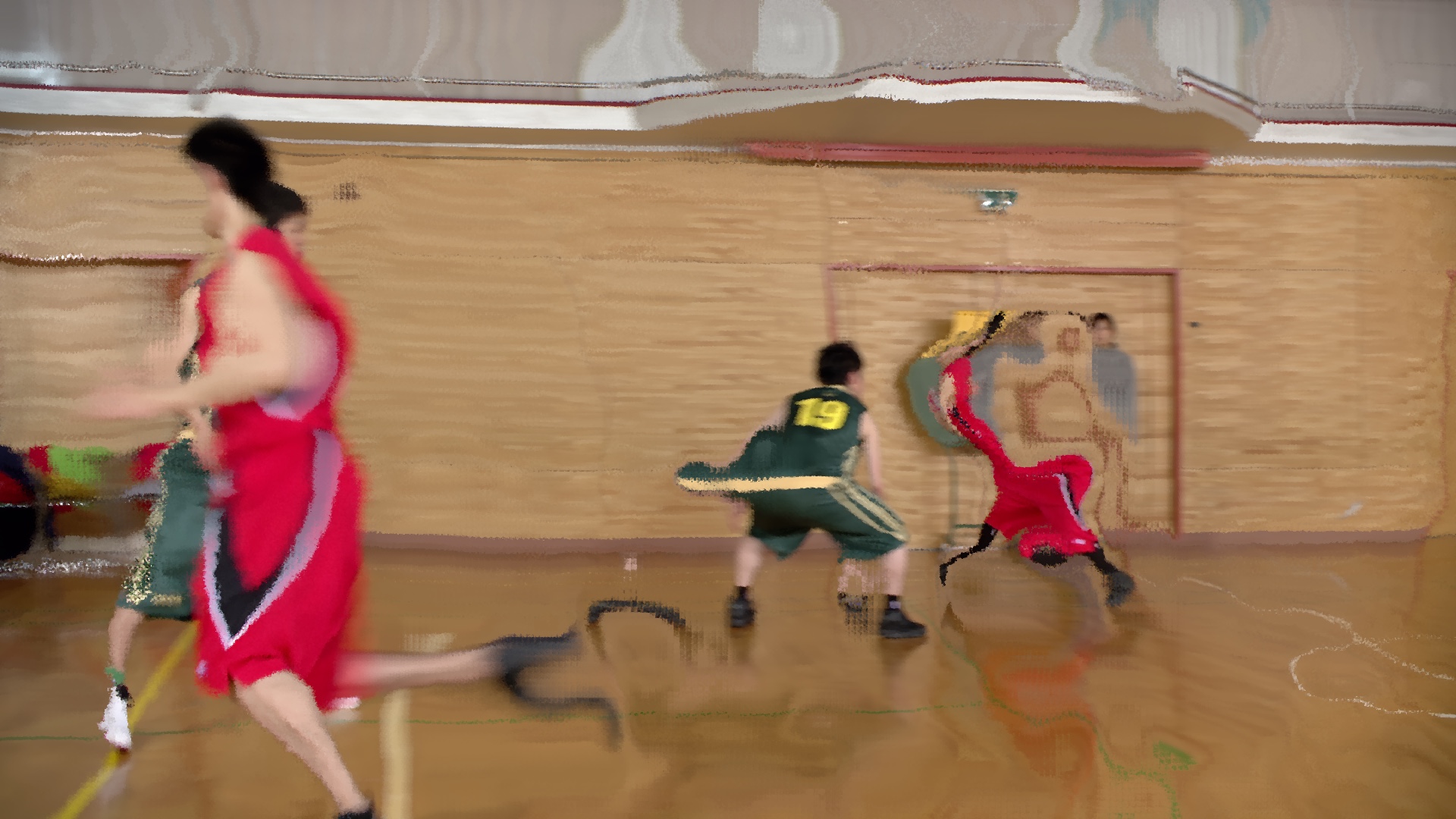} \\

 &
$\boldsymbol{e}_{0\rightarrow16}^{(1)}$ &
$\boldsymbol{e}_{0\rightarrow16}^{(2)}$ &
$\boldsymbol{e}_{0\rightarrow16}^{(4)}$ &
$\boldsymbol{e}_{0\rightarrow16}^{(8)}$ &
$\boldsymbol{e}_{0\rightarrow16}^{d}$ &
$\boldsymbol{e}_{0\rightarrow16}^{a}$ &
$\boldsymbol{e}_{0\rightarrow16}$ \\

 &
\includegraphics[width=0.115\textwidth]{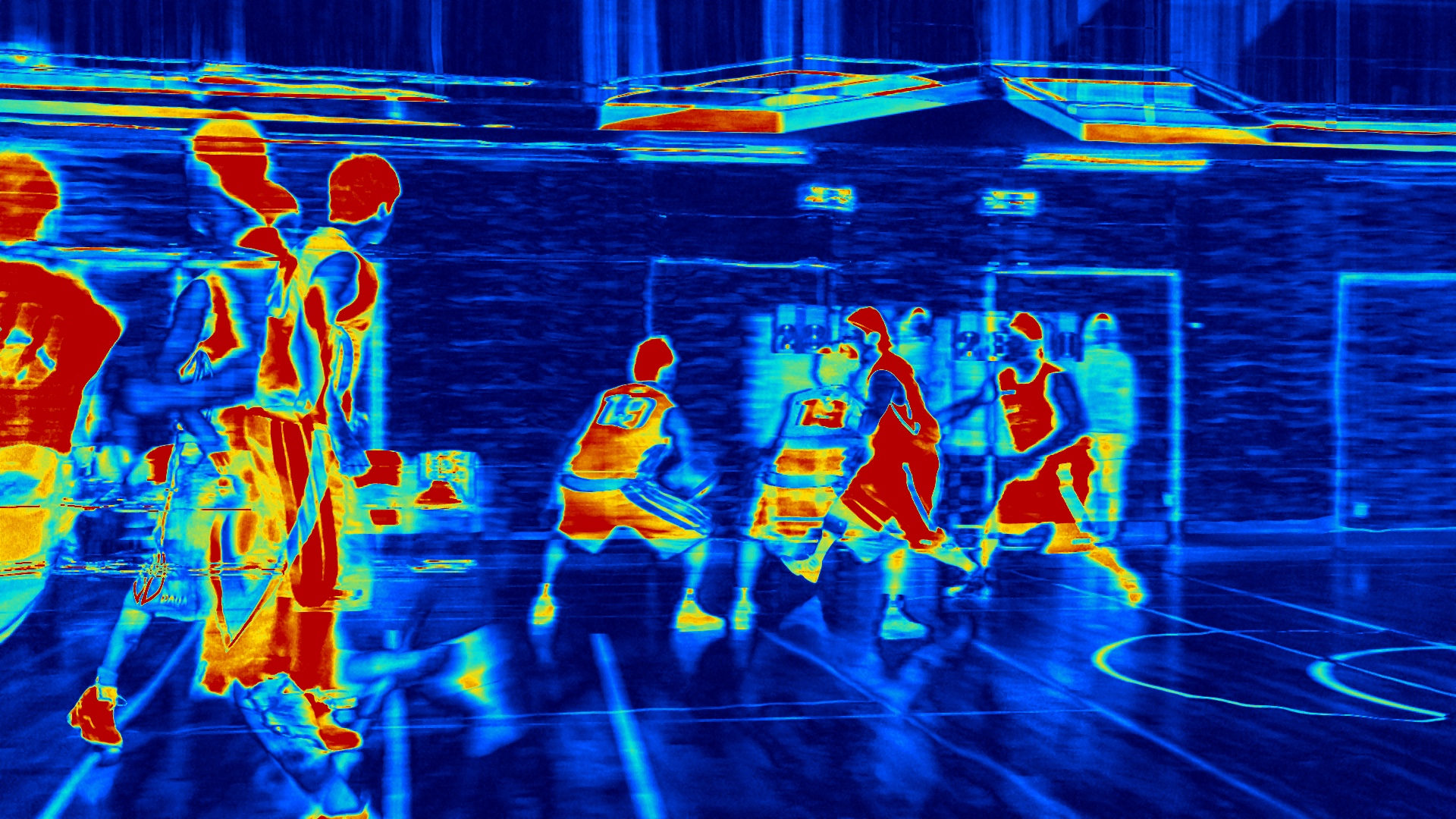} &
\includegraphics[width=0.115\textwidth]{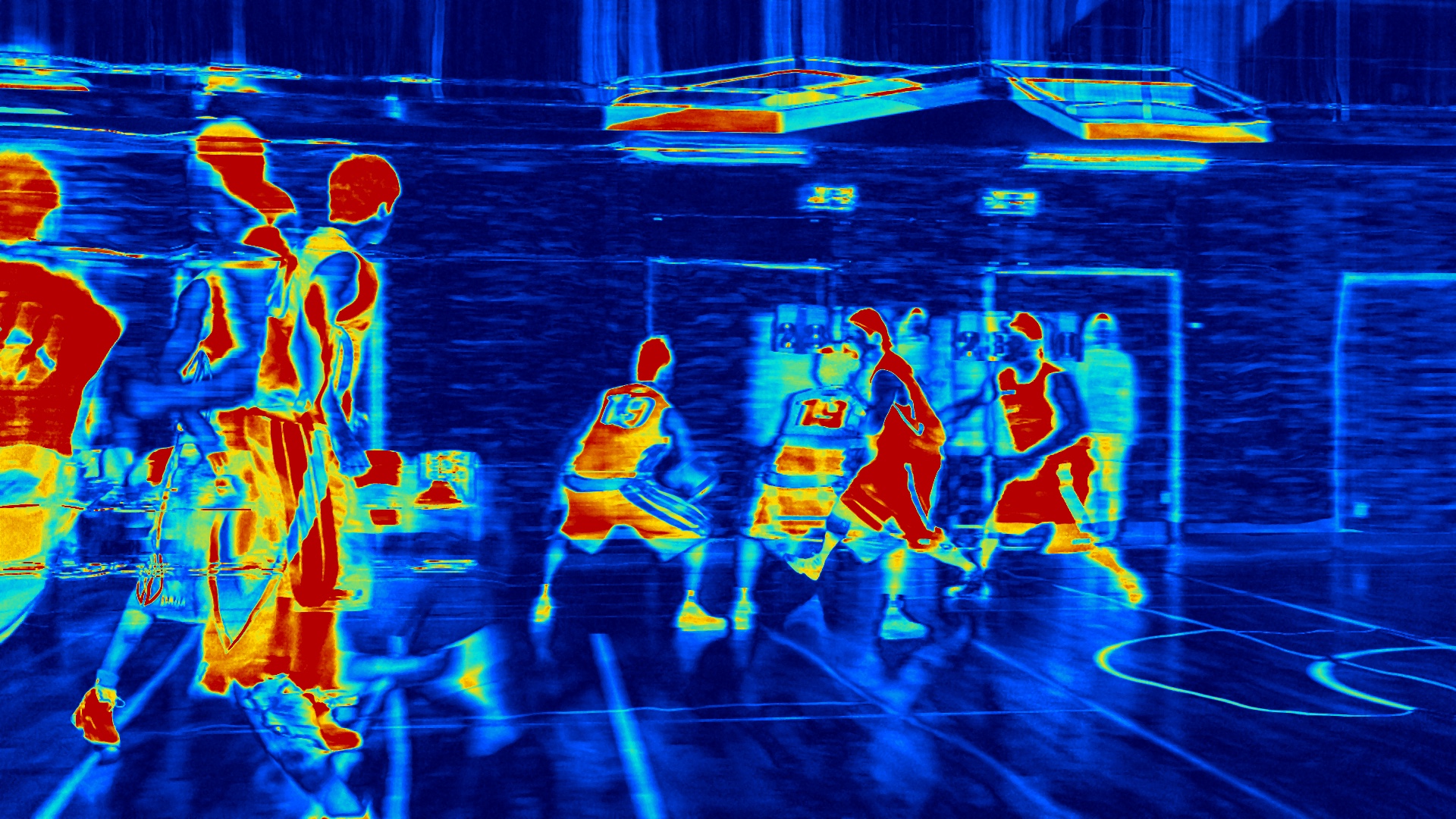} &
\includegraphics[width=0.115\textwidth]{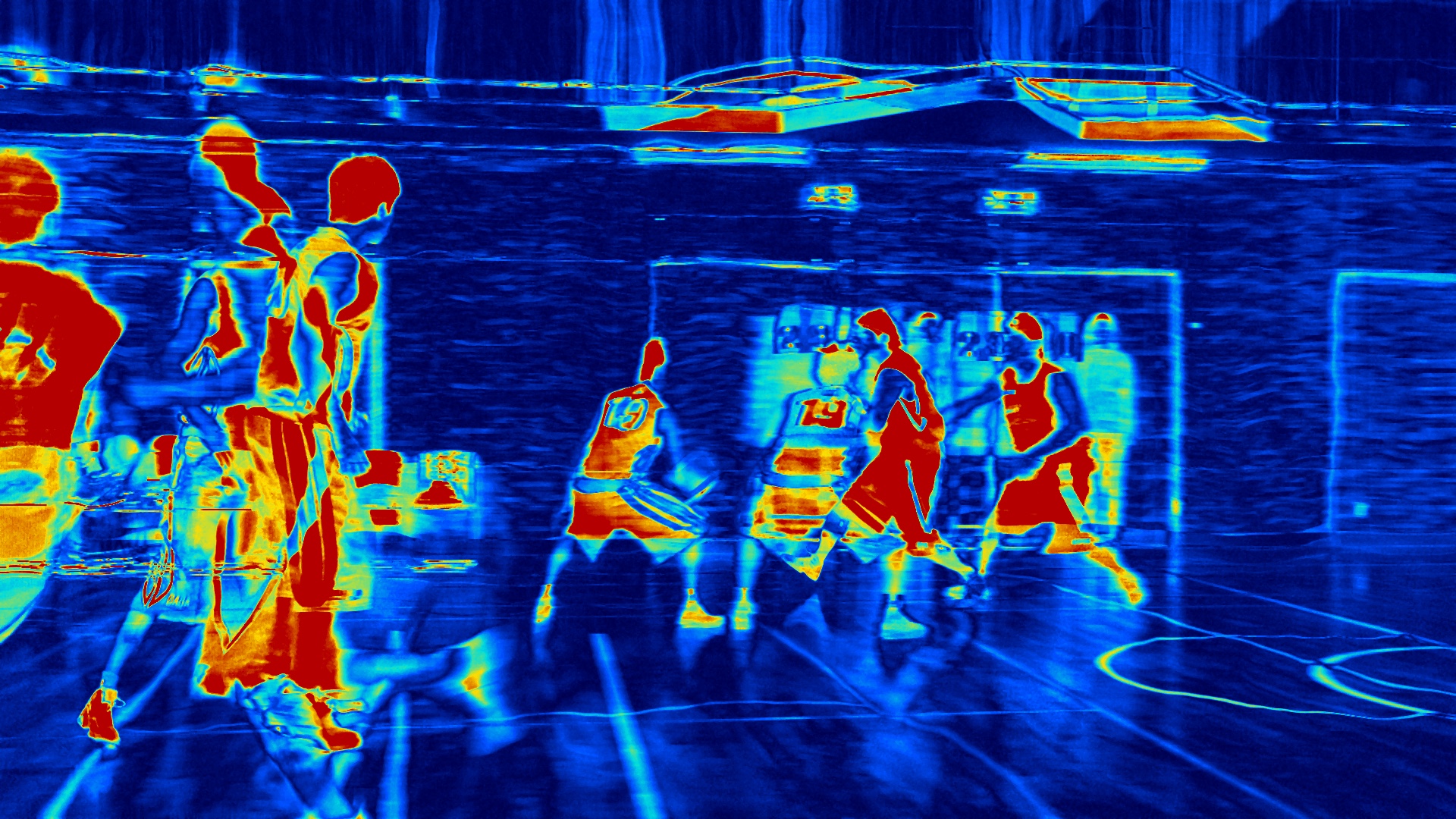} &
\includegraphics[width=0.115\textwidth]{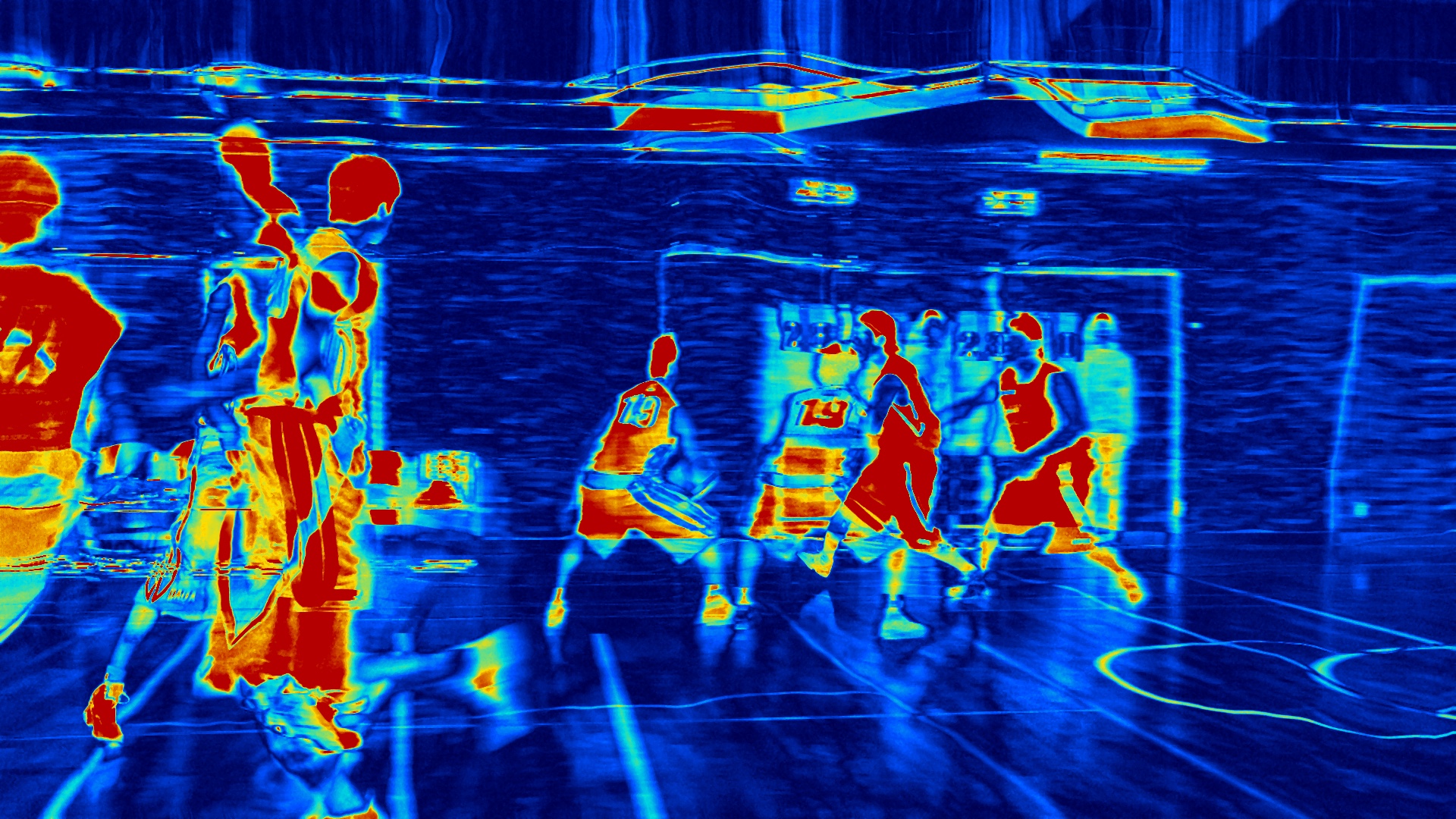} &
\includegraphics[width=0.115\textwidth]{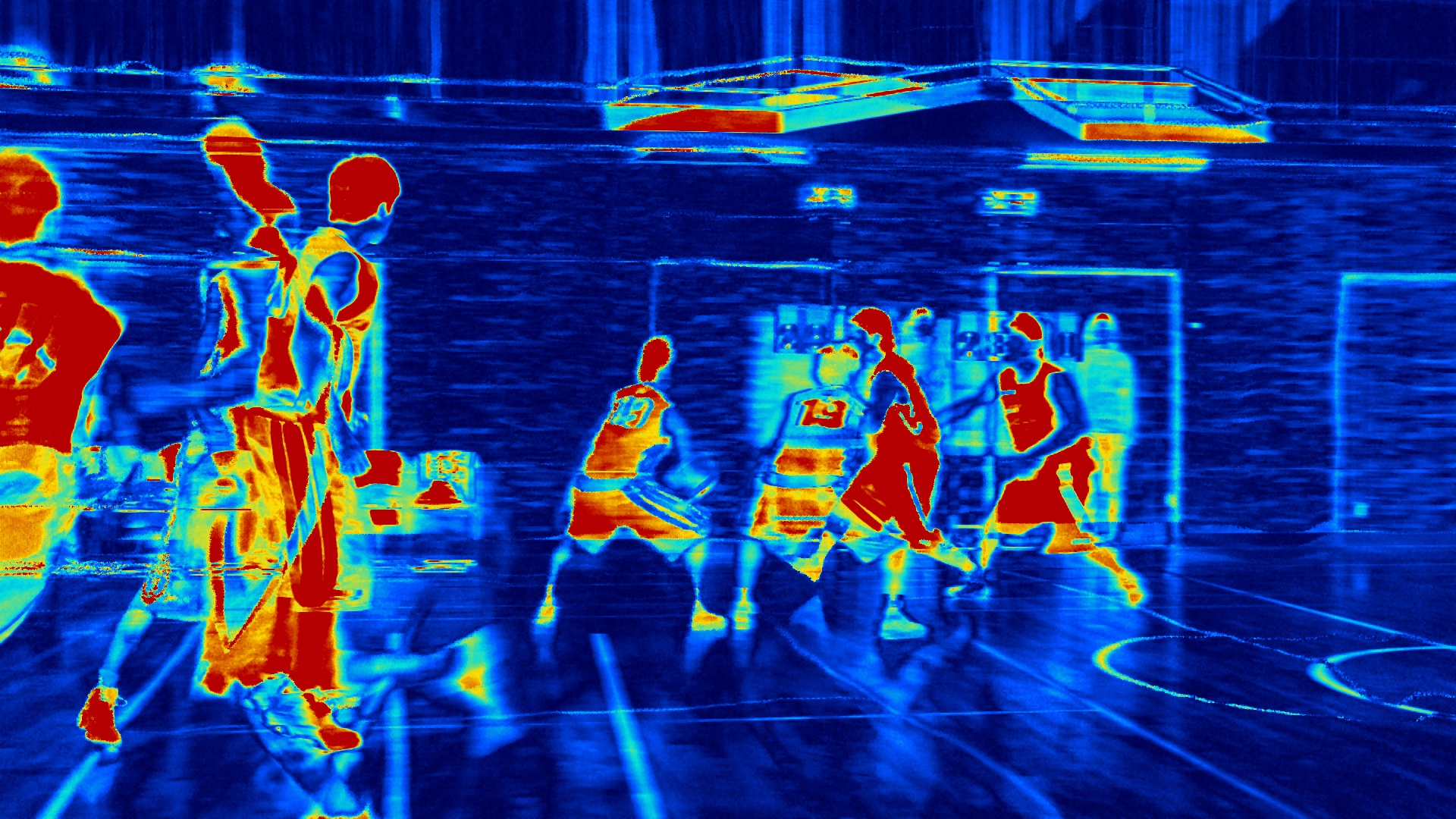} &
\includegraphics[width=0.115\textwidth]{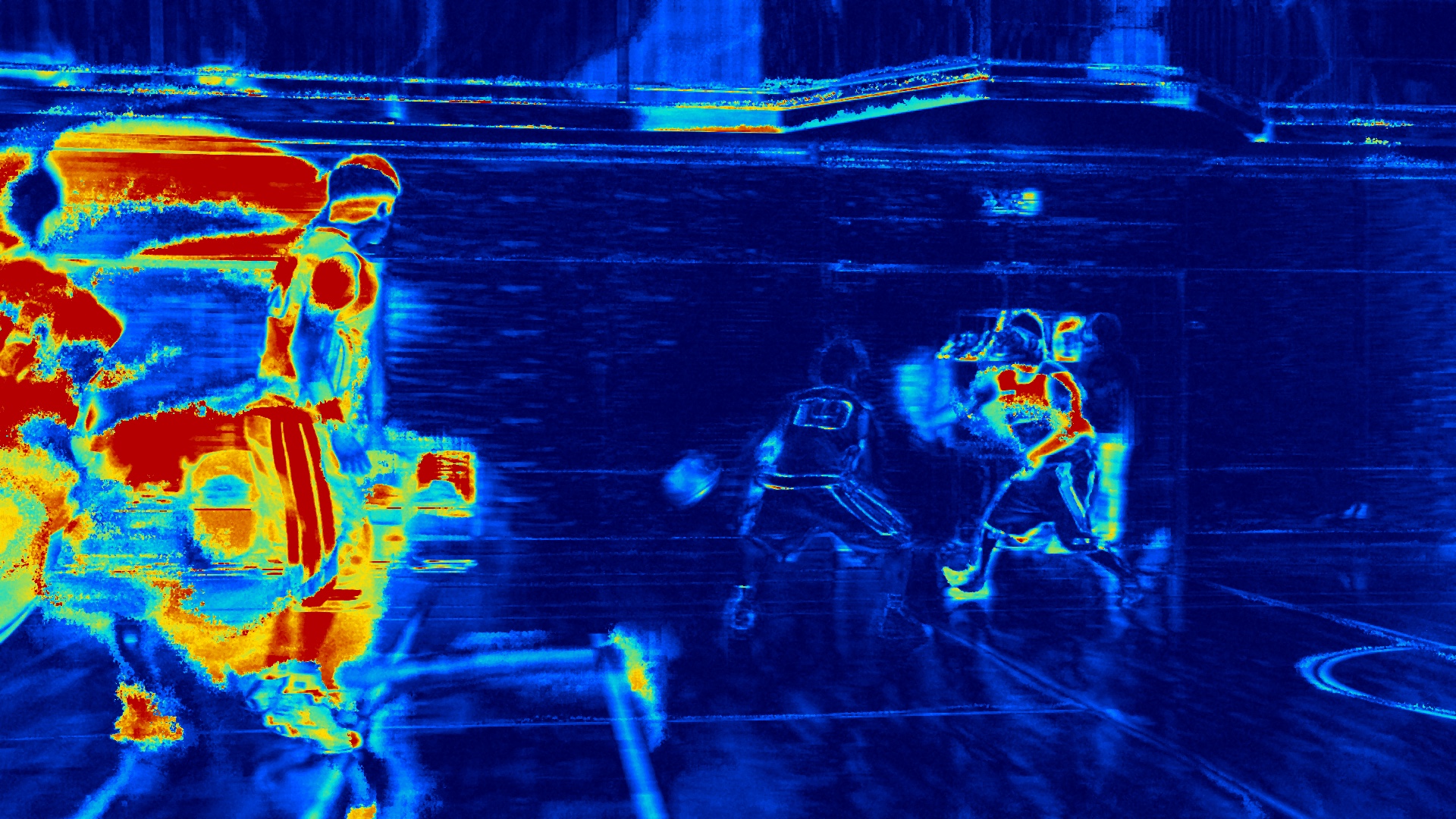} &
\includegraphics[width=0.115\textwidth]{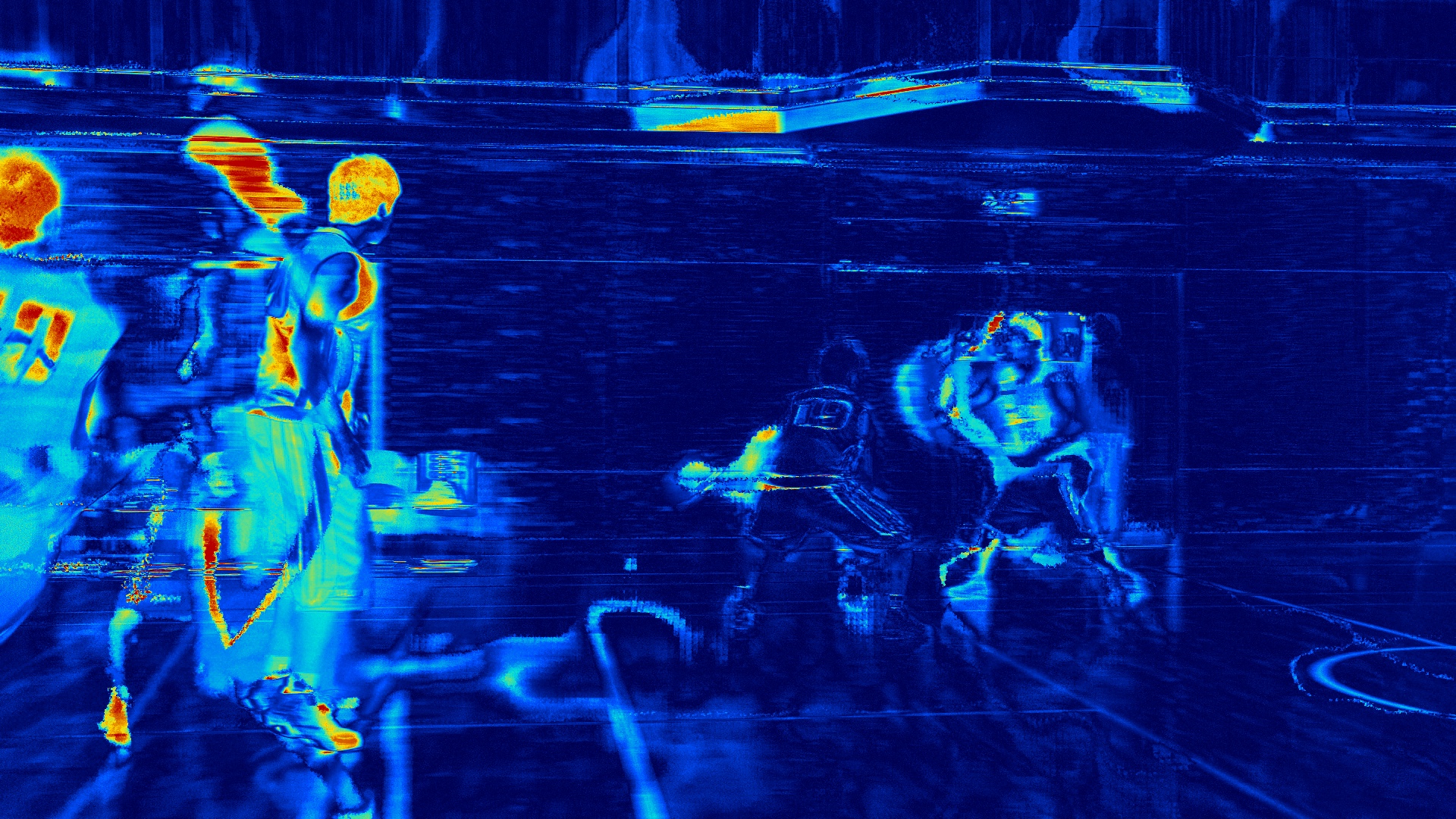} \\
\end{tabular}
\caption{Visualization of the proposed parallel accumulated motion estimation on BasketballDrive. The examples show motion fields, warped frames, and warping errors from POC 0 to POC 8 and POC 16.}
\label{fig:basketballdrive_pame_vis}
\end{figure*}
This design is enabled by the backward-flow formulation used in our motion estimator.
Specifically, \textit{the optical flow is estimated from the current frame to each reference frame, 
rather than from the reference frame to the current frame.}
During motion compensation, each current-frame pixel uses its backward flow to sample the corresponding pixel from the reference frame.
This backward warping avoids holes and, more importantly, \textit{defines all motion candidates on the coordinate
grid of the current frame}.
Let $\boldsymbol{p}$ denote a pixel location in the current frame.
For the $s$-th motion candidate, backward warping is written as
\begin{equation}
\tilde{\boldsymbol{x}}_{\tau\rightarrow t}^{(s)}(\boldsymbol{p})
=
\mathcal{W}\!\left(\boldsymbol{x}_\tau,\boldsymbol{v}_{t\rightarrow \tau}^{(s)}\right)(\boldsymbol{p})
=
\boldsymbol{x}_\tau\!\left(\boldsymbol{p}+\boldsymbol{v}_{t\rightarrow \tau}^{(s)}(\boldsymbol{p})\right),
\end{equation}
where bilinear interpolation is used for non-integer sampling locations.
The corresponding pixel-wise warping error is defined as
\begin{equation}
\boldsymbol{e}_{\tau\rightarrow t}^{(s)}(\boldsymbol{p})
=
\left\|
\boldsymbol{x}_t(\boldsymbol{p})
-
\tilde{\boldsymbol{x}}_{\tau\rightarrow t}^{(s)}(\boldsymbol{p})
\right\|_1.
\end{equation}
Therefore, the warping error at each pixel and the motion candidates at the same pixel are \textit{spatially aligned}.
We use this pixel-wise warping error as a reliability cue and feed it into a motion aggregation module to predict normalized weights for the multi-scale candidates.
This process is formulated as
\begin{equation}
\{\alpha_{t\rightarrow \tau}^{(s)}\}_{s=0}^{3}
=
\mathrm{Agg}\!\left(\{\boldsymbol{e}_{\tau\rightarrow t}^{(s)}\}_{s=0}^{3}\right),
\quad
\sum_{s=0}^{3}\alpha_{t\rightarrow \tau}^{(s)}(\boldsymbol{p})=1,
\end{equation}
where $\mathrm{Agg}(\cdot)$ denotes the motion aggregation module and
$\alpha_{t\rightarrow \tau}^{(s)}(\boldsymbol{p})$ is the normalized weight of the $s$-th candidate at pixel $\boldsymbol{p}$.
The aggregated motion field is obtained by weighted fusion:
\begin{equation}
\boldsymbol{v}_{t\rightarrow \tau}^{\mathrm{agg}}(\boldsymbol{p})
=
\sum_{s=0}^{3}
\alpha_{t\rightarrow \tau}^{(s)}(\boldsymbol{p})
\boldsymbol{v}_{t\rightarrow \tau}^{(s)}(\boldsymbol{p}).
\end{equation}
Benefiting from the aligned coordinate system, the aggregated motion field can be directly used to warp the reference frame.
The warped reference and the current frame are then fed into a motion vector refinement module, which predicts a residual correction to produce the final motion field.
This refinement step is written as
\begin{equation}
\Delta\boldsymbol{v}_{t\rightarrow \tau}^{d}
=
\mathrm{MVR}\!\left(
\boldsymbol{x}_t,
\mathcal{W}\!\left(\boldsymbol{x}_\tau,\boldsymbol{v}_{t\rightarrow \tau}^{\mathrm{agg}}\right)
\right),
\quad
\boldsymbol{v}_{t\rightarrow \tau}^d
=
\boldsymbol{v}_{t\rightarrow \tau}^{\mathrm{agg}}
+
\Delta\boldsymbol{v}_{t\rightarrow \tau}^{d},
\end{equation}
where $\mathrm{MVR}(\cdot)$ denotes the motion vector refinement module.
\subsubsection{Parallel-Accumulated Motion Estimation}
MSME enlarges the effective coverage of the optical flow network by estimating motion on downsampled inputs.
This improves direct flow estimation between the current frame and its references.
However, when the inter-frame motion is extremely large, the motion fields produced by MSME may still contain large errors, 
as illustrated in Fig.~\ref{fig:basketballdrive_pame_vis}.
A straightforward solution is to introduce more candidates by using coarser input scales.
This strategy, however, is not always effective, because stronger downsampling removes fine details that are important for accurate motion estimation.
Therefore, simply increasing the number of scales cannot fully resolve the large-motion problem.\par
To address the above issue, we propose parallel accumulated motion estimation (PAME).
The key idea is to estimate long-range motion in an indirect manner.
As a basic form of indirect estimation, one can decompose the motion from the current frame to
a distant reference into a set of shorter-range motions through intermediate frames.
When the motion changes smoothly over time, the motion from $t$ to $\tau$ can be approximated by
accumulating the motions along the temporal path between them.
Specifically, under the backward optical flow formulation, 
the motion fields cannot be directly summed 
because they are defined on different coordinate grids.
Let $\{t_n\}_{n=0}^{N}$ denote the temporal path from $\tau$ to $t$, where $t_0=\tau$ and $t_N=t$.
For each adjacent pair $(t_{n-1},t_n)$, we estimate a backward motion field $\boldsymbol{v}_{t_n\rightarrow t_{n-1}}$ from frame $t_n$ to frame $t_{n-1}$.
The accumulated motion from frame $t_n$ to the reference frame $\tau$ is then computed recursively as
\begin{equation}
\label{eq:serial_accumulation}
\begin{aligned}
\boldsymbol{v}_{t_1\rightarrow \tau}^{a}
&=
\boldsymbol{v}_{t_1\rightarrow t_0},\\
\boldsymbol{v}_{t_n\rightarrow \tau}^{a}
&=
\mathcal{W}\!\left(
\boldsymbol{v}_{t_{n-1}\rightarrow \tau}^{a},
\boldsymbol{v}_{t_n\rightarrow t_{n-1}}
\right)
+
\boldsymbol{v}_{t_n\rightarrow t_{n-1}}, \quad n=2,\ldots,N.
\end{aligned}
\end{equation}
Here, $\mathcal{W}(\cdot,\cdot)$ warps the previously accumulated motion field to the coordinate grid of frame $t_n$ using the current step motion $\boldsymbol{v}_{t_n\rightarrow t_{n-1}}$.
The final accumulated motion is $\boldsymbol{v}_{t\rightarrow \tau}^{a}=\boldsymbol{v}_{t_N\rightarrow \tau}^{a}$.
Although this accumulation strategy has been explored in prior work~\cite{zhai2025llbvc}, it is inherently sequential.
The motion field must be propagated step by step along the temporal path, which increases the encoding time for frames far from their references.
\par
\begin{figure*}
  \includegraphics[width=\textwidth]{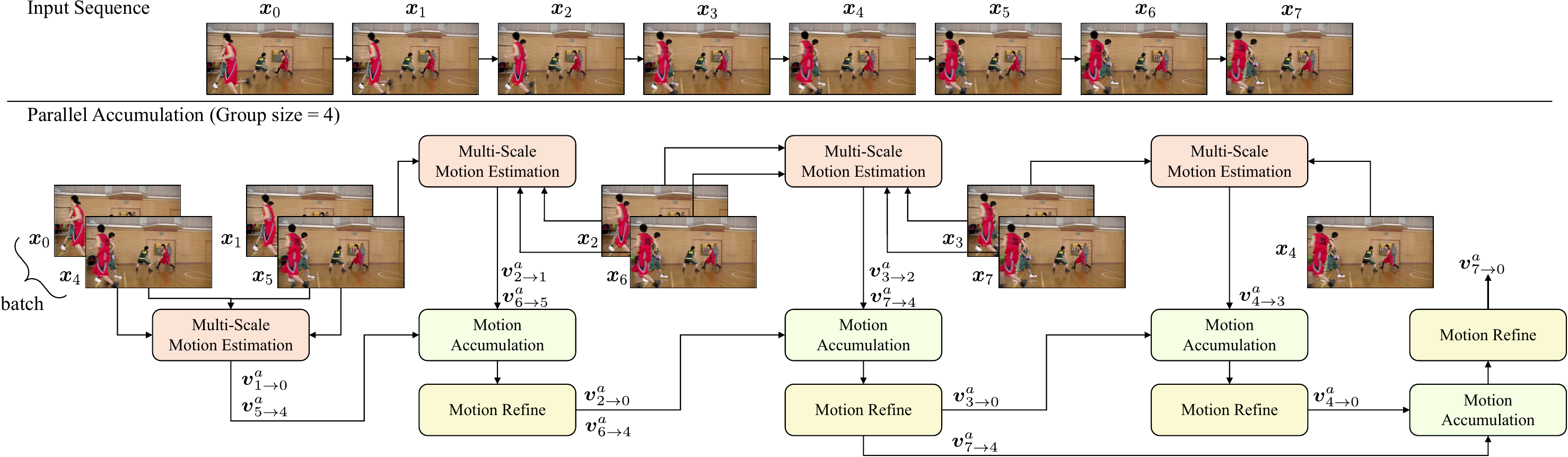}
  \caption{Architecture of the proposed parallel accumulated motion estimation.}
  \label{fig:abme}
\end{figure*}

To reduce this dependency, PAME adopts a group-wise parallel accumulation strategy, as illustrated in Fig.~\ref{fig:abme}.
Specifically, for a temporal path with $N$ adjacent intervals, we divide the intervals into $K=\lceil N/G\rceil$ groups, where $G$ is the group size.
The accumulations within different groups are independent and can therefore be concatenated along the batch dimension and evaluated in parallel.
This batched execution exposes more independent motion-estimation tasks to the GPU and reduces the latency caused by a purely step-by-step propagation.
If the last group contains fewer than $G$ intervals, it is simply treated as a shorter group and accumulated with its actual length.
Let the $k$-th group be defined by
\begin{equation}
a_k=t_{kG}, \quad
b_k=t_{\min((k+1)G,N)}, \quad
k=0,\ldots,K-1.
\end{equation}
The intra-group accumulated motion is computed as
\begin{equation}
\boldsymbol{u}_k
=
\mathcal{A}(b_k,a_k),
\end{equation}
where $\mathcal{A}(b_k,a_k)$ denotes the recursive accumulation from $b_k$ to $a_k$ using Eq.~(\ref{eq:serial_accumulation}).
All $\{\boldsymbol{u}_k\}_{k=0}^{K-1}$ are independent and can be evaluated in \textit{parallel}.
The final long-range motion is obtained by composing these group-level motions:
\begin{equation}
\begin{aligned}
\boldsymbol{V}_0
&=
\boldsymbol{u}_0,\\
\boldsymbol{V}_k
&=
\mathcal{W}\!\left(
\boldsymbol{V}_{k-1},
\boldsymbol{u}_k
\right)
+
\boldsymbol{u}_k,
\quad k=1,\ldots,K-1,\\
\boldsymbol{v}_{t\rightarrow \tau}^{a}
&=
\boldsymbol{V}_{K-1}.
\end{aligned}
\end{equation}
In this way, the expensive short-range accumulations are performed in parallel within groups, while only the lightweight group-level composition remains sequential.
This design approximately doubles the encoding throughput compared with serial accumulated-motion estimation on 1080p sequences.
\begin{figure*}
  \centering
  \includegraphics[width=\textwidth]{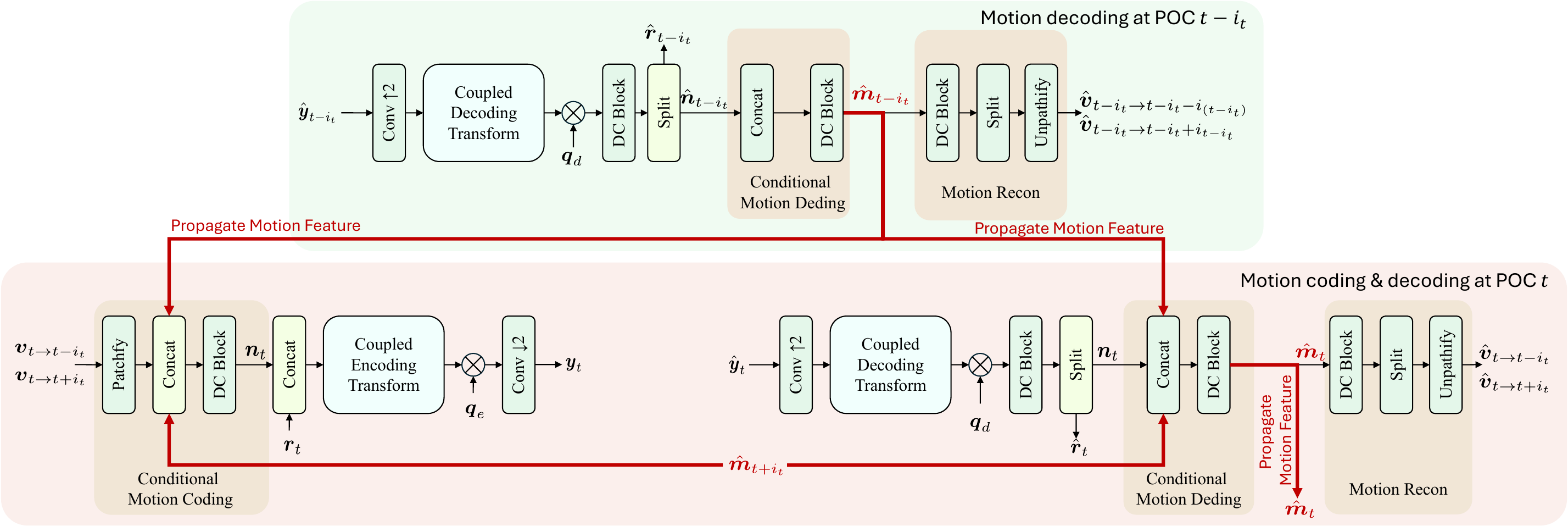}
  \caption{Architecture of the proposed bidirectional motion feature propagation.}
  \label{fig:mfp}
\end{figure*}
\subsubsection{Motion Vector Merging}
MSME and PAME produce two complementary motion candidates, denoted by
$\boldsymbol{v}_{t\rightarrow \tau}^{d}$ and $\boldsymbol{v}_{t\rightarrow \tau}^{a}$, respectively.
Since only one motion field is used for compensation, we merge them with a pixel-wise candidate fusion strategy, as illustrated in Fig.~\ref{fig:basketballdrive_pame_vis}.
Specifically, we compute the warping errors of the two candidates and use them to predict normalized weights
$\beta_{t\rightarrow \tau}^{d}$ and $\beta_{t\rightarrow \tau}^{a}$.
The merged motion field is obtained by weighted fusion:
\begin{equation}
\boldsymbol{v}_{t\rightarrow \tau}^{mg}(\boldsymbol{p})
=
\beta_{t\rightarrow \tau}^{d}(\boldsymbol{p})
\boldsymbol{v}_{t\rightarrow \tau}^{d}(\boldsymbol{p})
+
\beta_{t\rightarrow \tau}^{a}(\boldsymbol{p})
\boldsymbol{v}_{t\rightarrow \tau}^{a}(\boldsymbol{p}).
\end{equation}
Here, the weights are normalized at each pixel, i.e.,
$\beta_{t\rightarrow \tau}^{d}(\boldsymbol{p})+\beta_{t\rightarrow \tau}^{a}(\boldsymbol{p})=1$.
We then refine the merged motion field with a motion vector refinement module.
The merged motion is first used to warp the reference frame, and the warped reference, the current frame, and the merged motion are fed into MVR to predict a residual correction:
\begin{equation}
\begin{aligned}
\Delta\boldsymbol{v}_{t\rightarrow \tau}
&=
\mathrm{MVR}\!\left(
\boldsymbol{x}_t,
\mathcal{W}\!\left(\boldsymbol{x}_\tau,\boldsymbol{v}_{t\rightarrow \tau}^{mg}\right),
\boldsymbol{v}_{t\rightarrow \tau}^{mg}
\right),\\
\boldsymbol{v}_{t\rightarrow \tau}
&=
\boldsymbol{v}_{t\rightarrow \tau}^{mg}
+
\Delta\boldsymbol{v}_{t\rightarrow \tau}.
\end{aligned}
\end{equation}
\subsection{Bidirectional Motion Feature Propagation}
\label{sec:bmfp}
After obtaining more accurate motion vectors from MCME, an efficient motion compression strategy is still needed.
In BiCRVC, motion information is transmitted to the decoder through coupled representation coding.
The quality of motion compression directly affects the accuracy of decoder-side context alignment and thus the final reconstruction quality.
Moreover, the accumulated motion used in BiCRVC can have a larger value range than directly estimated motion, which makes motion compression more challenging.
\par
To address these issues, we propose bidirectional motion feature propagation (BMFP), which reuses motion priors from previously coded layers.
The architecture of BMFP is illustrated in Fig.~\ref{fig:mfp}.
For the current POC $t$, BMFP uses two propagated motion features, denoted by
$\hat{\boldsymbol{m}}_{t-i_t}$ and $\hat{\boldsymbol{m}}_{t+i_t}$.
They are produced when the motions of the two reference POCs are decoded, and are reused as motion priors for the current motion coding.
For the first coding layer, the two references are intra frames.
Since intra frames do not have decoded motion features, no motion prior is available at this layer.
We therefore set the two motion priors to zero, i.e., $\hat{\boldsymbol{m}}_{t-i_t}=\hat{\boldsymbol{m}}_{t+i_t}=\boldsymbol{0}$.
At the encoder, conditional motion encoding (CME) first patchifies the two motion vectors and concatenates them with the two propagated motion features:
\begin{equation}
\boldsymbol{n}_t
=
\mathrm{CME}\!\left(
\mathrm{Concat}\!\left(
\mathcal{P}(\boldsymbol{v}_{t\rightarrow t-i_t}),
\mathcal{P}(\boldsymbol{v}_{t\rightarrow t+i_t}),
\hat{\boldsymbol{m}}_{t-i_t},
\hat{\boldsymbol{m}}_{t+i_t}
\right)
\right).
\end{equation}
No spatial alignment is applied between the propagated motion features and the patchified motion representation.
This keeps the motion coding path simple, which is important because motion usually occupies a relatively small portion of the total bit rate.
Unlike DCVC-MIP~\cite{Qi_2023_CVPR}, which propagates motion information for forward prediction through complex motion-frame interactions, 
BMFP directly reuses decoded motion features from two reference POCs for bidirectional random-access coding, 
leading to a simpler and lower-complexity motion coding path, which is important for practical neural video compression systems.
The resulting $\boldsymbol{n}_t$ is then concatenated with $\boldsymbol{r}_t$ and fed into the coupled encoding transform for unified coding.
\par
At the decoder, BMFP decodes motion in two steps.
First, the decoded motion representation $\hat{\boldsymbol{n}}_t$ is split from the coupled representation.
It is then decoded by conditional motion decoding (CMD) with the same two propagated motion features to produce the current motion feature:
\begin{equation}
\begin{aligned}
\hat{\boldsymbol{r}}_t,\hat{\boldsymbol{n}}_t
&=
\mathrm{Split}\!\left(\mathrm{CRD}(\hat{\boldsymbol{y}}_t)\right),\\
\hat{\boldsymbol{m}}_t
&=
\mathrm{CMD}\!\left(
\mathrm{Concat}\!\left(
\hat{\boldsymbol{n}}_t,
\hat{\boldsymbol{m}}_{t-i_t},
\hat{\boldsymbol{m}}_{t+i_t}
\right)
\right).
\end{aligned}
\end{equation}
The motion feature $\hat{\boldsymbol{m}}_t$ is propagated to subsequent coding layers.
It is also fed into a motion reconstruction module to obtain the decoded bidirectional motion vectors:
\begin{equation}
\left(
\hat{\boldsymbol{v}}_{t\rightarrow t-i_t},
\hat{\boldsymbol{v}}_{t\rightarrow t+i_t}
\right)
=
\mathrm{MRec}\!\left(\hat{\boldsymbol{m}}_t\right).
\end{equation}
\subsection{Optimized Training Strategy}
\subsubsection{Coupled Distortion Training}
The coupled representation in BiCRVC makes training more challenging than 
in conventional BVC frameworks~\cite{sheng2025bi,jiang2025biecvc}.
The difficulty comes from the shared latent representation.
A single model must encode both frame and motion information,
and decode them jointly from the same latent.
Therefore, the common strategy of optimizing motion coding and frame coding separately is not well suited to BiCRVC.
Instead, the training objective should jointly supervise 
frame reconstruction and motion reconstruction.
To this end, we introduce a multi-stage coupled distortion 
training strategy.
In the first stage, we train the model with a coupled objective that jointly considers frame distortion and motion distortion.
Since BiCRVC adopts multi-candidate motion estimation, we supervise motion by the warping quality of different motion fields.
For a reference POC $\tau\in\{t-i_t,t+i_t\}$ and a motion field $\boldsymbol{u}_{t\rightarrow \tau}$, the warping loss is defined as
\begin{equation}
\mathcal{D}_{w}\!\left(\boldsymbol{u}_{t\rightarrow \tau}\right)
=
\left\|
\boldsymbol{x}_t
-
\mathcal{W}\!\left(\boldsymbol{x}_{\tau},\boldsymbol{u}_{t\rightarrow \tau}\right)
\right\|_2^2,
\end{equation}
where the squared $\ell_2$ norm is averaged over all pixels and color channels.
The motion distortion includes the four MSME scale candidates, the direct motion field $\boldsymbol{v}_{t\rightarrow \tau}^{d}$, the accumulated motion field $\boldsymbol{v}_{t\rightarrow \tau}^{a}$, the final estimated motion field $\boldsymbol{v}_{t\rightarrow \tau}$, and the reconstructed motion field $\hat{\boldsymbol{v}}_{t\rightarrow \tau}$.
We collect them as
\begin{equation}
\mathcal{V}_{t\rightarrow \tau}
=
\left\{
\boldsymbol{v}_{t\rightarrow \tau}^{(s)}
\right\}_{s=0}^{3}
\cup
\left\{
\boldsymbol{v}_{t\rightarrow \tau}^{d},
\boldsymbol{v}_{t\rightarrow \tau}^{a},
\boldsymbol{v}_{t\rightarrow \tau},
\hat{\boldsymbol{v}}_{t\rightarrow \tau}
\right\}.
\end{equation}
The motion distortion is then written compactly as
\begin{equation}
\mathcal{D}_{m}
=
\sum_{\tau\in\{t-i_t,t+i_t\}}
\sum_{\boldsymbol{u}\in\mathcal{V}_{t\rightarrow \tau}}
\mathcal{D}_{w}\!\left(\boldsymbol{u}\right).
\end{equation}
The frame distortion is defined between the reconstructed frame $\hat{\boldsymbol{x}}_t$ and the current frame $\boldsymbol{x}_t$:
\begin{equation}
\mathcal{D}_{f}
=
\left\|
\boldsymbol{x}_t
-
\hat{\boldsymbol{x}}_t
\right\|_2^2,
\end{equation}
where the squared $\ell_2$ norm is also averaged over all pixels and color channels.
The first-stage loss is then given by
\begin{equation}
\mathcal{L}_{1}
=
\mathcal{D}_{f}+\mathcal{D}_{m}.
\end{equation}
In the second stage, we add the rate term of the coupled latent $\hat{\boldsymbol{y}}_t$ to learn joint compression of motion and frame information.
For variable-rate training, we randomly sample a quality level $q\in\{0,\ldots,63\}$ and use the corresponding Lagrange multiplier $\lambda_q$.
Following DCVC-FM~\cite{li2024neural}, $\lambda_q$ is interpolated between $\lambda_{\min}=1$ and $\lambda_{\max}=768$.
The second-stage loss is written as
\begin{equation}
\mathcal{L}_{2}
=
\mathcal{R}\!\left(\hat{\boldsymbol{y}}_t\right)
+
\lambda_q
\left(
\mathcal{D}_{f}+\mathcal{D}_{m}
\right),
\end{equation}
where $\mathcal{R}(\hat{\boldsymbol{y}}_t)$ denotes the estimated bit rate of the coupled representation.
In the final stage, we remove the motion distortion term and optimize the model only with respect to the rate-distortion tradeoff of frame reconstruction.
At this point, the model has already learned to estimate and reconstruct useful motion fields, and removing the explicit motion constraint allows the final objective to focus on compression performance.
The final-stage loss is
\begin{equation}
\mathcal{L}_{3}
=
\mathcal{R}\!\left(\hat{\boldsymbol{y}}_t\right)
+
\lambda_q
\mathcal{D}_{f}.
\end{equation}
\subsubsection{Random GOP Structure Training}
BiCRVC follows the VTM-like GOP32 structure~\cite{bross2021overview} in Fig.~\ref{fig:gop32}, which contains six hierarchical coding layers.
Lower-layer frames are referenced more often, so their reconstruction quality has a larger impact on later frames.
We therefore assign each layer $\ell$ a quality adapter and use a layer-dependent weight $w_\ell$ to 
modulate the Lagrange multiplier~\cite{sheng2025bi,jiang2025biecvc}:
\begin{equation}
\lambda_{q,\ell}=w_\ell\lambda_q.
\end{equation}
The weighted multiplier $\lambda_{q,\ell}$ replaces $\lambda_q$ in the second- and final-stage losses.
This gives lower layers stronger rate-distortion constraints while keeping all layers in the same variable-rate framework.
\par
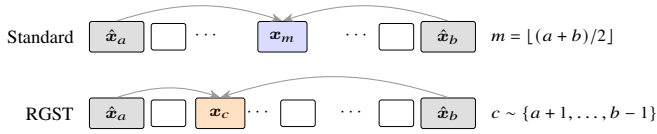
\begin{figure}[t]
  \centering
  \footnotesize
  \resizebox{\columnwidth}{!}{
  \begin{tikzpicture}[x=0.78cm,y=0.72cm]
    \tikzset{
      ref/.style={draw, rounded corners=1pt, fill=gray!25, minimum width=0.82cm, minimum height=0.46cm, inner sep=1pt, font=\scriptsize},
      mid/.style={draw, rounded corners=1pt, fill=blue!14, minimum width=0.74cm, minimum height=0.46cm, inner sep=1pt, font=\scriptsize},
      rnd/.style={draw, rounded corners=1pt, fill=orange!24, minimum width=0.74cm, minimum height=0.46cm, inner sep=1pt, font=\scriptsize},
      cand/.style={draw, rounded corners=1pt, fill=white, minimum width=0.52cm, minimum height=0.40cm, inner sep=1pt, font=\scriptsize},
      arr/.style={->, gray!80, line width=0.45pt, >=Stealth}
    }

    \node[anchor=east] at (-0.7,1.6) {Standard};
    \node[ref] (sa) at (0,1.6) {$\hat{\boldsymbol{x}}_a$};
    \node[cand] at (1.0,1.6) {};
    \node[font=\scriptsize] at (1.75,1.6) {$\cdots$};
    \node[mid] (sm) at (3.2,1.6) {$\boldsymbol{x}_m$};
    \node[font=\scriptsize] at (4.65,1.6) {$\cdots$};
    \node[cand] at (5.4,1.6) {};
    \node[ref] (sb) at (6.4,1.6) {$\hat{\boldsymbol{x}}_b$};
    \draw[arr] (sa.north) to[bend left=18] (sm.north);
    \draw[arr] (sb.north) to[bend right=18] (sm.north);
    \node[anchor=west, font=\scriptsize] at (7.1,1.6) {$m=\lfloor(a+b)/2\rfloor$};

    \node[anchor=east] at (-0.7,0.0) {RGST};
    \node[ref] (ra) at (0,0.0) {$\hat{\boldsymbol{x}}_a$};
    \node[cand] at (1.0,0.0) {};
    \node[rnd] (rc) at (2.0,0.0) {$\boldsymbol{x}_c$};
    \node[font=\scriptsize] at (2.75,0.0) {$\cdots$};
    \node[cand] at (3.5,0.0) {};
    \node[font=\scriptsize] at (4.55,0.0) {$\cdots$};
    \node[cand] at (5.4,0.0) {};
    \node[ref] (rb) at (6.4,0.0) {$\hat{\boldsymbol{x}}_b$};
    \draw[arr] (ra.north) to[bend left=18] (rc.north);
    \draw[arr] (rb.north) to[bend right=18] (rc.north);
    \node[anchor=west, font=\scriptsize] at (7.1,0.0) {$c\sim\{a+1,\ldots,b-1\}$};

  \end{tikzpicture}}
  \caption{Illustration of random GOP structure training (RGST). The standard hierarchy uses a fixed midpoint in each reference interval, while RGST randomly samples the next coded frame between two decoded reference frames and recursively partitions the two resulting sub-intervals.}
  \label{fig:rgst}
\end{figure}

Pretraining, however, is usually performed on short clips.
For example, Vimeo90K~\cite{xue2019video} contains only seven frames per clip, and training with long sequences from the beginning is costly and less stable.
With a fixed midpoint partition, such short clips cover fewer coding layers than GOP32 and cannot expose the model to all layer-quality cases.
To reduce this mismatch, we adopt random GOP structure training (RGST).
Starting from the reference interval $(0,N-1)$, the standard hierarchy chooses the next B frame as the midpoint of each interval.
In RGST, for an interval bounded by decoded reference frames $\hat{\boldsymbol{x}}_a$ and $\hat{\boldsymbol{x}}_b$ at POCs $a$ and $b$, we instead sample the next coded POC $c$ from $\{a+1,\ldots,b-1\}$ and code $\boldsymbol{x}_c$ with $\hat{\boldsymbol{x}}_a$ and $\hat{\boldsymbol{x}}_b$ as references.
The same rule is then applied recursively to the two sub-intervals, as illustrated in Fig.~\ref{fig:rgst}.
In this way, even seven-frame clips can cover the six hierarchical quality layers used by GOP32 during pretraining.
\begin{figure*}
\centering
\includegraphics[width=\textwidth]{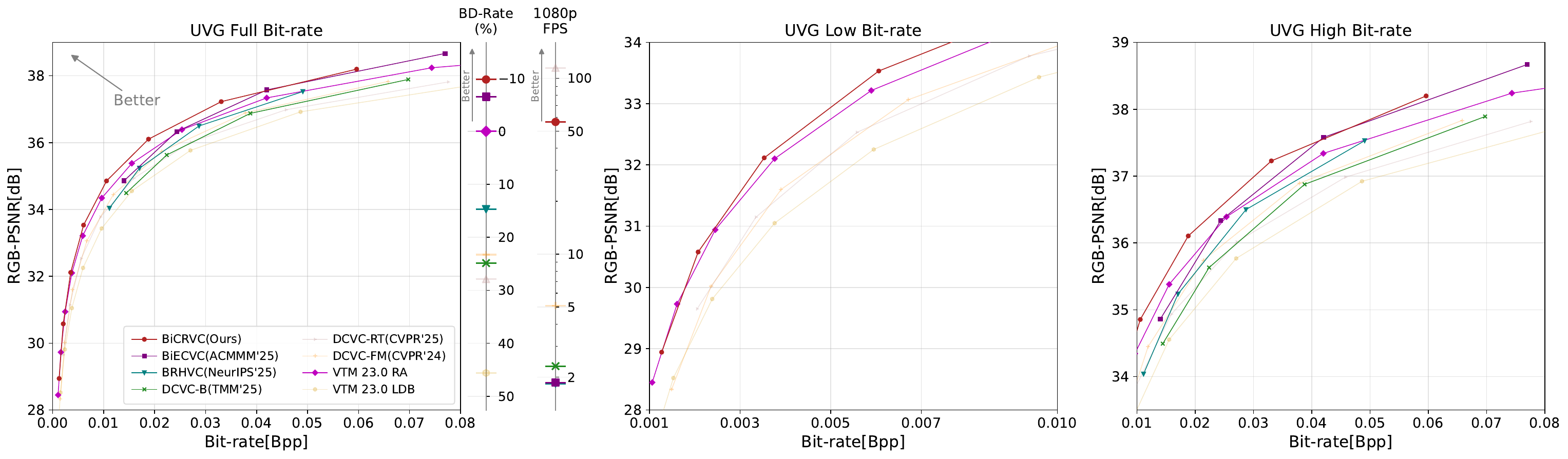}\\
\includegraphics[width=\textwidth]{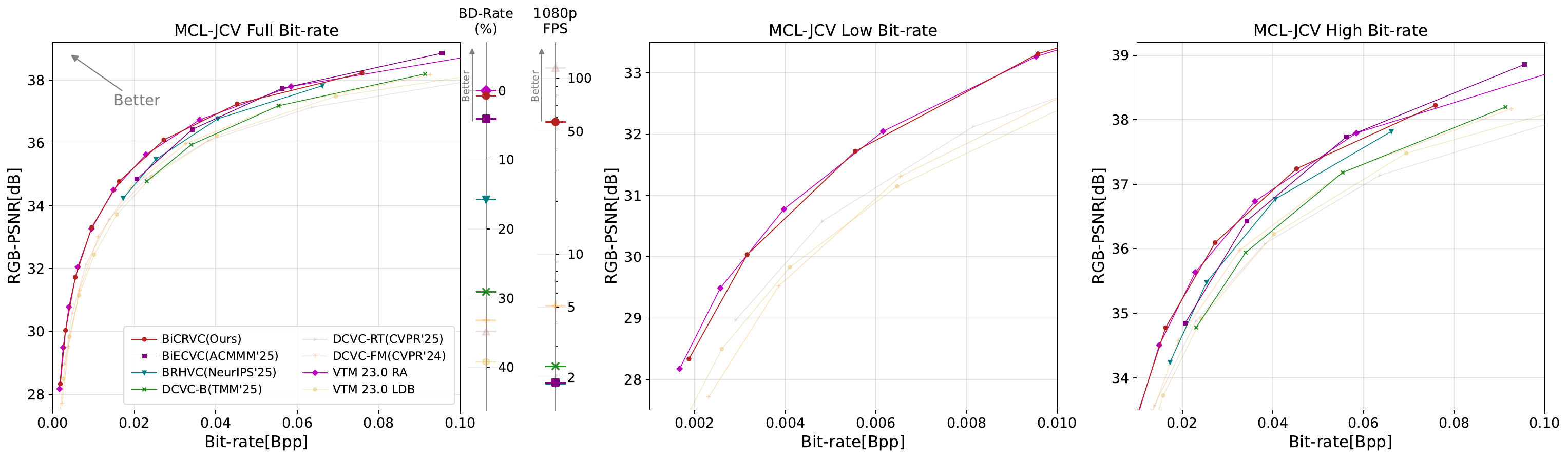}\\
\includegraphics[width=\textwidth]{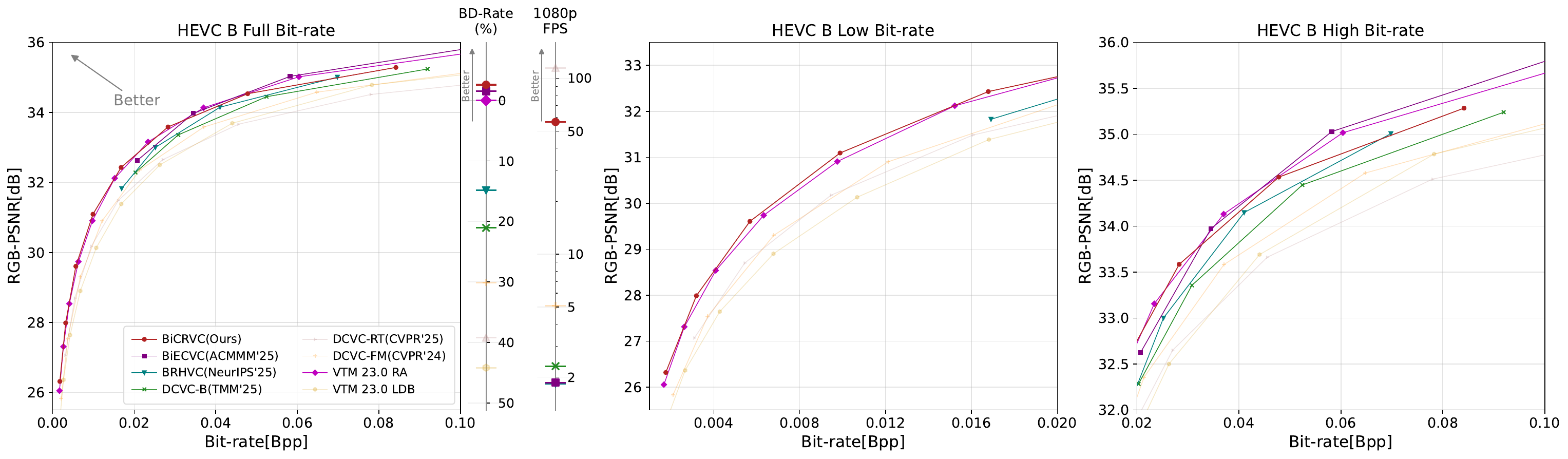}
\caption{Rate-distortion curves for HEVC B, UVG, and MCL-JCV.
96 frames are tested in RGB color space with intra-period=32.}
\label{fig:rd_ip32_f96}
\end{figure*}

\begin{figure*}
\centering
\includegraphics[width=\textwidth]{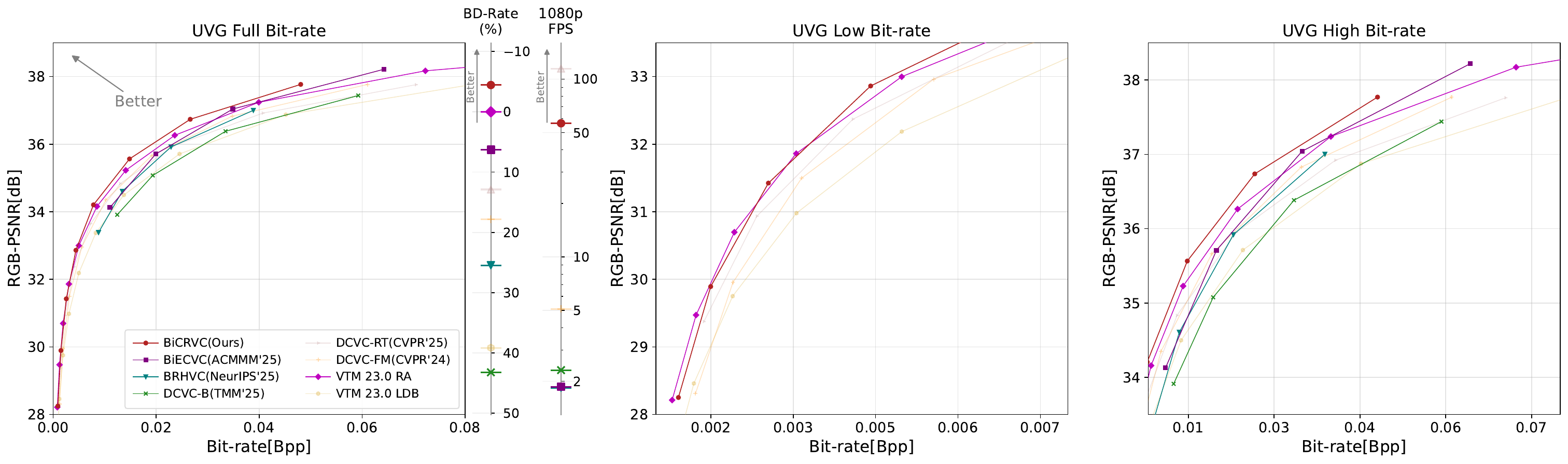}\\
\includegraphics[width=\textwidth]{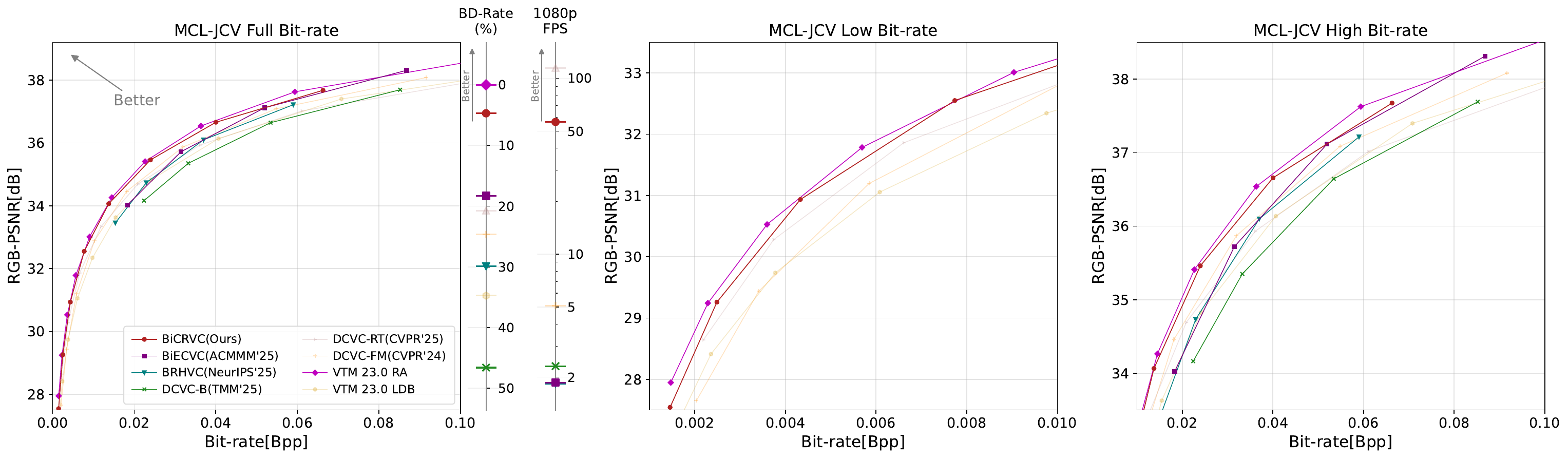}\\
\includegraphics[width=\textwidth]{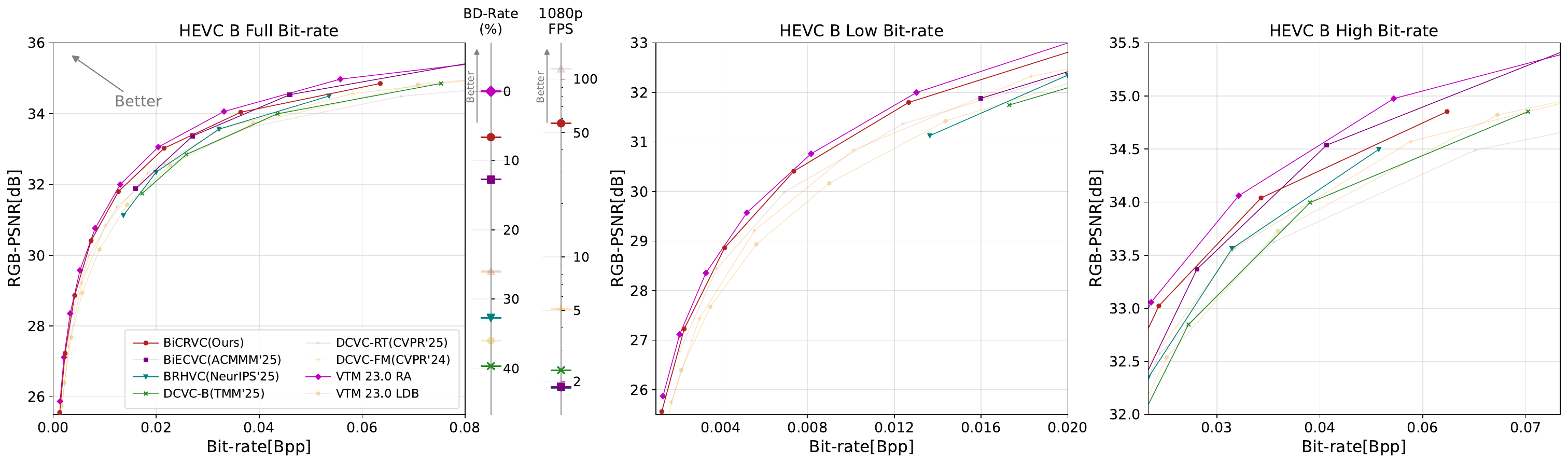}
\caption{Rate-distortion curves for HEVC B, UVG, and MCL-JCV.
64 frames are tested in RGB color space with intra-period=64.}
\label{fig:rd_ip64_f64}
\end{figure*}

\begin{table*}[t]
  \centering
  \setlength{\tabcolsep}{1.4mm}
  \tiny
  \caption{BD-Rate $(\%)$~\cite{bjontegaard2001calculation} comparison for \textbf{RGB PSNR} (dB) under different frame numbers (FN) and intra periods (IP). The anchor is \textbf{VTM 23.0 RA (GOP 32)}~\cite{bross2021overview}.}
  \renewcommand{\arraystretch}{1.12}
  \begin{threeparttable}
  \resizebox{0.98\textwidth}{!}{
  \begin{tabular}{@{}cccccccccccc@{}}
  \toprule
  Method & Venue & Type & FN & IP & HEVC B & HEVC C & HEVC D & HEVC E & UVG & MCL-JCV & Average \\
  \midrule
  \textcolor{gray!65}{VTM 23.0 LDB~\cite{bross2021overview}} & \textcolor{gray!65}{---} & \textcolor{gray!65}{LD} & \textcolor{gray!65}{96} & \textcolor{gray!65}{32} & \textcolor{gray!65}{44.2} & \textcolor{gray!65}{35.9} & \textcolor{gray!65}{35.7} & \textcolor{gray!65}{36.6} & \textcolor{gray!65}{45.6} & \textcolor{gray!65}{39.2} & \textcolor{gray!65}{39.5}  \\
  \rowcolor{lightgray}\textcolor{gray!65}{DCVC-DC~\cite{li2023neural}} & \textcolor{gray!65}{CVPR'23} & \textcolor{gray!65}{LD} & \textcolor{gray!65}{96} & \textcolor{gray!65}{32} & \textcolor{gray!65}{24.9} & \textcolor{gray!65}{25.1} & \textcolor{gray!65}{0.1} & \textcolor{gray!65}{14.4} & \textcolor{gray!65}{4.6} & \textcolor{gray!65}{12.6} & \textcolor{gray!65}{13.6} \\
  \textcolor{gray!65}{DCVC-FM~\cite{li2024neural}} & \textcolor{gray!65}{CVPR'24} & \textcolor{gray!65}{LD} & \textcolor{gray!65}{96} & \textcolor{gray!65}{32} & \textcolor{gray!65}{30.1} & \textcolor{gray!65}{11.0} & \textcolor{gray!65}{-7.5} & \textcolor{gray!65}{6.8} & \textcolor{gray!65}{23.3} & \textcolor{gray!65}{33.2} & \textcolor{gray!65}{16.1} \\
  \rowcolor{lightgray}\textcolor{gray!65}{DCMVC~\cite{tang2025neural}} & \textcolor{gray!65}{CVPR'25} & \textcolor{gray!65}{LD} & \textcolor{gray!65}{96} & \textcolor{gray!65}{32} & \textcolor{gray!65}{13.4} & \textcolor{gray!65}{13.5} & \textcolor{gray!65}{-10.1} & \textcolor{gray!65}{-6.4} & \textcolor{gray!65}{0.9} & \textcolor{gray!65}{9.0} & \textcolor{gray!65}{3.4} \\
  \textcolor{gray!65}{SEVC~\cite{bian2025augmented}} & \textcolor{gray!65}{CVPR'25} & \textcolor{gray!65}{LD} & \textcolor{gray!65}{96} & \textcolor{gray!65}{32} & \textcolor{gray!65}{22.8} & \textcolor{gray!65}{16.3} & \textcolor{gray!65}{-2.4} & \textcolor{gray!65}{2.6} & \textcolor{gray!65}{-0.2} & \textcolor{gray!65}{2.1} & \textcolor{gray!65}{6.9} \\
  \rowcolor{lightgray}\textcolor{gray!65}{ECVC~\cite{jiang2025ecvc}} & \textcolor{gray!65}{CVPR'25} & \textcolor{gray!65}{LD} & \textcolor{gray!65}{96} & \textcolor{gray!65}{32} & \textcolor{gray!65}{11.6} & \textcolor{gray!65}{7.2} & \textcolor{gray!65}{-12.7} & \textcolor{gray!65}{17.2} & \textcolor{gray!65}{-5.5} & \textcolor{gray!65}{3.0} & \textcolor{gray!65}{3.5} \\
  \textcolor{gray!65}{DCVC-RT~\cite{jia2025practical}} & \textcolor{gray!65}{CVPR'25} & \textcolor{gray!65}{LD} & \textcolor{gray!65}{96} & \textcolor{gray!65}{32} & \textcolor{gray!65}{38.3} & \textcolor{gray!65}{32.4} & \textcolor{gray!65}{15.9} & \textcolor{gray!65}{19.2} & \textcolor{gray!65}{27.9} & \textcolor{gray!65}{34.8} & \textcolor{gray!65}{28.1} \\
  \rowcolor{lightgray}\textcolor{gray!65}{HyTIP~\cite{chen2025hytip}} & \textcolor{gray!65}{ICCV'25} & \textcolor{gray!65}{LD} & \textcolor{gray!65}{96} & \textcolor{gray!65}{32} & \textcolor{gray!65}{13.8} & \textcolor{gray!65}{15.4} & \textcolor{gray!65}{-6.5} & \textcolor{gray!65}{27.7} & \textcolor{gray!65}{-2.4} & \textcolor{gray!65}{7.9} & \textcolor{gray!65}{9.3} \\
  \textcolor{gray!65}{EHVC~\cite{liao2025ehvc}} & \textcolor{gray!65}{ACMMM'25} & \textcolor{gray!65}{LD$^\ast$} & \textcolor{gray!65}{96} & \textcolor{gray!65}{32} & \textcolor{gray!65}{0.3} & \textcolor{gray!65}{15.1} & \textcolor{gray!65}{-5.7} & \textcolor{gray!65}{-11.0} & \textcolor{gray!65}{-23.9} & \textcolor{gray!65}{-3.5} & \textcolor{gray!65}{-4.8} \\
  \rowcolor{lightgray}UCVC~\cite{yang2024ucvc} & DCC'24 & RA & 96 & 32 & 47.1 & 47.7 & 18.1 & 24.6 & 35.9 & 65.9 & 39.9 \\
  DCVC-B~\cite{sheng2025bi} & TMM'25 & RA & 96 & 32 & 22.0 & 28.0 & -6.9 & -8.1 & 25.8 & 29.9 & 15.1 \\
  \rowcolor{lightgray}L-LBVC~\cite{zhai2025llbvc} & DCC'25 & RA & 96 & 32 & 26.7 & 31.5 & -10.6 & -4.4 & 15.6 & 38.1 & 16.1 \\
  BiECVC~\cite{jiang2025biecvc} & ACMMM'25 & RA & 96 & 32 & -1.5 & -2.0 & -26.5 & -20.3 & -6.5 & 4.1 & -8.8 \\
  \rowcolor{lightgray}BRHVC~\cite{liuneural} & NeurIPS'25 & RA & 96 & 32 & 14.8 & 9.3 & -17.1 & -5.2 & 14.7 & 15.8 & 5.4 \\
  BiCRVC & Ours & RA & 96 & 32 &  \textbf{-2.6} & \textbf{-4.8} & \textbf{-21.4} & \textbf{-17.9} & \textbf{-9.8} & \textbf{0.7} & \textbf{-9.3} \\
  \midrule
  DCVC-B~\cite{sheng2025bi} & TMM'25 & RA & 97 & 32 & 15.5 & 22.5 & -11.0 & -13.8 & 20.3 & 26.1 & 9.9 \\
  \rowcolor{lightgray}BiECVC~\cite{jiang2025biecvc} & ACMMM'25 & RA & 97 & 32 & -7.0 & -6.2 & -29.5 & -24.9 & -9.2 & 1.4 & -12.6 \\
  BRHVC~\cite{liuneural} & NeurIPS'25 & RA & 97 & 32 & 8.4 & 5.3 & -20.3 & -9.6 & 9.8 & 12.1 & 1.0 \\
  \rowcolor{lightgray}BiCRVC & Ours & RA & 97 & 32 & \textbf{-7.7} & \textbf{-8.7} & \textbf{-24.8} & \textbf{-22.6} & \textbf{-12.8} & \textbf{-1.2} & \textbf{-13.0} \\
  \midrule
  \textcolor{gray!65}{VTM 23.0 LDB~\cite{bross2021overview}} & \textcolor{gray!65}{---} & \textcolor{gray!65}{LD} & \textcolor{gray!65}{64} & \textcolor{gray!65}{64} & \textcolor{gray!65}{36.0} & \textcolor{gray!65}{30.5} & \textcolor{gray!65}{29.6} & \textcolor{gray!65}{33.5} & \textcolor{gray!65}{39.2} & \textcolor{gray!65}{34.8} & \textcolor{gray!65}{33.9} \\
  \rowcolor{lightgray}\textcolor{gray!65}{DCVC-DC~\cite{li2023neural}} & \textcolor{gray!65}{CVPR'23} & \textcolor{gray!65}{LD} & \textcolor{gray!65}{64} & \textcolor{gray!65}{64} & \textcolor{gray!65}{22.6} & \textcolor{gray!65}{26.4} & \textcolor{gray!65}{-0.1} & \textcolor{gray!65}{19.5} & \textcolor{gray!65}{5.5} & \textcolor{gray!65}{12.0} & \textcolor{gray!65}{14.3} \\
  \textcolor{gray!65}{DCVC-FM~\cite{li2024neural}} & \textcolor{gray!65}{CVPR'24} & \textcolor{gray!65}{LD} & \textcolor{gray!65}{64} & \textcolor{gray!65}{64} & \textcolor{gray!65}{26.1} & \textcolor{gray!65}{8.9} & \textcolor{gray!65}{-12.4} & \textcolor{gray!65}{7.2} & \textcolor{gray!65}{17.8} & \textcolor{gray!65}{24.7} & \textcolor{gray!65}{12.1} \\
  \rowcolor{lightgray}\textcolor{gray!65}{SEVC~\cite{bian2025augmented}} & \textcolor{gray!65}{CVPR'25} & \textcolor{gray!65}{LD} & \textcolor{gray!65}{64} & \textcolor{gray!65}{64} & \textcolor{gray!65}{17.9} & \textcolor{gray!65}{13.2} & \textcolor{gray!65}{-6.9} & \textcolor{gray!65}{-1.8} & \textcolor{gray!65}{-4.8} & \textcolor{gray!65}{-2.7} & \textcolor{gray!65}{2.5} \\
  \textcolor{gray!65}{DCMVC~\cite{tang2025neural}} & \textcolor{gray!65}{CVPR'25} & \textcolor{gray!65}{LD} & \textcolor{gray!65}{64} & \textcolor{gray!65}{64} & \textcolor{gray!65}{19.3} & \textcolor{gray!65}{17.0} & \textcolor{gray!65}{-6.6} & \textcolor{gray!65}{16.4} & \textcolor{gray!65}{3.0} & \textcolor{gray!65}{7.6} & \textcolor{gray!65}{9.5} \\
  \rowcolor{lightgray}\textcolor{gray!65}{ECVC~\cite{jiang2025ecvc}} & \textcolor{gray!65}{CVPR'25} & \textcolor{gray!65}{LD} & \textcolor{gray!65}{64} & \textcolor{gray!65}{64} & \textcolor{gray!65}{11.4} & \textcolor{gray!65}{3.8} & \textcolor{gray!65}{-17.4} & \textcolor{gray!65}{19.0} & \textcolor{gray!65}{-7.7} & \textcolor{gray!65}{-1.0} & \textcolor{gray!65}{1.4} \\
  \textcolor{gray!65}{DCVC-RT~\cite{jia2025practical}} & \textcolor{gray!65}{CVPR'25} & \textcolor{gray!65}{LD} & \textcolor{gray!65}{64} & \textcolor{gray!65}{64} & \textcolor{gray!65}{25.9} & \textcolor{gray!65}{24.6} & \textcolor{gray!65}{4.6} & \textcolor{gray!65}{8.2} & \textcolor{gray!65}{12.9} & \textcolor{gray!65}{20.7} & \textcolor{gray!65}{16.2} \\
  \rowcolor{lightgray}\textcolor{gray!65}{HyTIP~\cite{chen2025hytip}} & \textcolor{gray!65}{ICCV'25} & \textcolor{gray!65}{LD} & \textcolor{gray!65}{64} & \textcolor{gray!65}{64} & \textcolor{gray!65}{10.0} & \textcolor{gray!65}{12.0} & \textcolor{gray!65}{-10.2} & \textcolor{gray!65}{37.7} & \textcolor{gray!65}{-5.6} & \textcolor{gray!65}{2.8} & \textcolor{gray!65}{7.8} \\
  \textcolor{gray!65}{EHVC~\cite{liao2025ehvc}} & \textcolor{gray!65}{ACMMM'25} & \textcolor{gray!65}{LD$^\ast$} & \textcolor{gray!65}{64} & \textcolor{gray!65}{64} & \textcolor{gray!65}{4.3} & \textcolor{gray!65}{20.1} & \textcolor{gray!65}{-1.5} & \textcolor{gray!65}{10.1} & \textcolor{gray!65}{-20.3} & \textcolor{gray!65}{-3.5} & \textcolor{gray!65}{1.5} \\
  \rowcolor{lightgray}DCVC-B~\cite{sheng2025bi} & TMM'25 & RA & 64 & 64 & 40.7 & 39.4 & 0.0 & 2.4 & 44.2 & 47.6 & 29.0 \\
  BiECVC~\cite{jiang2025biecvc} & ACMMM'25 & RA & 64 & 64 & 12.7 & 4.1& -22.7 & -20.6 & 6.3 & 18.3 & -0.3 \\
  \rowcolor{lightgray}BRHVC~\cite{liuneural} & NeurIPS'25 & RA & 64 & 64 & 32.7 & 13.5 & -12.0 & 5.0 & 25.5 & 29.9 & 15.8 \\
  BiCRVC & Ours & RA & 64 & 64 & \textbf{6.6} & \textbf{0.8} & \textbf{-19.9} & \textbf{-22.1} & \textbf{-4.4} & \textbf{4.7} & \textbf{-5.7} \\
  \midrule
  DCVC-B~\cite{sheng2025bi} & TMM'25 & RA & 65 & 64 & 22.7 & 24.4 & -7.8 & -17.6 & 27.0 & 33.8 & 13.8 \\
  \rowcolor{lightgray}BiECVC~\cite{jiang2025biecvc} & ACMMM'25 & RA & 65 & 64 & -2.3 & -4.5 & -29.3 & -32.3 & -2.1 & 8.9 & -10.3 \\
  BRHVC~\cite{liuneural} & NeurIPS'25 & RA & 65 & 64 & 15.1 & 5.4 & -19.1 & -12.0 & 12.8 & 17.9 & 3.4 \\
  \rowcolor{lightgray}BiCRVC & Ours & RA & 65 & 64 & \textbf{-6.7} & \textbf{-8.1} & \textbf{-26.2} & \textbf{-31.3} & \textbf{-11.6} & \textbf{0.0} & \textbf{-14.0}\\
  \bottomrule
  \end{tabular}}
  \par\vspace{2pt}
  \begin{minipage}{0.98\textwidth}
    \footnotesize\raggedright
    LD$^\ast$ indicates methods that require look-ahead.\\Some BD-Rate values may differ slightly from those reported in the original papers because we compute BD-Rate over a wider quality range.
  \end{minipage}
  \end{threeparttable}
  \label{tab:rd_rgb}
\end{table*}
\section{Experiments}
\subsection{Settings}
\subsubsection{Training}
Our training procedure consists of two stages: pretraining and fine-tuning.
In the pretraining stage, we train BiCRVC on the Vimeo90k training split~\cite{xue2019video}, gradually increase the number of input frames from 3 to 5 and then to 7, and follow the coupled-distortion curriculum $\mathcal{L}_{1}\rightarrow\mathcal{L}_{2}\rightarrow\mathcal{L}_{3}$.
In the fine-tuning stage, we filter and process the original Vimeo videos~\cite{ori_vimeo} to obtain 27,649 high-quality sequences, set the number of input frames to 33, and directly optimize the final $\mathcal{L}_{3}$ objective.
During both pretraining and fine-tuning, the batch size is set to 16.
\subsubsection{Evaluation}
Following previous methods~\cite{jiang2025ecvc, jiang2025biecvc, zhai2025llbvc, sheng2025bi, liuneural,li2023neural,li2022hybrid,li2021deep},
we evaluate the rate-distortion performance on the HEVC common test sequences, including Classes B, C, D, and E.
We also report results on the UVG~\cite{mercat2020uvg} and MCL-JCV~\cite{wang2016mcl} datasets.\par
To provide a comprehensive evaluation, we compare BiCRVC with both low-delay and random-access methods.
The low-delay baselines include DCVC-DC~\cite{li2023neural}, DCVC-FM~\cite{li2024neural}, ECVC~\cite{jiang2025ecvc}, DCMVC~\cite{tang2025neural}, SEVC~\cite{bian2025augmented}, DCVC-RT~\cite{jia2025practical}, HyTIP~\cite{chen2025hytip}, and EHVC~\cite{liao2025ehvc}.
The random-access baselines include DCVC-B~\cite{sheng2025bi}, UCVC~\cite{yang2024ucvc}, L-LBVC~\cite{zhai2025llbvc}, BiECVC~\cite{jiang2025biecvc}, and BRHVC~\cite{liuneural}.
Since the original BiECVC bitrate range has limited overlap with those of other random-access baselines~\cite{sheng2025bi,liuneural}, we train a bitrate-extended variant aligned with the compared random-access baselines for more reliable BD-Rate comparisons.
We further compare BiCRVC with VTM 23.0, a strong traditional codec, under the low delay B and random access configurations.\par 
We evaluate performance using PSNR in the RGB domain.
All input frames and reconstructed frames are evaluated in RGB color space.
For comparison with low-delay methods, we report results with an intra period of 32 on 96 frames and an intra period of 64 on 64 frames.
For random-access evaluation, we additionally report results on 97 and 65 frames.
The intra periods of 32 and 64 follow the JVET common test conditions~\cite{bossen2019jvet}, where the intra period is set according to the sequence frame rate and roughly corresponds to one intra frame per second.
We use the Bjontegaard Delta Rate (BD-Rate) metric~\cite{bjontegaard2001calculation} for comparison.
The rate is calculated using actual bits per pixel (BPP) rather than estimated BPP.
\subsection{Rate-Distortion Performance}
Table~\ref{tab:rd_rgb} summarizes the BD-Rate results under multiple frame-number and intra-period settings, and Figs.~\ref{fig:rd_ip32_f96} and~\ref{fig:rd_ip64_f64} show representative RD curves. Overall, BiCRVC consistently outperforms VTM 23.0 RA on average, achieving $-9.3\%$ BD-Rate under IP32/FN96 and $-5.7\%$ under IP64/FN64. Compared with VTM 23.0 LDB, the gap is larger, showing that the proposed random-access neural codec provides a clear coding-efficiency advantage over the low-delay traditional configuration.
\par
Compared with low-delay neural codecs, BiCRVC is especially stronger than the real-time-oriented DCVC-RT~\cite{jia2025practical}. 
DCVC-RT is highly efficient in runtime, but it only supports low-delay prediction and has higher BD-Rate under the same IP/FN test settings, with average BD-Rates of 28.1\% and 16.2\% under IP32/FN96 and IP64/FN64, respectively.
BiCRVC reduces the average BD-Rate to $-9.3\%$ and $-5.7\%$ in the two settings, improving over DCVC-RT by 37.4 and 21.9 percentage points, respectively.
This improvement comes from adapting the efficient coding backbone to random-access coding with efficient bidirectional prediction and coupled representation coding.
\par
Among random-access neural codecs, BiCRVC gives competitive performance across different GOP lengths and datasets. 
It substantially improves over DCVC-B~\cite{sheng2025bi}, L-LBVC~\cite{zhai2025llbvc}, UCVC~\cite{yang2024ucvc}, and BRHVC~\cite{liuneural} on average.
Compared with the strong RA baseline BiECVC~\cite{jiang2025biecvc}, BiCRVC also achieves lower average BD-Rate across all evaluated IP/FN settings, with gains of 0.5, 0.4, 5.4, and 3.7 percentage points under IP32/FN96, IP32/FN97, IP64/FN64, and IP64/FN65, respectively.
This indicates that the proposed compact coupled representation can deliver stronger average compression efficiency while keeping a more deployment-friendly architecture, as discussed in the complexity analysis.
BiECVC improves compression with stronger multi-reference prediction and non-local context mining, whereas BiCRVC focuses on compact coupled representation, accurate motion estimation, broad quality-range support, and fast decoding.
Since the stronger prediction and context tools in BiECVC are complementary to the proposed coupled representation, integrating them may further improve the compression performance of BiCRVC.
The RD curves further confirm these trends on UVG, MCL-JCV, and HEVC Class B: 
BiCRVC stays near the upper-left region across the bit-rate range, showing 
stable behavior rather than improvements limited to a single bit-rate point.
\subsection{Subjective Quality Comparison}
The subjective quality comparison is shown in Fig.~\ref{fig:subjective_basketballdrive}.
On POC12 of BasketballDrive, BiCRVC achieves the lowest bit cost and the highest RGB-PSNR among the compared random-access neural codecs, using 0.0266 bpp and reaching 34.85 dB.
Compared with BiECVC, BRHVC, and DCVC-B, BiCRVC preserves clearer local structures in the selected regions while consuming fewer bits.
This visual comparison is consistent with the frame-level rate-distortion numbers annotated below each reconstruction.
\subsection{Model Complexity}
For complexity evaluation, we include open-source low-delay methods
DCVC-DC~\cite{li2023neural}, DCVC-FM~\cite{li2024neural}, DCVC-RT~\cite{jia2025practical},
SEVC~\cite{bian2025augmented}, DCMVC~\cite{tang2025neural}, ECVC~\cite{jiang2025ecvc},
and EHVC~\cite{liao2025ehvc} as reference baselines.
Since BiCRVC targets random-access coding, the main comparison is made with random-access methods,
including DCVC-B~\cite{sheng2025bi}, BiECVC~\cite{jiang2025biecvc}, and BRHVC~\cite{liuneural}.
Table~\ref{tab:runtime_fps} reports kMACs/pixel and the encoding/decoding FPS at 1080p, 720p, and 480p resolutions,
measured on a single 80GB GPU.
\begin{table*}[t]
  \centering
  \setlength{\tabcolsep}{1.0mm}
  \fontsize{9.2pt}{9.2pt}\selectfont
  \caption{Encoding and decoding complexity comparison of different methods on a single 80GB GPU at multiple resolutions.}
  \label{tab:runtime_fps}
  \renewcommand{\arraystretch}{1.12}
  \resizebox{0.995\textwidth}{!}{
  \begin{tabular}{@{}>{\centering\arraybackslash}m{0.18\textwidth}
                  >{\centering\arraybackslash}m{0.07\textwidth}
                  >{\centering\arraybackslash}m{0.09\textwidth}
                  >{\centering\arraybackslash}m{0.09\textwidth}
                  >{\centering\arraybackslash}m{0.09\textwidth}
                  *{6}{>{\centering\arraybackslash}m{0.10\textwidth}}@{}}
  \toprule
  & & & & &
  \multicolumn{6}{c}{Encoding \& Decoding FPS} \\
  \cmidrule(l){6-11}
  \multicolumn{1}{c}{Method} & Type & Params (M) &
  \shortstack{Enc.\\kMACs/pixel} & \shortstack{Dec.\\kMACs/pixel} &
  \multicolumn{2}{c}{1080p} &
  \multicolumn{2}{c}{720p} &
  \multicolumn{2}{c}{480p} \\
  \cmidrule(lr){6-7}\cmidrule(lr){8-9}\cmidrule(lr){10-11}
  & & & & & Enc. & Dec. & Enc. & Dec. & Enc. & Dec. \\
  \midrule
  \textcolor{gray!65}{DCVC-DC~\cite{li2023neural}} & \textcolor{gray!65}{LD} & \textcolor{gray!65}{50.8} & \textcolor{gray!65}{1297.37} & \textcolor{gray!65}{885.16} & \textcolor{gray!65}{3.14} & \textcolor{gray!65}{3.91} & \textcolor{gray!65}{6.80} & \textcolor{gray!65}{8.13} & \textcolor{gray!65}{13.70} & \textcolor{gray!65}{15.63} \\
  \textcolor{gray!65}{DCVC-FM~\cite{li2024neural}} & \textcolor{gray!65}{LD} & \textcolor{gray!65}{44.9} & \textcolor{gray!65}{1099.81} & \textcolor{gray!65}{835.54} & \textcolor{gray!65}{4.31} & \textcolor{gray!65}{5.10} & \textcolor{gray!65}{9.17} & \textcolor{gray!65}{10.75} & \textcolor{gray!65}{16.13} & \textcolor{gray!65}{19.61} \\
  \textcolor{gray!65}{DCVC-RT~\cite{jia2025practical}} & \textcolor{gray!65}{LD} & \textcolor{gray!65}{66.3} & \textcolor{gray!65}{\textbf{163.68}} & \textcolor{gray!65}{\textbf{179.01}} & \textcolor{gray!65}{\textbf{123.20}} & \textcolor{gray!65}{\textbf{114.71}} & \textcolor{gray!65}{\textbf{202.31}} & \textcolor{gray!65}{\textbf{175.07}} & \textcolor{gray!65}{\textbf{217.20}} & \textcolor{gray!65}{\textbf{212.13}} \\
  \textcolor{gray!65}{SEVC~\cite{bian2025augmented}} & \textcolor{gray!65}{LD} & \textcolor{gray!65}{76.1} & \textcolor{gray!65}{1511.11} & \textcolor{gray!65}{1405.56} & \textcolor{gray!65}{2.03} & \textcolor{gray!65}{2.16} & \textcolor{gray!65}{3.95} & \textcolor{gray!65}{4.23} & \textcolor{gray!65}{7.07} & \textcolor{gray!65}{7.69} \\
  \textcolor{gray!65}{DCMVC~\cite{tang2025neural}} & \textcolor{gray!65}{LD} & \textcolor{gray!65}{52.0} & \textcolor{gray!65}{1907.69} & \textcolor{gray!65}{1495.47} & \textcolor{gray!65}{2.02} & \textcolor{gray!65}{2.30} & \textcolor{gray!65}{4.41} & \textcolor{gray!65}{5.03} & \textcolor{gray!65}{9.17} & \textcolor{gray!65}{10.10} \\
  \textcolor{gray!65}{ECVC~\cite{jiang2025ecvc}} & \textcolor{gray!65}{LD} & \textcolor{gray!65}{61.9} & \textcolor{gray!65}{2242.68} & \textcolor{gray!65}{1403.59} & \textcolor{gray!65}{2.16} & \textcolor{gray!65}{2.23} & \textcolor{gray!65}{2.31} & \textcolor{gray!65}{4.52} & \textcolor{gray!65}{6.17} & \textcolor{gray!65}{8.93} \\
  \textcolor{gray!65}{EHVC~\cite{liao2025ehvc}} & \textcolor{gray!65}{LD} & \textcolor{gray!65}{51.4} & \textcolor{gray!65}{1518.93} & \textcolor{gray!65}{1085.86} & \textcolor{gray!65}{2.52} & \textcolor{gray!65}{3.01} & \textcolor{gray!65}{5.52} & \textcolor{gray!65}{6.49} & \textcolor{gray!65}{11.49} & \textcolor{gray!65}{12.46} \\
  \midrule
  DCVC-B~\cite{sheng2025bi} & RA & 55.4 & 2796.05 & 2000.55 & 1.87 & 2.31 & 4.08 & 4.95 & 8.47 & 9.71 \\
  BiECVC~\cite{jiang2025biecvc} & RA & 63.9 & 2870.08 & 1921.77 & 1.31 & 1.86 & 2.84 & 4.06 & 5.52 & 8.33 \\
  BRHVC~\cite{liuneural} & RA & 60.4 & 3718.48 & 2444.12 & 1.30 & 1.84 & 2.82 & 4.00 & 5.46 & 8.13 \\
  BiCRVC & RA & 80.8 & \textbf{1275.60} & \textbf{342.74} & \textbf{4.37} & \textbf{56.47} & \textbf{7.33} & \textbf{92.27} & \textbf{9.76} & \textbf{119.23} \\
  \bottomrule
  \end{tabular}}
  \vspace{0.3ex}
  \begin{tablenotes}
    \footnotesize
    \item Params (M) includes the parameters of both I-frame and B/P-frame codecs.
    \item kMACs/pixel is measured at 1080p.
  \end{tablenotes}
\end{table*}
\par
As shown in Table~\ref{tab:runtime_fps}, BiCRVC uses 80.8M parameters, which is moderately higher than DCVC-B, BiECVC, and BRHVC (55.4M--63.9M). However, parameter count is not the main deployment bottleneck once model weights are loaded and reused for video streams. The kMACs/pixel and end-to-end FPS provide a more direct view of runtime complexity.
\par
Among random-access codecs, BiCRVC has the lowest computational cost, especially on the decoding side. Its decoding cost is only 342.74 kMACs/pixel, much lower than DCVC-B, BiECVC, and BRHVC (2000.55, 1921.77, and 2444.12 kMACs/pixel), which explains its large decoding-speed advantage. At 1080p, BiCRVC achieves 56.47 decoding FPS, while these random-access baselines decode at only 2.31, 1.86, and 1.84 FPS.
\par
The encoding cost of BiCRVC is higher than its decoding cost because several tools are placed on the encoder side, including motion estimation and accumulated motion/flow candidate construction. Its decoding kMACs/pixel is still about twice that of the real-time low-delay codec DCVC-RT, mainly because BiCRVC performs bidirectional random-access prediction from two references. Further optimizing these encoder-side tools and the bidirectional decoding path may reduce kMACs/pixel and improve FPS in future versions.
\begin{figure*}[t]
\centering
\setlength{\tabcolsep}{0.6mm}
\begin{tabular}{ccccc}
\footnotesize Original &
\footnotesize BiCRVC &
\footnotesize BiECVC &
\footnotesize BRHVC &
\footnotesize DCVC-B\\
\includegraphics[width=0.19\textwidth]{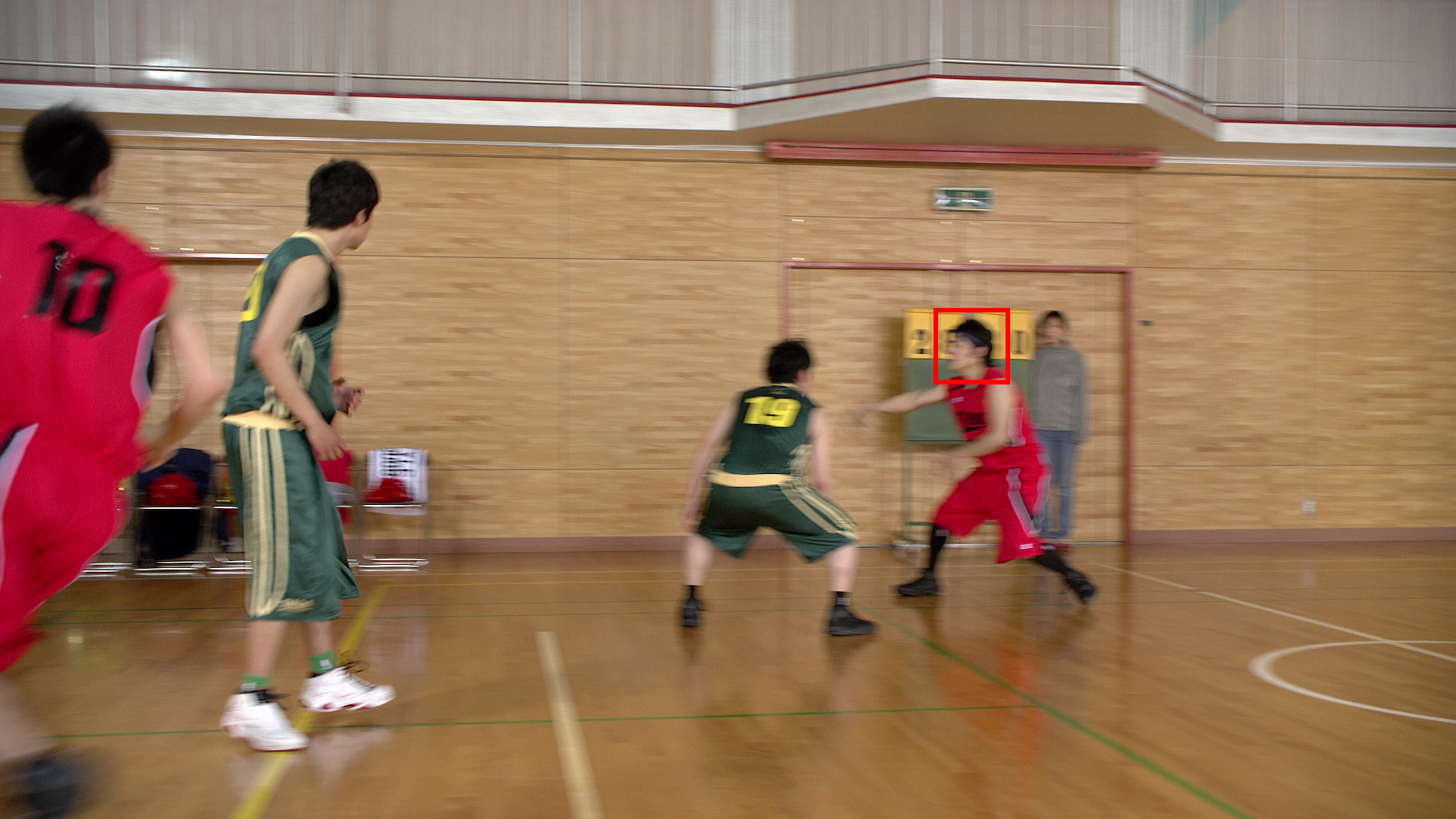} &
\includegraphics[width=0.19\textwidth]{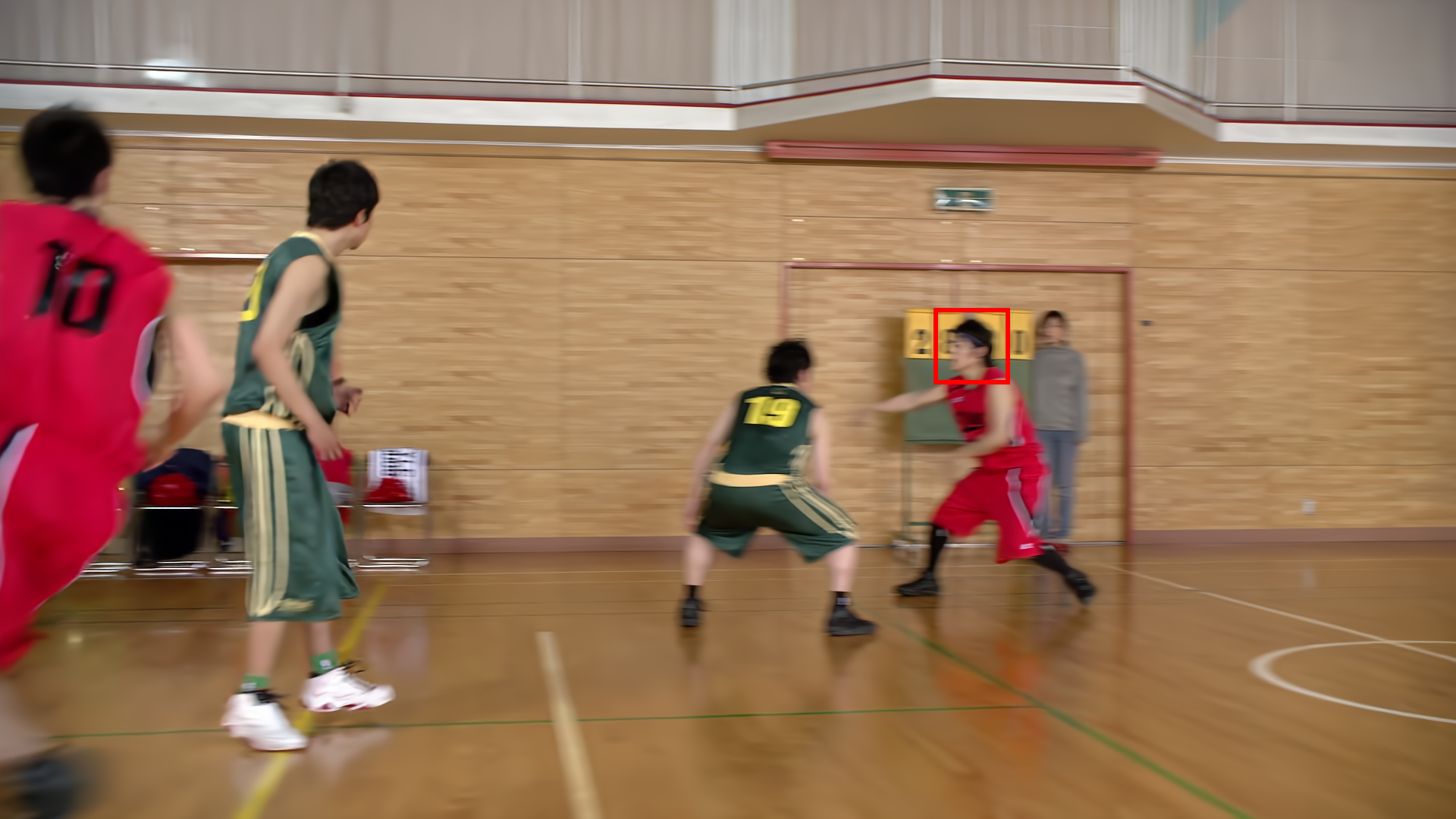} &
\includegraphics[width=0.19\textwidth]{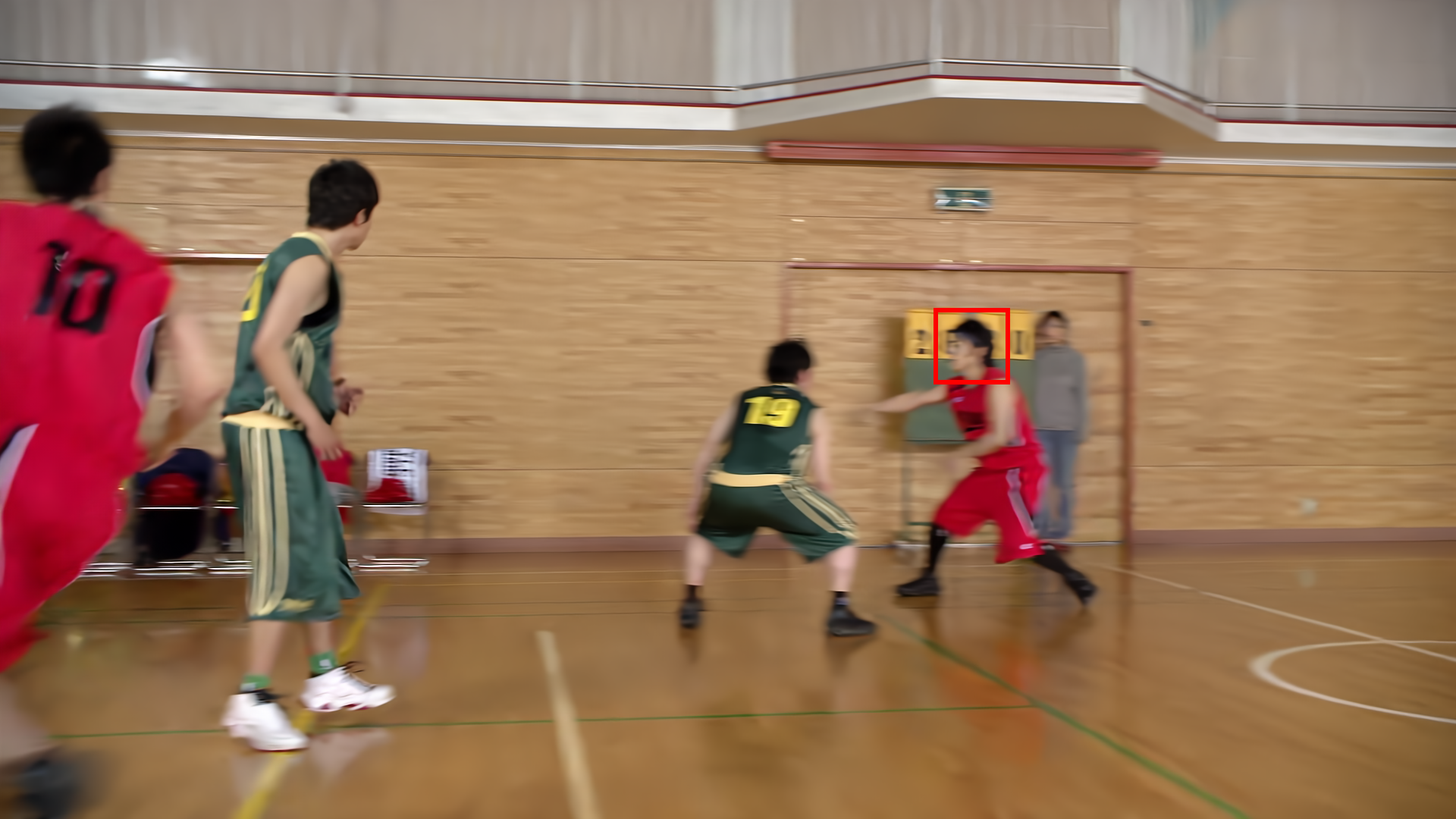} &
\includegraphics[width=0.19\textwidth]{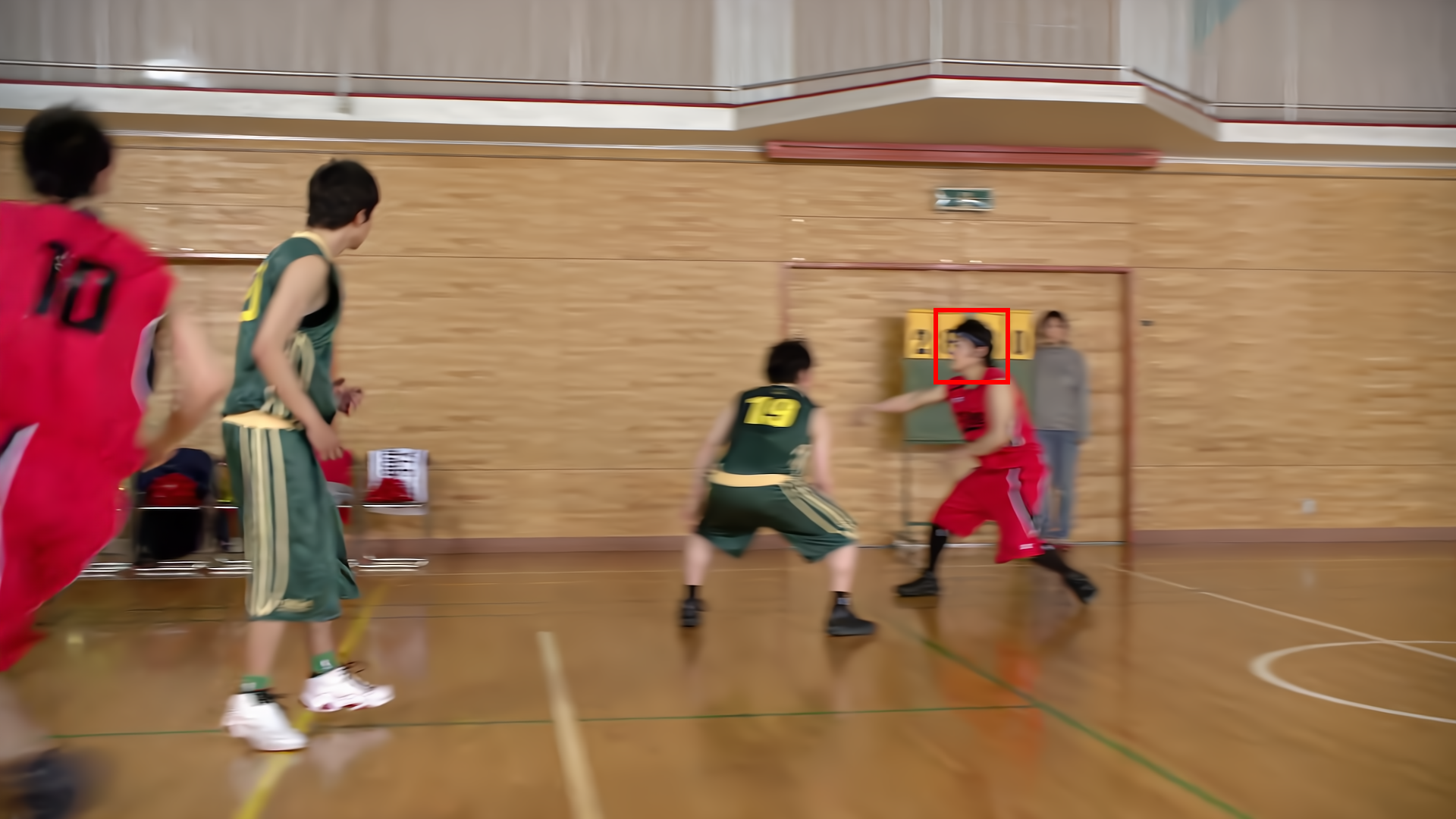} &
\includegraphics[width=0.19\textwidth]{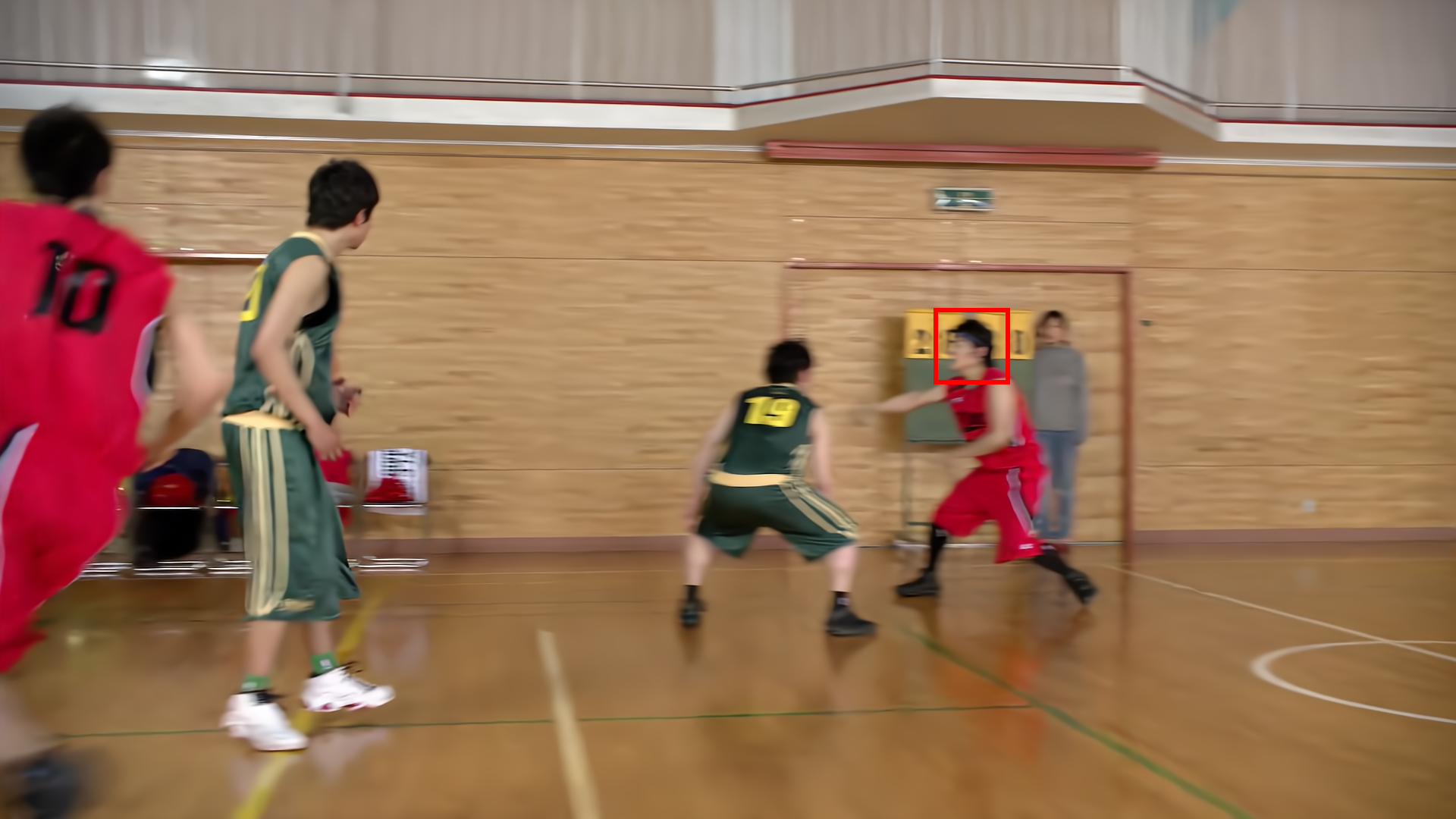} \\
&
\footnotesize \shortstack{0.0266 bpp / 34.85 dB} &
\footnotesize \shortstack{0.0380 bpp / 34.40 dB} &
\footnotesize \shortstack{0.0394 bpp / 34.54 dB} &
\footnotesize \shortstack{0.0458 bpp / 34.05 dB}
\end{tabular}
\vspace{0.5mm}
\setlength{\tabcolsep}{0.6mm}
\begin{tabular}{ccccc}
\makebox[0.19\textwidth][c]{\includegraphics[width=0.14\textwidth]{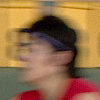}} &
\makebox[0.19\textwidth][c]{\includegraphics[width=0.14\textwidth]{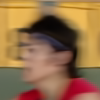}} &
\makebox[0.19\textwidth][c]{\includegraphics[width=0.14\textwidth]{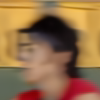}} &
\makebox[0.19\textwidth][c]{\includegraphics[width=0.14\textwidth]{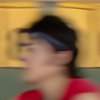}} &
\makebox[0.19\textwidth][c]{\includegraphics[width=0.14\textwidth]{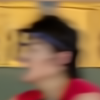}} \\
\end{tabular}
\caption{Subjective reconstruction comparison on POC12 of BasketballDrive. Red boxes indicate the selected local regions for visual inspection.}
\label{fig:subjective_basketballdrive}
\end{figure*}
\begin{figure}[t]
  \centering
  \includegraphics[width=0.49\columnwidth]{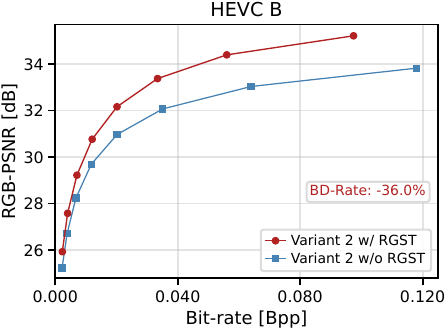}\hfill
  \includegraphics[width=0.49\columnwidth]{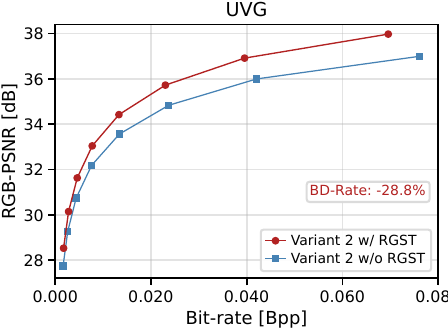}
  \caption{RD comparison of Variant 2 with and without random GOP structure training (RGST) on HEVC B and UVG.}
  \label{fig:rgst_rd}
\end{figure}
\subsection{Ablation Studies}
In the ablation study, we build a series of model variants by progressively enabling different modules. Each variant is trained to convergence so that the contribution of each module can be evaluated fairly. We report results on the HEVC common test sequences (Classes B, C, D, and E), UVG, and MCL-JCV under an intra period of 32 with 96 frames. RGB PSNR is used as the distortion metric, and BD-Rate is computed relative to the final full model. The results are summarized in Table~\ref{tab:ablation}.
\par
As shown in Table~\ref{tab:ablation}, DCVC-RT itself gives an average BD-Rate gap of 43.2\%. The basic bidirectional conditional coding (BCC) variant, a straightforward bidirectional adaptation that keeps the DCVC-RT-style implicit motion modeling (Variant 1), does not improve the result; instead, its average gap increases to 48.3\%, with particularly large degradation on HEVC Class B and UVG. This indicates that simply relying on implicit bidirectional contexts without explicitly transmitted motion is not sufficient for random-access coding. After introducing coupled representation coding (Variant 2), which explicitly couples motion and frame representations in a coupled coded representation, the average gap decreases to 25.2\%, with clear improvements on HEVC Class B and UVG.
\par
To further improve motion estimation accuracy, we introduce multi-candidate motion estimation, which includes multi-scale motion estimation and parallel accumulated motion estimation. Adding MSME (Variant 3) reduces the average BD-Rate gap from 25.2\% to 16.7\%. The gain is more evident on sequences with more complex motion, including HEVC Classes B, C, and D, as well as UVG and MCL-JCV. Based on Variant 3, adding PAME in Variant 4 further reduces the average gap to 10.6\%. The improvement is especially clear on challenging sequences, with additional gains of 8.4\% on HEVC Class B and 14.0\% on UVG. HEVC E is an exception in this setting, where the gap increases from 7.6\% to 9.0\%. This is expected because HEVC E mainly contains smooth and small motions; direct estimation is already reliable, while accumulated motion may introduce composition error and extra motion information to encode.
\par
The effect of PAME on HEVC E is therefore not determined by motion accuracy alone, but by the rate-distortion tradeoff between the additional accumulated-motion candidate and its coding cost. When BMFP and QCM are enabled, the same PAME operation becomes beneficial on HEVC E: comparing Variant 7 and Variant 9, the gap decreases from 6.6\% to 2.8\%. BMFP provides temporal priors for motion coding, and QCM improves context modeling efficiency, so the extra motion candidate can be represented with lower overhead. After adding SSC, Variant 8 is already slightly better than the final model on HEVC E (-0.4\% vs. 0.0\%), indicating that accumulated motion is almost unnecessary for this slow-motion class under the strongest coding configuration. However, PAME still improves the average gap from 3.4\% to 0.0\% and is consistently useful on the other test sets. This confirms that PAME mainly targets large or complex motion, while its benefit on simple-motion sequences depends on whether the motion representation is compressed efficiently enough.
\par
MSME and PAME affect only motion estimation, rather than motion reconstruction or motion compensation. As a result, Variants 3 and 4 have the same decoder complexity as Variant 2.
\par
To reduce motion compression overhead, we further introduce bidirectional motion feature propagation and conduct this ablation based on Variant 3. As shown by Variant 5, bidirectional motion feature propagation reduces the average BD-Rate gap from 16.7\% to 11.9\%.
\par
In BiCRVC, we further employ quadtree-based context modeling (QCM)~\cite{li2023neural} and scale-based selective coding (SSC)~\cite{shi2022alphavc,jia2025practical} to improve compression efficiency. Starting from Variant 3, QCM alone (Variant 6) reduces the average BD-Rate gap to 11.2\%. Combining bidirectional motion feature propagation and QCM in Variant 7 brings a larger gain, reducing the gap to 7.5\%, which suggests that the two components are complementary. Adding SSC in Variant 8 further reduces the average gap to 3.4\%. Finally, combining PAME with BMFP, QCM, and SSC gives the full model (Variant 10), which is used as the anchor. The comparison between Variants 8 and 10 also shows the intended tradeoff: PAME slightly hurts HEVC E by 0.4\%, but it improves HEVC B, HEVC C, HEVC D, UVG, and MCL-JCV by 3.5\%, 0.9\%, 1.8\%, 13.0\%, and 1.5\%, respectively.
\par
We also isolate the effect of random GOP structure training on Variant 2 while keeping the network architecture unchanged. As shown in Fig.~\ref{fig:rgst_rd}, RGST reduces BD-Rate by 36.0\% on HEVC Class B and 28.8\% on UVG compared with training without RGST. This shows that exposing the model to diverse reference intervals improves its adaptation to random-access coding.
\begin{table*}
  \setlength{\tabcolsep}{1.2mm}
  \renewcommand\arraystretch{1.2}
  \centering
  \footnotesize
  \caption{Ablation study on HEVC B$\sim$E, UVG, and MCL-JCV. The metric is BD-Rate (\%) for RGB PSNR (dB), computed relative to the final full model.}
  \resizebox{\textwidth}{!}{
  \begin{tabular}[t]{c|ccccccc|ccccccc}
  \toprule
  \multirow{2}{*}{Variant} & \multirow{2}{*}{BCC} & \multirow{2}{*}{CRC} & \multicolumn{2}{c}{MCME} & \multirow{2}{*}{BMFP} & \multirow{2}{*}{QCM} & \multirow{2}{*}{SSC} & \multirow{2}{*}{HEVC B} & \multirow{2}{*}{HEVC C} & \multirow{2}{*}{HEVC D} & \multirow{2}{*}{HEVC E} & \multirow{2}{*}{UVG} & \multirow{2}{*}{MCL-JCV} & \multirow{2}{*}{Average}\\
  &  &  & MSME & PAME &  &  &  &  &  &  &  &  &  & \\
  \midrule
  DCVC-RT &  &  &  &  &  &  & & 42.7 & 43.3 & 52.7 & 43.8 & 42.2 & 34.3 & 43.2 \\
  Variant 1 & \ding{51} &  &  &  &  &  &  & 64.4 & 38.8 & 49.7 & 33.7 & 64.2 & 38.6 & 48.3 \\
  Variant 2 & \ding{51} & \ding{51} &  &  &  &  &  & 29.5 & 20.6 & 16.1 & 11.1 & 45.1 & 29.1 & 25.2 \\
  Variant 3 & \ding{51} & \ding{51} & \ding{51} &  &  &  &  & 19.4 & 11.1 & 8.9 & 7.6 & 34.3 & 18.8 & 16.7 \\
  Variant 4 & \ding{51} & \ding{51} & \ding{51} & \ding{51} &  &  &  & 11.0 & 6.9 & 4.3 & 9.0 & 20.3 & 12.1 & 10.6 \\
  Variant 5 & \ding{51} & \ding{51} & \ding{51} &  & \ding{51} &  &  & 12.2 & 8.5 & 7.1 & 7.3 & 23.8 & 12.2 & 11.9 \\
  Variant 6 & \ding{51} & \ding{51} & \ding{51} &  &  & \ding{51} &  & 13.9 & 6.0 & 4.0 & 2.2 & 28.1 & 13.1 & 11.2 \\
  Variant 7 & \ding{51} & \ding{51} & \ding{51} &  & \ding{51} & \ding{51} &  & 8.7 & 4.0 & 4.4 & 6.6 & 16.5 & 4.8 & 7.5\\
  Variant 8 & \ding{51} & \ding{51} & \ding{51} & & \ding{51} & \ding{51} & \ding{51} & 3.5 & 0.9 & 1.8 & -0.4 & 13.0 & 1.5 & 3.4\\
  Variant 9 & \ding{51} & \ding{51} & \ding{51} & \ding{51} & \ding{51} & \ding{51} &  & 4.6 & 2.9 & 1.9 & 2.8 & 10.2 & 4.4 & 4.5 \\
  Variant 10 & \ding{51} & \ding{51} & \ding{51} & \ding{51} & \ding{51} & \ding{51} & \ding{51} & 0.0 & 0.0 & 0.0 & 0.0 & 0.0 & 0.0 & 0.0\\
  \bottomrule
  \end{tabular}}
  \label{tab:ablation}
\end{table*}
\section{Conclusion}
This paper presents BiCRVC, an efficient bidirectional neural video compression framework for random-access coding.
By jointly coding motion and frame information through coupled representation coding, BiCRVC preserves explicit motion-aligned contexts while avoiding the motion-first decoding pipeline of existing bidirectional neural codecs.
Together with multi-candidate motion estimation, bidirectional motion feature propagation, and random-GOP-aware training, BiCRVC consistently outperforms VTM 23.0 RA and recent random-access neural codecs on average.
It also provides about 30$\times$ faster 1080p decoding than recent random-access neural codecs~\cite{sheng2025bi,zhai2025llbvc,yang2024ucvc,jiang2025biecvc,liuneural}, showing a favorable balance between compression efficiency and practical runtime.
Future work will further improve compression performance by incorporating stronger bidirectional prediction tools, such as the multi-reference mechanism and non-local context mining~\cite{jiang2022mlic,jiang2025mlicpp,jiang2026mlicv2} used in BiECVC~\cite{jiang2025biecvc}, and will also explore cross-platform consistency~\cite{le2022mobilecodec,van2024mobilenvc,jia2025practical} for more robust practical deployment.

\bibliographystyle{IEEEtran}
\bibliography{main}

@article{xue2019video,
  title={Video Enhancement with Task-Oriented Flow},
  author={Xue, Tianfan and Chen, Baian and Wu, Jiajun and Wei, Donglai and Freeman, William T},
  journal={IJCV},
  volume={127},
  number={8},
  pages={1106--1125},
  year={2019},
  publisher={Springer}
}

@misc{ori_vimeo,
title={{Original Vimeo links}}, 
note={\url{https://github.com/anchen1011/toflow/blob/master/data/original_vimeo_links.txt}.}
}

@inproceedings{le2022mobilecodec,
  title={Mobilecodec: neural inter-frame video compression on mobile devices},
  author={Le, Hoang and Zhang, Liang and Said, Amir and Sautiere, Guillaume and Yang, Yang and Shrestha, Pranav and Yin, Fei and Pourreza, Reza and Wiggers, Auke},
  booktitle={ACM MMSys},
  pages={324--330},
  year={2022}
}

@inproceedings{van2024mobilenvc,
  title={Mobilenvc: Real-time 1080p neural video compression on a mobile device},
  author={Van Rozendaal, Ties and Singhal, Tushar and Le, Hoang and Sautiere, Guillaume and Said, Amir and Buska, Krishna and Raha, Anjuman and Kalatzis, Dimitris and Mehta, Hitarth and Mayer, Frank and others},
  booktitle={WACV},
  pages={4323--4333},
  year={2024}
}

@misc{ye2025ecm,
  author       = {Yan Ye},
  title        = {Enhanced Compression Model for Beyond-VVC Capability},
  year         = {2025},
  month        = jan,
  howpublished = {Presentation at the ITU and ISO/IEC JTC 1/SC 29 Joint Workshop on ``Future Video Coding -- Advanced Signal Processing, AI and Standards'', Geneva, Switzerland, 17 January 2025},
  institution  = {ITU and ISO/IEC JTC 1/SC 29},
  url          = {https://www.itu.int/en/ITU-T/Workshops-and-Seminars/2025/0117/Documents/Yan%20Ye.pdf},
  note         = {Accessed: 2026-05-07}
}

@inproceedings{chen2025hytip,
  title={Hytip: Hybrid temporal information propagation for masked conditional residual video coding},
  author={Chen, Yi-Hsin and Yao, Yi-Chen and Ho, Kuan-Wei and Wu, Chun-Hung and Phung, Huu-Tai and Benjak, Martin and Ostermann, J{\"o}rn and Peng, Wen-Hsiao},
  booktitle={ICCV},
  pages={17889--17898},
  year={2025}
}

@inproceedings{liuneural,
  title={Neural B-frame Video Compression with Bi-directional Reference Harmonization},
  author={Liu, Yuxi and Jin, Dengchao and Huo, Shuai and Gu, Jiawen and Zhou, Chao and Bai, Huihui and Lu, Ming and Ma, Zhan},
  booktitle={NeurIPS},
  year={2025}
}

@inproceedings{tang2025neural,
  title={Neural video compression with context modulation},
  author={Tang, Chuanbo and Li, Zhuoyuan and Bian, Yifan and Li, Li and Liu, Dong},
  booktitle={CVPR},
  pages={12553--12563},
  year={2025}
}

@inproceedings{liao2025ehvc,
  title={Ehvc: Efficient hierarchical reference and quality structure for neural video coding},
  author={Liao, Junqi and Wu, Yaojun and Lin, Chaoyi and Deng, Zhipin and Li, Li and Liu, Dong and Sun, Xiaoyan},
  booktitle={ACM MM},
  pages={12083--12091},
  year={2025}
}

@inproceedings{bian2025augmented,
  title={Augmented deep contexts for spatially embedded video coding},
  author={Bian, Yifan and Tang, Chuanbo and Li, Li and Liu, Dong},
  booktitle={CVPR},
  pages={2094--2104},
  year={2025}
}

@inproceedings{jiang2025biecvc,
  title={Biecvc: Gated diversification of bidirectional contexts for learned video compression},
  author={Jiang, Wei and Li, Junru and Zhang, Kai and Zhang, Li},
  booktitle={ACM MM},
  pages={7248--7257},
  year={2025}
}

@article{sheng2025prediction,
  title={Prediction and reference quality adaptation for learned video compression},
  author={Sheng, Xihua and Li, Li and Liu, Dong and Li, Houqiang},
  journal={IEEE TIP},
  year={2025},
  publisher={IEEE}
}

@article{sheng2024vnvc,
  title={Vnvc: A versatile neural video coding framework for efficient human-machine vision},
  author={Sheng, Xihua and Li, Li and Liu, Dong and Li, Houqiang},
  journal={IEEE TPAMI},
  volume={46},
  number={7},
  pages={4579--4596},
  year={2024},
  publisher={IEEE}
}

@article{yang2024adaptive,
  title={Adaptive Prediction Structure for Learned Video Compression},
  author={Yang, Jiayu and Zhai, Yongqi and Jiang, Wei and Yang, Chunhui and Gao, Feng and Wang, Ronggang},
  journal={ACM TOMM},
  volume={21},
  number={2},
  pages={1--23},
  year={2024}
}

@inproceedings{djelouah2019neural,
  title={Neural inter-frame compression for video coding},
  author={Djelouah, Abdelaziz and Campos, Joaquim and Schaub-Meyer, Simone and Schroers, Christopher},
  booktitle={ICCV},
  pages={6421--6429},
  year={2019}
}

@inproceedings{yang2024ucvc,
  title={UCVC: A Unified Contextual Video Compression Framework with Joint P-frame and B-frame Coding},
  author={Yang, Jiayu and Jiang, Wei and Zhai, Yongqi and Yang, Chunhui and Wang, Ronggang},
  booktitle={DCC},
  pages={382--391},
  year={2024},
  organization={IEEE}
}

@inproceedings{zhai2024hybrid,
  title={Hybrid Local-Global Context Learning for Neural Video Compression},
  author={Zhai, Yongqi and Yang, Jiayu and Jiang, Wei and Yang, Chunhui and Tang, Luyang and Wang, Ronggang},
  booktitle={DCC},
  pages={322--331},
  year={2024},
  organization={IEEE}
}

@inproceedings{zhai2025llbvc,
  title={L-LBVC: Long-Term Motion Estimation and Prediction for Learned Bi-Directional Video Compression},
  author={Zhai, Yongqi and Tang, Luyang and Jiang, Wei and Yang, Jiayu and Wang, Ronggang},
  booktitle={DCC},
  year={2025},
  organization={IEEE}
}

@article{lu2020end,
  title={An end-to-end learning framework for video compression},
  author={Lu, Guo and Zhang, Xiaoyun and Ouyang, Wanli and Chen, Li and Gao, Zhiyong and Xu, Dong},
  journal={IEEE TPAMI},
  volume={43},
  number={10},
  pages={3292--3308},
  year={2020},
  publisher={IEEE}
}

@article{bjontegaard2001calculation,
  title={Calculation of average PSNR differences between RD-curves},
  author={Bjontegaard, Gisle},
  journal={ITU-T SG16 Q},
  volume={6},
  year={2001}
}

@article{guo2023learning,
  title={Learning cross-scale weighted prediction for efficient neural video compression},
  author={Guo, Zongyu and Feng, Runsen and Zhang, Zhizheng and Jin, Xin and Chen, Zhibo},
  journal={IEEE TIP},
  volume={32},
  pages={3567--3579},
  year={2023},
  publisher={IEEE}
}

@article{bossen2019jvet,
  title={JVET common test conditions and software reference configurations for SDR video},
  author={Bossen, Frank and Boyce, Jill and Li, Xiang and Seregin, Vadim and S{\"u}hring, Karsten},
  journal={JVET},
  volume={16},
  pages={19--27},
  year={2019}
}

@inproceedings{shi2022alphavc,
  title={Alphavc: High-performance and efficient learned video compression},
  author={Shi, Yibo and Ge, Yunying and Wang, Jing and Mao, Jue},
  booktitle={ECCV},
  pages={616--631},
  year={2022},
  organization={Springer}
}

@ARTICLE{sheng2024spatial,
  author={Sheng, Xihua and Li, Li and Liu, Dong and Li, Houqiang},
  journal={TCSVT}, 
  title={Spatial Decomposition and Temporal Fusion Based Inter Prediction for Learned Video Compression}, 
  year={2024},
  volume={34},
  number={7},
  pages={6460-6473}}

@inproceedings{yang2021hierarchical,
  title={Hierarchical Autoregressive Modeling for Neural Video Compression},
  author={Yang, Ruihan and Yang, Yibo and Marino, Joseph and Mandt, Stephan},
  booktitle={ICLR},
  year={2021}
}

@inproceedings{jiang2024lvc,
  title={LVC-LGMC: Joint Local and Global Motion Compensation for Learned Video Compression},
  author={Jiang, Wei and Li, Junru and Zhang, Kai and Zhang, Li},
  booktitle={ICASSP},
  pages={2955--2959},
  year={2024},
  organization={IEEE}
}

@article{sheng2025bi,
  title={Bi-Directional Deep Contextual Video Compression},
  author={Sheng, Xihua and Li, Li and Liu, Dong and Wang, Shiqi},
  journal={IEEE TMM},
  year={2025},
  publisher={IEEE}
}

@article{bross2021overview,
  title={Overview of the versatile video coding (VVC) standard and its applications},
  author={Bross, Benjamin and Wang, Ye-Kui and Ye, Yan and Liu, Shan and Chen, Jianle and Sullivan, Gary J and Ohm, Jens-Rainer},
  journal={TCSVT},
  volume={31},
  number={10},
  pages={3736--3764},
  year={2021},
  publisher={IEEE}
}

@inproceedings{alexandre2023hierarchical,
  title={Hierarchical B-frame video coding using two-layer CANF without motion coding},
  author={Alexandre, David and Hang, Hsueh-Ming and Peng, Wen-Hsiao},
  booktitle={CVPR},
  pages={10249--10258},
  year={2023}
}

@ARTICLE{chen2024bcanf,
  author={Chen, Mu-Jung and Chen, Yi-Hsin and Peng, Wen-Hsiao},
  journal={TCSVT}, 
  title={B-CANF: Adaptive B-Frame Coding With Conditional Augmented Normalizing Flows}, 
  year={2024},
  volume={34},
  number={4},
  pages={2908-2921}}

@inproceedings{li2024neural,
  title={Neural video compression with feature modulation},
  author={Li, Jiahao and Li, Bin and Lu, Yan},
  booktitle={CVPR},
  pages={26099--26108},
  year={2024}
}

@inproceedings{jia2025practical,
  title={Towards Practical Real-Time Neural Video Compression},
  author={Jia, Zhaoyang and Li, Bin and Li, Jiahao and Xie, Wenxuan and Qi, Linfeng and Li, Houqiang and Lu, Yan},
  booktitle={CVPR},
  year={2025}
}

@inproceedings{hu2020improving,
  title={Improving deep video compression by resolution-adaptive flow coding},
  author={Hu, Zhihao and Chen, Zhenghao and Xu, Dong and Lu, Guo and Ouyang, Wanli and Gu, Shuhang},
  booktitle={ECCV},
  pages={193--209},
  year={2020},
  organization={Springer}
}

@inproceedings{qi2024long,
  title={Long-term temporal context gathering for neural video compression},
  author={Qi, Linfeng and Jia, Zhaoyang and Li, Jiahao and Li, Bin and Li, Houqiang and Lu, Yan},
  booktitle={ECCV},
  pages={305--322},
  year={2024},
  organization={Springer}
}

@inproceedings{hu2022coarse,
  title={Coarse-to-fine deep video coding with hyperprior-guided mode prediction},
  author={Hu, Zhihao and Lu, Guo and Guo, Jinyang and Liu, Shan and Jiang, Wei and Xu, Dong},
  booktitle={CVPR},
  pages={5921--5930},
  year={2022}
}

@InProceedings{Qi_2023_CVPR,
    author    = {Qi, Linfeng and Li, Jiahao and Li, Bin and Li, Houqiang and Lu, Yan},
    title     = {Motion Information Propagation for Neural Video Compression},
    booktitle = {CVPR},
    month     = {June},
    year      = {2023},
    pages     = {6111-6120}
}

@InProceedings{jiang2025ecvc,
  title={ECVC: Exploiting Non-Local Correlations in Multiple Frames for Contextual Video Compression},
  author={Jiang, Wei and Li, Junru and Zhang, Kai and Zhang, Li},
  booktitle = {CVPR},
  year={2025}
}

@article{hu2022fvc,
  title={Fvc: An end-to-end framework towards deep video compression in feature space},
  author={Hu, Zhihao and Xu, Dong and Lu, Guo and Jiang, Wei and Wang, Wei and Liu, Shan},
  journal={IEEE TPAMI},
  volume={45},
  number={4},
  pages={4569--4585},
  year={2022},
  publisher={IEEE}
}

@article{tang2026nvc,
  title={NVC-1B: Scaling up Neural Video Coding Models},
  author={Tang, Chuanbo and Sheng, Xihua and Li, Li and Liu, Dong and Wu, Feng},
  journal={IEEE TPAMI},
  year={2026},
  publisher={IEEE}
}

@article{sullivan2012overview,
  title={Overview of the high efficiency video coding (HEVC) standard},
  author={Sullivan, Gary J and Ohm, Jens-Rainer and Han, Woo-Jin and Wiegand, Thomas},
  journal={TCSVT},
  volume={22},
  number={12},
  pages={1649--1668},
  year={2012},
  publisher={IEEE}
}

@article{jiang2026mlicv2,
  title={Mlicv2: Enhanced multi-reference entropy modeling for learned image compression},
  author={Jiang, Wei and Zhai, Yongqi and Yang, Jiayu and Gao, Feng and Wang, Ronggang},
  journal={ACM TOMM},
  volume={22},
  number={4},
  pages={1--23},
  year={2026},
  publisher={ACM New York, NY}
}

@inproceedings{yang2020learning,
  title={Learning for video compression with hierarchical quality and recurrent enhancement},
  author={Yang, Ren and Mentzer, Fabian and Gool, Luc Van and Timofte, Radu},
  booktitle={CVPR},
  pages={6628--6637},
  year={2020}
}

@InProceedings{lin2020mlvc,
author = {Lin, Jianping and Liu, Dong and Li, Houqiang and Wu, Feng},
title = {M-LVC: Multiple Frames Prediction for Learned Video Compression},
booktitle = {CVPR},
month = {June},
year = {2020}
}

@article{li2025ustc,
  title={Ustc-td: A test dataset and benchmark for image and video coding in 2020s},
  author={Li, Zhuoyuan and Liao, Junqi and Tang, Chuanbo and Zhang, Haotian and Li, Yuqi and Bian, Yifan and Sheng, Xihua and Feng, Xinmin and Li, Yao and Gao, Changsheng and others},
  journal={IEEE TMM},
  year={2025},
  publisher={IEEE}
}

@article{chen2024maskcrt,
  title={Maskcrt: Masked conditional residual transformer for learned video compression},
  author={Chen, Yi-Hsin and Xie, Hong-Sheng and Chen, Cheng-Wei and Gao, Zong-Lin and Benjak, Martin and Peng, Wen-Hsiao and Ostermann, J{\"o}rn},
  journal={TCSVT},
  year={2024},
  publisher={IEEE}
}

@inproceedings{ranjan2017optical,
  title={Optical flow estimation using a spatial pyramid network},
  author={Ranjan, Anurag and Black, Michael J},
  booktitle={CVPR},
  pages={4161--4170},
  year={2017}
}

@inproceedings{lu2024deep,
  title={Deep Hierarchical Video Compression},
  author={Lu, Ming and Duan, Zhihao and Zhu, Fengqing and Ma, Zhan},
  booktitle={AAAI},
  volume={38},
  number={8},
  pages={8859--8867},
  year={2024}
}

@inproceedings{wu2018video,
  title={Video compression through image interpolation},
  author={Wu, Chao-Yuan and Singhal, Nayan and Krahenbuhl, Philipp},
  booktitle={ECCV},
  pages={416--431},
  year={2018}
}

@inproceedings{ho2022canf,
  title={Canf-vc: Conditional augmented normalizing flows for video compression},
  author={Ho, Yung-Han and Chang, Chih-Peng and Chen, Peng-Yu and Gnutti, Alessandro and Peng, Wen-Hsiao},
  booktitle={ECCV},
  pages={207--223},
  year={2022},
  organization={Springer}
}

@article{vctmentzer2022,
  title={VCT: A Video Compression Transformer},
  author={Mentzer, Fabian and Toderici, George D and Minnen, David and Caelles, Sergi and Hwang, Sung Jin and Lucic, Mario and Agustsson, Eirikur},
  journal={NeurIPS},
  volume={35},
  pages={13091--13103},
  year={2022}
}

@inproceedings{rippel2021elf,
  title={Elf-vc: Efficient learned flexible-rate video coding},
  author={Rippel, Oren and Anderson, Alexander G and Tatwawadi, Kedar and Nair, Sanjay and Lytle, Craig and Bourdev, Lubomir},
  booktitle={ICCV},
  pages={14479--14488},
  year={2021}
}

@inproceedings{pourreza2021extending,
  title={Extending neural p-frame codecs for b-frame coding},
  author={Pourreza, Reza and Cohen, Taco},
  booktitle={ICCV},
  pages={6680--6689},
  year={2021}
}

@inproceedings{agustsson2020scale,
  title={Scale-space flow for end-to-end optimized video compression},
  author={Agustsson, Eirikur and Minnen, David and Johnston, Nick and Balle, Johannes and Hwang, Sung Jin and Toderici, George},
  booktitle={CVPR},
  pages={8503--8512},
  year={2020}
}

@article{wiegand2003overview,
  title={Overview of the H. 264/AVC video coding standard},
  author={Wiegand, Thomas and Sullivan, Gary J and Bjontegaard, Gisle and Luthra, Ajay},
  journal={TCSVT},
  volume={13},
  number={7},
  pages={560--576},
  year={2003},
  publisher={IEEE}
}

@inproceedings{wang2016mcl,
  title={MCL-JCV: a JND-based H. 264/AVC video quality assessment dataset},
  author={Wang, Haiqiang and Gan, Weihao and Hu, Sudeng and Lin, Joe Yuchieh and Jin, Lina and Song, Longguang and Wang, Ping and Katsavounidis, Ioannis and Aaron, Anne and Kuo, C-C Jay},
  booktitle={ICIP},
  pages={1509--1513},
  year={2016},
  organization={IEEE}
}

@inproceedings{mercat2020uvg,
  title={UVG dataset: 50/120fps 4K sequences for video codec analysis and development},
  author={Mercat, Alexandre and Viitanen, Marko and Vanne, Jarno},
  booktitle={ACM MMSys},
  pages={297--302},
  year={2020}
}

@inproceedings{jiang2022mlic,
author = {Jiang, Wei and Yang, Jiayu and Zhai, Yongqi and Ning, Peirong and Gao, Feng and Wang, Ronggang},
title = {MLIC: Multi-Reference Entropy Model for Learned Image Compression},
year = {2023},
booktitle = {ACM MM},
pages={7618--7627},
}

@article{jiang2025mlicpp,
  title={MLIC++: Linear complexity multi-reference entropy modeling for learned image compression},
  author={Jiang, Wei and Yang, Jiayu and Zhai, Yongqi and Gao, Feng and Wang, Ronggang},
  journal={ACM TOMM},
  volume={21},
  number={5},
  pages={1--25},
  year={2025},
  publisher={ACM New York, NY}
}

@article{li2021deep,
  title={Deep contextual video compression},
  author={Li, Jiahao and Li, Bin and Lu, Yan},
  journal={NeurIPS},
  volume={34},
  pages={18114--18125},
  year={2021}
}

@ARTICLE{sheng2022temporal,
  author={Sheng, Xihua and Li, Jiahao and Li, Bin and Li, Li and Liu, Dong and Lu, Yan},
  journal={IEEE TMM}, 
  title={Temporal Context Mining for Learned Video Compression}, 
  year={2022},
  volume={25},
  pages={7311-7322}}

@inproceedings{li2022hybrid,
  title={Hybrid spatial-temporal entropy modelling for neural video compression},
  author={Li, Jiahao and Li, Bin and Lu, Yan},
  booktitle={ACM MM},
  pages={1503--1511},
  year={2022}
}

@inproceedings{lu2019dvc,
  title={Dvc: An end-to-end deep video compression framework},
  author={Lu, Guo and Ouyang, Wanli and Xu, Dong and Zhang, Xiaoyun and Cai, Chunlei and Gao, Zhiyong},
  booktitle={CVPR},
  pages={11006--11015},
  year={2019}
}

@inproceedings{hu2021fvc,
  title={FVC: A new framework towards deep video compression in feature space},
  author={Hu, Zhihao and Lu, Guo and Xu, Dong},
  booktitle={CVPR},
  pages={1502--1511},
  year={2021}
}

@inproceedings{li2023neural,
  title={Neural video compression with diverse contexts},
  author={Li, Jiahao and Li, Bin and Lu, Yan},
  booktitle={CVPR},
  pages={22616--22626},
  year={2023}
}

\end{document}